\PassOptionsToPackage{svgnames,table}{xcolor}
\documentclass[11pt,reqno]{article}
\usepackage{amsmath,hyperref,amssymb, amsthm}
\usepackage{authblk}
\usepackage[all]{xy}
\usepackage{graphicx,url}
\usepackage{color}
\usepackage{slashed}
\usepackage{tikz}
\usepackage{lipsum}
\usepackage{caption}
\usepackage{ytableau}
\usepackage{longtable}
\usepackage{multirow}

\usepackage{caption}

\usepackage[style=ieee,citestyle=numeric-comp]{biblatex}
\newcommand{\be}{\begin{eqnarray}}
\newcommand{\ee}{\end{eqnarray}}
\newcommand{\eeq}{\end{equation}}
\newcommand{\beq}{\begin{equation}}

\allowdisplaybreaks \numberwithin{equation}{section}

\DeclareSymbolFont{AMSa}{U}{msa}{m}{n}
\DeclareSymbolFont{AMSb}{U}{msb}{m}{n}
\DeclareMathSymbol{\fieldR}{\mathalpha}{AMSb}{"52}

\renewcommand{\Im}{\imag}
\DeclareMathOperator{\imag}{Im}

\newcommand{\CA}{{\cal A}}

\newcommand{\CS}{{\cal S}}

\newcommand{\CH}{\mathcal{H}}
\newcommand{\Hh}{\mathcal{H}}

\newcommand{\CI}{{\cal I}}
\newcommand{\CJ}{{\cal J}}
\newcommand{\CL}{{\cal L}}
\newcommand{\CN}{\mathcal{N}}
\newcommand{\CM}{\mathcal{M}}
\newcommand{\CO}{{\cal O}}

\newcommand{\CQ}{\mathcal{Q}}

\newcommand{\calC}{\mathcal{C}}
\newcommand{\calR}{\mathcal{R}}

\newcommand{\CV}{\mathcal{V}}
\newcommand{\CZ}{\mathcal{Z}}

\newcommand{\HRR}{\text{RR}}
\newcommand{\NSNS}{\text{NSNS}}
\newcommand{\R}{\text{R}}
\newcommand{\NS}{\text{NS}}
\newcommand{\GTVW}{\text{GTVW}}

\DeclareMathOperator{\Tr}{Tr}
\DeclareMathOperator{\Aut}{Aut}

\newcommand{\NN}{\mathbb{N}}
\newcommand{\ZZ}{\mathbb{Z}}
\newcommand{\RR}{\mathbb{R}}
\newcommand{\CC}{\mathbb{C}}
\newcommand{\QQ}{\mathbb{Q}}
\newcommand{\FF}{\mathbb{F}}

\newcommand{\HH}{\mathbb{H}}

\newcommand{\cTop}{\mathsf{Top}}

\newcommand{\ii}{\mathbf{i}}
\newcommand{\jj}{\mathbf{j}}
\newcommand{\kk}{\mathbf{k}}
\newcommand{\ssi}{\boldsymbol{\sigma}}

\def\beq{\begin{equation}}
\def\eeq{\end{equation}}
\def\bea{\begin{eqnarray}}
\def\eea{\end{eqnarray}}

\def\<{\langle}

\newtheorem{theorem}{Theorem}
\newtheorem{claim}[theorem]{Claim}
\newtheorem{conjecture}[theorem]{Conjecture}
\newtheorem{corollary}[theorem]{Corollary}

\DeclareMathOperator{\ch}{ch}
\DeclareMathOperator{\diag}{diag}

\usepackage{tcolorbox}
\tcbuselibrary{skins,breakable}
\usetikzlibrary{shadings,shadows}
    {\endtcolorbox}

    {\endtcolorbox}

    {\endtcolorbox}

\title{Non-invertible defects from the Conway SCFT to K3 sigma models II: duality and Fibonacci defects}
\author[1]{Roberta Angius\thanks{roberta.angius@uni-hamburg.de}}
\author[2]{Stefano Giaccari}
\author[3]{Sarah M. Harrison}
\author[4,5]{Roberto Volpato\thanks{volpato@pd.infn.it}}
{\small \affil[1]{\small  II. Institut f\"ur Theoretische Physik, Universit\"at Hamburg, Notkestrasse 9, 22607 Hamburg, Germany}
\affil[2]{\small Istituto Nazionale di Ricerca Metrologica, Strada delle Cacce 91, I-10135 Torino, Italy}
\affil[3]{\small  Department of Physics and Department of Mathematics,  Northeastern University, Boston, MA 02115, USA}
\affil[4]{\small Dipartimento di Fisica e Astronomia, Universit\`a di Padova, Via Marzolo 8, 35131, Padova, Italy}
\affil[5]{\small INFN, sez. di Padova, Via Marzolo 8, 35131, Padova, Italy}}

\begin{document}

\maketitle

\abstract{We continue the study, initiated in \cite{Angius:2025zlm}, of topological defect lines (TDLs) in the Conway  module $V^{f\natural}$ 
and K3 non-linear sigma models (NLSMs). In the  case of $V^{f\natural}$, we fully classify the potential $\CN=1$ (and $\CN=4$)--preserving duality defects for cyclic Tambara--Yamagami categories TY$(\ZZ_N)$, noting a curious relation to genus zero groups of monstrous moonshine. We use the correspondence with Leech lattice endomorphisms, discovered in \cite{Angius:2025zlm}, to construct a number of non-trivial examples of TDLs in $V^{f\natural}$, including examples of irrational quantum dimension.
 In particular, we fully classify and construct defects for the TY$(\ZZ_2)$ and TY$(\ZZ_3)$ cases, and provide examples of duality defects for TY$(\ZZ_2\times \ZZ_2)$ and Fibonacci fusion categories as well. In the case of K3 NLSMs, we describe a duality defect of irrational quantum dimension $\sqrt 2$ for the category TY$(\ZZ_2, -1)$ in a particular torus orbifold, which exists on a 16-dimensional slice of the moduli space. We also provide a detailed analysis of spectral flow--preserving TDLs in Gepner models of K3, of independent interest, and use this to construct non-invertible defects for Fibonacci and $Rep(S_3)$ categories in particular examples. Finally we provide evidence for our conjecture in \cite{Angius:2025zlm} that  special subcategories of such TDLs in $V^{f\natural}$ correspond to $\CN=(4,4)$ and spectral flow--preserving defect lines in a corresponding K3 NLSM. In particular, we compute defect--twined elliptic genera for all non-invertible defects constructed in this article, demonstrating that for each defect found in a K3 NLSM, there is a corresponding defect in $V^{f\natural}$ with coincident twining genus, and making a prediction for a number of TDLs in K3 NLSMs yet to be found.}

\tableofcontents

\section{Introduction and outlook}

The reinterpretation of conventional symmetries in terms of topological defects supported on codimension-$1$ submanifolds of the spacetime, acting on local operators and obeying group-like fusion rules has enabled multiple generalizations of the symmetry concept. These include higher form symmetries, obtained relaxing the condition on the dimension of the support manifold \cite{Gaiotto:2014kfa}, and non-invertible symmetries,  arising from more general (non-invertible) fusion rules among the operators. Such developments have driven extensive research across a wide range of physical settings. Nevertheless, systematic methods for a complete classification remain elusive, even in special frameworks like two-dimensional conformal field theories (CFTs),  where many key results have been established \cite{Verlinde:1988sn, Petkova:2000ip, Frohlich:2004ef,Frohlich:2006ch,Frohlich:2009gb,Chang_2019,Bhardwaj:2017xup}.

For rational CFTs, the topological defects preserving the full chiral algebra can be systematically classified by imposing a Cardy-like condition on the defects, as described in \cite{Petkova:2000ip}. This approach relies crucially on rational algebras admitting only finitely many irreducible representations.
An analogous construction for defects preserving non-rational chiral algebras does not exist, due to the presence of infinitely many irreducible representations, the only exception being the case of free bosons \cite{Fuchs:2007tx,Bachas:2012bj,Thorngren:2021yso}. Since rational theories constitute a measure-zero subset of the moduli space of physically relevant CFTs, understanding defects in non-rational settings represents a broadly encountered and unresolved challenge, and is one of the primary motivations of our study. Our article focuses on two particular theories--the Conway moonshine module $V^{f\natural}$ and K3 non-linear sigma models (NLSMs)--each of which is of interest in its own right.

On the one hand, the present work continues the investigation, initiated in \cite{Angius:2025zlm}, of the Conway moonshine module $V^{f \natural}$ \cite{Frenkel:1988flm,Duncan:2006}, the unique holomorphic superconformal field theory (SCFT) with $c=12$ and no fields of weight $1/2$.  By studying topological defect lines preserving the non-rational $\mathcal{N}=1$ superVirasoro algebra, and imposing certain technical constraints, in the companion article we proposed a generalization of the  Cardy-like condition adapted to this setting. This approach enables us to place nontrivial constraints on the fusion ring structure of the corresponding category $\cTop$. In particular, in \cite{Angius:2025zlm}, we proved that every TDL $\CL$ in $\cTop$ can be associated with a $\ZZ$-linear map $\rho(\CL):\Lambda\to \Lambda$ from the Leech lattice $\Lambda$ into itself. The association $\rho$ is compatible with the basic operations on defects, namely the fusion product ($\CL_1\CL_2)$ of two lines, their superposition $\CL_1+\CL_2$, and with duality $\CL^*$ defined by the change of line orientation. Compatibility here means that:
\be \rho(\CL_1\CL_2)=\rho(\CL_1)\rho(\CL_2)\ ,\qquad \rho(\CL_1+\CL_2)=\rho(\CL_1)+\rho(\CL_2),\qquad \rho(\CL^*)=\rho(\CL)^t\ ,
\ee where transposition $\rho(\CL)^t$ is defined with respect to an orthonormal basis of $\Lambda\otimes \RR$. Physically, $\Lambda\otimes \RR$ is identified with the $24$-dimensional real space ${}^\RR V^{f\natural}_{tw}(1/2)$ of CPT-self-conjugate ground states (conformal weight $1/2$) in the Ramond sector $V^{f\natural}_{tw}$, and $\rho(\CL)$ is defined by the action of $\CL$ on such states. Knowing the lattice endomorphism $\rho(
\CL)$ is not sufficient to fully reconstruct the topological defect $\CL$: indeed, there can be (infinitely!) many non-isomorphic TDLs $\CL$ associated with the same $\rho(\CL)$. Although these constraints do not provide a complete determination of the underlying tensor category $\cTop$, they are sufficient to establish a set of general results on the admissible defects, including a correspondence between these elements and automorphisms of the Leech lattice, as well as constraints on their quantum dimensions.

The restriction of $\rho$ to lines $\CL$ that are invertible, i.e. that satisfy $\CL\CL^*=1$, defines an isomorphism between the group $\Aut_\tau(V^{f\natural})$ of symmetries of $V^{f\natural}$ preserving the $\CN=1$ supercurrent $\tau(z)$, and the Conway group $\Aut(\Lambda)\cong Co_0$ of automorphisms of the Leech lattice $\Lambda$. In fact, this group isomorphism was proved earlier in \cite{Duncan:2014eha}, and its existence was used in \cite{Angius:2025zlm} to extend $\rho$ to non-invertible defects.

On the other hand, the same question, i.e. the classification of non-invertible symmetries in non-rational CFTs,  was the motivation in \cite{Angius:2024evd} for the study of TDLs in K3 non-linear sigma models (NLSM) preserving the $\mathcal{N}=(4,4)$ superconformal algebra and spectral flow.  In that context, rather that employing Cardy-like conditions, the analysis focused on the action of defects on the boundary states of the theory.  In particular, examining how defects act on these boundary states, corresponding to $1/2$ BPS D-branes charged under the $U(1)^{24}$ gauge group of the R-R ground states of the theory, enabled the derivation of a set of general constraints on the admissible defects and the structure of their associated fusion rings. The key insight underpinning the analysis is the identification of a correspondence between topological defects satisfying the above conditions and endomorphisms of the lattice of R-R D-brane charges in the theory.
The formal analogy between the results obtained in \cite{Angius:2024evd} and \cite{Angius:2025zlm}, together with the still mysterious connection between symmetry groups and twining functions of K3 NLSMs and $V^{f\natural}$ \cite{Gaberdiel:2011fg,Duncan:2015xoa,Cheng:2016org} constitutes a second major motivation for this work. We aim to investigate and provide evidence for a potential generalization of this connection at the level of tensor categories.

A detailed proposal for the conjectural relationship  between categories of TDLs in $V^{f\natural}$ and in NLSMs on K3 was described in \cite{Angius:2025zlm} (see section \ref{s:conj} for a review). In $V^{f\natural}$, one considers subcategories $\cTop_\Pi$ of $\cTop$, whose defects act trivially on the Ramond ground states in some fixed $4$-dimensional subspace $\Pi\subset \Lambda\otimes \RR$. According to the conjecture in \cite{Angius:2025zlm}, every such $\cTop_\Pi$ should be equivalent to a category $\cTop^{K3}_\calC$ of topological defects in some NLSM $\calC$ on K3, preserving the $\CN=(4,4)$ superconformal algebra and the spectral flow. The representations of these categories on the space of states in the two CFTs are also expected to be closely related. For example, for every $\CL\in \cTop_\Pi$, the endomorphism $\rho(\CL)$ acting on $V^{f\natural}_{tw}(1/2)$ matches with the action of the corresponding defect in $\cTop^{K3}_\calC$ on the RR ground states on the K3 model $\calC$, 
upon a suitable identifications of these spaces in the two CFTs. An even stronger conjecture involves the so called $\CL$-twining genera $\phi^{\CL}(\calC,\tau,z)$ in the K3 model, some $\CL$-equivariant version of the elliptic genus of K3 (see section \ref{s:K3defects}). The conjecture is that all such genera can be reproduced by analogous functions $\phi^{\CL}(V^{f\natural},\tau,z)$ defined in $V^{f\natural}$:
\be \phi^{\CL}(\calC,\tau,z)=\phi^{\CL}(V^{f\natural},\tau,z) \qquad \forall \CL\in \cTop_\Pi\ .
\ee Again, these conjectures were motivated by previous analogous observations concerning groups of invertible symmetries in the two CFTs \cite{Duncan:2015xoa}.

The goal of the present article is to apply the methods developed in \cite{Angius:2025zlm} to discover a number of new non-trivial examples of topological defect lines and their corresponding fusion categories in both $V^{f\natural}$ and K3 NLSMs. Furthermore, we use these examples to provide further evidence for the conjectural relation between fusion categories in $V^{f\natural}$ and K3 NLSMs. This article contains many new results not seen before in the literature, which we summarize here.
\begin{enumerate}

\item In Section \ref{sec:TY_categories} we review the classification of cyclic orbifolds $V^{f\natural}/\langle g \rangle$ for all conjugacy classes $g\in Co_0$. We use this to classify the potential supersymmetry--preserving duality defects corresponding to Tambara-Yamagami categories TY$(\mathbb{Z}_N)$, where $N=o(g)$, the order of $g$; these are the 42 classes with balanced Frame shapes listed in Table \ref{tbl:selforb}.\footnote{As explained in \S \ref{section:TDLs_Vfnat}, though there exist (generically multiple) duality defects $\CN_g$ for TY$(\mathbb{Z}_N)$ corresponding to each of these conjugacy classes, it is much more difficult to determine whether these defects preserve the $\CN=1$ superconformal algebra. We manage to fully classify the possibilities only in examples of order 2 and 3.} Among these, seven classes, 2B, 3B, 4B, 4E, 5B, 6K, 9C (decorated with an underscore in the table), are those for which the corresponding duality defects may preserve a four-plane in $V^{f \natural}_{tw}(1/2)$, and for which analogous defects are expected to exist in certain K3 sigma models.
    
\item In Section \ref{s:Monster}, we describe an interesting connection between cyclic orbifolds of $V^{f\natural}$ and the genus zero groups appearing in monstrous moonshine. Roughly speaking, when $V^{f\natural}/\langle g \rangle$ is isomorphic to $V^{f\natural}$ itself, the orbifold theory can either preserve the fermion number in the Ramond sector (Table \ref{tbl:selforb}), or have the opposite sign (Table \ref{tbl:VfnatnoSUSY}). The conjugacy classes in these two cases correspond, respectively, to genus zero groups appearing in monstrous moonshine with or without the Fricke involution. The other two possibilities, that $V^{f\natural}/\langle g\rangle$ is isomorphic to $V^{fE_8}$ (Table \ref{tbl:VfE8}) or is anomalous (Table \ref{tbl:anomalous}), also correspond to distinguished classes of genus zero groups.

\item We explicitly classify and construct all $\CN=1$ supersymmetry--preserving duality defects in $V^{f\natural}$ corresponding to TY categories for cyclic groups $\ZZ_2$ and $\ZZ_3$ arising from conjugacy classes 2B and 3B in $Co_0$. These results are contained in \S \ref{s:Z2duality} and \S \ref{sec:Z3duality}, respectively. The conclusion is that defects for both $\ZZ_2$ Tambara--Yamagami  categories (TY$(\mathbb{Z}_2, +1)$ and TY$(\mathbb{Z}_2,-1)$) exist; these are constructed in \S \ref{s:freefermion}. Moreover, in the $\ZZ_3$ case, we find duality defects for three of the four possible Tambara--Yamagami categories, given by TY$(\mathbb{Z}_3, \chi_+,-1 )$, TY$(\mathbb{Z}_3, \chi_-,+1)$ and TY$(\mathbb{Z}_3,\chi_-,-1)$; these are constructed in \S \ref{s:Z3defects}. We verify the fusion category data in all cases by considering the spin selection rules (see \S \ref{s:spinselect}) arising from the corresponding defect-twisted partition functions.

\item Moreover, as each of the $\ZZ_2$ and $\ZZ_3$ duality defects mentioned in the previous point preserves a four--plane in $Co_0$, and thus a copy of the $c=6, \mathcal N=4$ superconformal algebra in $V^{f\natural}$, according to the conjecture proposed in \cite{Angius:2025zlm}, each of these defects should admit a realization in a corresponding K3 NLSM. We compute the corresponding defect twining genus \eqref{Vfnat:L_twined_graded_part_func} in $V^{f\natural}$ for each of these defects. See Table \ref{t:summary} for a summary of the equation numbers. Thus in all cases, we make a prediction for the defect--twined elliptic genus in any K3 NLSM with a  $\ZZ_2$ or $\ZZ_3$ duality defect for the corresponding TY fusion category. It remains for most of these predictions to be verified in explicit examples; however see the next point. 

\item In Section \ref{s:Z2dualityK3} we construct an explicit realization of the category TY$(\mathbb{Z}_2,-1)$ in a specific K3 model obtained as a torus orbifold $T^4/ \calR$, where $T^4$ is given by the product of four mutually orthogonal circles, three at the self-dual radius $R_{sd}$ and one with radius $\sqrt{2}R_{sd}$.  We then compute the corresponding duality-twining genus in \eqref{eq:K3Z2genus}, verifying its agreement with \eqref{twinZ2dual-}. This defect is present on a 16--dimensional sublocus of the moduli space of K3 NLSMs, of which we expect a generic point to be irrational. These are the first examples (as far as we know) of K3 NLSMs with a cyclic Tambara--Yamagami category and duality defect of irrational dimension.

    \item In Section \ref{s:duality_def_noncyclic} we review the construction of duality defects in orbifold theories obtained by gauging a subgroup $H$ of the (generally non-abelian) symmetry group $G$ of the parent theory. Implementing the procedure in $V^{f \natural}$ with $G \cong D_8$ and $H \cong \mathbb{Z}_2$, we explicitly construct five duality defects 
    (see Table \ref{t:summary}), which among themselves realize the four possible $\ZZ_2\times \ZZ_2$ Tambara–Yamagami categories, i.e. TY$(\mathbb{Z}_2 \times \mathbb{Z}_2, \chi_a , \pm 1)$ and TY$(\mathbb{Z}_2 \times \mathbb{Z}_2, \chi_s , \pm 1)$, whose spin selection rules appear in Table \ref{t:spinselectZ2Z2}. In section \ref{s:Z2Z2_GTVW} we construct duality defects for the two categories TY$(\mathbb{Z}_2 \times \mathbb{Z}_2, \chi_a, +1)$ and TY$(\mathbb{Z}_2 \times \mathbb{Z}_2, \chi_s, +1)$  in the K3 model $\mathcal{C}_{GTVW}$ of \cite{Gaberdiel:2013psa}, and we compute the corresponding defect twining genera in \eqref{twinN3456} and \eqref{GTVWRepH8}. These respectively match the twining genera \eqref{VfnatRepD8} and \eqref{eq:Z2Z2-+} in $V^{f \natural}$, in agreement with the conjecture.
    \item Going beyond duality defects, for which systematic constructions are known\footnote{At least if one only requires them to preserve Virasoro; some ad hoc methods are needed to determine whether a given duality defect preserves the $\CN=1$ supercurrent.}, even if their implementation is often technically involved, in Section \ref{s:Fibonacci} we construct two commuting Fibonacci defects $W^A$ and $W^B$, which are related by conjugation in $Co_0$, and commute with a subgroup $ 2.(A_5 \times J_2):2 \subset Co_0$ of invertible defects. In equation \eqref{eq:FibTwining} we compute the $W^A$-twining partition function. 
    \item In Section \ref{s:Gepner} we discuss the construction of supersymmetry-preserving TDLs in Gepner models obtained as orbifolds of tensor products of diagonal $\mathcal{N}=2$ minimal models. We describe the conditions for TDLs arising as Verlinde lines in the component minimal models to preserve spectral flow in the corresponding Gepner model in \S \ref{s:Gepnerdefects}. In \S \ref{sec:338} we use these methods to construct Fibonacci fusion categories in the $(3)^2(8)$ and $(2)(3)(18)$ Gepner models. The twining elliptic genus for these Fibonacci defects is given in equation \eqref{eq:GepnerFibgenus} and exactly matches the Fibonacci twining genus in $V^{f \natural}$ in equation \eqref{eq:FibTwining}, providing further evidence for the conjecture. In the same section we provide explicit realizations of $Rep (S_3)$ in the Gepner models $(4)^3$, $(1)^2(4)^2$ and $(2)(4)(10)$ which twining defect functions \eqref{RepS3twining} match the results of the $Rep(S_3)$ category in $V^{f \natural}$ that we computed in the companion article \cite{Angius:2025zlm}.
\end{enumerate}
Table \ref{t:summary} gives a summary of all of the non-invertible TDLs constructed in this article, in both $V^{f\natural}$ and K3 NLSMs, and their corresponding defect twining genera, with references. In all examples considered in this article, the defect--twined elliptic genus of K3 matches the twining genus for a corresponding non-invertible TDL in $V^{f\natural}$ arising in the same fusion category.
\begin{center}
\begin{table}[htb!]\begin{center}
\begin{tabular}{|c|c|c|c|c|}\hline
Fusion Category & Defect in $V^{f\natural}$& $V^{f\natural}$ twining  &  K3 NLSM & K3 twining  \\\hline\hline
TY$(\ZZ_2,+1)$ & $\CN_g^{(1)}$, \S \ref{s:freefermion} & Eq. \eqref{twinZ2dual+}& ?? & \\\hline
TY$(\ZZ_2,-1)$ & $\CN_g^{(2)}$, \S \ref{s:freefermion} &Eq. \eqref{twinZ2dual-}& $T^4/\ZZ_2$, \S \ref{s:K3Z2duality} & Eq. \eqref{eq:K3Z2genus}\\\hline
TY$(\ZZ_3,\chi_+,-1)$ & $\CN^{(b_\pm)}$, \S \ref{s:Z3defects} & Eq. \eqref{Nbpmgenus}& ??&\\\hline
TY$(\ZZ_3,\chi_-,+1)$ & $\CN^{(a'_\pm)}$, \S \ref{s:Z3defects}& Eq. \eqref{Na'pmgenus} & ??&\\\hline
TY$(\ZZ_3,\chi_-,-1)$ & $\CN^{(a'_0)}$, \S \ref{s:Z3defects}& Eq. \eqref{Na0genus} & ??&\\\hline
\shortstack{TY$(\ZZ_2\times \ZZ_2,\chi_a,+1)$\\$\cong Rep(D_8)$} & \shortstack{$\CN_{p_2}$, \S\ref{s:Z2Z2Vfnat}\\$\CN_{p_4}$, \S\ref{s:Z2Z2Vfnat}} &\shortstack{Eq. \eqref{eq:Z2Z2++} \\ Eq. \eqref{VfnatRepD8}} &\shortstack{?? \\GTVW, \S \ref{s:Z2Z2_GTVW}} &\shortstack{\\Eq.\eqref{twinN3456}}\\\hline
\begin{minipage}{0.2\textwidth}\shortstack[c]{TY$(\ZZ_2\times \ZZ_2,\chi_a,-1)$\\$\cong Rep(Q_8)$} \end{minipage}& \shortstack[c]{$\CN_{p_5}$, \S\ref{s:Z2Z2Vfnat}} & \shortstack[c]{Eq. \eqref{VfnatRepQ8}} & \shortstack[c]{??} &\\\hline
\begin{minipage}{0.2\textwidth}\shortstack{TY$(\ZZ_2\times \ZZ_2,\chi_s,+1)$\\$\cong Rep(H_8)$}\end{minipage} & \shortstack[c]{$\CN_{p_1}$, \S\ref{s:Z2Z2Vfnat}} &Eq. \eqref{eq:Z2Z2-+}& \shortstack[c]{GTVW, \S \ref{s:Z2Z2_GTVW}}  &  Eq.\eqref{GTVWRepH8} \\\hline
TY$(\ZZ_2\times \ZZ_2,\chi_s,-1)$ &$\CN_{p_3}$, \S\ref{s:Z2Z2Vfnat} &Eq.\eqref{eq:Z2Z2--}& ?? &\\\hline
Fibonacci & \shortstack[c]{$W$, \S \ref{s:FibDef}} &Eq. \eqref{eq:FibTwining}& \begin{minipage}{0.1\textwidth}\shortstack{
$(3)^2(8)$, \\$(2)(3)(18)$} \end{minipage}
\S \ref{sec:338}  & Eq.\eqref{eq:GepnerFibgenus}\\\hline
$Rep(S_3)$ & \shortstack[c]{$\CL_\rho,\CL_{\cal X}$, \S 5.2 in \cite{Angius:2025zlm}} & \shortstack[c]{Eq. (5.33) \cite{Angius:2025zlm}} & \begin{minipage}{0.13\textwidth}\shortstack{$(4)^3,
(1)^2(4)^2$,\\  $(2)(4)(10)$,  }\end{minipage}\S \ref{sec:338} & Eq.\eqref{RepS3twining}\\\hline
\end{tabular}
\end{center}
\caption{{\small Summary of results. The first column lists the fusion categories that we explicitly construct throughout the article. For each category, the second column specifies the non-invertible topological defects it contains together with a reference to the section where their explicit construction is carried out. The third column provides the reference to the equation that computes the corresponding twining genus in $V^{f \natural}$. The last two columns respectively indicate the specific K3 model in which we realize each fusion category and the reference to the equation computing the associated twining genus in that model. The question marks ?? mean that we have not yet found a TDL in  a K3 model matching the corresponding defect in $V^{f\natural}$. }} \label{t:summary}
\end{table}
\end{center}

We end with some discussion of our results and open problems which arise naturally from our findings.
\begin{enumerate}
    \item In this work we exploit the description of $V^{f \natural}$ as a vertex operator algebra and its relation with other $c=12$ holomorphic theories  to give an explicit characterization of defects in the category $\cTop$. Assuming Conjecture \ref{conj:K3relation}, one expects that any defect lying in the subcategory $\cTop_{\Pi^{\natural}}$, i.e. defects that preserve a fixed 4-plane inside the $24$-dimensional space of Ramond ground states, should correspond to a defect in the category $\cTop_{\mathcal{C}}^{K3}$ of a suitable K3 NLSM. In particular, for each fusion category that we realize explicitly in $V^{f \natural}$ (see Table \ref{t:summary}) we predict that there should exist a K3 NLSM (or family of K3 NLSMs) realizing the same category and with coincident defect--twined elliptic genera. Can we construct the corresponding examples on the K3 side of the relation?
    
    As Table \ref{t:summary} shows, in the present work we were able to identify such a K3 model and to construct the corresponding category only for a subset of examples. The analysis in this article is not meant to be complete, and we certainly expect that many more examples of TDLs in K3 sigma models could be discovered using the  currently known methods. Nevertheless, one can interpret this as evidence of a larger difficulty in constructing the K3-side defects explicitly. This is not surprising: while we have several explicit constructions of the Conway module, only a few special families of exactly solvable K3 models (such as torus orbifolds, or Gepner models) are known, where the topological defects can be studied in detail. Aside from the potential computational challenges, this observation raises the question of whether the difficulty of finding realizations on the K3 side can be due to the fact that K3 models admitting defects satisfying the conditions of the conjecture are rare in the K3 moduli space and, if so, how the loci admitting nontrivial $\cTop_{\mathcal{C}}^{K3}$ are distributed within that moduli space. In \cite{Angius:2024evd} it has been shown that K3 models potentially admitting a nontrivial category $\cTop_{\mathcal{C}}^{K3}$ form a dense subset of null measure in the $K3$ moduli space $\mathcal{M}_{K3}$. It would be interesting to determine whether a nontrivial category is actually realized at all such points or only at a (possibly finite) subset. And if these points are enough to realize all the topological defects that we can find in $V^{f \natural}$, including those we have constructed in this article. In this sense, we could use Conjecture \ref{conj:K3relation} to predict the existence of K3 models in which such topological defects are actually realized.
    
    \item  Another interesting direction, still related to understanding the distribution of generalized symmetries across the K3 moduli space, is to study for any given defect $\mathcal{L}$, present in a particular K3 model, the locus in moduli space where $\mathcal{L}$ is actually preserved. In other words, one would like to analyze deformations of K3 models that preserve a given category of defects. In a similar spirit of \cite{Cordova:2023qei}, one may start from a K3 model in which $\mathcal{L}$ is realized and study which marginal deformations preserve the defect; these directions should span a codimension-$n$ sublocus (with $n \geq 1$) of $\mathcal{M}_{K3}$. Exactly marginal operators are obtained as superconformal descendants of the $80$ $\left( \frac{1}{2}, \frac{1}{2}; \frac{1}{2}, \frac{1}{2} \right)$ NS–NS ground fields that are related by spectral flow to the twenty R-R ground fields labeled by $\left( \frac{1}{4}, 0 ; \frac{1}{4}, 0 \right)$. Since the defects we consider preserve the supercurrents and the spectral-flow operators, the deformation directions that leave the defect invariant can be read off from the action of the defect on these twenty R-R ground states. Equivalently, in the lattice language used in this work and in \cite{Angius:2024evd, Angius:2025zlm}, where any defect induces an endomorphism of the D-brane charge lattice $\Gamma^{4,20}$, the preserved directions are given by the eigenspace of that endomorphism whose eigenvalue equals the quantum dimension of the defect; the codimension of the preserved locus is then determined by the multiplicity of that eigenvalue. 
    
    For example, the $\ZZ_2$ duality defect in the torus orbifold $T^4/\langle \calR\rangle$, described in section \ref{s:Z2dualityK3}, preserves $8$ RR ground fields, namely the four spectral flow operators and four more states in $\left( \frac{1}{4}, 0 ; \frac{1}{4}, 0 \right)$ representations, coming from the $\calR$-twisted sector. This means that it preserves the $4\cdot 4=16$ NSNS exactly marginal operators related by spectral flow to the latter four RR states. In turn, this implies that all deformations generated by such marginal operators leave the duality defect topological, so that there is a $16$-dimensional family of K3 sigma models, within the $80$-dimensional moduli space, where this TDL exists. Furthermore, since all the $16$ marginal operators come from the twisted sector, any such deformation will move us away from the $T^4/\ZZ_2$ orbifold locus. A similar reasoning applies for the Fibonacci defect of the $(3)(3)(8)$ Gepner model, considered in section \ref{s:Gepner}. In this case, there are $8$ RR ground fields in $\left( \frac{1}{4}, 0 ; \frac{1}{4}, 0 \right)$ representations that are preserved, besides the four spectral flow operators. This means that there is a $4\cdot 8=32$-dimensional family of K3 sigma models that admit such a Fibonacci defect, and the Gepner model we considered is just one rational point in this family. It is reasonable to expect the generic element in this family to be non-rational.
    
    \item  A correspondence analogous to the one between the Conway module $V^{f \natural}$ and K3 sigma models also exists between $V^{fE_8}$, the $\mathcal{N}=1$ supersymmetrized version of the VOA $V^{E_8}$ based on the lattice $E_8$, and the $\mathcal{N}=(4,4)$ non-linear sigma models on $T^4$. In \cite{Anagiannis:2020hkk} the authors show that any twining elliptic genus of $T^4$ with respect to a supersymmetry-preserving symmetry, see \cite{Volpato:2014zla} for a complete classification, can be reproduced by a supersymmetry-preserving twined elliptic genus of $V^{f E_8}$. This correspondence is fully consistent with the characterization of discrete symmetry groups of $T^4$ as subgroups of the group of even-determinant Weyl transformations of $E_8$ that fix pointwise a sublattice of rank at least four. As in the case of the $V^{f \natural}/K3$ correspondence, it would be interesting to investigate whether a similar relationship holds between supersymmetry-preserving topological defects in $V^{fE_8}$ and those in $T^4$ sigma models. Moreover, since the theories $V^{f \natural}$ and $V^{f E_8}$ are related by cyclic orbifolds that preserve the supercurrent, one can obtain topological defects in $V^{f \natural}$ from those on $V^{f E_8}$, and vice versa, provided the defects preserve the corresponding gauging groups. Indeed, we have verified that some of the $\mathbb{Z}_2$ and the $\mathbb{Z}_3$ duality defects, as well as the Fibonacci defect identified in $V^{f \natural}$ have a correspondent in $V^{f E_8}$ with the same twining genus. Since certain K3 models admit a torus orbifold description, it would be interesting to explore whether, in these cases, their topological defects exhibit relations analogous to those found between $V^{f \natural}$ and $V^{f E_8}$. 
    
    \item A topological defect $\CL$ in $V^{f\natural}$ can be described, using a standard `doubling trick', as a boundary state in the holomorphically factorized CFT $V^{f\natural}\otimes\overline{V^{f\natural}}$, where $\overline{V^{f\natural}}$ is a anti-holomorphic copy of $V^{f\natural}$. In other words, one can interpret the defects $\CL\in \cTop$ as ($\CN=1$-preserving) D-branes in type II superstring compactified on $V^{f\natural}\otimes\overline{V^{f\natural}}$. The lattice of R-R charges in this string model is $\Lambda\otimes\Lambda^*\cong {\rm End}(\Lambda)$, and the charge of a D-brane corresponding to $\CL$ is $\rho(\CL)$. What does the condition that the defect $\CL$ preserves a four-plane $\Pi\subset V^{f\natural}_{tw}(1/2)$ correspond to, in this context?  It is reasonable to speculate that massless open string states on one of these D-branes would correspond to continuous deformations of the defect, thus signaling that there is a  moduli space of D-branes with the same charge $\rho(\CL)$. What is the geometry of these moduli spaces? These kind of type II compactifications were considered in \cite{Harrison:2021gnp}, where it was shown that the $1+1$ dimensional space-time theory has either $(24,24)$ or $(0,48)$ supersymmetry, depending on the GSO projection. The space-time supersymmetry algebra can also be seen as the zero degree subalgebra in an infinite dimensional Borcherds-Kac-Moody superalgebra \cite{Harrison:2018joy}, which acts on certain spaces of BPS states. Do the D-branes preserve any of this space-time supersymmetry? Does the BKM superalgebra act on the degrees of freedom of a D-brane?
    
    \item While it is quite easy to construct examples of topological defect lines in $V^{f\natural}$ that preserve some rational subalgebra, showing that a certain topological defect in $V^{f\natural}$ preserves the $\CN=1$ superVirasoro algebra is a very challenging task.  For each non-invertible defect constructed in this paper, we had to make use of sometimes complicated and \emph{ad hoc} arguments that cannot be easily generalized to other cases. This is true even if one knows the corresponding endomorphism of the Leech lattice that determines how the defect acts on the Ramond ground states. The main problem is that the $\CN=1$ supercurrent $\tau(z)$ is a very particular element in the $2048$-dimensional space of spin $3/2$ operators, with no simple characterization (as far as we know).  Given an endomorphism of the Leech lattice, is there a simple and systematic way to construct a corresponding topological defect (or defects) preserving $\tau(z)$ (or to rule out their existence)? The next two points might be relevant in this respect.
    
    \item Our construction implies that the choice of a supercurrent $\tau(z)$ in $V^{f\natural}$ determines an embedding of the Leech lattice $\Lambda$ in the space of Ramond ground fields and vice versa. However, the only way we know to relate these two objects is via the subgroup $\Aut(V^{f\natural})\cong Co_0$ of $Spin(24)$ that preserves $\tau(z)$, and that coincides with the group of automorphisms of the embedded lattice $\Lambda$. Is there a more direct and explicit connection? The space of spin $3/2$ fields in the NS sector of $V^{f\natural}$ is an irreducible module for a Clifford algebra based on the space $V^{f\natural}_{tw}(1/2)$ of Ramond ground fields. Thus, it seems that a characterization of $\tau(z)$ in terms of the `Clifford ring' generated by elements in the Leech lattice should be possible; this would be helpful in proving that a given defect preserves the $\CN=1$ supercurrent. These arguments suggest the existence of some special integral form of $V^{f\natural}$, i.e. a definition of the SVOA over the ring of integers rather than the field of complex numbers. Such an integral form is defined as follows. Suppose there exists a set of generators for $V^{f\natural}$ with respect to which all OPE coefficients are integral. Then, one could replace the vector space $V^{f\natural}$ with the $\ZZ$-module $V^{f\natural}_\ZZ$ (in fact, a lattice) generated by such operators, and the OPE would provide a well-defined `vertex operator ring' structure on this module, satisfying axioms analogous to the ones of a VOA. We would be particularly interested in an integral form $V^{f\natural}_\ZZ$ containing (a multiple of) the $\CN=1$ supercurrent $\tau(z)$, and the corresponding Leech lattice of Ramond ground fields. Which defects in $\cTop$ would act on such a $V^{f\natural}_\ZZ$ by endomorphisms? All such defects must have integral quantum dimension; in particular, we would expect all invertible lines in $\Aut_\tau(V^{f\natural})$ to be included.  The same question could be asked for the module $V^{f\natural}_\ZZ\otimes_\ZZ R$ defined over a more general ring $R$.
    
    \item There are many different constructions of the Leech lattice based on rings of integers in some algebraic extension $K$ of $\QQ$.  These constructions are  useful for defining endomorphisms of the Leech lattice with eigenvalues in the number field $K$. Examples include the icosian ring in the number field $K=\QQ(\zeta)$, considered in section \ref{s:Fibonacci}, or the ring $\ZZ[\omega]$ of Eisenstein integers, which is mentioned in section \ref{sec:Z3duality}. Can one use the other constructions of the Leech lattice to define interesting rings of lattice endomorphisms? Can one show that a corresponding fusion category of $\CN=1$-preserving topological defects exists? A suggestive example is a construction, due to Craig \cite{Craig}, of the Leech lattice in terms of the ring $\ZZ[e^{\frac{2\pi i}{39}}]$. The corresponding number field $K=\QQ(e^\frac{2\pi i}{39})$ contains the eigenvalues of the Haagerup fusion ring \cite{AsaedaHaagerup}. Can one use Craig's construction to define a Haagerup fusion ring of Leech lattice endomorphisms? In this case, is there an associated Haagerup fusion category of topological defects?

    \item In section \ref{s:Gepner} we describe the topological defects in Gepner models that preserve the full chiral algebra and spectral flow. This particular category of TDLs is very interesting because it corresponds to (possibly non-invertible) symmetries that commute with space-time supersymmetry. However, the analysis we present in this article is not complete. First of all, we mostly focus on models with $c=6$ (K3 models);  the generalization to other cases, in particular to sigma models on Calabi-Yau $3$-folds ($c=9$) should be straightforward. The most challenging issue concerns the treatment of topological defects in models where the resolution of fixed points is necessary. As explained in section \ref{s:Gepner}, this problem arises when some of the topological defects induced from the Verlinde lines in the tensor product of $\CN=2$ minimal models turn out not to be simple defects in the Gepner model, after the orbifold ensuring the integrality of the $U(1)$ charge. In all such cases, we know that the resulting TDL in the Gepner model is the superposition of two simple defects. We expect to be able to determine the action of such simple defects in various specific examples, but a systematic treatment of this issue is, as far as we know, still an open problem. Completing this analysis and explicitly constructing the categories of topological defects in Gepner models would pave the way to the investigation of a number of interesting questions: the study of exactly marginal deformations preserving certain defects, along the lines of \cite{Cordova:2023qei} (see also point 2 above); the action of defects on boundary states, especially the ones preserving some space-time supersymmetry, providing new examples   and  generalizing the analysis of \cite{Angius:2024evd} for K3 models; the relation between the existence of non-trivial topological defects and geometric properties of the target space; and the consequences of the existence of TDLs on the worldsheet for the space-time theory (at least in a perturbative limit). We hope to come back to these questions in future work.
\end{enumerate}

\subsection*{Notation}
\begin{itemize} \item $\Lambda$: the Leech lattice ($24$-dim. even unimodular, positive definite, with no roots) \cite{ConwaySloane}.
    \item $\Aut(\Lambda)\cong Co_0$: the Conway group of order $2^{22}\cdot 3^9\cdot 5^4\cdot 7^2\cdot 11\cdot 13\cdot 23$ \cite{Atlas}.
    \item $\CL$: a topological defect line (TDL) in a CFT.
    \item $\hat\CL$: the linear operator on the Hilbert space of states, associated with inserting $\CL$ along the $S^1$ circle, when the CFT is defined on a cylinder $S^1\times \RR$.
    \item $\langle\CL\rangle$: the quantum dimension of $\CL$ (defined as the vacuum eigenvalue of $\hat\CL$).
    \item $V^{f\natural}$: the Conway module SCFT (NS sector) (see section \ref{s:ConwayModule}).
    \item $V^{f\natural}_{tw}$: the canonically twisted module (Ramond sector) of $V^{f\natural}$.
    \item $V^{f\natural}_{tw}(1/2)$: the $24$-dimensional space of Ramond ground states (conformal weight $1/2$) in $V^{f\natural}_{tw}$;
    \item ${}^\RR V^{f\natural}_{tw}(1/2)$: the real space of CPT self-conjugate operators in $V^{f\natural}_{tw}(1/2)$.
    \item $V^{f\natural}_{\CL}$, $V^{f\natural}_{tw,\CL}$: $\CL$-twisted spaces of $V^{f\natural}$ on $S^1\times \RR$, defined by inserting $\CL$ along the `time' direction (NS and R sector, respectively).
    \item $\tau(z)$: the $\CN=1$ supercurrent of $V^{f\natural}$.
    \item $\Aut_\tau(V^{f\natural})\cong \Aut(\Lambda)\cong Co_0$: group of (invertible) symmetries of $V^{f\natural}$ fixing $\tau(z)$.
    \item $\cTop$: tensor category of TDLs $\CL$ in $V^{f\natural}$ preserving $\tau(z)$ and `well-behaved' with respect to fermion number (see section \ref{s:ConwayModule}); the invertible lines in $\cTop$ generate $\Aut_\tau(V^{f\natural})$.
    \item $\cTop_{\Pi^\natural}$: given a four-dimensional subspace $\Pi^\natural\subset {}^\RR V^{f\natural}_{tw}(1/2)$, the subcategory of of lines $\CL\in \cTop$ that preserve $\Pi^\natural$.
    \item $Z_{\NS}^{\CL,\pm}$, $Z_{\R}^{\CL,\pm}$: torus partition functions with insertion of $\CL$ along the `space' circle ($\CL$-twined), for various choices of spin structures.
    \item $Z_{\CL,\NS}^\pm$, $Z_{\CL,\R}^\pm$: torus partition functions with insertion of $\CL$ along the `time' circle ($\CL$-twisted), for various choices of spin structures.
    \item $\CZ^{\CL}(\tau)$, $\CZ_{\CL}(\tau)$: vector-valued partition functions with components $(Z_{\NS}^{\CL,+},Z_{\NS}^{\CL,-},Z_{\R}^{\CL,+},Z_{\R}^{\CL,-})^t$ and $(Z_{\CL,\NS}^{+},Z_{\CL,\NS}^{-},Z_{\CL,\R}^{+},Z_{\CL,\R}^{-})^t$, respectively.
    \item $TY(A,\chi,\epsilon)$: a Tambara-Yamagami category with abelian group $A$, bicharacter $\chi$, Frobenius-Schur indicator $\epsilon$ (see section \ref{sec:TY_categories}).
    \item $\calC$: a supersymmetric non-linear sigma model (NLSM) on K3.
    \item $\Gamma^{4,20}$: lattice of R-R charges in a K3 NLSM (even unimodular with signature $(4,20)$);
    \item $\Pi$: four-dimensional space of spectral-flow operators in the R-R sector of a NLSM on K3, often seen as a subspace of $\Gamma^{4,20}\otimes \RR$ (see section \ref{s:K3defects}).
    \item $\cTop^{K3}_{\calC}$: tensor category of TDLs in the K3 NLSM $\calC$, preserving the $\CN=(4,4)$ superconformal algebra and the spectral flow operators.
    \item $\phi^\CL(V^{f\natural},\tau,z)$, $\phi^\CL(\calC,\tau,z)$: $\CL$-twined elliptic genus in $V^{f\natural}$ and in the K3 NLSM $\calC$, respectively (see section \ref{s:conj}).
\end{itemize}
\section{Summary of previous results}

Two-dimensional CFTs are typically defined in terms of a set of local operators equipped with operator product expansions (OPEs) that satisfy associativity and determine correlation functions obeying modular invariance. To incorporate generalized symmetries in this framework, one introduces extended operators ${\mathcal{L}}_{i}(\gamma_i)$, known as topological defect lines (TDLs), which are supported on oriented lines $\gamma_i$ in the spacetime manifold $\Sigma$. These TDLs preserve correlation functions under any smooth deformation of $\gamma_i$, provided that the lines do not cross any other operator insertions. As TDLs generally satisfy non-invertible fusion algebras, they do not admit a standard group structure, as for ordinary symmetries, but rather they can be naturally described into the language of tensor categories (see \cite{Etingof:2015} for an introduction to the subject). One concrete way to define these extended defects is through their action on local operators, namely, by specifying how correlation functions change when a TDL is moved across a local operator insertion. If moving the line $\mathcal{L}_i(\gamma_i)$  across a subset of local operators, corresponding to the holomorphic fields $\left\lbrace T(z), \phi_1(z), \phi_2(z) , ... \right\rbrace$, leaves all correlators invariant, then the TDL $\mathcal{L}_i$ is said to preserve the full chiral algebra $\mathcal{A}$ generated by these fields. An analogous statement holds for an anti-chiral algebra $\bar{\mathcal{A}}$. The complete preserved algebra $\mathcal{A} \times \bar{\mathcal{A}}$ necessarily contains the Virasoro algebra $Vir_c \times Vir_{\bar{c}}$, generated by the stress-energy tensors $T(z)$ and $\bar{T}(\bar{z})$, and it is typically non-rational. As mentioned in the introduction, when $\mathcal{A} \times \bar{\mathcal{A}}$ is rational, there exists a method to classify TDLs preserving the algebra, based on imposing Cardy-like conditions on the defects, \cite{Petkova:2000ip}.  Indeed, rational algebras possess only finitely  many irreducible representations, and the action of a defect over the corresponding primary fields can be encoded in a finite set of complex parameters. By computing the torus partition function with such a defect $\mathcal{L}$ inserted in the trace, i.e. the $\mathcal{L}-$twined partition function, one obtains a function depending on these complex parameters. Requiring that the modular $S-$transformation of this function yields a well-defined partition function on the $\mathcal{L}-$twisted Hilbert space, which still decomposes into the finitely many irreducible representations of $\mathcal{A} \times \bar{\mathcal{A}}$, imposes (discrete) lattice-like conditions on the parameter space, with the simple defects corresponding to an appropriate basis of this lattice.

However, because of the presence of infinitely many irreducible representations, no general analogue of this procedure exists for defects preserving non-rational chiral algebras. In \cite{Angius:2025zlm}, we proposed a Cardy-like construction for TDLs in $V^{f \natural}$ that preserve the non-rational $\mathcal{N}=1$ superVirasoro algebra, where the 24 Ramond ground states of the theory play the role of primary states in the rational case. Although the construction does not provide a complete determination of the underlying tensor category $\cTop$, it allows us to impose conditions on the admissible defects, to establish a correspondence between them and the automorphisms of the Leech lattice, as well as to derive constraints on their quantum dimensions.

Similar results were obtained in \cite{Angius:2024evd} for TDLs in K3 non-linear sigma models preserving the $\mathcal{N}=(4,4)$ superconformal algebra and spectral flow. Instead of imposing Cardy-like conditions, the analysis in that case exploits the action of TDLs on the boundary states of the theory, corresponding to R–R charged D-branes. These branes span a 24-dimensional charge lattice $\Gamma^{4,20}$ with signature $(4,20)$ on which TDLs act as endomorphisms. The analogue of this lattice in $V^{f \natural}$ is the Leech lattice $\Lambda$, embedded in the space of Ramond ground states.

In the rest of this section we review the main results obtained by implementing these two constructions and show how the similarities between them suggest a conjectured correspondence between topological defects in K3 that preserve the $\mathcal{N}=(4,4)$ superconformal algebra and spectral flow, and defects of $V^{f \natural}$ that preserve supersymmetry and a chosen four-plane in the space of Ramond ground states. This conjecture extends the relationship established in \cite{Duncan:2015xoa} linking the symmetries of the two theories.

\subsection{Topological defects in $V^{f \natural}$}\label{s:ConwayModule}
The Conway moonshine module $V^{f \natural}$ \cite{Frenkel:1988flm,Duncan:2006} is one of the only three  holomorphic superconformal field theories (SCFTs) with $c=12$, the other two being $F(24)$, the theory of $24$ free chiral fermions, and $V^{f E_8}$, the supersymmetrized version of the VOA $V^{E_8}$ based on the lattice $E_8$. Among these three, $V^{f \natural}$ is uniquely distinguished by having a minimal fermions conformal weight of $3/2$.

There exist various equivalent descriptions for this theory. The simplest one is in terms of the self-dual SVOA of $12$ chiral free bosons on the lattice
\begin{equation}\label{D12plus}
    D_{12}^+= \left\lbrace \frac{1}{2} (x_1, x_2, ..., x_{12}) \in \left( \frac{1}{2} \mathbb{Z}\right)^{12} \, \vert \, x_i \equiv x_j \, \,  \text{mod} \, \, 2 \, , \sum x_i \in 4 \mathbb{Z} \right\rbrace
\end{equation}
containing the vacuum $V_0$ and the spinorial $V_s$ irreducible modules of the affine algebra $\hat{so}(24)_1$: 
\begin{equation}
    V^{f \natural} = \,  V_0 \,  \oplus \,  V_s \, .
\end{equation}
$V^{f \natural }$ provides an irreducible NS module for $\mathcal{N}=1$ superVirasoro $sVir_{c=12}$, whose unique  supercurrent $\tau(z)$ is fixed by the action of a finite automorphism group $\Aut_\tau(V^{f\natural})$ isomorphic to the Conway group
\begin{equation}
    Co_0 \simeq \Aut_\tau(V^{f\natural}):=\left\lbrace g \in Aut(V^{f \natural}) \vert \, g(\tau)=\tau\right\rbrace \subset Aut (V^{f \natural}) \, \simeq \, Spin(24),
\end{equation}
the automorphism group of the $24$-dimensional Leech lattice $\Lambda$. 

The Ramond sector can be realized as the $(-1)^F$-twisted module of $V^{f \natural}$, where the fermion number symmetry $\langle (-1)^F\rangle\cong\mathbb{Z}_2 \subset Aut(V^{f \natural})$ acts trivially on $V_0$ and by $-1$ on $V_s$. It corresponds to the remaining vectorial and conjugate spinorial irreducible modules of $\hat{so}(24)_1$:
\begin{equation}
    V^{f \natural}_{tw} = \, V_v \, \oplus \, V_c \, . 
\end{equation}
We assign fermion number $(-1)^F=+1$ to $V_v$ and $(-1)^F=-1$ to $V_c$; the opposite assignment would also be consistent. 
Other notable constructions for $V^{f \natural}$ highlight its connection to the other two self-dual SVOAs at $c=12$. In particular, $V^{f \natural}$ can be obtained as a $\mathbb{Z}_2$ orbifold of $F(24)$, acting by reversing the signs of the $24$ chiral fermions of weight $1/2$. Similarly, $V^{f \natural}$ can be realized as the $\mathbb{Z}_2$ orbifold of $V^{f E_8}$, where the involution flips the sign of the $8$ free chiral bosons compactified on $E_8$ and their fermionic superpartners, while preserving the $\mathcal{N}=1$ supercurrent. These descriptions are particularly useful for constructing TDLs in $V^{f \natural}$ that manifestly preserve the $\mathcal{N}=1$ supercurrent, as we will see, for example, in section \ref{s:freefermion}.

Within this theory, we consider the category $\cTop$, whose objects are TDLs $\mathcal{L}$ satisfying the following three properties:
\begin{itemize}
    \item[1.] With each $\CL\in \cTop$ are associated two linear maps
    \be \hat{\mathcal{L}}: V^{f \natural} \mapsto V^{f \natural}\qquad \hat{\mathcal{L}}: V^{f \natural}_{tw} \mapsto V^{f \natural}_{tw}\ ,\ee that commute with $(-1)^F$ and with the action of the $\mathcal{N}=1$ superVirasoro algebra.
    \item[2.] With each $\CL\in \cTop$ are associated the $\mathcal{L}$-twisted sector $V^{f \natural}_{\mathcal{L}}$ and the $(-1)^F \mathcal{L}$-twisted sector $V^{f \natural}_{ \mathcal{L}, tw}$, that are unitary representations of, respectively, the NS and Ramond $\mathcal{N}=1$ superVirasoro algebras, with a well-defined and compatible $\mathbb{Z}_2$-grading by $(-1)^F$. Furthermore, the spectrum of $L_0$ on $V^{f \natural}_{\mathcal{L}}$ and $V^{f \natural}_{ \mathcal{L}, tw}$ is discrete, and each $L_0$-eigenspace is a finite dimensional $\mathbb{Z}_2$-graded vector space. 
    \item[3.] For each defect $\mathcal{L}$, we can define a vector valued $\mathcal{L}-$twined partition function $\mathcal{Z}^{\mathcal{L}}(\tau)= \left( Z^{\mathcal{L},+}_{NS}, Z^{\mathcal{L},-}_{NS}, Z^{\mathcal{L},+}_{R}, Z^{\mathcal{L},-}_{R} \right)^t$ with components
    \begin{equation}\label{ZLtwinNS}
        Z^{\mathcal{L},\pm}_{NS} = \Tr_{V^{f \natural}} \left( q^{L_0 - \frac{1}{2}} (\pm 1)^F \hat{\mathcal{L}} \right)
    \end{equation}
   \begin{equation}\label{ZLtwinR}
        Z^{\mathcal{L},\pm}_{R} = \Tr_{V^{f \natural}_{tw}} \left( q^{L_0 - \frac{1}{2}} (\pm 1)^F \hat{\mathcal{L}} \right)
    \end{equation}
    and a vector valued $\mathcal{L}-$twisted partition function $\mathcal{Z}_{\mathcal{L}}(\tau)= \left( Z^{+}_{\mathcal{L}, NS}, Z^{-}_{\mathcal{L}, NS}, Z^{+}_{\mathcal{L}, R}, Z^{-}_{\mathcal{L}, R} \right)^t$ with
        \begin{equation}\label{ZLtwistNS}
        Z^{\pm}_{\mathcal{L}, NS} = \Tr_{V^{f \natural}_{\mathcal{L}}} \left( q^{L_0 - \frac{1}{2}} (\pm 1)^F  \right)
    \end{equation}
   \begin{equation}\label{ZLtwistR}
        Z^{\pm}_{\mathcal{L}, R} = \Tr_{V^{f \natural}_{\mathcal{L},tw}} \left( q^{L_0 - \frac{1}{2}} (\pm 1)^F  \right)
    \end{equation}
    which are related by a modular $S$-transformation:
    \begin{equation}
        \CZ_{\mathcal{L}}(-1/ \tau) \, = \, S \, \CZ^{\mathcal{L}}(\tau) 
    \end{equation}
    where $S$ is the appropriate S-matrix for a holomorphic $c=12$ SCFT, namely
    \begin{equation}
        S = \left( \begin{matrix} 1 & 0 & 0 & 0 \\ 0 & 0 & 1 & 0 \\ 0 & 1 & 0 & 0 \\ 0 & 0 & 0 & 1 \end{matrix} \right).
    \end{equation}
\end{itemize}
By property 2, for each $\mathcal{L} \in \cTop$, the sector $V^{f \natural}_{\mathcal{L},tw}$ furnishes a unitary representation of the Ramond superVirasoro algebra $sVir_{c=12}$ with a discrete $L_0$ spectrum. Due to the presence of supersymmetry, apart from the ground state, each excited level of $L_0$ contains an equal number of bosonic and fermionic states. Hence the function $Z^-_{\mathcal{L},R}$ is constant and integral, a supersymmetric index counting the difference between the number of bosonic and fermionic ground states in the twisted Hilbert space $V^{f \natural}_{\mathcal{L},tw}$. Furthermore, property 3 implies that the $\mathcal{L}-$twined function $Z^{\mathcal{L},-}_R$ must be the same constant, and therefore it also takes integral values:
\begin{equation}
    Z^{\mathcal{-}}_{\mathcal{L},R} = Z^{\mathcal{L},-}_R \in \mathbb{Z}.
\end{equation}

All invertible topological defect lines in the category $\cTop$ are given by $\mathcal{L}_g$ for each $g \in Co_0$, the symmetry group of the theory. Using fusion, for every defect $\mathcal{L} \in \cTop$, one can define a set of integer invariants through the trace functions:
\begin{equation}
    Z^{\mathcal{-}}_{\CL_g\mathcal{L},R} = Z^{\CL_g\mathcal{L},-}_R=Tr_{V^{f \natural}_{tw} (1/2)} \left( (-1)^F \hat{\mathcal{L}_g} \hat{\mathcal{L}}\right) \in \mathbb{Z} \, \, , \quad \quad \forall \, g \, \in Co_0,
    \label{L:set_integers}
\end{equation}
where $V^{f \natural}_{tw}(1/2) \simeq \mathbb{C}^{24}$ denotes the 24 dimensional Hilbert space of R ground states of conformal weight $1/2$. The action of $\Aut_\tau(V^{f\natural})\cong Co_0$ on the $24$-dimensional real space ${}^{\mathbb{R}}V^{f \natural}_{tw}(1/2) \subset V^{f \natural}_{tw}(1/2)$ of CPT self-conjugate states, fixes (setwise) a copy of the Leech lattice $\Lambda\subset {}^{\mathbb{R}}V^{f \natural}_{tw}(1/2)$. For each given $\CL\in \cTop$, integrality of the trace functions \eqref{L:set_integers} for all $g\in \Aut_\tau(V^{f\natural})\cong Co_0$ implies that the linear operator $\hat\CL$, when restricted to ${}^{\mathbb{R}}V^{f \natural}_{tw}(1/2)$, acts linearly on $\Lambda$ as a $\mathbb{Z}$-endomorphism. This result is formally established in the following theorem, proved in Section 3 of \cite{Angius:2025zlm}:
\begin{theorem}\label{th:main}
    Let $\cTop$ be a tensor category of topological defects $\mathcal{L}$ of $V^{f\natural}$ containing all invertible defects $\mathcal{L}_g$, $g\in Co_0$, and such that all objects $\CL\in \cTop$ satisfy properties 1, 2 and 3 above. Then there is an embedding $\Lambda\hookrightarrow {}^\RR V^{f\natural}_{tw}(1/2)$ of the Leech lattice $\Lambda$ into ${}^\RR V^{f\natural}_{tw}(1/2)\subset V^{f\natural}_{tw}(1/2)\cong \mathbb{C}^{24}$, such that for every $\mathcal{L}\in \cTop$, the $\CC$-linear map $\hat\CL_{\rvert V^{f\natural}_{tw}(1/2)}:V^{f\natural}_{tw}(1/2)\to V^{f\natural}_{tw}(1/2)$ is contained in ${\rm End}(\Lambda)$. The assignment \begin{align} \label{rho:Vfnat_map}
       \rho:\cTop &\to {\rm End}(\Lambda)\ ,\\ \CL &\mapsto \rho(\CL):=\hat\CL_{\rvert V^{f\natural}_{tw}(1/2)} 
    \end{align} defines a surjective, non-injective ring homomorphism from the Grothendieck ring $Gr(\cTop)$ to ${\rm End}(\Lambda)$, such that $\rho(\CL^*)=\rho(\CL)^\dag=\rho(\CL)^t$.
\end{theorem}
Among the main implications of the theorem, there are  the following two corollaries, proved in Section 3 of \cite{Angius:2025zlm}.

 \begin{corollary}\label{th:integralqdim}  If $\CL\in \cTop$ of quantum dimension $\langle \CL\rangle$ preserves a field $\lambda\in\Lambda\subset {}^\RR V^{f\natural}_{tw}(1/2)$, i.e. if $\hat\CL$ acts by $\hat\CL|\lambda\rangle=\langle\CL\rangle |\lambda\rangle$, then necessarily $\langle \CL\rangle\in \ZZ_{\ge 1}$. More generally, if $\CL$ preserves any field in $V^{f\natural}_{tw}(1/2)$ (not necessarily in $\Lambda$), then the quantum dimension $\langle \CL\rangle$ must be an algebraic integer of degree at most $24$. 
  \end{corollary} 
  
  \begin{corollary}\label{identity} Suppose $\hat\CL$ acts trivially  on some $\psi\in V^{f\natural}_{tw}(1/2)$ (i.e. $\hat\CL|\psi\rangle=\langle\CL\rangle|\psi\rangle$) such that $\psi^\perp \cap (\Lambda\otimes \bar\QQ)=0$, where $\bar\QQ$ is the algebraic closure of $\QQ$. Then, $\CL$ is multiple of the identity defect, i.e. $\CL=d\CI$, $d\in \NN$ (and in particular $\langle \CL\rangle=d$ is integral).\end{corollary}
The first assertion follows directly from the homomorphism \eqref{rho:Vfnat_map}, which ensures that the linear operators $\hat{\mathcal{L}}_{\vert V^{f \natural}_{tw}(1/2)}$ admit integral $24 \times 24$ matrix representations in the lattice basis. Consequently, their minimal polynomial is monic with integral coefficients, and if the quantum dimension is one of its roots, it must be an algebraic integer. In the special case in which the corresponding eigenvector is aligned with the lattice $\Lambda$, the eigenvalue is exactly an integer.\\ The second statement follows from the fact that if $\hat{\mathcal{L}}$ acts trivially (i.e., only by multiplication by its quantum dimension) on a sufficiently generic vector $\psi$ in $V^{f \natural}_{tw}(1/2) \cong \Lambda \otimes \mathbb{C}$, then $\mathcal{L}$ itself must be proportional to the trivial defect. The hypothesis that $\psi$ is generic is enforced by the condition $\psi^{\perp} \cap (\Lambda \otimes \overline{\mathbb{Q}})=0$, where $\overline{\mathbb{Q}}$ denotes the algebraic closure of $\mathbb{Q}$. Concretely, since $\Lambda \otimes \bar{\mathbb{Q}}$ is a countable subset of zero-measure in $\Lambda \otimes \mathbb{C}$ with respect to the Lebesgue measure in $\mathbb{C}^{24}$, the condition above holds for almost every $\psi$, so $\psi$ is generic.

\subsection{Topological defects in K3 non-linear sigma models}\label{s:K3defects}
Non-linear sigma models (NLSMs) on K3 \cite{Aspinwall:1996mn,Nahm:1999ps}, together with those on $T^4$, are the only known unitary SCFTs with $\mathcal{N}=(4,4)$ superconformal algebra at $(c, \bar{c})=(6,6)$ whose spectra are invariant under spectral flow. These models play a central role both in physics, where they describe the worldsheet dynamics of perturbative type II superstrings on a target K3, and in mathematics, for example, through their connection with certain sporadic simple groups and moonshine phenomena \cite{Mukai,Eguchi:2010ej}.

Both the holomorphic and anti-holomorphic copy of the $\mathcal{N}=4$ superconformal algebra are generated by the stress-energy tensor, an $\hat{su}(2)_1$ triplet of currents, and the four supercurrents. Their unitary irreducible representations are labeled by the pairs $\left( h , q ; \tilde{h}, \tilde{q}\right)$ associated to the highest weight ground states in the representation, where $h$ ($\tilde{h}$) is the conformal weight and $q$ ($\tilde{q}$) is the eigenvalue with respect to the zero-mode $J^3_0$ ($\tilde{J}^3_0$) Cartan current of $\hat{su}(2)_1$, normalized so as $q,\tilde q\in \frac{1}{2}\ZZ$. These representations are organized into the standard RR, RNS, NSR, NSNS sectors related one to the other by spectral flow transformations.

A special feature of these theories is that every K3 NLSM exhibits a topological sector of massless RR ground states = $\CH_{RR,gr}$ containing the lowest weight fields of a single $\left( \frac{1}{4}, \frac{1}{2} ; \frac{1}{4}, \frac{1}{2}\right)$ $\CN=(4,4)$ representation, spanning a four-dimensional real space $\Pi$, and of $20$ $\left( \frac{1}{4}, 0; \frac{1}{4},0 \right)$ representations spanning a twenty-dimensional real space $\Pi^{\perp}$. The four RR ground state fields in $\Pi$ transform under the bifundamental representation of the $SU(2) \times SU(2)$ R-symmetry group. Crucially, the OPEs with these fields generate the spectral flow transformations, which motivates referring to these fields as spectral flow generators.

The exactly marginal deformations of these models preserving the $\mathcal{N}=(4,4)$ superconformal symmetry span the $80$ dimensional moduli space
\begin{equation}
    \mathcal{M}_{K3} = O(4,20; \mathbb{Z}) \backslash O(4,20; \mathbb{R})/ \left( O(4) \times O(20) \right)
    \label{K3:moduli_space}
\end{equation}
where $O(4,20; \mathbb{Z})$ is the T-duality group acting via automorphisms on the charge lattice $\Gamma^{4,20}$ of RR D-brane charges, which is an even unimodular lattice of signature $(4,20)$. Considering the embedding of this lattice  in the 24--dimensional real space of RR ground states $V= \Gamma^{4,20} \otimes \mathbb{R} = \Pi \oplus \Pi^{\perp} \simeq \mathbb{R}^{4,20}$, the space \eqref{K3:moduli_space} parameterizes the possible ways to choose a positive definite four-dimensional subspace $\Pi \simeq \mathbb{R}^{4,0}$ within $V$, modulo T-dualities. The choice of a specific K3 model $\mathcal{C} \equiv \mathcal{C}_{\Pi}$, then corresponds to a specific choice of $\Pi \subset V$.

Given a K3 NLSM $\mathcal{C}$, we consider the category $\cTop_{\mathcal{C}}^{K3}$, whose objects are TDLs $\mathcal{L}$ satisfying the following two properties:
\begin{itemize}
    \item[1.] They commute with the full $\mathcal{N}=(4,4)$ superconformal algebra.
    \item[2.] They commute with spectral flow generators. 
\end{itemize}
It was shown in \cite{Angius:2024evd} that these two properties imply that for every defect $\mathcal{L} \in \cTop^{K3}_{\mathcal{C}}$, the restriction of the corresponding linear operator $\hat{\mathcal{L}}$ to the RR ground state Hilbert space yields an $\mathbb{R}$-linear extension of a lattice endomorphism:
\begin{equation}
\begin{split}
    \rho \, \, : \,\ \, \cTop_{\mathcal{C}}^{K3} \,\,  &\mapsto \, \, End \left( \Gamma^{4,20} \right) 
\end{split}
\label{rho:K3_map}
\end{equation}
which acts trivially on the spectral flow generators, i.e., by multiplication by the quantum dimension, $\hat{\mathcal{L}} \vert_{\Pi} = \langle \mathcal{L} \rangle \, \text{id}_{\Pi} $. For generic non-invertible TDLs $\mathcal{L} \in \cTop^{K3}_{\mathcal{C}}$ the map \eqref{rho:K3_map} defines a ring homomorphism from the fusion ring of $Top^{K3}_{\mathcal{C}}$ to $End(\Gamma^{4,20})$. When restricted to invertible defects $\mathcal{L}_g$ with $g \in G_{\Pi} \simeq Aut \left( \Gamma^{4,20} \right) $, where $G_{\Pi}\simeq \left\lbrace g \in O (4,20 ; \mathbb{Z}) \vert \, g \vert_{\Pi} = \text{id}\vert_{\Pi} \right\rbrace \subset Aut (\Gamma^{4,20})$ is the symmetry group of $\mathcal{C}$ preserving the $\mathcal{N}=(4,4)$ superalgebra and spectral flow, the map $\rho$ enhances to an isomorphism between invertible defects and automorphisms of the lattice $\Gamma^{4,20}$ \cite{Gaberdiel:2011fg}. 
The key results established in \cite{Angius:2024evd} through the map \eqref{rho:K3_map} and properties 1 and 2, which are particularly relevant for our current analysis, can be summarized in the following two central claims (corresponding to Claim 2 and 3 in \cite{Angius:2024evd}).
\begin{claim}\label{claim:quantum_dim_K3} The quantum dimension $\langle \mathcal{L} \rangle$ of a defect $\mathcal{L} \in \cTop^{K3}_{\mathcal{C}}$ is an algebraic integer of degree at most 6. Furthermore, if $\Pi \cap \Gamma^{4,20} \neq 0$, then $\langle \mathcal{L} \rangle$ is an integral for all $\mathcal{L} \in \cTop^{K3}_{\mathcal{C}}$.
\end{claim}
\begin{claim}\label{claim:identity_K3} For a generic K3 sigma model $\mathcal{C}$, the only topological defects in $\cTop^{K3}_{\mathcal{C}}$ are integral multiples of the identity.
\end{claim}
The first claim follows directly from the fact that $\hat{\mathcal{L}}$, when acting on the RR ground-state space $V= \Gamma^{4,20} \otimes \mathbb{R}$, admits an integral $24 \times 24$ matrix representation in the lattice basis, which minimal polynomial is monic, of degree $24$ with integral coefficients. Since by property 2 the quantum dimension $\langle \mathcal{L} \rangle$ is a root of this polynomial with multiplicity at least $4$, it must be an algebraic integer of degree at most $6$.  Furthermore, when the corresponding eigenvector aligns with a vector in the lattice $\Gamma^{4,20}$, which physically corresponds to the presence of D-brane configurations in the model minimizing the BPS mass, the quantum dimensions of all the defects in $\cTop^{K3}_{\mathcal{C}}$ are genuine integers. The second assertion tells us that the category $\cTop^{K3}_{\mathcal{C}}$ is trivial for a generic K3 NLSM in $\mathcal{M}_{K3}$. This generalizes the expectation that the most generic K3 NLSM possesses no additional (generalized) symmetries beyond its $\mathcal{N}=(4,4)$ superconformal symmetry.

\subsection{The Conjecture}\label{s:conj}
The first observation \cite{Duncan:2015xoa} connecting the theory $V^{f \natural}$ with K3 NLSMs was based on the connection between their symmetry groups. Specifically, it was noted that all symmetry groups of K3 sigma models preserving the $\mathcal{N}=(4,4)$ superconformal algebra and the spectral flow can be realized as subgroups of the symmetry group of $V^{f \natural}$ \cite{Gaberdiel:2011fg}, which is the Conway group $Co_0$. This group is the automorphism group of the Leech lattice, the unique $24$-dimensional even unimodular lattice without roots.  

For each K3 NLSM, the corresponding symmetry group $G_{\Pi^{\natural}} \subset Co_0$ fixes a subspace $\Pi^{\natural}$ of dimension at least $4$ in the $\mathbf{24}$ representation of $Co_0$. Conversely, every such fixed subspace $\Pi^{\natural}$ in $\mathbf{24}$ defines at least a K3 NLSM whose symmetry group is precisely $G_{\Pi^{\natural}}$. This connection between $V^{f \natural}$ and K3 NLSMs extends beyond the abstract group structure to include the action of these symmetries on the physical states. In particular, the action of any element in $G_{\Pi^{\natural}}$ on the topological sector of $1/4$ BPS states in a corresponding K3 NLSM is exactly reproduced (with few exceptions) by the action of the corresponding symmetry in $Co_0$ on the canonically-twisted module $V^{f \natural}_{tw}$ \cite{Duncan:2015xoa,Cheng:2016org}.

Since $Co_0 \subset SO(24)$ acts on the space ${}^{\mathbb{R}}V^{f \natural}_{tw} (1/2) \simeq \mathbb{R}^{24}$ of Ramond ground states in the standard vectorial representation, then a subgroup $G_{\Pi^{\natural}}\subset Co_0$ that fixes a four-plane $\Pi^{\natural}\subset \mathbb{R}^{24}$ also  fixes a subalgebra $so(4)_1 \simeq su(2)_1 \oplus su(2)_1 \subset so(24)_1$. These two $su(2)_1$ algebras are the R-symmetries of two commuting copies of the $\CN=4$ superconformal algebra at $c=6$, in which the $\CN=1$ superVirasoro at $c=12$ generated by $\tau(z)$ is diagonally embedded \cite{Cheng:2015kha}. Choosing one of the two $su(2)_1$ determines a graded partition function in the theory computed over the Ramond sector of $V^{f \natural}$ furnished by the twisted module:
\begin{equation}
    \phi \left( V^{f \natural}, \tau, z \right) = \Tr_{V^{f \natural}_{tw}} \left( (-1)^F q^{L_0 - 1/2} y^{J^3_0}\right)
    \label{Vfnat:graded_part_func}
\end{equation}
where $y=e^{2 \pi i z}$ and $J^3_0$ is the Cartan generator of $su(2)_1$. This function exactly reproduces the elliptic genus $\phi \left( \mathcal{C}, \tau, z \right)$ of a K3 NLSM $\mathcal{C}$:
\begin{equation}
    \phi \left( \mathcal{C}, \tau, z \right) = \Tr_{RR} \left( (-1)^{F + \bar{F}} q^{L_0-1/4} \bar{q}^{\bar{L}_0 -1/4} y^{J^3_0}\right)
    \label{K3:elliptic_genus}
\end{equation}
which is a purely holomorphic function preserved by marginal deformations preserving supersymmetry, and is a signed count of 1/4-BPS states in the RR sector. Notably the evaluation of the two functions \eqref{Vfnat:graded_part_func} and \eqref{K3:elliptic_genus} provide the same result \cite{Duncan:2015xoa}.

To understand the action of the symmetry $G_{\Pi^{\natural}}$ on these same sectors of the Hilbert spaces of the two theories, one can consider $g$--twined versions of the functions \eqref{Vfnat:graded_part_func} and \eqref{K3:elliptic_genus}. In particular, for any $g \in G_{\Pi^{\natural}}$, one can define, respectively, the $g$-twined partition function:
\begin{equation}
    \phi^g \left( V^{f \natural}, \tau, z \right) = \Tr_{V^{f \natural}_{tw}} \left( g \,(-1)^F q^{L_0 - 1/2} y^{J^3_0}\right)
    \label{Vfnat:g_twined_graded_part_func}
\end{equation}
in $V^{f \natural}$, and the $g$-twined elliptic genus
\begin{equation}
    \phi^g \left( \mathcal{C}, \tau, z \right) = \Tr_{RR} \left(g \, (-1)^{F + \bar{F}} q^{L_0-1/4} \bar{q}^{\bar{L}_0 -1/4} y^{J^3_0}\right)
    \label{K3:g_twined_elliptic_genus}
\end{equation}
in the K3 NLSM $\mathcal{C}$ which symmetry group is $G_{\Pi^{\natural}}$. Comparing these functions for any suitable symmetry $g \in Co_0$, that fixes a four-plane $\Pi^{\natural}$ in $\mathbf{24}$, one (almost \cite{Cheng:2016org}) always recovers the same result \cite{Duncan:2015xoa}. Despite this evidence for a connection between the two theories, a full understanding is still lacking.

Motivated by the formal analogy among the properties of topological defects in $V^{f \natural}$, stated in Theorem \ref{th:main} and the Corollaries \ref{th:integralqdim} and \ref{identity}, and the ones in K3 NLSMs, articulated in Claims \ref{claim:quantum_dim_K3} and \ref{claim:identity_K3} and proven through the existence of the map \eqref{rho:K3_map}, in \cite{Angius:2025zlm} we proposed a generalization of their relationship that extends the previous one from the level of symmetry groups to the level of their tensor categories. In particular, for any defect $\mathcal{L} \in \cTop_{\Pi^{\natural}}$, where $\cTop_{\Pi^{\natural}}$ is the subcategory of $\cTop$ of objects in $V^{f \natural}$ that fix a four-dimensional space $\Pi^{\natural} \subset V^{f\natural}_{tw}(1/2)$, and therefore the corresponding affine subalgebra $\hat{so}(4)_1 \subset \hat{so} (24)_1$, one can define a generalized $\mathcal{L}$-twined partition function (or ``twining genus"),
\begin{equation}
    \phi^\CL \left( V^{f \natural}, \tau, z \right) = \Tr_{V^{f \natural}_{tw}} \left( \hat{\mathcal{L}} \,(-1)^F q^{L_0 - 1/2} y^{J^3_0}\right)
    \label{Vfnat:L_twined_graded_part_func}
\end{equation}
by taking the trace of the action of $\mathcal{L}$ on the states of the twisted module $V^{f \natural}_{tw}$. Similarly, given the K3 model $\mathcal{C}$, for each topological defect $\mathcal{L} \in \cTop^{K3}_{\mathcal{C}}$, one can define the $\mathcal{L}$-twined elliptic genus:
\begin{equation}
    \phi^\CL \left( \mathcal{C}, \tau, z \right) = \Tr_{RR} \left(\hat{\mathcal{L}} \, (-1)^{F + \bar{F}} q^{L_0-1/4} \bar{q}^{\bar{L}_0 -1/4} y^{J^3_0}\right).
    \label{K3:L_twined_elliptic_genus}
\end{equation}

The expectation of an exact correspondence between these classes of topological defects in the two theories is formalized in the following conjecture (Conjecture 4 in \cite{Angius:2025zlm}).
\begin{conjecture}\label{conj:K3relation}
    Let $\Pi^{\natural} \subset \Lambda\otimes\RR\subset V^{f\natural}_{tw}$ be a subspace of dimension $4$ and $\cTop_{\Pi^{\natural}}$ the full subcategory of $\cTop$ such that
    \begin{equation}
        Obj \left( \cTop_{\Pi^{\natural}}\right) \, = \, \left\lbrace \mathcal{L} \in Obj (\cTop) \, \vert \, \hat{\mathcal{L}}\vert_{\Pi^{\natural}} = \langle \mathcal{L} \rangle \text{id}_{\Pi^{\natural}}\right\rbrace.
    \end{equation}
     Then, there exists a non-linear sigma model $\calC$ on K3 such that:\begin{enumerate}
        \item[\textit{(i)}] there is an equivalence $F:\cTop_{\Pi^{\natural}}\longrightarrow \cTop^{K3}_\calC$ of tensor categories between $\cTop_{\Pi^{\natural}}$ and the category $\cTop^{K3}_\calC$ of topological defects of $\calC$ preserving $\CN=(4,4)$ and spectral flow;
        \item[\textit{(ii)}] for every $\CL\in \cTop_{\Pi^{\natural}}$, the $\CL$-twining genus $\phi^\CL(V^{f\natural},\tau,z)$  computed in $V^{f\natural}$ (eq.\eqref{Vfnat:L_twined_graded_part_func}) coincides with the $F(\CL)$-twining genus in the K3 model $\calC$ (eq.\eqref{K3:L_twined_elliptic_genus}):
 \be        \phi^\CL(V^{f\natural},\tau,z)=\phi^{F(\CL)}(\calC,\tau,z).
        \ee 
\item[\textit{(iii)}] there is an isometry of Hilbert spaces $\varphi:V^{f\natural}_{tw}(1/2)\stackrel{\cong}{\longrightarrow} \Hh^{K3}_{\HRR,gr}$ such that for each $\CL\in \cTop_{\Pi^{\natural}}$
\be \varphi\circ \hat\CL_{\rvert V^{f\natural}_{tw}(1/2)} =\widehat{F(\CL)}_{\rvert \Hh^{K3}_{\HRR,gr}}\circ \varphi\ .
\ee
    \end{enumerate}
\end{conjecture}\noindent

In \cite{Angius:2025zlm}, we provided a conditional argument supporting this conjecture for the subcategory of $\mathsf{Top}$ in $V^{f \natural}$ whose objects have integral quantum dimension. In that special case, one can prove the existence of the isomorphism $\varphi: V^{f \natural}_{tw}(1/2) \mapsto \mathcal{H}^{K3}_{RR,gr}$, and show that the induced endomorphisms $\varphi \, \circ \, \hat{\mathcal{L}}_{\vert V^{f \natural}_{tw} (1/2)} \, \circ \varphi^{-1}$ establish an exact correspondence with topological defects in a specific K3 NLSM $\mathcal{C}$, consistent with the known properties of the theory.

This argument only works for the tensor (sub)categories of certain special K3 NLSMs that only contain defects with integral quantum dimension, and that are realized at particular points in moduli space where the geometric condition $\Pi \cap \Gamma^{4,20} \neq 0$ is satisfied, as stated in Claim \ref{claim:quantum_dim_K3}. Besides these special cases, the check of the conjecture on other K3 models remains an open problem. In the companion article we initiated the explicit construction of non-invertible topological defects in the K3 model $\mathcal{C}' = \mathcal{C}_{\text{GTVW}}/S_3$, obtained as the orbifold of the rational SCFT $\mathcal{C}_{\text{GTVW}}$ introduced in \cite{Gaberdiel:2013psa} by the non-anomalous subgroup $S_3 \subset \mathbb{Z}_2^8 \rtimes M_{20}$ of the symmetry group of $\mathcal{C}_{GTVW}$, and we compared the result with defects in $\mathsf{Top}$.

Further concrete examples are crucial to test the conjecture and provide solid evidence for the correspondence between the objects in the tensor categories defined on both sides of the conjecture. In the following sections, we will provide explicit constructions of such objects in the two theories.

\section{Duality defects and Tambara-Yamagami categories in $V^{f\natural}$}\label{sec:TY_categories}

A Tambara-Yamagami (TY) category is a fusion category whose simple objects are given by the invertible defects $\CL_a$ associated with a finite group $A$, and a single non-invertible defect $\CN$ (a duality defect) that is unoriented $\CN^*=\CN$ and satisfies
\be\label{TYfusion}  \CN^2=\sum_{a\in A}\CL_a\ ,\qquad\qquad \CN \CL_a=\CL_a\CN=\CN\ ,\quad \forall a\in A\ .
\ee In particular, $\CN$ has dimension $\sqrt{|A|}$. Tambara and Yamagami \cite{Tambara:1998} proved that such a fusion category exists if and only if the group $A$ is abelian, and classified all possible such categories for every abelian finite group $A$. In physics, and in particular in two dimensional CFT, duality defects and TY categories occur when the CFT is self-orbifold with respect to a finite non-anomalous abelian group $A$. In this section, after reviewing the general properties of duality defects in CFT, we classify all possible duality defects associated with cyclic groups that may preserve the $\CN=1$ supercurrent in $V^{f\natural}$. In \S \ref{s:spinselect} we review the spin selection rules for the TY categories we consider in this paper. Finally, in \S \ref{s:Monster}, we describe a relation between cyclic orbifolds of $V^{f\natural}$ and genus zero groups appearing in monstrous moonshine. In the following sections, we will provide  explicit examples of duality defects for TY$(\ZZ_2)$, TY$(\ZZ_3)$, and TY$(\ZZ_2\times \ZZ_2)$ categories in $V^{f\natural}$.

\subsection{General properties of TY categories}

Let $A$ be a finite abelian group; we use additive notation for the group law. Then, the possible TY categories associated with $A$ are specified by a choice of Frobenius-Schur indicator $\epsilon\in \{\pm 1\}$ and a symmetric, bilinear, non-degenerate function called the bicharacter  $\chi:A\times A\to U(1)$ with the properties:
\bea \nonumber\chi(a,b)&=&\chi(b,a)\ ,\\
  \chi(a+b,c)&=&\chi(a,c)\chi(b,c)\ ,\qquad \forall a,b,c\in A\ ,\\\nonumber
 \chi(a,b)&=&1\ ,\forall b\in A\qquad \Leftrightarrow \qquad a=0\ .
\eea
 The data $\chi,\epsilon$ determine all fusion matrices of the category, up to gauge equivalences (see \cite{Tambara:1998} for explicit formulae); in particular, the subcategory generated by the invertible defects $\CL_a$ is always non-anomalous. We denote by TY$(A,\chi,\epsilon)$ the corresponding fusion category; sometimes, we use the notation TY$(A)$ if we do not want to specify the fusion matrix data. Two fusion categories TY$(A,\chi,\epsilon)$ and TY$(A,\chi',\epsilon')$ for the same group $A$ are equivalent  if and only if $\epsilon=\epsilon'$ and $\chi$ and $\chi'$ are related by some automorphism $\phi:A\to A$ of $A$, i.e.
 \be \chi'(a,b)=\chi(\phi(a),\phi(b))\ ,\qquad \forall a,b\in A\ .
 \ee

In the context of a (Euclidean) two dimensional CFT $\calC$  (that we always assume to be unitary and with a unique vacuum), a TY$(A)$ category of topological defects arises when $A$ is a non-anomalous symmetry group of $\calC$ such that the orbifold $\calC/A$ is isomorphic to the original theory, i.e. $\calC=\calC/A$. In this case, the duality defect $\CN$ can be obtained by `half-gauging', i.e. by the taking the orbifold of $\calC$ by $A$ on half of the two-dimensional space, and then using the isomorphism $\calC=\calC/A$ to map the theory back to itself. The domain wall separating the two regions of the space carries a copy of the duality defect $\CN$. 

Let us give a more explicit description of such a duality defect. Let $\Hh^A$ be the subspace of states/fields that are fixed by $A$, and let $\Hh_A$ be its orthogonal complement, i.e. the space of fields transforming in non-trivial irreducible representations of $A$. Notice that the space $\Hh^A$ is closed under OPE on the sphere. The fusion relations \eqref{TYfusion} imply that the operator $\hat\CN$ is zero on $\Hh_A$, while $\frac{1}{\sqrt{|A|}}\hat\CN$ acts on $\Hh^A$ as an involution
\be\label{dualinvo} \frac{1}{\sqrt{|A|}}\hat\CN:\Hh^A\to \Hh^A\ ,\qquad (\frac{1}{\sqrt{|A|}}\hat\CN)^2={\rm id}_{\Hh^A}\ ,
\ee that respects the OPE. When the CFT $\calC$ is a holomorphic VOA or SVOA $V$, this means that $\frac{1}{\sqrt{|A|}}\hat\CN$ is an automorphism of order $2$ of the fixed subVOA $V^A$.

In general, if a CFT $\calC$ has a non-anomalous finite group $A$ of symmetries, then the orbifold CFT $\calC/A$ admits a fusion category of topological defects (the Wilson lines for $A$)  isomorphic to the category $Rep(A)$ of finite dimensional representations of $A$ \cite{Frohlich:2009gb,Bhardwaj:2017xup}. In this category, the identity is the trivial representation, simple objects $\CL_R$ correspond to irreducible representations $R\in Irr(A)$, the dual of a representation $R$ is the conjugate representation $R^*$, and a representation $R$ is invertible if and only if it is $1$-dimensional. Furthermore, if one takes the generalized orbifold of $\calC/A$ by this category of Wilson lines, one gets back the original theory $\calC$. When the group $A$ is abelian, then all irreducible representations are $1$-dimensional (i.e. they are characters $A\to U(1)$), so that the category $Rep(A)$ is generated by invertible defects and has the structure of a group with respect to tensor product. The underlying group $\hat A:=Hom(A,U(1))$  is known as the Pontryagin dual of $A$ and is always (non-canonically) isomorphic to the abelian group $A$. A choice of isomorphism \begin{align}\label{PontrDual} A&\stackrel{\cong}{\longrightarrow} \hat A\\ \notag
a&\mapsto R_a\end{align} is equivalent to a choice of a non-degenerate symmetric bilinear form $\chi:A\times A\to U(1)$, as above. Indeed, given such a $\chi$, one can associate with each $a \in A$ the $1$-dimensional representation $R_a:A\to U(1)$ given by $b\mapsto R_a(b)=\chi(a,b)$. 

Recall that a CFT $\calC$ with (finite, abelian, non-anomalous) symmetry group $A$ can be decomposed as
\be \Hh_{\calC}=\oplus_{R\in Irr(A)} \Hh_R\ ,
\ee where $\Hh_R$ is generated by fields transforming in the irreducible representation $R$ of $A$. On the other hand, the orbifold theory $\calC/A$ decomposes as
\be \Hh_{\calC/A}=\oplus_{a\in A} \Hh_a\ ,
\ee where $\Hh_a$ is the space of $A$ invariant fields in the $a$-twisted sector. The two theories share the common subspace $\Hh^A\equiv \Hh_{R=1}\equiv \Hh_{a=0}$, which is the space of untwisted $A$-invariant fields. Furthermore, for each $a\in A$ and for each $R\in Irr(A)$, the spaces $\Hh_R$ and $\Hh_a$ are modules for $\Hh^A$, considered as an algebra with respect to OPE. 

When a duality defect for the group $A$ exists, there is an isomorphism \be\label{CFTiso} \calC\stackrel{\cong}{\longrightarrow} \calC/A\ee under which each $\Hh_{R}\subseteq \Hh_{\calC}$ is mapped to $\Hh_a\subseteq \Hh_{\calC/A}$. The correspondence between $\Hh^A$-modules induced by \eqref{CFTiso}  establishes a group isomorphism \eqref{PontrDual}
between $A$ and its Pontryagin dual $\hat A=Irr(A)$. The form $\chi$ associated with the isomorphism \eqref{PontrDual} determines (some of) the fusion matrices of the Tambara-Yamagami category TY$(A,\chi,\epsilon)$ where $\CN$ belongs. In particular, when the duality defect $\CN$ in this category is moved across a field $\phi \in \Hh_{R_a}$, it maps it to the corresponding $a$-twisted field $\Hh_a$, attached to a defect line $\CL_a$, determined by the isomorphism \eqref{CFTiso}. On the other hand, when $\CN$ is moved across a field $\phi$ in the $A$-fixed subspace $\Hh^A$, it is mapped to $\frac{1}{\sqrt{|A|}}\hat\CN(\phi)$. Consistency of the OPE on the two sides of $\CN$ implies that \eqref{dualinvo} must be an automorphism of $\Hh^A$ that exchanges each $\Hh^A$-module $\Hh_{R_a}$ with the $\Hh^A$-module $\Hh_a$. Vice versa, any such order $2$ automorphism of $\Hh^A$, exchanging $\Hh_{R_a}$ with $\Hh_a$, defines a duality defect $\CN$ that extends the group $A$ to a Tambara-Yamagami category  TY$(A,\chi,\epsilon)$. This argument provides a very effective method for determining explicitly a duality defect $\CN$. 
TY categories differing only for the value $\epsilon\in \{\pm 1\}$ can be related to each other by changes the sign of the fusion matrices where all the four `external' defects are duality defects. It can also be detected by the fact that the insertion of a circle with the duality defect $\CN$ around the identity operator (vacuum) on the sphere multiplies the correlation function by a factor $\epsilon\langle \CN\rangle=\epsilon\sqrt{|A|}$.

\subsection{Duality defects in $V^{f\natural}$}\label{section:TDLs_Vfnat}

Let us now specialize the previous general discussion on TY categories to the case where the CFT is the holomorphic SVOA $V^{f\natural}$. We would like to understand which TY categories TY$(A,\chi,\epsilon)$ can arise as subcategories in $\cTop$, i.e. such that all defects preserve the supercurrent and are `well-behaved' with respect to fermion number. Clearly, the group $A$ in such a TY$(A,\chi,\epsilon)$ must be a non-anomalous abelian subgroup in $\Aut_\tau(V^{f\natural})$, so there are only finitely many possibilities.

In this section we classify the potential  $\CN=1$--preserving duality defects  corresponding to the TY categories TY$(\ZZ_N, \chi,\epsilon)$ for cyclic subgroups $\langle g\rangle\cong \ZZ_N$ with $g\in Co_0$.  If a duality defect $\CN_A$ exists, then $g$ must be non-anomalous and the orbifold $V^{f\natural}/\langle g\rangle$ must be isomorphic to $V^{f\natural}$. It is well known which conjugacy classes in $Co_0$ give rise to non-anomalous symmetries (see, for example, \cite{Albert:2022gcs}). For such classes, the orbifold $V^{f\natural}/\langle g\rangle$ is a consistent holomorphic SVOA with $c=12$  and a distinguished $\CN=1$ supercurrent, inherited from the one in the parent theory.

There are three different holomorphic SCFTs at $c=12$: $V^{f\natural}$, $V^{fE_8}$, and the theory $F(24)$ with $24$ free fermions. In the latter case, there are $8$ inequivalent choices of the $\CN=1$ supercurrent. It was proved in \cite{Harrison:2020wxl} that an orbifold of $F(24)$  by a cyclic group preserving some choice of $\CN=1$ supercurrent cannot be $V^{f\natural}$. This implies that, vice versa, any consistent orbifold $V^{f\natural}/A$, where $A\cong \ZZ_N$ preserves $\tau(z)$, must be either $V^{fE_8}$ or $V^{f\natural}$ itself. These two possibilities can be  distinguished by computing the torus partition function with fully periodic boundary conditions (Witten index) for the orbifold. This is actually just a number, and it is equal to $\pm 24$ for $V^{f\natural}$ (corresponding to the $24$ ground states in the twisted module $V^{f\natural}_{tw}$) and to $0=8-8$ for $V^{fE_8}$. The sign in $\pm 24$ distinguishes the two inequivalent choices for the definition of the fermion number in the Ramond sector $V^{f\natural}_{tw}$.

The orbifold Witten index can be computed directly from the Frame shape of $g$, which provides a formal way to encode the information about the characteristic polynomial and, consequently, the conjugacy class of $g$ in $Co_0 \subset O(24,\RR)$. In particular, if $g\in O(24,\RR)$ is an element of order $N$, and there exists a suitable basis of $\mathbb{R}^{24}$ such that the matrix representation of $g$  has rational coefficients, then the characteristic polynomial factorizes uniquely as
\begin{equation}
    p(\lambda)= \det \left( \lambda -g \right)= \prod_{k \vert N} (1- \lambda^k)^{l_k}
\end{equation}
where the integers $l_k$ may in general be negative.  This data can be conveniently encoded in the Frame shape \cite{Frame:1970}, defined by the formal symbol:
\begin{equation}\label{eq:Frameshape}
    \pi_g \equiv \prod_{k \vert N} k^{l_k}.
\end{equation}
From the Frame shape, one can easily  determine the set of eigenvalues of $g$ in the $24$-dimensional representation. Since the $24$ eigenvalues always appear in complex-conjugate pairs, we can parameterize them as $\left\lbrace e^{-2\pi i \beta_{g,1}}, e^{2\pi i \beta_{g,1}}, \ldots,e^{-2\pi i \beta_{g,12}}, e^{2\pi i \beta_{g,12}} \right\rbrace$ with $0\le \beta_{g,i}\le \frac{1}{2}$.  
The $160$ distinct Frame shapes of $Co_0$, associated with its $167$ conjugation classes, were first computed in \cite{Kondo:1985}.

For each conjugacy class $\left[ g \right]$ with Frame shape $\pi_g$, we define the product of Dedekind eta functions
\begin{equation}\label{eq:Etaquotient}
    \eta_g (\tau) := \prod_{k \vert N} \eta (k \tau)^{l_k}
\end{equation}
as well as the product of eigenvalue factors
\begin{equation}
    \mathcal{C}_g := \Tr_{R_{spin}}(g)=\pm \prod_{i=1}^{12} \left( e^{\pi i \beta_{g,i}} - e^{- \pi i \beta_{g,i}} \right)\ ,
\end{equation} where $R_{spin}={\bf 1}+{\bf 276}+{\bf 1771}+{\bf 24}+{\bf 2024}$ is the $2^{12}=4096$-dimensional spinorial representation of $Co_0\subset Spin(24)$.\footnote{Note that from the Frame shapes $\pi_g$ and $\pi_{-g}$ one can calculate the traces $\calC_{g}$ and $\mathcal{C}_{-g}$ only up to a sign.} 
We denote by $\calC_{-g}$ and $\pi_{-g}$ the analogous trace associated with $-g\in SO(24)$.   
Given this data, we can now determine, for all $g\in Co_0$, a general formula for $\CZ^g$ given by 
\begin{align}\label{eq:NSgtwining}
 	Z^{g,\pm}_{\NS}(V^{f\natural},\tau)&={1\over 2}\left({\eta_g(\tau/2)\over \eta_g(\tau)} + {\eta_{-g}(\tau/2)\over \eta_{-g}(\tau)} \pm \calC_g\eta_g(\tau) \pm \calC_{-g}\eta_{-g}(\tau)\right )\\\label{eq:Rgtwining}
 	Z^{g,\pm}_{\R}(V^{f\natural},\tau)&={1\over 2}\left(-{\eta_g(\tau/2)\over \eta_g(\tau)} + {\eta_{-g}(\tau/2)\over \eta_{-g}(\tau)} \mp \calC_g\eta_g(\tau) \pm \calC_{-g}\eta_{-g}(\tau)\right ).
 \end{align}
  Note that \eqref{eq:NSgtwining} is symmetric under interchange of the Frame shape $\pi_g$ with its dual Frame shape $\pi_{-g}$, indicating that only $Co_0/\ZZ_2=Co_1$ acts faithfully on $V^{f\natural}$. In particular, the $g$-twined Witten index is a constant (and in particular, modular invariant)
  \be\label{TwinWitt} Z^{g,-}_{\R}(V^{f\natural},\tau)=\Tr_{\bf 24}(g)=l_1\ ,
  \ee where $l_1$ is the multiplicity of $k=1$ in the Frame shape $\pi_g=\prod_{k|N}k^{l_k}$.

  When $g$ is non-anomalous of order $N$ with Frame shape $\pi_g=\prod_{k|N}k^{l_k}$, the Witten index of the orbifold theory $V^{f\natural}/\langle g\rangle$ can be computed using \cite{Persson:2015jka,Albert:2022gcs}
  \begin{align} Z^{-}_{\R}(V^{f\natural}/\langle g\rangle,\tau)&=\frac{1}{N}\sum_{i,j=0}^{N-1}Z^{g^j,-}_{g^i,\R}(V^{f\natural},\tau)=\frac{1}{N}\sum_{i,j=0}^{N-1}\Tr_{\bf 24}(g^{\gcd(i,j,N)})=\sum_{d|N}\frac{J_2(N/d)}{N}\Tr_{\bf 24}(g^d)\notag
  \\ 
  &=\sum_{d|N}\frac{J_2(N/d)}{N}\sum_{k|d}kl_k=\sum_{k|N}kl_k\sum_{r|\frac{N}{k}}\frac{J_2(N/(rk))}{N}=\sum_{k|N}l_k\frac{N}{k}\ ,\label{OrbWittt}
  \end{align} where $J_2(n)$ is the Jordan totient function counting the number of pairs $(i,j)$ of integers with $1\le i,j\le n$ and such that $\gcd(i,j,n)=1$. In this derivation, we used that the $g^i$-twisted $g^j$-twined $Z^{g^j,-}_{g^i,\R}$ is related by modular transformations to the (untwisted) $g^{d}$-twined $Z^{g^d,-}_{\R}$ with $d=\gcd(i,j,N)$, and therefore it equals $\Tr_{\bf 24}(g^d)=\sum_{k|d}kl_k$ by \eqref{TwinWitt}; furthermore,  we used the identity $\sum_{r|n}J_2(r)=n^2$.

\begin{center}
\begin{table}
{\footnotesize
    \centering
    \[
        \begin{array}{|c|cccccccc|}
        \hline
            \text{Co}_0 & \text{1A}   & \underline{\text{2B}} &  \text{6A} & \underline{\text{3B}} & \underline{\text{4B}} & \underline{\text{4E}} & \text{10A} & \underline{\text{5B}} \\
            \text{Co}_1  & \text{1A} & \text{2A} &  \text{3A} & \text{3B}  & \text{4A} & \text{4C} & \text{5A} & \text{5B}    \\
            \pi_g & 1^{24} & 1^82^8 & {1^{12}6^{12}\over 2^{12}3^{12}} & 1^63^6& {1^84^8\over 2^8} & 1^42^24^4& {1^610^6\over 2^65^6} & 1^45^4  \\
            \Tr_{24} g  & 24 &  8  & 12 & 6 &  8 & 4  & 6 & 4      \\\hline
            \mathbb M & \text{1A}   & \text{2A} &  \text{6B} & \text{3A} & \text{4A} & \text{4A} & \text{10D} & \text{5A}    \\
            \Gamma & 1+ & 2+ &  6+6 & 3+  & 4+ & 4+ & 10+10 & 5+   \\
            \hline\hline\text{Co}_0  & \underline{\text{6K}} & \text{14A} & \text{7B} &  \text{8D} & \text{8G} &\text{18A} & \underline{\text{9C} } & \text{11A}   \\
            \text{Co}_1 & \text{6E} & \text{7A} & \text{7B} & \text{8C} & \text{8E}  & \text{9A} & \text{9C} & \text{11A}  \\
            \pi_g  & 1^22^23^26^2 & {1^414^4\over 2^47^4} & 1^37^3 & {1^48^4 \over 2^24^2} & 1^22.4.8^2 & {1^3 18^3 \over 2^39^3} & \frac{1^39^3}{3^2} & 1^2 11^2 \\
            \Tr_{24} g  & 2 & 4 & 3 & 4 & 2 & 3 & 3 & 2   \\\hline\mathbb M &   \text{6A} & \text{14C} & \text{7A} & \text{8A} & \text{8A} & \text{18E} & 9A & 11A   \\
            \Gamma &   6+ & 14+14  & 7+ & 8+ & 8+ & 18+18 & 9+ & 11+  \\\hline
            \hline\text{Co}_0  &  \text{12D} &  \text{12I}   & \text{12Q} & \text{26A} &  \text{14C} & \text{15A} & \text{30B} & \text{15D} \\
            \text{Co}_1   & \text{12A} & \text{12E} &  \text{12K} & \text{13A} &  \text{14B} & \text{15A}  & \text{15B}  & \text{15D}  \\
            \pi_g  & {1^412^4\over 3^44^4} & {1^23^24^212^2\over2^26^2} & {1^312^3\over 2.3.4.6} & {1^226^2\over 2^23^2} & 1.2.7.14 & {1^315^3\over 3^35^3}  & {1^25^26^230^2\over 2^23^210^215^2} & 1.3.5.15 \\
            \Tr_{24}g  & 4 & 2 &  3 & 2 & 1 & 3  & 2 & 1  \\\hline
            \mathbb M & \text{12H} & \text{12A} & \text{12H}   & \text{26B} &  \text{14A} & \text{15C} & \text{30D} & \text{15A} \\
            \Gamma  & 12+12 & 12+ & 12+12 & 26+26 &  14+ & 15+15  & 30+5,6,30 & 15+  \\\hline
            \hline\text{Co}_0  & \text{15E} &  \text{16B} &  \text{20D}  & \text{20G} & \text{21A}  & \text{42B}& \text{23AB}  & \text{24E}    \\
            \text{Co}_1 &  \text{15E} & \text{16B} & \text{20A}  & \text{20C} & \text{21A}  & \text{21B}& \text{23AB}  & \text{24B}   \\
            \pi_g  & {1^215^2\over3.5}  & {1^216^2\over2.8} & {1^220^2\over4^25^2} &{1.2.10.20\over 4.5} & {1^221^2\over 3^27^2}  & {1.3.14.42\over2.6.7.21} & 1.23  & {1^24.6.24^2\over 2.3^28^212} \\
            \Tr_{24}g  & 2  & 2 &  2 & 1 & 2  & 1 & 1  & 2 \\\hline
            \mathbb M & \text{15C} & \text{16C} &   \text{20F} & \text{20F} & \text{21D} & \text{42D} & \text{23A} & \text{24I}  \\
            \Gamma & 15+15 & 16+ & 20+20 & 20+20  & 21+21 & 42+3,14,42 & 23+ & 24+24  \\\hline
            \hline\text{Co}_0  & \text{24J} & \text{28B} & \text{30F} & \text{66A}& \text{35A} & \text{36B} & \text{39AB} & \text{60C} \\
            \text{Co}_1  & \text{24F}   & \text{28A} & \text{30A}  & \text{33A} & \text{35A} & \text{36A} & \text{39AB}  & \text{60A}\\
            \pi_g & {1.4.6.24\over 3.8} & {1.4.7.28\over2.14} & {1.2.15.30\over3.5.6.10}  & {1.6.11.66\over2.3.22.33} & {1.35\over 5.7} & {1.36\over 4.9} & {1.39\over 3.13} & {1.4.6.10.15.60\over2.3.5.12.20.30} \\
            \Tr_{24}g  & 1  & 1 & 1  & 1 & 1 & 1 & 1 & 1  \\\hline
            \mathbb M & \text{24I}   & \text{28B} &  \text{30F} & \text{66B} & \text{35B} & \text{36D}  & \text{39CD}   & \text{60C}  \\
            \Gamma & 24+24 & 28+ &  30+2,15,30 & 66+6,11,66 & 35+35 & 36+36 & 39+39   & 60+4,15,60  \\\hline
        \end{array}
    \]
    \caption{\small{The 42 conjugacy classes under which $V^{f\natural}$ is self-orbifold, corresponding to 40 different Frame shapes (the pairs 39A,39B and 23A,23B have the same Frame shape). All Frame shapes are balanced and all conjugacy classes have positive $ \Tr_{24}g$. All these classes correspond to genus zero groups containing the Fricke involution (see \S \ref{s:Monster}).  Finally, there are $7$ non-trivial classes 2B, 3B, 4B, 4E, 5B, 6K, 9C (decorated by an underscore) for which the corresponding duality defect might potentially preserve a $4$-dimensional subspace of $V^{f\natural}_{tw}(1/2)$, and therefore might be related to duality defects in K3 sigma models. }}
    \label{tbl:selforb}}
\end{table}
\end{center}

\begin{center}
\begin{table}
{\footnotesize
    \centering
    \[
        \begin{array}{|c|cc|cc|cc|cc|cc|}
        \hline
            \text{Co}_0 & \text{6H} & \text{6I}  & \text{10G} & \text{10H} & \text{12L} & \text{12O} & \text{18F}  &  \text{18I}  & \text{30H} & \text{30J}\\
            \text{Co}_1 & \text{6C} & \text{6D}  & \text{10D} & \text{10E} & \text{12H} & \text{12I} & \text{18B} & \text{18C} &\text{30D} & \text{30E}\\
            \pi_g & {1^42.6^5\over 3^4} & {1^53.6^4\over 2^4} & \frac{1^2 2 . 10^3}{5^2} & \frac{1^3 5. 10^2}{2^2} & \frac{1.2^2 3. 12^2}{4^2} & \frac{1^2 4. 6^2 12}{3^2} & \frac{1^2 9.18}{2.3} & \frac{1.2.18^2}{6.9}& {1.6.10.15\over3.5} & {2.3.5.30\over 6.10}\\
            \Tr_{24} g & 4 & 5 & 2 & 3 & 1 & 2 & 2 & 1 & 1 & 0 
             \\\hline
        \end{array}
    \]
    \caption{\small{The 10 conjugacy classes with unbalanced Frame shapes under which $V^{f\natural}$ is self-orbifold but a duality defect $\mathcal{N}$ does not exist. They form five pairs of conjugacy classes of the same order, whose Frame shapes $\pi_g, \pi_{g'}$ are related by \eqref{balance}. All these classes have positive $ \Tr_{24}g$. None of these classes satisfies relation \eqref{MCo0relation}, therefore, we cannot associate to any of them a class $\left[ g' \right]$ on $\mathbb{M}$ and a genus zero group $\Gamma$. }}
    \label{tbl:selforb_no_balanced}}
\end{table}
\end{center}

\begin{table}
\small
    \centering
    \[
        \begin{array}{|c|cccccccc|}
            \hline\text{Co}_0 & \text{2C}   & \text{3C} &  \text{6C} & \text{4D} & \text{4F} & \text{5C} & \text{10C} & \text{6F}   \\
            \text{Co}_1 & \text{2A}   & \text{3C} &  \text{3C} & \text{4B} & \text{4C} & \text{5C} & \text{5C} & \text{6A}    \\
            \pi_g & {2^{16}\over 1^8} & {3^9\over 1^3} & {1^36^9\over2^33^9} & {4^8\over 2^4} & {2^64^4\over1^4} & {5^5\over1}& {1.10^5\over2.5^5}& {1^46^8\over2^83^4} \\
            \Tr_{24} g  & -8 &  -3  & 3 & 0 &  -4 & -1  & 1 & 4    \\
            \hline
            \mathbb M & \text{2B} & \text{3B} &  \text{6E} & \text{4C}  & \text{4C} & \text{5B} & \text{10E} & \text{6E}    \\
            \Gamma_{g'} & 2- & 3- &  6- & 4-  & 4- & 5- & 10- & 6-    \\\hline
            \hline
            \text{Co}_0 & \text{6G} & \text{6L} & \text{6M}   & \text{6N} & \text{8F} &  \text{8H} & \text{9B} &\text{18B}   \\
            \text{Co}_1 & \text{6C} & \text{6E} & \text{6F}  & \text{6F} & \text{8D} &  \text{8E} & \text{9B} &\text{9B}    \\
            \pi_g & {2^53^46\over1^4} & {2^46^4\over1^23^2} & {3^36^3\over1.2} & {1.6^6\over2^23^3} & {8^4\over4^2} & {2^34.8^2\over1^2} & {9^3\over3} & {3.18^3\over6.9^3} \\
            \Tr_{24} g & -4 & -2 & -1  & 1 & 0 & -2 & 0 & 0  \\\hline
            \mathbb M & \times & \text{6C} &  \text{6D} & \text{6E}  & \text{8D} & \text{8E} & \text{9B} & \text{18D}    \\
            \Gamma_{g'} & \times & 6+3 &  6+2 & 6-  & 8|2- & 8- & 9- & 18-    \\\hline
            \hline\text{Co}_0 & \text{10E} & \text{10F} & \text{12E} & \text{12G} & \text{12H} & \text{12K}  & \text{12M} &  \text{12N}     \\
            \text{Co}_1  & \text{10A} & \text{10D} & \text{12B} & \text{12D} & \text{12D} & \text{12G}   & \text{12H} &  \text{12I}    \\
            \pi_g & {1^210^4\over2^45^2} & {2^35^210\over1^2} & {2^212^4\over 4^46^2}  & {2.3^312^3\over1.4.6^3} & {1.12^3\over3^34} &  {4^212^2 \over 2.6} & {2^36.12^2\over1.3.4^2} & {2^23^24.12\over1^2}  \\
            \Tr_{24}g & 2 & -2 & 0 & -1 & 1 & 0 & -1 & -2  \\\hline
            \mathbb M & \text{10E} & \times &  \text{12I} & \text{12B}  & \text{12I} & \text{12E} & \text{12I} & \times    \\
            \Gamma_{g'} & 10- & \times &  12- & 12+4  & 12- & 12+3 & 12- & \times    \\\hline
            \hline\text{Co}_0 & \text{14D}  & \text{18E} & \text{18H} & \text{20F}  & \text{24I} & \text{30G} & \text{30I} & \text{30K} \\
            \text{Co}_1 & \text{14B}  & \text{18A} & \text{18C} & \text{20C}  & \text{24F} & \text{30A} & \text{30D}& \text{30E} \\
             \pi_g & {2^214^2\over1.7}  & {1.18^2\over 2^29} & {2^29.18\over1.6} & {2^25.20\over1.4} & {2.3.4.24\over 1.8} & {2^23.5.30^2\over1.6^210^215} & {2.3.5.30\over1.15} & {2.30\over3.5}\\
             \Tr_{24}g & -1 & 1 & -1 & -1 & -1 & -1 & -1 & 0  
            \\\hline
            \mathbb M & \text{14B} & \text{18D} &  \times & \text{20C}  & \text{24C} & \text{30G} & \times & \text{30G}    \\
            \Gamma_{g'} & 14+7 & 18- &  \times & 20+4  & 24+8 & 30+15 & \times & 30+15    \\\hline
        \end{array}
    \]
    \caption{\small{The 32 conjugacy classes under which the orbifold of $V^{f\natural}$ yields $V^{fE_8}$. Note that none of these Frame shapes is balanced. Moreover, all but 5 classes satisfy the relation to monster conjugacy classes described in \S \ref{s:Monster}, and correspond to non-Fricke genus zero groups.}}
    \label{tbl:VfE8}
\end{table}

\begin{table}
\small
    \centering
    \[
        \begin{array}{|c|cccccccc|}
            \hline\text{Co}_0 & \text{2A}   & \text{3A} &  \text{6B} & \text{4C} & \text{5A} & \text{10B} & \text{6E} & \text{6J}      \\
            \text{Co}_1 & \text{1A} & \text{3A} &  \text{3B} & \text{4A}  & \text{5A} & \text{5B} & \text{6A} & \text{6D}    \\
            \pi_g & {2^{24}\over 1^{24}} & {3^{12}\over 1^{12}} & {2^66^6\over 1^63^6} & {4^8\over 1^8} & {5^6\over 1^6} & {2^410^4\over 1^45^4} & {3^46^4\over1^42^4} & {2.6^5\over1^53}  \\
            \Tr_{24} g  &  -24 & -12 & -6 & -8 & -6 & -4 & -4 & -5    \\\hline
            \mathbb M & \text{2B} & \text{3B} &  \text{6C} & \text{4C}  & \text{5B} & \text{10B} & \text{6D} & \text{6E}    \\
            \Gamma_{g'} & 2- & 3- &  6+3 & 4-  & 5- & 10+5 & 6+2 & 6-    \\\hline
            \hline\text{Co}_0 & \text{7A} & \text{14B} & \text{8E} & \text{9A} & \text{18C} &  \text{10D} & \text{10I} & \text{22A}    \\
            \text{Co}_1 & \text{7A} & \text{7B}  & \text{8C} & \text{9A} & \text{9C} & \text{10A} & \text{10E}  & \text{11A}      \\
            \pi_g & {7^4\over1^4} & {2^314^3\over1^37^3} & {2^28^4\over 1^44^2} & {9^3\over1^3} & {2^33^218^3\over1^36^29^3} & {5^210^2\over1^22^2} & {2.10^3\over1^35} & {2^222^2\over1^211^2}  \\
            \Tr_{24} g & -4 & -3  & -4 & -3 & -3 & -2 & -3 & -2   \\\hline
            \mathbb M & \text{7B} & \text{14B} &  \text{8E} & \text{9B}  & \text{18C} & \text{10C} & \text{10E} & \text{22B}    \\
            \Gamma_{g'} & 7- & 14+7 &  8- & 9-  & 18+9 & 10+2 & 10- & 22+11    \\\hline
            \hline\text{Co}_0 &\text{12C}  & \text{12J} & \text{12R} & \text{13A}  & \text{30A} & \text{15B} & \text{30D} & \text{30E}  \\
            \text{Co}_1 & \text{12A}  & \text{12E} &  \text{12K} & \text{13A}  & \text{15A} & \text{15B} &  \text{15D} &  \text{15E} \\
            \pi_g & {2^43^412^4\over1^44^46^4} & {4^212^2\over1^23^2} & {2^23.12^3\over 1^34.6^2} & {13^2\over 1^2}  & {2^33^35^330^3\over1^36^310^315^3} & {3^215^2\over1^25^2} & {2.6.10.30\over1.3.5.15} & {2^23.5.30^2\over 1^26.10.15^2} \\
            \Tr_{24}g  & -4 & -2 & -3 & -2  & -3 & -2 & -1 & -2   \\\hline
            \mathbb M & \text{12B} & \text{12E} &  \text{12I} & \text{13B}  & \text{30A} & \text{15B} & \text{30C} & \text{30G}    \\
            \Gamma_{g'} & 12+4 & 12+3 &  12- & 13-  & 30+6,10,15 & 15+5 & 30+3,5,15 & 30+15    \\\hline
             \hline\text{Co}_0 & \text{16C} &  \text{18D}  & \text{18G} & \text{20C}  & \text{42A} & \text{21B} & \text{46AB}  & \text{24D} \\
             \text{Co}_1 & \text{16B} & \text{18A}   & \text{18B} & \text{20A} & \text{21A} & \text{21B}  & \text{23AB}  & \text{24B} \\
             \pi_g & {2.16^2\over 1^28} & {9.18\over1.2} & {2.3.18^2\over 1^26.9} & {2^25^220^2\over1^24^210^2} & {2^23^27^242^2\over1^26^214^221^2} & {7.21\over 1.3} & {2.46\over 1.23} & {2.3^24.24^2\over1^26.8^212}  \\
             \Tr_{24}g & -2 &  -1 & -2 & -2 & -2 & -1 & -1 & -2 \\\hline
             \mathbb M & \text{16B} & \text{18A} &  \text{18D} & \text{20C}  & \text{42B} & \text{21B} & \text{46AB} & \text{24C}    \\
            \Gamma_{g'} & 16- & 18+2 &  18- & 20+4  & 42+6,14,21 & 21+3 & 46+23 & 24+8    \\\hline
             \hline\text{Co}_0 & \text{28C} & \text{33A} & \text{70A} & \text{36A} & \text{78AB}  &  \text{60D} && \\
             \text{Co}_1 &  \text{28A} & \text{33A} & \text{35A} & \text{36A} & \text{39AB}   &  \text{60A} && \\
             \pi_g & {4.28\over1.7} & {3.33\over 1.11} & {2.5.7.70 \over 1.10.14.35} & {2.9.36\over1.4.18} & {2.3.13.78\over1.6.26.39}  & {3.4.5.60\over1.12.15.20} &&\\
             \Tr_{24}g &  -1 & -1 & -1 & -1 & -1 & -1 &&
            \\\hline
            \mathbb M & \text{28C} & \text{33A} &  \text{70B} & \text{36B}  & \text{78BC} & \text{60D} &  &     \\
            \Gamma_{g'} & 28+7 & 33+11 &  70+10,14,35 & 36+4  & 78+6,26,39 & 60+12,15,20 &  &     \\\hline
        \end{array}
    \]
    \caption{\small{The 40 conjugacy classes associated with 38 different Frame shapes under which the orbifold of $V^{f\natural}$ is again $V^{f\natural}$, but with the opposite fermion number in the Ramond sector. Note these are precisely the conjugacy classes which have balanced Frame shape and negative $\Tr_{24}g$. The 38 corresponding genus zero groups are precisely those of type $\Gamma_0(N)$, extended by Atkin-Lehner involutions but without Fricke involutions (see \S \ref{s:Monster}).} }
    \label{tbl:VfnatnoSUSY}
\end{table}

\begin{table}
\small
    \centering
    \[
        \begin{array}{|c|cccccccc|}
            \hline\text{Co}_0 &  \text{4A} &  \text{2D} &   \text{3D} &  \text{6D}  &  \text{4G} &  \text{8A} &  \text{4H} &  \text{12A}\\
            \text{Co}_1 &  \text{2B} &  \text{2C} &  \text{3D} &  \text{3D} &  \text{4D} &  \text{4E} &  \text{4F}  &  \text{6B}\\
            \pi_g & {4^{12}\over 2^{12}} & 2^{12} & 3^8 & {6^8\over 3^8} & 2^44^4 & {8^6\over 4^6} & 4^6 & {2^612^6\over 4^66^6} \\\hline
            \mathbb M &  \text{4D} &  \text{4A} &   \text{3C} &  \text{6F}  &  \text{4B} &  \text{8F} &  \text{8B} &  \text{12F}\\
            \Gamma_{g'} &  4|2- &  4+ &   3|3 &  6|3-  &  4|2+ &  8|4- &  8|2+ &  12|2+6\\\hline
            \hline\text{Co}_0  &  \text{6O} &  \text{12B} &  \text{6P} &  \text{8B} &  \text{8C}&  \text{8I} &  \text{20A} &  \text{20B} \\
            \text{Co}_1  &  \text{6G} &  \text{6H} &  \text{6I} &  \text{8A} &  \text{8B} &  \text{8F} &  \text{10B} &  \text{10C}  \\
            \pi_g & 2^36^3 & {12^4\over 6^4} & 6^4 & {8^4\over 2^4} & {2^48^4\over 4^4} & 4^28^2 & {2^320^3\over 4^310^3} & {4^220^2\over 2^210^2}  \\\hline
            \mathbb M &  \text{12A} &  \text{12J} &   \text{12D} &  \text{8D}  &  \text{8B} &  \text{8C} &  \text{20E} &  \text{20D}\\
            \Gamma_{g'} &  12+ &  12|6- &   12|3+ &  8|2-  &  8|2+ &  8|4+ &  20|2+10 &  20|2+5\\\hline
           \hline\text{Co}_0  &  \text{10J} &  \text{12F} &  \text{24A}&  \text{12P} &  \text{24B} &  \text{12S} &  \text{28A} &  \text{15C} \\
            \text{Co}_1 &  \text{10F} &  \text{12C} &  \text{12F} &  \text{12J}  &  \text{12L} &  \text{12M} &  \text{14A} &  \text{15C}  \\
            \pi_g & 2^210^2 & {6^212^2\over 2^24^2} & {4^324^3\over 8^312^3} & 2.4.6.12 & {24^2\over 12^2} & 12^2 & {2^228^2\over 4^214^2} & {15^2\over 3^2} \\\hline
            \mathbb M &  \text{20A} &  \text{12G} &   \text{24F} &  \text{12C}  &  \text{24J} &  \text{24B} &  \text{28D} &  \text{15D}\\
            \Gamma_{g'} &  20+ &  12|2+2 &   24|4+6 &  12|2+  &  24|12- &  24|6+ &  28|2+14 &  15|3-\\\hline
            \hline\text{Co}_0 &  \text{30C} &  \text{16A} &  \text{20E}  &  \text{21C} &  \text{42C} &  \text{22BC} &   \text{24C} &  \text{24F}\\
            \text{Co}_1 &  \text{15C} &  \text{16A} &  \text{20B} &  \text{21C} &  \text{21C} &  \text{22A}  &  \text{24A} &  \text{24C} \\
            \pi_g & {3^230^2\over 6^215^2} & {2^216^2\over 4.8} & 4.20 & 3.21 & {6.42\over 3.21} & 2.22 & {2^224^2\over 6^28^2} & {8.24\over 2.6}  \\\hline
            \mathbb M &  \text{30E} &  \text{16A} &   \text{40B} &  \text{21C}  &  \text{42C} &  \text{44AB} &  \text{24H} &  \text{24D}\\
            \Gamma_{g'} &  30|3+10 &  16|2+ &   40|2+ &  21|3+  &  42|3+7 &  44+ &  24|2+12 &  24|2+3\\\hline
             \hline\text{Co}_0 &  \text{24G} &  \text{24H} &  \text{52A}  &  \text{56AB}  &  \text{60A} &  \text{60B} &  \text{40AB}  &  \text{84A} \\
              \text{Co}_1 &  \text{24D} &  \text{24E}  &  \text{26A} &  \text{28B}  &  \text{30B} &  \text{30C} &  \text{40A} &  \text{42A} \\
              \pi_g & {12.24\over 4.8}  & {2.6.8.24\over 4.12} &{2.52\over 4.26} & {4.56\over 8.28} & {2.10.12.60\over 4.6.20.30} & {6.60\over 12.30} & {2.40\over 8.10} & {4.6.14.84\over2.12.28.42}\\\hline
              \mathbb M &  \text{24G} &  \text{24A} &   \text{52B} &  \text{56B}  &  \text{60E} &  \text{60F} &  \text{40CD} &  \text{84B}\\
            \Gamma_{g'} &  24|4+2 &  24|2+ &   52|2+26 &  56|4+14  &  60|2+5,6,30 &  60|6+10 &  40|2+20 &  84|2+6,14,21
         \\\hline
        \end{array}
    \]
    \caption{{\small{The 43 anomalous conjugacy classes of $Co_0$ associated with 40 different Frame shapes. For all such conjugacy classes, $\Tr_{24}g=0$. Note these all have Frame shapes which are balanced. Except for the five $Co_0$ conjugacy classes 2D, 6O, 10J and 22BC, these all correspond to genus zero groups $\Gamma_{g'}$ of $n|h$--type, with or without Atkin-Lehner involutions. (See \S \ref{s:Monster}.)} }}
    \label{tbl:anomalous}
\end{table}
The results of the orbifold Witten indices for non-anomalous Frame shapes of $Co_0$ are reported in Tables \ref{tbl:selforb}, \ref{tbl:selforb_no_balanced}, \ref{tbl:VfE8},  and \ref{tbl:VfnatnoSUSY}. Table \ref{tbl:selforb} and \ref{tbl:selforb_no_balanced} contain all $52$ conjugacy classes for which the orbifold Witten index is $24$, so that $V^{f\natural}$ is self-orbifold $V^{f\natural}/\langle g\rangle\cong V^{f\natural}$. In particular, table \ref{tbl:selforb} contains the $42$ such classes for which the Frame shape $\prod_{k|N} k^{l_k}$, $l_k\in \ZZ$, where $N$ is the order of $g$, is \emph{balanced}, i.e. it is invariant under the transformation
\be\label{balance} \prod_{k|N} k^{l_k} \quad \to\quad  \prod_{k|N} (N/k)^{l_k}\ .
\ee Table \ref{tbl:selforb_no_balanced} contains the remaining $10$ classes $[g]$ such that $V^{f\natural}/\langle g\rangle\cong V^{f\natural}$ but the Frame shape $\pi_g$ is not balanced. Table \ref{tbl:VfE8} contains all classes for which $V^{f\natural}/\langle g\rangle$ is isomorphic to $V^{fE_8}$. Table \ref{tbl:VfnatnoSUSY} contains non-anomalous classes for which the orbifold Witten index is $-24$. This means that the orbifold theory is isomorphic to $V^{f\natural}$, but with the opposite choice of fermion number for the Ramond sector; in particular, the $24$ Ramond ground states in $V^{f\natural}/\langle g\rangle $ have negative fermion number.  For such groups $\langle g\rangle$, even if a duality defect $\CN$ exists, it does not commute with the fermion number in the Ramond sector, so it is not contained in the category $\cTop$.
Finally, Table \ref{tbl:anomalous} contains all anomalous Frame shapes. In all such cases, by applying the formula for the orbifold Witten index, one obtains a number that does not match any of the holomorphic SVOA at $c=12$; this confirms that the orbifold $V^{f\natural}/\langle g\rangle$ is inconsistent, as expected.

We will now argue that the only cyclic subgroups of $Co_0$ that can be potentially  associated with a duality defect in $\cTop$ are the ones in Table \ref{tbl:selforb}. This means that not only the orbifold theory $V^{f\natural}/\langle g\rangle$ must be consistent and isomorphic to $V^{f\natural}$ with the same fermion number in the Ramond sector, but the Frame shape $\pi_g$ must also be balanced.

Indeed, the existence of a duality defect implies not only the isomorphism $V^{f\natural}/\langle g\rangle\cong V^{f\natural}$, but also that each $g$-invariant $g^k$-twisted sector $(V_{g^k})^g$ in $V^{f\natural}/\langle g\rangle$ is isomorphic to the $g$-eigenspace with $g=e^{\frac{2\pi i a}{N}}$ in the parent theory $V^{f\natural}$, for some $a\in \ZZ/N\ZZ$. The space $(V_{g^k})^g$ in $V^{f\natural}/\langle g\rangle$ is the eigenspace of the quantum symmetry $Q$ with eigenvalue $Q= e^{\frac{2\pi i k}{N}}$. Thus, the existence of $\CN$ implies that $g$ and $Q$ must have the same eigenvalues at each conformal weight, and in particular the same Frame shape. As proved in \cite{Persson:2015jka}, if $g$ has Frame shape $\prod_{k|N} k^{l_k}$, then the the Frame shape of the quantum symmetry $Q$ is given by the formula $\prod_{k|N} (N/k)^{l_k}$ in \eqref{balance}. Thus, in the $10$ unbalanced cases of Table \ref{tbl:selforb_no_balanced}, a duality defect $\CN$ cannot exist, because $g$ and $Q$ have different Frame shapes. These $10$ elements can be organized into five pairs $\pi_g,\pi_{Q}$ of the same order, such that if $g$ has Frame $\pi_g$, then the quantum symmetry $Q$ has Frame shape $\pi_{Q}$ given by the transformation \eqref{balance}.

Therefore, we are left to study the $42$ conjugacy classes of $Co_0$ with balanced Frame shape in Table \ref{tbl:selforb}. Suppose that, for some of these cyclic groups $\langle g\rangle$, the duality defect $\CN$ exists as a topological defect in $V^{f\natural}$. Then, as discussed in the previous section $\frac{1}{\sqrt{N}}\hat \CN$ must act on the $g$-fixed subSVOA $(V^{f\natural})^g$ by an order $2$ automorphism that exchanges suitable $(V^{f\natural})^g$-modules. When the subSVOA $(V^{f\natural})^g$ contains a rational affine algebra, all such automorphisms can be determined using a theorem by Kac \cite{Kac}; this technique was recently applied to determine the  duality defects in the bosonic $E_8$ lattice VOA $V^{E_8}$  in \cite{Burbano:2021loy}.
 In order for $\CN$ to be an element of $\cTop$, one needs to show that such an automorphism $\frac{1}{\sqrt{N}}\hat \CN$ can be chosen so that it acts trivially on the supercurrent $\tau$; as we will show in the next sections, this is usually the most difficult task in determining the duality defects in $\cTop$.

Finally, to make contact with non-invertible symmetries in K3 NLSMs, we are interested in determining which duality defects in $V^{f\natural}$ preserve a choice of four-plane $\Pi^\natural$ in $V^{f\natural}_{tw}(1/2)$ as described in the introduction. Finding corresponding SUSY--preserving duality defects in K3 NLSMs will provide evidence for our main conjecture \ref{conj:K3relation}. As reported in Table \ref{tbl:selforb}, there are 7 non-trivial $Co_0$ conjugacy classes: 2B, 3B, 4B, 4E, 5B, 6K, 9C, corresponding to the Frame shapes $1^82^8$, $1^63^6$, $1^84^8/2^8$, $1^42^24^4$, $1^45^4$, $1^22^23^26^2$, and $1^39^3/3^2$, resp., which may potentially preserve a choice of $\CN=4$ superconformal algebra.

Let $\CN_g$ be the duality defect corresponding to the TY category TY$(\langle g\rangle)$ if it exists. For those classes where the order $o(g)$ is not a perfect square, we can determine from the Frame shape the eigenvalues of ${1\over \sqrt{o(g)}}\hat \CN_g$ acting on $V^{f\natural}_{tw}(1/2)$. Indeed, we know that the possible eigenvalues of $\hat\CN_g$ on the $g$-fixed subspace are $\pm\sqrt{o(g)}$, while on the orthogonal complement the eigenvalue is necessarily $0$. The trace of $\hat\CN_g$ on $V^{f\natural}_{tw}(1/2)$ must be an integer, and if $o(g)$ is not a perfect square the only possibility is that the trace is $0$, i.e. $+\sqrt{o(g)}$ and $-\sqrt{o(g)}$ must have the same multiplicity. Thus, the dimension of the subspace of $V^{f\natural}_{tw}(1/2)$ that is preserved by $\CN_g$, i.e. the multiplicity of the eigenvalue $\langle \CN_g\rangle=+\sqrt{o(g)}$,  is half of the dimension of the $g$-fixed subspace. As a consequence, $\hat\CN_g$ can fix a $4$-dimensional subspace $\Pi^\natural$ only if $g$ fixes a subspace of dimension at least $8$.

We identify four-plane-preserving duality defects corresponding to TY categories for cyclic groups generated by $g=2B$ ($\pi_g=1^82^8$) and $g=3B$ ($\pi_g=1^63^6$) and prove they preserve $\CN=4$ supersymmetry in sections \ref{s:Z2duality} and \ref{sec:Z3duality}, respectively. We leave it as an open question whether one can construct $\CN=4$--preserving duality defects for $\ZZ_N$ where $N=4,5,6,$ and 9.

\subsection{Spin selection rules}\label{s:spinselect}

In this article, we will determine the $\CN$-twining and $\CN$-twisted partition functions $\CZ^\CN$, $\CZ_\CN$ for various duality defects in TY categories TY$(A,\chi,\epsilon)$ in $V^{f\natural}$. We will now review how the precise TY category, i.e. the values of $\chi$  and $\epsilon$, determine the possible conformal weights (or, more generally, the spin in a non-holomorphic CFT) of the $\CN$-twisted sector; see for example \cite{Chang_2019,Thorngren:2021yso} for more details. In the following sections, we will use these spin selection rules to determine the precise Tambara-Yamagami category of a given defect $\CN$, once the $\CN$-twisted partition function is known.

Consider an amplitude on the cylinder $S^1\times \RR$ with the line $\CN$ extending along the Euclidean time direction $\RR$. As incoming and outgoing states we take the same CPT self-conjugate normalized state $\psi$ in the $\CN$-twisted sector, so that we simply get $\langle \psi|\psi\rangle=1$. Now, we insert an operator $e^{4\pi i L_0}$ in the amplitude (in a non-holomorphic CFT, we would insert $e^{4\pi i(L_0-\bar L_0)}$\footnote{To be precise, the generator of translations along $S^1$ on the cylinder is $L_0-\frac{c}{24}-(\bar L_0-\frac{\bar c}{24})$. However, if $c\neq \bar c$, then on the RHS of eq.\eqref{spinselect1} one should include a factor $e^{-4\pi i (c-\bar c)/24}$ due to the gravitational anomaly. This is equivalent to simply inserting $e^{4\pi i(L_0-\bar L_0)}$ on the LHS of \eqref{spinselect1}.}). This is equivalent to the defect line $\CN$ wrapping twice the $S^1$ circle, while going from $t=-\infty$ to $t=+\infty$. Using the fusion rule $\CN^2=\sum_{a\in A} \CL_a$ and the suitable fusion matrices, one gets the identity \cite{Chang_2019,Thorngren:2021yso}
\be\label{spinselect1} \langle \psi|e^{4\pi iL_0}|\psi\rangle=\frac{\epsilon}{\sqrt{|A|}}\sum_{a\in A} \langle \psi|\hat\CL_a^{(\CN)}|\psi\rangle\ .
\ee
 
The operators $\hat\CL_a^{(\CN)}$, $a\in A$, are (unitary) linear operators mapping the $\CN$-twisted space into itself, and define a \emph{projective} representation of the abelian group $A$, obeying the relation
\be \hat\CL_a^{(\CN)}\hat\CL_b^{(\CN)}=\chi(a,b)\hat\CL^{(\CN)}_{a+b}\ ,\qquad a,b\in A\ .
\ee Alternatively, one can think of the operators $\hat\CL^{(\CN)}_a$ on the $\CN$-twisted sector as defining a central extension of the abelian group $A$. Notice that the central extension is still abelian, thanks to the symmetry of the bicharacter $\chi(a,b)=\chi(b,a)$. In fact, there is an ambiguity by a phase in the very definition of each $\hat\CL^{(\CN)}_a$, which is related to the choice of $3$-way topological junction operators between one $\CL_a$ and two $\CN$ defects, and their ordering; the formulae we write correspond to a standard conventional choice, where $\hat\CL^{(\CN)}_a$ corresponds to $\widehat{( r_a)}_-$ in the notation of \cite{Thorngren:2021yso}. 

In order to see how the spin selection rules arise, let us consider the example of $A\cong \ZZ_3$, that will be useful in \S \ref{sec:Z3duality}. 
 For $A\cong \ZZ_3$, there are only two choices of non-degenerate symmetric bicharacter, namely $\chi(a,b)=\chi_+(a,b)$ or $\chi(a,b)=\chi_-(a,b)$, with
\be\label{bicharZ3} \chi_\pm(1,1)=\chi_\pm(2,2)=\omega^{\pm 1}\ ,\quad 
\chi_\pm(2,1)=\chi_\pm(1,2)=\omega^{\mp 1}\ ,\quad \chi_\pm(0,b)=\chi_\pm(a,0)=1\ ,
\ee and $\omega =e^{\frac{2\pi i}{3}}$. Let us pass to the multiplicative notation for the group $\langle g\rangle\cong \ZZ_3$, with the identification $\CL_a\equiv \CL_{g^a}$. For this group, \eqref{spinselect1} becomes
\be\label{spinselect2} \langle \psi|e^{4\pi iL_0}|\psi\rangle=\frac{\epsilon}{\sqrt{3}}\langle \psi|1+\hat\CL_g^{(\CN)}+\hat\CL_{g^2}^{(\CN)}|\psi\rangle=\frac{\epsilon}{\sqrt{3}}\langle \psi|1+\hat\CL_g^{(\CN)}+\chi(1,1)^{-1}(\hat\CL_{g}^{(\CN)})^2|\psi\rangle\ .
\ee Because $L_0$ and $\hat\CL^{(\CN)}_g$ commute, we can simultaneously diagonalize them on the $\CN$-twisted sector. One can see that the operator $\hat\CL^{(\CN)}_{g}$, acting on $\CN$-twisted sector, has still order $3$, i.e.
\be (\hat\CL^{(\CN)}_g)^3=\chi(1,1)\hat\CL^{(\CN)}_{g^2}\hat\CL^{(\CN)}_{g}=\chi(1,1)\chi(2,1)\hat\CL^{(\CN)}_{g^3}=\mathbf{1}\ ,
\ee where we used $\hat\CL^{(\CN)}_{g^3}=\hat\CL^{(\CN)}_{1}=\mathbf{1}$ and $\chi(1,1)\chi(2,1)=\chi(0,1)=1$ by bilinearity. Thus, the possible eigenvalues of $\hat\CL^{(\CN)}_g$ are third roots of unity.  If we take $|\psi\rangle$ to be an eigenstate with $L_0|\psi\rangle=h|\psi\rangle$ and $\hat\CL^{(\CN)}_g|\psi\rangle=\zeta|\psi\rangle$, with $\zeta^3=1$, then \eqref{spinselect2} becomes
\be\label{spinselection} e^{4\pi i h}=\frac{\epsilon}{\sqrt{3}}(1+\zeta+\chi(1,1)^{-1}\zeta^2)\ .
\ee  We conclude that, for a given category TY$(\ZZ_3,\chi,\epsilon)$, an $L_0$-eigenvalue $h$ in the $\CN$-twisted sector is possible only if there exists a third root of unity $\zeta$ such that \eqref{spinselection} holds.
In Table \ref{t:spinselectZ3} we report the possible values of $\epsilon$, $\chi(1,1)$ and eigenvalues $\zeta$, and the corresponding allowed values for $h$ satisfying \eqref{spinselection}.
\begin{table}[h]
    \centering
    
    \begin{tabular}{|c|c|c|c|c|}\hline
    $\epsilon$     & $\chi(1,1)$ & $\zeta$ & $\frac{\epsilon}{\sqrt{3}}(1+\zeta+\chi(1,1)^{-1}\zeta^2)$ & $h$  \\
    \hline $+1$ & $\omega$ &  $1,\omega^{-1}$ & $e^{-\frac{2\pi i}{12}}$ &$-\frac{1}{24}+\frac{1}{2}\ZZ$\\
         $+1$ & $\omega$ &   $\omega$ & $e^{\frac{2\pi i}{4}}$ & $\frac{1}{8}+\frac{1}{2}\ZZ$\\
         \hline $+1$ & $\omega^{-1}$ &  $1,\omega$ & $e^{\frac{2\pi i}{12}}$ &$+\frac{1}{24}+\frac{1}{2}\ZZ$\\
         $+1$ & $\omega^{-1}$ &   $\omega^{-1}$ & $e^{-\frac{2\pi i}{4}}$ & $-\frac{1}{8}+\frac{1}{2}\ZZ$\\
         \hline $-1$ & $\omega$ &  $1,\omega^{-1}$ & $e^{2\pi i\frac{5}{12}}$ &$\frac{5}{24}+\frac{1}{2}\ZZ$\\
         $-1$ & $\omega$ &   $\omega$ & $e^{-\frac{2\pi i}{4}}$ & $-\frac{1}{8}+\frac{1}{2}\ZZ$\\
         \hline $-1$ & $\omega^{-1}$ &  $1,\omega$ & $e^{-2\pi i\frac{5}{12}}$ &$-\frac{5}{24}+\frac{1}{2}\ZZ$\\
         $-1$ & $\omega^{-1}$ &   $\omega^{-1}$ & $e^{\frac{2\pi i}{4}}$ & $\frac{1}{8}+\frac{1}{2}\ZZ$\\ \hline
    \end{tabular}
    \caption{The spin selection rules for $A=\ZZ_3$.}\label{t:spinselectZ3}
\end{table}

 In the case $A= \mathbb{Z}_2$, which we will analyze in section \ref{s:Z2duality}, there exists a single non-degenerate symmetric bicharacter,
\begin{equation}
    \chi(1,1)=-1 \, \, , \quad \quad  \chi (1,0)= \chi (0,1) = \chi (0,0) =1 \, .
\end{equation}
Accordingly, one can distinguish two distinct Tambara–Yamagami categories, TY$(\mathbb{Z}_2, \chi, \pm 1) \equiv {\rm TY}(\mathbb{Z}_2 , \pm 1)$, which differ only by the sign of the Frobenius–Schur indicator $\epsilon= \pm 1$. Because $(\hat{\mathcal{L}}_g^{(\CN)})^2=\chi(1,1)=-1$, the eigenvalues $\zeta$ of $\hat{\mathcal{L}}_g^{(\CN)}$  in the $\mathcal{N}$-twisted sector must satisfy the condition $\zeta= \pm i$. By \eqref{spinselect1}, we recover the spin selection rules summarized in Table \ref{t:spinselectZ2}.
\begin{table}[h]
    \centering
    
    \begin{tabular}{|c|c|c|c|}\hline
    $\epsilon$      & $\zeta$ & $\frac{\epsilon}{\sqrt{2}}(1+\zeta)$ & $h$  \\
    \hline $+1$  &  $i$ & $e^{\frac{2\pi i}{8}}$ &$\frac{1}{16}+\frac{1}{2}\ZZ$\\
    \hline $+1$  &  $-i$ & $e^{-\frac{2\pi i}{8}}$ &$-\frac{1}{16}+\frac{1}{2}\ZZ$\\
    \hline $-1$  &  $i$ & $e^{-2\pi i\frac{3}{8}}$ &$-\frac{3}{16}+\frac{1}{2}\ZZ$\\
    \hline $-1$  &  $-i$ & $e^{2\pi i\frac{3}{8}}$ &$\frac{3}{16}+\frac{1}{2}\ZZ$\\ \hline
    \end{tabular}
    \caption{\small{The spin selection rules for $A=\ZZ_2$.}}\label{t:spinselectZ2}
\end{table}

In \S \ref{s:Z2Z2Vfnat} we will consider duality defects for the group $A=\langle a,Q\rangle\cong \ZZ_2\times\ZZ_2$.
In this case, the bicharacter $\chi$ satisfies $\chi(x,y)\in\{\pm 1\}$ and $\chi(aQ,aQ)=\chi(a,a)\chi(Q,Q)\chi(a,Q)^2=\chi(a,a)\chi(Q,Q)$. This implies that $\chi(x,x)=+1$ for at least one of the three non-trivial elements $x\in \{a,Q,aQ\}$. Up to automorphisms in $\Aut(\ZZ_2\times\ZZ_2)\cong S_3$, which correspond to equivalences of the Tambara-Yamagami category, we can set $\chi(Q,Q)=1$. Non-degeneracy of $\chi$ then implies that $\chi(a,Q)=-1$. Thus, the only two possibilities for $\chi$ are $\chi=\chi_a$ or $\chi=\chi_s$, with 
\be \chi_a(a,a)=+1\ ,\qquad \chi_s(a,a)=-1\ .
\ee
Eq. \eqref{spinselect1} implies that 
$$ \langle \psi|e^{4\pi i L_0}|\psi\rangle=\frac{\epsilon}{2}\langle \psi|(1+\hat\CL_{a}^{(\CN)}+\hat\CL_{Q}^{(\CN)}+\hat\CL_{aQ}^{(\CN)})|\psi\rangle=\frac{\epsilon}{2}\langle \psi|(1+\hat\CL_{a}^{(\CN)}+\hat\CL_{Q}^{(\CN)}-\hat\CL_{a}^{(\CN)}\hat\CL_{Q}^{(\CN)})|\psi\rangle\ ,
 $$ where we used $\chi(a,Q)=-1$. If $|\psi\rangle$ in the $\CN$-twisted sector is such that $L_0|\psi\rangle=h|\psi\rangle$, $\hat\CL_{Q}^{(\CN)}|\psi\rangle=\zeta_Q|\psi\rangle$ and $\hat\CL_{a}^{(\CN)}|\psi\rangle=\zeta_a|\psi\rangle$, we get the selection rule
 $$ e^{4\pi i h}=\frac{\epsilon}{2}(1+\zeta_a+\zeta_Q-\zeta_a\zeta_Q)\ .
 $$ Since $\chi(Q,Q)=1$, one has $(\hat\CL_{Q}^{(\CN)})^2=1$, so $\hat\CL_{Q}^{(\CN)}$ has eigenvalues $\zeta_Q\in \{+1,-1\}$. On the other hand, $(\hat\CL_{a}^{(\CN)})^2=\chi(a,a)$, so $\hat\CL_{Q}^{(\CN)}$ has eigenvalues $\zeta_a\in \{\pm 1\}$ if $\chi(a,a)=+1$ or $\zeta_a\in\{\pm i\}$ if $\chi(a,a)=-1$.
  The resulting possibilities are summarized in Table \ref{t:spinselectZ2Z2}.
  
  \begin{table}[h!]
  \centering
 \begin{tabular}{|c|c|c|c|}
 \hline
      $\chi(a,a)$ & $\epsilon$ & $(\zeta_a,\zeta_Q)$ & $h$  \\ \hline
  $+1$    &  $+1$  & $(-1,-1)$ & $ \frac{1}{4}+\frac{1}{2}\ZZ$\\
  $+1$    &  $+1$  & $(1,-1),\ (-1,1),\ (1,1)$ & $\frac{1}{2}\ZZ$\\ \hline
  $+1$    &  $-1$ & $(-1,-1)$ & $ \frac{1}{2}\ZZ$\\
  $+1$    &  $-1$  & $(1,-1),\ (-1,1),\  (1,1)$ & $ \frac{1}{4}+\frac{1}{2}\ZZ$\\
  \hline
  $-1$    &  $+1$ & $(+i,-1)$ & $ \frac{1}{8}+\frac{1}{2}\ZZ$\\
  $-1$    &  $+1$ & $(-i,-1)$ & $ \frac{3}{8}+\frac{1}{2}\ZZ$\\
   $-1$    &  $+1$ & $(\pm i,+1)$ & $ \frac{1}{2}\ZZ$\\\hline
   $-1$    &  $-1$ & $(+i,-1)$ & $ \frac{3}{8}+\frac{1}{2}\ZZ$\\
  $-1$    &  $-1$ & $(-i,-1)$ & $ \frac{1}{8}+\frac{1}{2}\ZZ$\\
   $-1$    &  $-1$ & $(\pm i,+1)$ & $ \frac{1}{4}+\frac{1}{2}\ZZ$\\\hline
 \end{tabular}\caption{\small{The spin selection rules for $A=\ZZ_2\times\ZZ_2$.}}\label{t:spinselectZ2Z2}\end{table}
 
\subsection{Cyclic orbifolds of $V^{f\natural}$ and the monster group}\label{s:Monster}

In this section we describe an interesting relationship between cyclic orbifolds $V^{f\natural}/\langle g\rangle$ for $g\in Co_0$ and monstrous moonshine. 

We begin with a brief review of the relevant definitions. In \cite{CN}, as part of their monstrous moonshine conjectures, for each conjugacy class $[g']\in \mathbb M$, the monster group, Conway and Norton associate  a McKay--Thompson series $T_{g'}(\tau)$, which is a weight zero modular function for a genus zero subgroup $\Gamma_{g'}<SL(2,\mathbb R)$. 
We now know each McKay-Thompson series is given by a trace in the moonshine module $V^\natural$ constructed by Frenkel, Lepowsky, and Meurman \cite{Frenkel:1988flm}:
\be
T_{g'}(\tau):= \Tr_{V^\natural}g' q^{L_0-1}.
\ee

On the other hand, to each conjugacy class $[g]\in Co_0$, we can associate a Frame shape $\pi_g$ and an eta quotient $\eta_g(\tau)$, defined in \eqref{eq:Frameshape} and \eqref{eq:Etaquotient}, respectively. We can also define a $g$-twined lattice theta series $\Theta_g(\tau)$ via the natural action of $g\in Co_0$ on the Leech lattice. This function is essentially a trace over the $g$--invariant vectors in the Leech lattice; see \cite{10.2969/jmsj/04120263} for explicit formulas.

In \cite{CN} Conway and Norton make the interesting conjecture that there is a correspondence between conjugacy classes in $Co_0$ and in $\mathbb M$: they predict that to each conjugacy class $[g]\in Co_0$, one can  assign a conjugacy class $[g']$  in $\mathbb M$ such that the monstrous McKay-Thompson series $T_{g'}(\tau)$ can be expressed as
\be\label{MCo0relation}
T_{g'}(\tau)={\Theta_g(\tau) \over \eta_g(\tau)} + c_g,
\ee
where  $c_g$ is a constant.
In  \cite{10.2969/jmsj/04120263}, Lang proves that this is true for all but precisely 15 conjugacy classes $[g]\in Co_0$. Therefore, in the cases where this correspondence holds, one can also associate a genus zero group to $[g]\in Co_0$, which is the genus zero group $\Gamma_{g'}$ associated to the hauptmodul $T_{g'}(\tau)$.

Now we need to briefly review the characterization of the genus zero groups $\Gamma_{g'}$ which arise in monstrous moonshine. Firstly, let $\Gamma_0(N)$  be the congruence subgroup of $SL(2,\mathbb Z)$ defined by
\be
\Gamma_0(N):=\left \{\begin{pmatrix} a & b\\c & d\end{pmatrix} \in SL(2,\mathbb Z) \Bigg| c\equiv 0 \pmod N \right \}.
\ee
Moreover, for all exact divisors  $e$ of $N$, define the following Atkin-Lehner involutions of $\Gamma_0(N)$ by the following set of determinant $e$ matrices,
\be
W_{e}:=\left \{\begin{pmatrix} ae & b\\cN & de\end{pmatrix}  \Bigg| a,b,c,d\in \ZZ ~\& ~e||N ~\& ~ade^2-bcN=e \right \},
\ee
where $e||N$ means that $e$ is an exact divisor of $N$, i.e. $(e, N/e)=1$.
Following \cite{CN}, we introduce the following notation
\begin{itemize}
\item $N-=\Gamma_0(N)$, i.e. the congruence subgroup of $SL(2, \ZZ)$
\item $N+e_1,e_2,\ldots= \langle \Gamma_0(N), W_{e_1}, W_{e_2}, \ldots \rangle$, i.e. the group generated by $\Gamma_0(N)$ and a subset $e_1, e_2, \ldots$ of its Atkin-Lehner involutions
\item $N+$, the group generated by $\Gamma_0(N)$ and {\it all} of its Atkin-Lehner involutions
\end{itemize}
Similarly instead of $\Gamma_0(N)$ itself, for any $h$ which divides $n$, we can consider  the group
\be
\Gamma_0(n|h):= \left \{\begin{pmatrix} a & b/h\\cn & d\end{pmatrix} \in SL(2,\mathbb R) \Bigg| a,b,c,d\in \ZZ \right \},
\ee
which is constructed by conjugating the group $\Gamma_0(n/h)$ by $\begin{pmatrix} h & 0 \\ 0 & 1\end{pmatrix}$.
In this context, for any Atkin-Lehner involution $W_e$ of $\Gamma_0(n/h)$, we define a corresponding involution of $\Gamma_0(n|h)$ by
\be
w_e = \begin{pmatrix} h^{-1} & 0 \\ 0 & 1\end{pmatrix}W_e\begin{pmatrix} h & 0 \\ 0 & 1\end{pmatrix}.
\ee
For these groups we use the notation
\begin{itemize}
    \item $n|h-=\Gamma_0(n|h)$
    \item $n|h+e_1, e_2, \ldots= \langle \Gamma_0(n|h), w_{e_1}, w_{e_2}, \ldots \rangle$.
\end{itemize}
Note that the matrix $\begin{pmatrix} 0 & -1 \\ N & 0 \end{pmatrix}$ is called the Fricke involution, and is present in the set $W_N$.
All genus zero groups $\Gamma_{g'}$ present in monstrous moonshine are of the type $N-, n|h-$ or have additional Atkin-Lehner involutions.

Now we would like to describe the relation to the cyclic orbifolds  $V^{f\natural}/\langle g\rangle$ discussed in \S \ref{section:TDLs_Vfnat}. For each conjugacy class $[g]\in Co_0$ where the relation \eqref{MCo0relation} holds, we can also associate a corresponding genus zero group $\Gamma_{g'}$. Moreover, we can divide the cyclic orbifolds into the following four classes corresponding the Tables \ref{tbl:selforb}-\ref{tbl:anomalous}: $(-1)^F$--preserving self-orbifolds, orbifolds which yield $V^{fE_8}$, $(-1)^F$--breaking self-orbifolds, and anomalous orbifolds. In each case, we see a correspondence with genus zero groups of a different type.

\begin{enumerate}
    \item All the $42$ conjugacy classes $[g]\in Co_0$ such that $V^{f\natural}/\langle g \rangle \cong V^{f\natural}$ and a duality defect might potentially exist, which are listed in Table \ref{tbl:selforb}, satisfy the relation \eqref{MCo0relation} for some conjugacy class $[g']\in \mathbb M$. Moreover, all corresponding genus zero groups $\Gamma_{g'}$ are of the type $\Gamma_0(o(g))+e_1, e_2,\ldots$, which contain the Fricke involution. 
    \item These properties do not hold for the $10$ classes with unbalanced Frame shape that are listed in Table \ref{tbl:selforb_no_balanced}, for which $V^{f\natural}/\langle g \rangle \cong V^{f\natural}$, but a duality defect cannot exist, since the symmetry $g$ in $V^{f\natural}$ and the corresponding quantum symmetry $Q$ in $V^{f\natural}/\langle g \rangle$ have different Frame shapes.
    \item For conjugacy classes $[g]\in Co_0$ such that $V^{f\natural}/\langle g \rangle \cong V^{fE_8}$, which are listed in Table \ref{tbl:VfE8}, all but 5 conjugacy classes satisfy the relation \eqref{MCo0relation} for some $[g']\in \mathbb M$. In all but one case, the corresponding genus zero groups $\Gamma_{g'}$ are all groups of non--Fricke type, i.e. of the form $\Gamma_0(o(g))$, possibly extended by a set of Atkin-Lehner involutions which do not contain the Fricke involution.\footnote{The one exception is the $Co_0$ conjugacy class 8F, which corresponds to the genus zero group $8|2-$.}
\item There are 40 conjugacy classes for which $V^{f\natural}/\langle g\rangle \cong V^{f\natural}$, but with opposite sign of $(-1)^F$ in the Ramond sector. As shown in Table \ref{tbl:VfnatnoSUSY}, all such classes satisfy the relation \eqref{MCo0relation}, leading to 38 conjugacy classes\footnote{This is because the conjugacy classes 46AB and 78AB in $Co_0$ lead to the same trace functions.} $[g']\in \mathbb M$ whose corresponding genus zero groups $\Gamma_{g'}$ are all non--Fricke extensions of $\Gamma_0(o(g))$ by a (possibly trivial) set of Atkin--Lehner involutions. Notably, these correspond to 38 of the 39 non--Fricke genus zero groups of this type, the only one missing being the group $25-$, which also does not play a role in monstrous moonshine, as there is no conjugacy class of order 25 in $\mathbb M$. Interestingly, these 39 genus zero groups precisely label the set of so--called ``optimal mock Jacobi forms" \cite{Cheng:2016klu}, 23 of which are related to the umbral moonshine phenomenon \cite{Cheng:2013cdh}. Is there a connection between these cyclic orbifolds of $V^{f\natural}$ and the optimal mock Jacobi forms?

    \item All 43 anomalous conjugacy classes of $Co_0$ listed in Table \ref{tbl:anomalous} satisfy the relation \eqref{MCo0relation} for some $[g']\in \mathbb M$. Moreover, for all but 5 of the 43 classes,\footnote{The five exceptions are 2D, 6O, 10J, and 22BC. These correspond to the Fricke--type genus zero groups $4+, 12+, 20+$, and $44+$, respectively.} the corresponding genus zero groups $\Gamma_{g'}$ are of $n|h$--type, with or without additional Atkin--Lehner involutions. 
    
\end{enumerate}
It would be interesting to find a deeper explanation for these facts, and we leave this as an open question.

\section{Duality defects for $\ZZ_2$}\label{s:Z2duality}

As mentioned in \S \ref{section:TDLs_Vfnat}, $V^{f\natural}$ is self-orbifold under cyclic groups $G$ generated by elements of conjugacy classes listed in Table \ref{tbl:selforb} that have balanced Frame shape. Thus, one expects a duality defect to be associated with each of these conjugacy classes. We are particularly interested in duality defects which preserve the $\mathcal N=1$ supercurrent. Moreover, for the purposes of connections to K3 NLSMs, we would like to understand which duality defects preserve a $c=6$, $\CN=4$ superconformal algebra.

In this section, we discuss this question for the case of duality defects for $G= \mathbb{Z}_2 = \langle g \rangle$, where $g$ has Frame shape $1^82^8$, i.e. class 2B in $Co_0$, as listed in table \ref{tbl:selforb}.  
Let us consider the orbifold of $V^{f\natural}$ by $\langle g\rangle$. The subgroup of $SO(24)=SO(\Lambda\otimes\RR)$ commuting with $g$ is
\be\label{Z2pres_subgroup} SO(24)\supset SO(16)\times SO(8)\ ,
\ee
and the corresponding $g$-preserved subalgebra of $so(24)_1$  is:
\begin{equation}
so(24)_1 \quad  \supset  \quad so(16)_1 \oplus so(8)_1.
\label{Z2:preserved_subalgebra}
\end{equation}
In describing the details of the orbifold procedure, we will use the descriptions of $V^{f\natural}$ both as a lattice VOA based on the odd lattice $D_{12}^+$, and as a rational CFT with chiral algebra $so(24)_1$.
Up to conjugation in $Spin(24)$, the $\mathbb{Z}_2$ symmetry $g$ can be taken to act on the vertex operators $V_{(x_1,\ldots,x_{12})}(z)$, $(x_1,\ldots,x_{12})\in D_{12}^+$ by multiplication by a $U(1)$ phase
\begin{equation}
    V_{\left( x_1 , ..., x_{12}\right)} \quad \quad \mapsto \quad \quad e^{2 \pi i \alpha_k x_k} V_{\left( x_1 , ..., x_{12}\right)}
    \label{Sym_action:lattice}
\end{equation}
with 
\begin{equation}
    \alpha = ( \underbrace{0, ..., 0 }_8, \underbrace{\frac{1}{2}, \frac{1}{2}, \frac{1}{2}, \frac{1}{2}}_4 ).
    \label{Z2:shift_vector}
\end{equation}

Let us now consider the orbifold of $V^{f\natural}$ by $\langle g\rangle\cong \ZZ_2$. To project the spectrum onto the $\mathbb{Z}_2$-invariant subset of states, we need to decompose the representations in $V^{f \natural}$ with respect to the preserved subalgebra    \eqref{Z2:preserved_subalgebra}, as shown in table \ref{tab:2}.
\begin{table}[h!]
	\centering
	\begin{tabular}{|c|c|c|l|c|}
		\hline
		&$\widehat{so}(24)_1$  & $g=+1$  & $g=-1$   \\
		\hline
		$NS_+$ & $0$ &  $(0,0)$ & $(v, v)$ 
        \\
		\hline
		$NS_-$ & s & $(s,s)$ & $(c, c)$ 
        \\
		\hline
		$R_+$ & v & $(v,0)$ & $(0, v)$ 
        \\
		\hline
		$R_-$ & c & $(c,s)$ & $(s, c)$ 
        \\
		\hline
	\end{tabular}
	\caption{\small{Decomposition of the $so(24)_1$ representations under the $\mathbb{Z}_2$-preserved subalgebra $so(16)_1 \oplus so(8)_1$, and the corresponding $\mathbb{Z}_2 = \langle g \rangle$ action. }}
	\label{tab:2}
\end{table}

From the third column of Table \ref{tab:2}, we can read off the representations preserved by the $\mathbb{Z}_2$ action. Specifically, they are:
\begin{equation}
\left( V^{f \natural} / \langle g\rangle\right)_{NS,untw} = (0,0)_+ \oplus (s,s)_- \qquad\left( V^{f \natural} / \langle g\rangle\right)_{R,untw} = (v,0)_+ \oplus (c,s)_-
\label{Z2:preserved_untwisted}
\end{equation}  where the subscript $\pm$ denotes the corresponding eigenvalue $\pm 1$ of the fermion number operator $(-1)^F$.
Note that the $\mathcal{N}=1$ supercurrent of $V^{f \natural}$ is contained in the representation $(s,s)$, which is preserved by the orbifold procedure. 

Define $\ch^{(n)}_\rho(\tau):=\Tr_{\rho^{(n)}} q^{L_0-c/24}$ to be the characters of $\widehat{so}(n)_1$ representations, where $\rho\in \{0,v,s,c\}$. They are explicitly given by \eqref{ch_0_so(n)}--\eqref{ch_s_so(n)} in appendix \ref{a:minimal}.

From table \ref{tab:2}, the $g$-twining partition functions can be written in terms of the characters of the $so(16)_1 \oplus so(8)_1$ algebra as (the argument $\tau$ is everywhere omitted):
	\begin{align}
Z_{NS}^{g,\pm}&=\ch^{(16)}_0\ch^{(8)}_0-\ch^{(16)}_v\ch^{(8)}_v\pm \ch^{(16)}_s\ch^{(8)}_s\mp \ch^{(16)}_c\ch^{(8)}_c\ ,\\
Z_{R}^{g,\pm}&=\ch^{(16)}_v\ch^{(8)}_0-\ch^{(16)}_0\ch^{(8)}_v\pm \ch^{(16)}_c\ch^{(8)}_s\mp \ch^{(16)}_s\ch^{(8)}_c\ .
\end{align} The S-transformations of these formulae provide the $g$-twisted partition functions:
\begin{align}Z_{g,NS}^{\pm}&=\ch^{(16)}_0\ch^{(8)}_s+\ch^{(16)}_v\ch^{(8)}_c\pm (\ch^{(16)}_s\ch^{(8)}_0+\ch^{(16)}_c\ch^{(8)}_v)\\
Z_{ g,R}^{\pm}&=\ch^{(16)}_0\ch^{(8)}_c+ \ch^{(16)}_v\ch^{(8)}_s\pm (\ch^{(16)}_s\ch^{(8)}_v+\ch^{(16)}_c\ch^{(8)}_0)\ .
\end{align}
The representations in the $g$-twisted sector can also be derived from those in the untwisted sector by shifting the corresponding lattices using the vector $\alpha$ defined in \eqref{Z2:shift_vector}. At the level of the algebra \eqref{Z2:preserved_subalgebra}, $\alpha$ is a vector in $(0,s)$.

The action of $g$ on the $g$-twisted sector is determined by its action on the original theory up to a phase factor, which is identical for all twisted states. This phase is fixed by requiring the OPEs between twisted operators to be local and ensuring the appropriate modularity properties of the partition function.  In particular,  since $Z^{\pm}_{g,NS}(\tau+1)=-Z^{g,\mp}_{g,NS}(\tau)$ and $Z^{\pm}_{g,R}(\tau+1)=Z^{g,\pm}_{g,R}(\tau)$, we obtain the actions
described in Table \ref{tab:3}.
\begin{table}[h!]
	\centering
	\begin{tabular}{|c|c|c|}
		\hline $g$-twisted
		&  $g=+1$  & $g=-1$  \\
		\hline
		$NS_+$ & $(v, c)$ &  $(0,s)$  
        \\
		\hline
		$NS_-$ & $(c, v)$ & $(s,0)$ 
        \\
		\hline
		$R_+$ &  $(0, c)$ & $(v,s)$  
        \\
		\hline
		$R_-$ & $(s, v)$ & $(c,0)$  
        \\
		\hline
	\end{tabular}
	\caption{\small{Decomposition of the representations in the $g$-twisted sector with respect to the preserved subalgebra $so(16)_1 \oplus so(8)_1$ and the corresponding $\ZZ_2=\langle g\rangle$ action.}}
	\label{tab:3}
\end{table}

Consequently, the representations preserved under the $g$ action in the $g$-twisted sector are:
\begin{equation}
\left( V^{f \natural} /\langle g\rangle\right)_{NS,g-tw} = (v,c)_+ \oplus (c,v)_- \qquad \left( V^{f \natural} /\langle g\rangle\right)_{R,g-tw}= (0,c)_+ \oplus (s,v)_-\ ,
\label{Z2:preserved_twisted}
\end{equation} where the subscript still denotes the fermion number $(-1)^F$.
The orbifold $V^{f\natural}/\langle g\rangle$ is given by the direct sum of the two preserved subsectors \eqref{Z2:preserved_untwisted} and \eqref{Z2:preserved_twisted}:
\begin{equation}
\left(V^{f \natural} /\langle g\rangle\right)_{NS}= \underbrace{(0,0)_+ \oplus (s,s)_-}_{\text{untwisted}} \oplus \underbrace{(v,c)_+ \oplus (c,v)_- }_{g-\text{twisted}}\end{equation}
\begin{equation} \left(V^{f \natural} /\langle g\rangle\right)_{R}= \underbrace{ (v,0)_+ \oplus (c,s)_-}_{\text{untwisted}} \oplus \underbrace{(0,c)_+ \oplus (s,v)_-}_{g-\text{twisted}}.        \end{equation}

It is straightforward to verify that the construction is self-orbifold by computing its fully periodic torus partition functions. This property is already encoded in the Witten index listed in the tables of Section \ref{section:TDLs_Vfnat}. Consequently, there must exist a duality defect $\CN_g$ that maps the parent theory to the daughter orbifold model and vice versa. 

Comparing the representations appearing in the daughter theory with the ones of the parent theory, we can deduce that the operator $\frac{1}{\sqrt{2}}\hat\CN_g$ must act as an external automorphism on $\widehat{so}(8)_1$ exchanging the representations $c \leftrightarrow v$ of the two models and map $s$ into itself by a non-trivial action. Every such automorphism can be represented by a $8\times 8$ real orthogonal matrix $O^{(8)}_\CN\in O(8,\RR)$ that describes its action on the $8$ ground states of the $s$ representation; the condition that the automorphism is outer and exchanges $v$ and $c$ is equivalent to requiring $\det O^{(8)}_\CN=-1$. 
In order to preserve the $\mathcal{N}=1$ supercurrent the operator $\frac{1}{\sqrt{2}}\hat\CN_g$ has to act as an inner automorphism on $\widehat{so}(16)_1$ compensating the change of the representation $s$ of $\widehat{so}(8)_1$ in $(s,s)$. 
Such an inner automorphism correspond to an involution  $S^{(16)}_\CN\in Spin(16)$ in the spin group generated by $\widehat{so}(16)_1$. Thus, every duality defect $\CN_g$ is determined by a pair
\be\label{O16O8} \CN_g \quad \leftrightarrow\quad  (S_{\CN}^{(16)},O_{\CN}^{(8)})\in Spin(16)\times O(8)\ ,
\ee such that $(S_{\CN}^{(16)})^2=1=(O_\CN^{(8)})^2$ and $\det O_\CN^{(8)}=-1$.

In section \ref{s:freefermion} we will show that the duality defect $\CN_g$ can be chosen in such a way that it preserves the $\CN=1$ superconformal symmetry. Even with this requirement, the duality defect $\CN_g$ for a given symmetry $g$ of Frame shape $1^82^8$ is not unique. Indeed, given any such $\CN_g$, one can obtain another one by considering a fusion product  with invertible defects
\be\label{allZ2duality} \CL_{h_1}\CN_g \CL_{h_2}\ ,
\ee where $h_1,h_2\in C_{Co_0}(g)\subset Co_0$ are any elements in the centralizer $C_{Co_0}(g)$ of $g$, i.e. that commute with $g$, and such that
\be (\CL_{h_1}\CN_g \CL_{h_2})^2=\CN_g^2=\CI+\CL_g\ .
\ee
In particular, for each $g$ of Frame shape $1^82^8$, we will find two examples $\CN_g^{(1)}$ and $\CN_g^{(2)}$ of duality defects generating the Tambara-Yamagami categories $TY(\ZZ_2,+)$ and $TY(\ZZ_2,-)$. Furthermore,  any other duality defect \eqref{allZ2duality} for $g$ is related to either $\CN_g^{(1)}$ or $\CN_g^{(2)}$ by conjugation in the group $SO(16)\times SO(8)$ of $SO(24)$ commuting with $g$, see the discussion in section \eqref{s:freefermion}. 

\subsection{The lattice endomorphism associated with the $\ZZ_2$-duality defect}\label{s:Z2endom}

Recall that the group $\Aut_\tau(V^{f\natural})\cong Co_0$ of automorphisms of $V^{f\natural}$ fixing the $\CN=1$ supercurrent $\tau(z)$ acts on the $24$-dimensional space   $V^{f\natural}_{tw}(1/2)$ of Ramond ground states by automorphisms of the Leech lattice $\Lambda\subset V^{f\natural}_{tw}(1/2)$. Our general results on $\CN=1$ preserving duality defects imply that the action of $\hat\CN_g$ on $V^{f\natural}_{tw}(1/2)$  maps vectors of $\Lambda \subset V^{f\natural}_{tw}(1/2)$ into vectors of $\Lambda$. In the following, we will show how this property imposes constraints to determine the pair of automorphisms $\left( \CS_{\CN}^{(16)}, \CO_{\CN}^{(8)}\right)$ defining $\CN_g$. \\
In $V^{f\natural}_{tw}(1/2)$, the group $\langle g\rangle\cong \mathbb{Z}_2$ acts fixing a $16-$dimensional subspace. The sublattice of $\Lambda$ fixed by an element $g\in Co_0$ of Frame shape $1^82^8$ is the $16-$dimensional Barnes-Wall lattice $\Lambda_{16} \subset \Lambda$ \cite{ConwaySloane}.  By definition we have:
\begin{equation}
    \hat{\mathcal{N}}^2_{g} \vert v \rangle = \left( \mathcal{L}_e + \mathcal{L}_{g} \right) \vert v \rangle, \quad \quad \quad \forall \vert v \rangle \in V^{f\natural}_{tw}(1/2) 
\end{equation}
where $\mathcal{L}_e$ is the trivial defect and $\mathcal{L}_g$ is the invertible defect associated to the symmetry $g \in \mathbb{Z}_2$. If $v \in \Lambda_{16}$ we have:
\begin{equation}
    \hat{\mathcal{N}}^2_{g} \vert v \rangle = 2 \vert v \rangle,
\end{equation}
which means that $\hat{\mathcal{N}}_{g}$ has eigenvalues $\pm\sqrt{2}$ on the space $\Lambda_{16}\otimes\RR$ and kills the vectors of the complementary orthogonal $8-$dimensional lattice $\Lambda_8 = \Lambda_{16}^{\perp} \cap \Lambda$. 
 While the duality defect $\CN_g$ preserving the $\CN=1$ supercurrent is not unique (see eq.\eqref{allZ2duality}in section \ref{s:freefermion}), this analysis, together with the requirement that $\Tr_{V^{f\natural}_{tw}(1/2)}(\hat\CN_g)\in \ZZ$, implies that the eigenvalues of $\hat\CN_g$ on $V^{f\natural}_{tw}(1/2)$ are always
\be\label{Z2_lattice_eigen} 8\times \sqrt{2}\ ,\qquad 8\times (-\sqrt{2})\ ,\qquad 8\times 0\ .
\ee  This also implies that the subalgebra of $\widehat{so}(16)_1$ that is preserved by $\CN_g$ is $\widehat{so}(8)_1^+\oplus \widehat{so}(8)_1^-$, corresponding to the eigenspaces of $\hat\CN_g$ on $V^{f\natural}_{tw}(1/2)$ with eigenvalues $+\sqrt{2}$ and $-\sqrt{2}$.

Now consider a vector $v=v_{\parallel}+v_\perp \in \Lambda$ with a leg $v_\parallel$ extending on the invariant $16-$dimensional space $\Lambda_{16}\otimes \RR$ and another leg $v_\perp$ extending on its $8-$dimensional complementary orthogonal space $\Lambda_8\otimes\RR$. For all $\lambda\in \Lambda_{16}$, one has $v_\perp\cdot \lambda=0$ and $v_\parallel\cdot \lambda=v\cdot\lambda\in\ZZ$, so that $v_\parallel\in \Lambda_{16}^\star$, where
\be \Lambda_{16}^\star=\{v\in \Lambda_{16}\otimes\RR\mid v\cdot\lambda\in \ZZ,\ \forall\lambda\in\Lambda_{16}\}
\ee is the dual lattice. Furthermore, because $\Lambda$ is self-dual, for each primitive vector $\lambda\in\Lambda_{16}\subset \Lambda$, there must exist a vector $v\in \Lambda$ such that $v\cdot\lambda=v_\parallel\cdot\lambda=1$, so that the vectors $v_\parallel$ span the whole $\Lambda_{16}^\star$. The action of $\hat{\mathcal{N}}_{g}$ on $v=v_\parallel+v_\perp$ kills its component $v_\perp$ along $\Lambda_8\otimes\RR$ and produces a vector $\hat\CN_g(v)=\hat\CN_g(v_\parallel)$ that must be contained in $\Lambda_{16}$, so that
\be \hat\CN_g(v_\parallel)\in \Lambda_{16}\ ,\qquad \forall v_\parallel\in \Lambda_{16}^\star\ .
\ee
The lattice $\Lambda_{16}$ is a $2$-modular lattice, i.e. it is isomorphic to its dual $\Lambda_{16}^\star$ rescaled by $\sqrt{2}$,
\be\label{NgLambda16} \Lambda_{16}=\sqrt{2} O^{(16)}_\CN\Lambda_{16}^\star\ ,
\ee where $O^{(16)}_\CN\in SO(16)$ is a suitable orthogonal matrix, which is unique up to acting from the left or from the right by lattice automorphisms in $Aut(\Lambda_{16})=Aut(\Lambda^\star_{16})$. An explicit description of the lattice $\Lambda_{16}$ and of a matrix $O^{(16)}_\CN$ providing the isomorphism \eqref{NgLambda16} is given in appendix \ref{a:Leechbasis}.

Thus, up to automorphisms of $\Lambda_{16}$, the only possible action of $\hat\CN_g$ on $v_\parallel\in \Lambda_{16}^\star$ compatible with \eqref{NgLambda16} and with  eigenvalues $\pm\sqrt{2}$ is
\be\hat\CN_g(v_\parallel)=\sqrt{2}O^{(16)}_\CN v_\parallel \in\Lambda_{16}\ ,\qquad \forall v_\parallel\in \Lambda_{16}^\star\ .
\ee
Compatibility of the OPE between $g$-invariant Ramond ground states and the NS currents in the $\widehat{so}(16)_1$ algebra fixed by $g$ implies that $O^{(16)}_\CN\in SO(16)$ in \eqref{NgLambda16} must be the image under the projection $Spin(16)\to SO(16)$ of the involution $S^{(16)}_\CN\in Spin(16)$ determining the inner automorphism of $\widehat{so}(16)_1$ currents in \eqref{O16O8}.
Given the automorphism \eqref{NgLambda16}, there are two possible lifts of $O^{(16)}_\CN\in SO(16)$ to the spin group $Spin(16)$, and they are related to each other by composition with the fermion number $(-1)^F$. As a consequence, at most one of these lifts can preserve the $\CN=1$ supercurrent.
Notice that the subgroup \be C_{Co_0}(g):=\{h\in Co_0\mid gh=hg\}\subset Co_0\ee of elements of $Co_0$ that commute with $g$ acts by automorphisms on the $g$-fixed sublattice $\Lambda_{16}$, so there must be a group homomorphism  $C_{Co_0}(g)\to \Aut(\Lambda_{16})$ with kernel $\langle g\rangle\cong\ZZ_2$ (no other element of $Co_0$ fixes the same sublattice). Because both $C_{Co_0}(g)/\langle g\rangle$ and $\Aut(\Lambda_{16})$ have the same  order $2^{21}\cdot 3^5\cdot 5^2\cdot 7=89181388800$ \cite{Atlas,ConwaySloane}, it follows that they must be isomorphic
\be C_{Co_0}(g)/\langle g\rangle\cong \Aut(\Lambda_{16})\ .
\ee An important consequence of this observation is the following. Suppose that, for a certain choice of the automorphism \eqref{NgLambda16} with $O^{(16)}_\CN\in SO(16)$ of order $2$, there exists a lift $S^{(16)}_\CN\in Spin(16)$ of $O^{(16)}_\CN$ and a matrix $O^{(8)}_\CN\in O(8)$ such that the duality defect $\CN_g$ associated with the pair $(S^{(16)}_\CN,O^{(8)}_\CN)$ preserves the $\CN=1$ supercurrent. Then, any other lattice automorphism \eqref{NgLambda16}
\be h_1 O^{(16)}_\CN h_2\ ,
\ee where $h_1,h_2\in \Aut(\Lambda_{16})\cong C_{Co_0}(g)/\langle g\rangle$ are such that $(h_1 O^{(16)}_\CN h_2)^2=1$, is also associated with a $\CN=1$-preserving duality defect $(h_1S^{(16)}_\CN h_2,h_1 O^{(8)}_\CN h_2)$.
\footnote{One might be worried about possible ambiguities in these definitions, arising from the choice of lifts of $h_1,h_2$ from the quotient group $C_{Co_0}(g)/\langle g\rangle$ to the group $C_{Co_0}(g)\subset Co_0\cong \Aut_\tau(V^{f\natural})$ that effectively acts on $V^{f\natural}$ and $V^{f\natural}_{tw}$. 
Notice, however, that the action of $S^{(16)}_\CN$ and $O^{(8)}_\CN$ is only defined on the $g$-invariant subspace of $V^{f\natural}$, while $\hat\CN$ is zero on states that are not $g$-invariant. Therefore, $h_1S^{(16)}_\CN h_2$ and $h_1 O^{(8)}_\CN h_2$ are perfectly well-defined for $h_1,h_2$ in the quotient $C_{Co_0}(g)/\langle g\rangle$, with no need to take any lifts.} 

As we will see in next section, some $\CN=1$-preserving duality defects $\CN_g$ do exist; the analysis of this section then shows that they are in one-to-one correspondence with lattice automorphisms \eqref{NgLambda16} such that $O^{(16)}_\CN$ is an involution. As a cross check of these statements, via a (tedious) computer calculation, we were able to find explicitly the pair $(S^{(16)}_\CN,O^{(8)}_\CN)$ preserving the $\CN=1$ supercurrent and corresponding to the matrix $O^{(16)}_\CN$ in eq.\eqref{O16Lattice}.

\subsection{Free fermion construction of the $\ZZ_2$ duality defect}\label{s:freefermion}

In this section, we will show explicitly that a duality defect $\CN_g$ for a symmetry $g\in Co_0$ with Frame shape $1^82^8$ can be chosen so as to preserve the $\CN=1$ supercurrent. The simplest route is to describe $V^{f\natural}$ as a $\ZZ_2\times \ZZ_2$ orbifold of the theory $F(24)$ of $24$ NS chiral free fermions $\psi^1,\ldots, \psi^{24}$, with a $\CN=1$ supercurrent
\be\label{F24SVir} \tau(z):=-i\sum_{k=1}^8 :\psi^i\psi^{i+8}\psi^{i+16}:(z)\ .
\ee This is one of the $8$ inequivalent choices of $\CN=1$ supercurrents in $F(24)$, and it is denoted by $A_1^8$ in \cite{Harrison:2020wxl}. In this theory, we consider the following $\CN=1$-preserving symmetries
\be Q_1:\begin{cases}
	\psi^i\to \psi^i & i=1,\ldots, 8\ ,\\	
	\psi^i\to -\psi^i & i=9,\ldots, 24\ ,
\end{cases} \qquad Q_2:\begin{cases}
\psi^i\to \psi^i & i=9,\ldots, 16\ ,\\	
\psi^i\to -\psi^i & i=1,\ldots,8,17,\ldots, 24\ ,
\end{cases}\ee
\be Q_3:\begin{cases}
\psi^i\to \psi^i & i=17,\ldots,24\ ,\\	
\psi^i\to -\psi^i & i=1,\ldots,16\ ,
\end{cases}
\ee so that $\langle Q_1,Q_2\rangle\cong \ZZ_2\times \ZZ_2$ with $Q_1Q_2=Q_3$. The symmetries are non-anomalous and we will  show below that $F(24)/\langle Q_1\rangle\cong V^{fE_8}$ and $F(24)/\langle Q_1,Q_2\rangle\cong V^{f\natural}$. 

The theory $F(24)$ has a $\widehat{so}(24)_1$ affine algebra generated by fermion biproducts $:\psi^i\psi^j:$, and the subalgebra that is invariant under $\langle Q_1,Q_2\rangle$ is 
\be\label{so8cubed} \widehat{so}(8)^a_1\oplus \widehat{so}(8)^b_1\oplus \widehat{so}(8)^c_1\ .
\ee Recall that the three representations $v,s,c$ of $\widehat{so}(8)_1$ are related to each other by the outer $S_3$ `triality' automorphism. For later convenience, it is useful to think of the free fermions $\psi^i$ as transforming into the $s$ representation of one of the $\widehat{so}(8)_1$ algebras. Therefore, the NS sector of $F(24)$ is given by the representations
\be \begin{aligned}  &(0,0,0)_+  \oplus (s,0,0)_-  \oplus (0,s,0)_-\oplus (0,0,s)_-  \oplus \\  &\oplus (s,s,0)_+\oplus  (0,s,s)_+\oplus (s,0,s)_+\oplus (s,s,s)_-\ , \\ \end{aligned}
\ee 
of $\widehat{so}(8)^a_1\oplus \widehat{so}(8)^b_1\oplus \widehat{so}(8)^c_1$, while the Ramond sector is
 \be  \begin{aligned} & (v,v,v)_+ \oplus (c,c,c)_-\oplus(v,c,c)_+ \oplus (c,v,v)_-\oplus \\ & \oplus (c,v,c)_+ \oplus (v,c,v)_-\oplus (c,c,v)_+ \oplus (v,v,c)_-\ . \\ \end{aligned}
 \ee Here, the subscripts $\pm$ denote the fermion number. For each of the three algebras $\widehat{so}(8)^a_1$, $\widehat{so}(8)^b_1$, and $\widehat{so}(8)^c_1$, let $\xi_8^a$, $\xi_8^b$, $\xi_8^c$ be the  central element in the corresponding spin group $Spin(8)$, acting trivially on the $0$ and $v$ representations of $\widehat{so}(8)_1$, and by $-1$ on the $s$ and $c$. Then, we can interpret $Q_1$, $Q_2$, and $Q_3$ as
\be Q_1=\xi_8^b\xi_8^c\ ,\qquad Q_2=\xi_8^a\xi_8^c\ ,\qquad Q_3=\xi_8^a\xi_8^b\ .
\ee
Let us now decompose the $Q_1$-, $Q_2$-, and $Q_3$-twisted sectors into representations of \eqref{so8cubed}. The results are summarized in table \ref{t:rep_decomposition_Z2}.

\begin{table}[htb]\begin{center}
\begin{tabular}{|c|c|c|c|c|c|} 
	\hline
	Sector & NS Reps & Ramond reps &$Q_1$ & $Q_2$ & $Q_3$\\
	\hline
	untwisted & $(0,0,0)_+\oplus (s,s,s)_-$ & $(v,v,v)_+ \oplus (c,c,c)_-$ & +& + & +\\
	& $(0,s,s)_+\oplus (s,0,0)_-$ & $(v,c,c)_+ \oplus (c,v,v)_-$ & +& - & -\\
	& $(s,0,s)_+\oplus (0,s,0)_-$ & $(c,v,c)_+ \oplus (v,c,v)_-$ & -& + & -\\
	& $(s,s,0)_+\oplus (0,0,s)_-$ & $(c,c,v)_+ \oplus (v,v,c)_-$ & -& - & +\\
	\hline
$Q_1$-twisted & $(0,v,v)_+\oplus (s,c,c)_-$ & $(v,0,0)_+ \oplus (c,s,s)_-$ & +& + & +\\
						 & $(0,c,c)_+\oplus (s,v,v)_-$ & $(v,s,s)_+ \oplus (c,0,0)_-$ & +& - & -\\
						  & $(0,v,c)_+\oplus (s,c,v)_-$ & $(c,s,0)_+ \oplus (v,0,s)_-$ & -& - & +\\
						    & $(0,c,v)_+\oplus (s,v,c)_-$& $(c,0,s)_+ \oplus (v,s,0)_-$  & -& + & -\\
\hline
$Q_2$-twisted & $(v,0,v)_+\oplus (c,s,c)_-$ & $(0,v,0)_+ \oplus (s,c,s)_-$ & +& + & +\\
& $(c,0,c)_+\oplus (v,s,v)_-$ & $(s,v,s)_+ \oplus (0,c,0)_-$ & -& + & -\\
& $(v,0,c)_+\oplus (c,s,v)_-$ & $(s,c,0)_+ \oplus (0,v,s)_-$  & -& - & +\\
& $(c,0,v)_+\oplus (v,s,c)_-$ & $(0,c,s)_+ \oplus (s,v,0)_-$ & +& - & -\\
\hline
$Q_3$-twisted & $(v,v,0)_+\oplus (c,c,s)_-$ & $(0,0,v)_+ \oplus (s,s,c)_-$ & +& + & +\\
& $(c,c,0)_+\oplus (v,v,s)_-$ & $(s,s,v)_+ \oplus (0,0,c)_-$ & -& - & +\\
& $(v,c,0)_+\oplus (c,v,s)_-$ & $(s,0,c)_+ \oplus (0,s,v)_-$ & -& + & -\\
& $(c,v,0)_+\oplus (v,c,s)_-$ & $(0,s,c)_+ \oplus (s,0,v)_-$ & +& - & -\\
\hline
\end{tabular}
\caption{\small{Decomposition of the untwisted, $Q_1$-, $Q_2$- and $Q_3$-twisted sectors into $\hat{so}(8)^a_1 \oplus \hat{so}(8)^b_1 \oplus \hat{so}(8)^c_1$ representations.}}
\label{t:rep_decomposition_Z2}
\end{center} \end{table}

The orbifold $F(24)/\langle Q_1\rangle$ includes the untwisted and the $Q_1$-twisted representations that are $Q_1$-invariant. The symmetry $Q_1$ acts trivially on the first $8$ NS free fermions $\psi^1,\ldots, \psi^8$, that generate a $F(8)$ subSVOA of $F(24)/\langle Q_1\rangle$, that includes the affine algebra $\widehat{so}(8)_1^a$. On the other hand, the other $16$ free fermions $\psi^9,\ldots, \psi^{24}$ are projected out, and the $\widehat{so}(8)_1^b\oplus \widehat{so}(8)_1^c$ algebra is enhanced by the currents in the sectors
\be (0,v,v)_+,\qquad (0,s,s)_+,\qquad (0,c,c)_+
\ee to the $\hat e_{8,1}$ affine algebra. Altogether, the orbifold theory is $F(24)/\langle Q_1\rangle\cong F(8)\otimes V^{E_8}\cong V^{fE_8}$.

On the other hand, the orbifold $F(24)/\langle Q_1,Q_2\rangle=V^{fE_8}/\langle Q_2\rangle$ includes the untwisted, $Q_1$-, $Q_2$- and $Q_3$-twisted representations that are invariant under $Q_1$, $Q_2$, and $Q_3$. Explicitly, the orbifold $F(24)/\langle Q_1,Q_2\rangle\cong V^{f\natural}$ contains the following representations of the algebra  $\widehat{so}(8)_1^a\oplus \widehat{so}(8)^b_1\oplus \widehat{so}(8)^c_1$ in \eqref{so8cubed}:
\be F(24)/\langle Q_1,Q_2\rangle\cong V^{f\natural}:\begin{cases}
	\text{NS}+: & (0,0,0)_+\oplus (v,v,0)_+\oplus (0,v,v)_+\oplus (v,0,v)_+\\
		\text{NS}-: &  (s,s,s)_-\oplus (c,c,s)_-\oplus (s,c,c)_-\oplus (c,s,c)_-\\
	\text{R}+: & (v,0,0)_+\oplus (0,v,0)_+\oplus (0,0,v)_+\oplus (v,v,v)_+\\
	\text{R}-: &  (c,s,s)_-\oplus (s,c,s)_-\oplus (s,s,c)_-\oplus (c,c,c)_-
\end{cases} 
\ee

 The algebra \eqref{so8cubed} is enhanced by the weight $1$ ground fields  in
$(0,v,v)_+$, $(v,0,v)_+$ $(v,v,0)_+$ to a $\widehat{so}(24)_1$ affine algebra, whose currents generate all bosonic NS operators in $F(24)/\langle Q_1,Q_2\rangle$. The fermionic $\text{NS}$ fields combine into a spinor representation of $\widehat{so}(24)_1$. This shows that $F(24)/\langle Q_1,Q_2\rangle\cong V^{f\natural}$. In particular, the $24$ dimensional space of Ramond ground states in $V^{f\natural}$ is contained in the representations
$(v,0,0)_+$ $(0,v,0)_+$ $(0,0,v)_+$ that combine into a vector representation of $\widehat{so}(24)_1$, as usual. 

In this description, we can identify the symmetry $g$ of $V^{f\natural}$ of Frame shape $1^82^8$ as a central element of the group $Spin(8)^c$ generated by the currents in the subalgebra $\widehat{so}(8)_1^c$, third components in \eqref{so8cubed}. More precisely, $g\in Spin(8)^c$ is the element acting by $-1$ on the $v$ and $c$  representations of $\widehat{so}(8)_1^c$ and trivially on the $0$ and $s$ representations. The NS affine subalgebra of $\widehat{so}(24)_1$ that is preserved by $g$ is given by \eqref{so8cubed} and its representation $(v,v,0)_+$, that combine into $\widehat{so}(16)_1\oplus \widehat{so}(8)^c_1$. The analysis at the beginning of section \ref{s:Z2duality} then shows that the $g$-orbifold $V^{f\natural}/\langle g\rangle$ contains the following representations of $\widehat{so}(8)_1^a\oplus \widehat{so}(8)^b_1\oplus \widehat{so}(8)^c_1$ in \eqref{so8cubed}:
\be V^{f\natural}/\langle g\rangle:\begin{cases}
	\text{NS untwisted:} & (0,0,0)_+\oplus (v,v,0)_+\oplus (s,s,s)_-\oplus (c,c,s)_-\\
	\text{NS $g$-twisted:} & (0,v,c)_+\oplus (v,0,c)_+\oplus (s,c,v)_-\oplus (c,s,v)_-\\
	\text{R untwisted:} & (v,0,0)_+\oplus (0,v,0)_+\oplus (c,s,s)_-\oplus (s,c,s)_-\\
	\text{R $g$-twisted:} & (0,0,c)_+\oplus (v,v,c)_+\oplus (s,s,v)_-\oplus (c,c,v)_-
\end{cases} 
\ee
Let us construct two different duality defects $\CN^{(1)}_g$ and $\CN^{(2)}_g$ for $g$ that explicitly preserve the $\CN=1$ supercurrent \eqref{F24SVir}. As discussed above, the operators $\frac{1}{\sqrt{2}}\hat\CN^{(i)}_g$ must act by an inner automorphism on the algebra $\widehat{so}(16)_1\supset \widehat{so}(8)^a_1\oplus \widehat{so}(8)^b_1$, and by an outer automorphism on $\widehat{so}(8)_1^c$ exchanging the $v$ and $c$ representations. Furthermore, its action on the space of $g$-fixed fields must have order $2$. It is easier to define such operators by describing their action on the $24$ free fermions $\psi^1,\ldots,\psi^{24}$  as follows
\be\label{Ngone} \frac{1}{\sqrt{2}}\hat\CN_g^{(1)}:\begin{cases}\begin{matrix}
	\psi^1\leftrightarrow \psi^2 &  \psi^3\to -\psi^3 & \psi^4\to -\psi^4& \psi^5\to -\psi^5 
			\\
	\psi^9\leftrightarrow \psi^{10}, &\psi^{11}\to -\psi^{11} & \psi^{12}\to -\psi^{12} &\psi^{13}\to -\psi^{13}\\
	\psi^{17}\leftrightarrow \psi^{18}& &&  \end{matrix}\\
    \psi^i\to \psi^i\quad \text{otherwise}
\end{cases}
\ee
\be\label{Ngtwo} \frac{1}{\sqrt{2}}\hat\CN_g^{(2)}:\begin{cases}\begin{matrix}
	\psi^1\leftrightarrow \psi^2, &  & \psi^4\to -\psi^4& \psi^5\to -\psi^5 & \psi^6\to -\psi^6 \ ,
			\\
	\psi^9\leftrightarrow \psi^{10}, &\psi^{11}\to -\psi^{11} & &\psi^{13}\to -\psi^{13} &\psi^{14}\to -\psi^{14} \\
	\psi^{17}\leftrightarrow \psi^{18},& \psi^{19}\to -\psi^{19} & \psi^{20}\to -\psi^{20} & & \end{matrix}\\
    \psi^i\to \psi^i\quad \text{otherwise}
\end{cases}
\ee
Strictly speaking, the $24$ fermions $\psi^1,\ldots,\psi^{24}$ are not fields in $V^{f\natural}$, so we really define $\frac{1}{\sqrt{2}}\hat\CN_g^{(i)}$ as the  automorphisms of $\widehat{so}(8)_1^a\oplus \widehat{so}(8)^b_1\oplus \widehat{so}(8)^c_1$ induced by this action on the fermions. The involutions $ \frac{1}{\sqrt{2}}\hat\CN_g^{(i)}$ are chosen in such a way that the supercurrent \eqref{F24SVir} is invariant. Let us show that these formulae provide well-defined duality defects for $g$ by relating the actions \eqref{Ngone} and \eqref{Ngtwo} to the pairs $(S^{(16)}_\CN,O^{(8)}_\CN)$ in \eqref{O16O8} and to the lattice automorphism \eqref{NgLambda16}.  The  eigenvalues of $\frac{1}{\sqrt{2}}\hat\CN_g^{(1)}$ and $\frac{1}{\sqrt{2}}\hat\CN_g^{(2)}$ on the three spaces of fermions spanned by $\{\psi^1,\ldots,\psi^8\}$, $\{\psi^9,\ldots,\psi^{16}\}$ and $\{\psi^{17},\ldots,\psi^{24}\}$ are reported in table \ref{t:Z2dual_autom}.
\begin{table}[h]\begin{center}
\begin{tabular}{|c|c|c|c|}\hline
	& $\{\psi^1,\ldots,\psi^8\}$ & $\{\psi^9,\ldots,\psi^{16}\}$ &$\{\psi^{17},\ldots,\psi^{24}\}$\\
	\hline
	$\frac{1}{\sqrt{2}}\hat\CN_g^{(1)}$ &  $(+1)^4,(-1)^4$ &  $(+1)^4,(-1)^4$ &  $(+1)^7,(-1)^1$ \\
	$\frac{1}{\sqrt{2}}\hat\CN_g^{(2)}$ &  $(+1)^4,(-1)^4$ &  $(+1)^4,(-1)^4$ &  $(+1)^5,(-1)^3$ \\ \hline
\end{tabular}    
\end{center}
\caption{\small{Eigenvalues of $\frac{1}{\sqrt{2}}\hat\CN_g^{(1)}$ and $\frac{1}{\sqrt{2}}\hat\CN_g^{(2)}$ on the three sets of free fermions.}}\label{t:Z2dual_autom}\end{table}

Recall that the three sets of free fermions transform in the $(s,0,0)$, $(0,s,0)$ and $(0,0,s)$ representations of $\widehat{so}(8)_1^a\oplus \widehat{so}(8)^b_1\oplus \widehat{so}(8)^c_1$. Thus, the action on $\psi^{17},\ldots,\psi^{24}$ provides directly the matrix $O^{(8)}_\CN$ in \eqref{O16O8} with $(O^{(8)}_\CN)^2=1$. Furthermore, the fact that the number of $-1$ eigenvalues on $\psi^{17},\ldots,\psi^{24}$ is odd means that $\det O^{(8)}_\CN=-1$, so that the induced automorphism on $\widehat{so}(8)_1^c$ is outer, and exchanges the $v$ and $c$ representation, as required for the duality defect. We could also consider operators $\frac{1}{\sqrt{2}}\hat\CN_g^{(1)}Q_1$ and $\frac{1}{\sqrt{2}}\hat\CN_g^{(2)}Q_1$ such that the multiplicity of the $(-1)$-eigenvalue on the space generated by  $\psi^{17},\ldots,\psi^{24}$ is $7$ and $5$, respectively. However, since $Q_1$ is trivial on $V^{f\natural}$, we would not get any new duality defects.

   On the other hand, the actions on $\{\psi^1,\ldots,\psi^8\}$ and $\{\psi^9,\ldots,\psi^{16}\}$ as in \eqref{Ngone} or \eqref{Ngtwo} determine elements in $Spin(8)^a/\langle(-1)^a_8\rangle$ and $Spin(8)^b/\langle(-1)^b_8\rangle$, respectively, where $(-1)^a_8$ and $(-1)^b_8$ are the central elements of $Spin(8)^a$ and $Spin(8)^b$ acting trivially on the $s$ representation.
   The fact that the number of $(-1)$-eigenvalues in the spaces spanned by these sets of free fermions  is a multiple of $4$ (see table \ref{t:Z2dual_autom}) ensures that these two involutions in $Spin(8)^a/\langle(-1)^a_8\rangle$ and $Spin(8)^b/\langle(-1)^b_8\rangle$ lift  to elements of order $2$ in the corresponding groups $Spin(8)^a$ and $Spin(8)^b$. On the contrary, if we had chosen the multiplicities of $(-1)$-eigenvalues on $ \{\psi^1,\ldots,\psi^8\}$ and $\{\psi^9,\ldots,\psi^{16}\}$  to be equal to $2\mod 4$, they would lift to elements order $4$ in $Spin(8)$. In general, one can verify that, given an element in $Spin(8)/\langle(-1)_8\rangle$ with eigenvalues $(+1)^4,(-1)^4$ on the $s$ representation, then both its lifts to $Spin(8)$ have the same eigenvalues multiplicities, namely $(-1)^4, (+1)^4$, when acting on the vector representation of $Spin(8)$; the two lifts are related to each other by exchanging the $+1$ and the $-1$ eigenspaces in the $v$ representation. Because they have the same eigenvalues both on the $s$ and on the $v$ representation, it follows that the two lifts are conjugate with each other within $Spin(8)$. 
   
   For each of the actions \eqref{Ngone} and \eqref{Ngtwo}  on $\{\psi^1,\ldots,\psi^{8}\}$ and $\{\psi^9,\ldots,\psi^{16}\}$, let us choose the lifts from $Spin(8)^a/\langle(-1)^a_8\rangle$ and $Spin(8)^b/\langle(-1)^b_8\rangle$ to $Spin(8)^a$ and $Spin(8)^b$. Via the homomorphism $Spin(8)^a\times Spin(8)^b\to Spin(16)$, this pair of lifts is mapped to an involution $S^{(16)}_\CN\in Spin(16)$ in \eqref{O16O8}, determining an inner automorphism of order $2$ on $\widehat{so}(16)_1\supset \widehat{so}(8)_1^a\oplus \widehat{so}(8)_1^b$. This argument completes the proof that the actions \eqref{Ngone} and \eqref{Ngtwo} determine two different pairs $(S^{(16)}_\CN,O^{(8)}_\CN)$ of automorphisms of $\widehat{so}(16)_1\supset \widehat{so}(8)_1$ preserving the $\CN=1$ supercurrent and satisfying all required properties for $\frac{1}{\sqrt{2}}\hat\CN_g$.
   
   As stressed in the previous section, the $\CN=1$-preserving duality defects $\CN_g$ constructed here are not unique. In fact, starting from any one of them, say $\CN_g^{(1)}\equiv (S^{(16)}_\CN,O^{(8)}_\CN)$, one can obtain all the other ones, including $\CN_g^{(2)}$, by composing $(h_1S^{(16)}_\CN h_2,h_1O^{(8)}_\CN h_2)$ with elements $h_1,h_2\in C_{Co_0}(g)/\langle g\rangle$, such that $h_1S^{(16)}_\CN h_2$ and $h_1O^{(8)}_\CN h_2$ are still involutions. On the other hand, we can argue that all such involutions $h_1S^{(16)}_\CN h_2$ and $h_1O^{(8)}_\CN h_2$ have the same eigenvalues on all the $g$-invariant representations of $Spin(16)\times O(8)$ as either $\CN^{(1)}_g$ or $\CN^{(2)}_g$, and therefore they are related to one of these two defects by conjugation within $Spin(16)\times Spin(8)$. For example, for any duality defect $\CN$, the corresponding involution $S^{(16)}_\CN$ must have eigenvalues $(+1)^8,(-1)^8$ in the vector representation of $Spin(16)$ in order to match the expected action \eqref{Z2_lattice_eigen} on $\Lambda^g\otimes \RR$.  Such an element of $Spin(16)$ necessarily acts on both the $s$ and $c$ representations with half eigenvalues $+1$ and half $-1$.  As for $O^{(8)}_\CN$, since the action on the $s$-representation must be an involution with negative determinant,  the multiplicity of $-1$ eigenvalues must be either $1$, $3$, $5$, or $7$. Notice that $(S^{(16)}_\CN,O^{(8)}_\CN)$ and $(\xi_{16}S^{(16)}_\CN,\xi_8O^{(8)}_\CN)$ act in the same way on all $g$-invariant fields, where $\xi_{16}\in Spin(16)$ and $\xi_{8}\in Spin(8)$ are the central elements acting non-trivially on the $s$ and $c$ representations of the respective spin group. Thus, without loss of generality, we can assume that the number of $-1$-eigenvalues of  $O^{(8)}_\CN$ on the $s$-representation is either $1$ or $3$, the same as for $\CN_g^{(1)}$ and $\CN_g^{(2)}$, respectively.     This analysis of the $(S^{(16)}_\CN,O^{(8)}_\CN)$ eigenvalues implies that all possible duality defects $\CN_g$ for $g$ are related to either $\CN_g^{(1)}$ or $\CN_g^{(2)}$ by conjugation within $Spin(16)\times SO(8)$.
    As a consequence, their twining partition functions and the fusion category they generate are the same as $\CN^{(1)}_g$ or $\CN^{(2)}_g$. By a computer calculation, we verified that the matrix $O^{(16)}_\CN$ in  eq.\eqref{O16Lattice} corresponds to a defect with the same eigenvalues as $\CN^{(2)}_g$. On the other hand, a defect with the same eigenvalues as $\CN^{(1)}_g$ can be represented by a matrix $h_1O^{(16)}_\CN$, where $h_1\in C_{Co_0}(g)$ is given by
   \be h_1=(\diag(1,1,1,1,1,1,1,1,1,1,-1,-1,1,1,-1,-1),\diag(1,1,-1,-1,1,1,-1,-1))\ ,
   \ee where the first component is an automorphism of the sublattice $\Lambda_{16}\subset \Lambda$ and the second an automorphism of its orthogonal complement $\Lambda_{16}^\perp \cap \Lambda$.

   Finally, let us determine the subspace of fields of $V^{f\natural}$ that is preserved by each of the two defects $\CN_g^{(1)}$ and $\CN^{(2)}_g$.
   The defect $\CN^{(1)}_g$ preserves the subalgebras $\widehat{so}(4)^{a+}_1\oplus \widehat{so}(4)^{a-}_1\subset \widehat{so}(8)^{a}_1$, $\widehat{so}(4)^{b+}_1\oplus \widehat{so}(4)^{b-}_1\subset \widehat{so}(8)^{b}_1$, and $\widehat{so}(7)^{c}_1\oplus \subset \widehat{so}(8)^{c}_1$. It also preserves $32$ out of $64$ currents in the $(v,v,0)$ representation of $\widehat{so}(8)_1^a\oplus \widehat{so}(8)^b_1\oplus \widehat{so}(8)^c_1$. In particular $16$ of them enhance $\widehat{so}(4)^{a+}_1\oplus \widehat{so}(4)^{b+}_1$ to $\widehat{so}(8)^{+}_1$, while  the other $16$ enhance $\widehat{so}(4)^{a-}_1\oplus \widehat{so}(4)^{b-}_1$ to $\widehat{so}(8)^{-}_1$. Furthermore, $\CN^{(1)}_g$ also preserves the difference between the Sugawara stress-tensor of $\widehat{so}(8)^{c}_1$ and the one of its subalgebra $\widehat{so}(7)^{c}_1$; this is a weight $2$ operator generating a copy of the $Vir_{c=1/2}$ algebra. Altogether, the bosonic subalgebra preserved by $\CN^{(1)}_g$ is
   \be V(\widehat{so}(8)^{+}_1\oplus \widehat{so}(8)^{-}_1\oplus \widehat{so}(7)^{c}_1)\otimes Vir_{c=1/2}\ . 
   \ee
   
   A similar argument shows that $\CN^{(2)}_g$ preserves the affine algebra
   \be \widehat{so}(8)^{+}_1\oplus \widehat{so}(8)^{-}_1\oplus \widehat{so}(5)^{c}_1\oplus \widehat{so}(3)^{c}_1
   \ee whose Sugawara stress-tensor coincides with the stress tensor of $V^{f\natural}$.

As a final remark, we notice that the symmetry $g$ and the duality defects $\hat\CN^{(i)}_g$ are well-defined also on the theory $F_{24}/\langle Q_1\rangle\cong V^{fE_8}$. On the original theory $F(24)$, $g$ corresponds to the symmetry acting trivially on the NS field, and by $-1$ on the Ramond sector, so that the orbifold $F(24)/\langle g\rangle$ is isomorphic to $F(24)$ but with the opposite choice of fermion number on the Ramond sector.

 \subsection{Duality defect partition functions and twining genera}

In this section, we compute the twining partition functions and twining genera for the defects $\CN_g^{(1)}$ and $\CN_g^{(2)}$ defined in the previous section
 
Using the formulas \eqref{ch_0_so(n)}, \eqref{ch_v_so(n)} and \eqref{ch_s_so(n)} and the decompositions in Table \ref{tab:2} in terms of the $g$-preserved affine algebra $\widehat{so}(16)_1\oplus \widehat{so}(8)_1$, it is straightforward to verify that the partition function $\CZ(V^{f\natural},\tau):=(Z^{+}_{\NS},Z^{-}_{\NS}, Z^{+}_{\R},Z^{-}_{\R})^t$ of $V^{f\natural}$, defined as in \eqref{ZLtwinNS} and \eqref{ZLtwinR} for $\CL$ the identity defect, can be reproduced by the formulas
\bea\label{eq:ZdecompNS}
Z^\pm_{\NS}(\tau)&=&\ch_{[0,0]}(\tau)+\ch_{[v,v]}(\tau)\pm\ch_{[s,s]}(\tau)\pm\ch_{[c,c]}(\tau)\\\label{eq:ZdecompR}
Z^\pm_{\R}(\tau)&=&\ch_{[v,0]}(\tau)+\ch_{[0,v]}(\tau)\pm\ch_{[c,s]}(\tau)\pm\ch_{[s,c]}(\tau)
\eea
where we have defined $\ch_{[i,j]}:=\ch^{(16)}_i(\tau)\ch^{(8)}_j(\tau)$ for $i,j\in\{0,v,s,c\}$.
 We now consider the defect twined partition functions $\CZ^\CN(V^{f\natural},\tau):=(Z^{\CN,+}_{\NS},Z^{\CN,+}_{\NS}, Z^{\CN,+}_{\R},Z^{\CN,-}_{\R})^t$ defined in equations \eqref{ZLtwinNS} and \eqref{ZLtwinR} 
 where $\CN=\CN_g^{(1)}$ or $\CN_g^{(2)}$. The duality defect $\CN$ annihilates all fields that are in the $v$ or $c$ representations of $\widehat{so}(8)_1$, so we only need to consider contributions from the $(0,0)$, $(s,s)$, $(v,0)$, and $(c,s)$ representations of $\widehat{so}(16)_1\oplus \widehat{so}(8)_1$. For $\CN$ defined by a pair $(S^{(16)}_{\CN},O^{(8)}_{\CN})$ as described in the previous section (see eq.\eqref{O16O8}), the corresponding defect-twined partition functions will schematically take the form 
 \bea\label{eq:ZdefectNS}
Z^{\CN,\pm}_{\NS}(\tau)&=&\sqrt 2\ch^\CN_{[0,0]}(\tau)\\\label{eq:ZdefectR}
Z^{\CN,\pm}_{\R}(\tau)&=&\sqrt 2 \ch^\CN_{[v,0]}(\tau)
\eea
where we have introduced $\ch_{[i,j]}^{\CN}(\tau):=\sqrt 2 \ch^{(16),S_{\CN}}_i(\tau)\ch^{(8),O_{\CN}}_j(\tau)$ with  \be \ch^{(16),S_\CN}_\rho(\tau):= \Tr_{\rho^{(16)}} S^{(16)}_\CN q^{L_0-1/3} \, , \quad \quad \ch^{(8),O_\CN}_\rho(\tau):= \Tr_{\rho^{(8)}} O^{(8)}_\CN q^{L_0-1/6} \, . \ee  

To understand why all the other sectors do not contribute to the defect-twining partition functions we need to explicitly compute the characters. Let us consider the general case of an $\hat{so}(2n)_1$ algebra described in terms $2n$ free fermions. And let $\CO \in SO(2n)$ be a $2n \times 2n$ matrix acting on the free fermions through the eigenvalues $(\zeta_{\CN,i},\zeta_{\CN,i}^{-1})$, $i=1,\ldots,n$, with $\zeta_{\CN,i}=e^{-2\pi i \beta_{\CN,i}}$ for $\beta_{\CN,i} \in [0,{1\over 2}]$. The twined characters in the different representations can be expressed as
\bea
\ch^{(2n),\CO}_0(\tau)&=&{1\over 2 \eta (\tau)^n} \left(\prod_{i=1}^n{\theta_3(\tau,\beta_{\CN,i})}+\prod_{i=1}^n{\theta_4(\tau,\beta_{\CN,i})}\right) \label{eq:TrO0}\\
\ch^{(2n),\CO}_v(\tau)&=&{1\over 2 \eta (\tau)^n}\left(\prod_{i=1}^n{\theta_3(\tau,\beta_{\CN,i})}-\prod_{i=1}^n{\theta_4(\tau,\beta_{\CN,i})}\right) \label{eq:TrOv}\\ \label{eq:TrOcs}
\ch^{(2n),\CO}_s(\tau)&=& \ch^{(2n),\CO_{\CN}}_c(\tau)=\pm {1\over 2}\prod_{i=1}^n{\theta_2(\tau,\beta_{\CN,i})\over\eta(\tau)^n}.
\eea
 Now, in our case of interest, we consider the algebra $\hat{so}(16)_1$ and the matrix $\CO = \CO^{(16)}_{\mathcal{N}}$ is the element of $SO(16)$ corresponding to the automorphism $S^{(16)}_{\CN}$ described in the previous section, for both $\CN^{(1)}_g$ and $\mathcal{N}^{(2)}_g$, under the projection $Spin(16) \to SO(16)$. $\CO^{(16)}_{\CN}$ acts on the fermions with eight eigenvalues $+1$ and eight eigenvalues $-1$. Equivalently we have $\beta_{\CN,i}=0$ for $i=1,\ldots,4$ and $\beta_{\CN,i}=1/2$ for $i=5,\ldots,8$, for both $O^{(16)}_{\CN^{(1)}_g}$ and $O^{(16)}_{\CN^{(2)}_g}$. The trace in equation \eqref{eq:TrOcs} then vanishes due to the factors $(e^{-\pi i \beta_{\CN,i}}-e^{\pi i \beta_{\CN,i}})$ in $\theta_2(\tau,\beta_{\CN,i})$ for $i=5,\ldots,8$, which arise from fermion zero modes. Moreover, for this set of eigenvalues it is clear that $\ch^{(16),\CO_{\CN}}_v(\tau)=0$ since the $\theta_3$ and $\theta_4$ contributions cancel exactly. Consequently, even the partition functions $Z^{\CN^{(1)}_g,\pm}_{\R}(\tau)=Z^{\CN^{(2)}_g,\pm}_{\R}(\tau)=0$ vanish identically.

Therefore, we are only left to determine the trace of $O^{(8)}_\CN$ on the $0$ representation of $\widehat{so}(8)_1$. There are two cases of outer automorphism of $\widehat{so}(8)_1$ as described in the previous section: $O^{(8)}_{\CN^{(1)}_g}$ preserves an $\widehat{so}(7)_1\oplus \widehat{so}(1)_1$, and $O^{(8)}_{\CN^{(2)}_g}$ preserves an $\widehat{so}(5)_1\oplus \widehat{so}(3)_1$.  

Through the fermionic representation of the $\hat{so}(8)_1$ algebra we can compute the corresponding twined traces. In the case of $\CO^{(8)}_{\CN_g^{(1)}}$ there is one eigenvalue equal to $-1$ and seven eigenvalues equal to $+1$, which leads to
\bea
\ch^{(8),O_{\CN^{(1)}_g}}_0(\tau)&=&{\sqrt{\theta_3(\tau)\theta_4(\tau)} (\theta_3^3(\tau) + \theta_4^3(\tau))\over 2\eta^4(\tau)}\\
&=& q^{-1/6}+14q^{5/6}+36 q^{11/6}+120 q^{17/6}+\ldots,
\eea
For $\CO^{(8)}_{\CN_g^{(3)}}$, there are three eigenvalues equal to $-1$ and five eigenvalues equal to $+1$. The corresponding character takes the form
\bea
\ch^{(8),O_{\CN^{(2)}_g}}_0(\tau)&=&{\theta_3(\tau)\theta_4(\tau) \sqrt{\theta_3(\tau)\theta_4(\tau)} (\theta_3(\tau) + \theta_4(\tau)) \over 2\eta^4(\tau)}\\
&=& q^{-1/6}-2q^{5/6}+4 q^{11/6}-8q^{17/6}+\ldots.
\eea

Combining these characters, we compute the defect-twined partition function in \eqref{eq:ZdecompNS} for $\CN^{(1)}_g$
\bea
Z^{\CN^{(1)}_g,\pm}_{\NS}(\tau)&=&\sqrt 2\ch^{(16),O_{\CN}}_0(\tau)\ch^{(8),\CO_{\CN_g^{(1)}}}_0(\tau)
\\ &=&\sqrt 2 {(\theta_3^3 + \theta_4^3)(\theta_3\theta_4)^{9/2}\over 2\eta^{12}}\\
&=&\sqrt 2 \left ({1\over \sqrt q} + 6 q^{1/2} -48 q^{3/2} + 160 q^{5/2} - 447 q^{7/2} + \ldots\right ) \, ,
\eea
and for $\CN^{(2)}_g$
\be
Z^{\CN^{(2)}_g,\pm}_{\NS}(\tau)&=&\sqrt 2\ch^{(16),O_{ \CN}}_0(\tau)\ch^{(8),\CO_{\CN_g^{(2)}}}_0(\tau)
\\
&=&\sqrt 2 {(\theta_3 + \theta_4)(\theta_3\theta_4)^{11/2}\over 2\eta^{12}}\\
&=&\sqrt 2 \left ({1\over \sqrt q} - 10 q^{1/2} +48 q^{3/2} - 160 q^{5/2} + 449 q^{7/2} + \ldots\right )
\ee
Using the modular properties of $\eta$ and $\theta_i$ under the S-transformation $\tau\to -1/\tau$ (see appendix \ref{a:minimal}), we find the following defect twisted partition functions
\bea
Z^\pm_{\CN^{(1)}_g,\NS}(\tau)&=&\sqrt 2 {(\theta_3^3 + \theta_2^3)(\theta_3\theta_2)^{9/2}\over 2\eta^{12}}\\
&=&16 q^{1/16}+128 q^{7/16}+240 q^{9/16}+1152 q^{15/16}+ 1824 q^{17/16} + \ldots
\\
Z^\pm_{\CN^{(2)}_g,\NS}(\tau)&=&\sqrt 2 {(\theta_3 + \theta_2)(\theta_3\theta_2)^{11/2}\over 2\eta^{12}}\\
&=&32 q^{3/16}+64 q^{5/16}+416 q^{11/16}+704 q^{13/16}+2848 q^{19/16}+\ldots.
\eea

Note that these both have positive integral degeneracies, as expected. As explained in \cite{Chang_2019}, the spin of the $\CN$-twisted sector states is directly related to the parameter $\epsilon\in \{\pm1\}$ determining the corresponding Tambara-Yamagami category TY($\ZZ_2$,$\epsilon$). From the spin selection rules summarized in table \ref{t:spinselectZ2}, we see that $\CN^{(1)}_g$ generates a TY($\ZZ_2$,$+1$) and $\CN^{(2)}_g$ generates a TY($\ZZ_2$,$-1$).

\medskip
 
Finally, let us compute the defect twining genera \eqref{Vfnat:graded_part_func} for the $\ZZ_2$ duality defects. We begin with the presentation of $V^{f\natural}$ in terms of $\widehat{so}(16)_1\oplus \widehat{so}(8)_1$ modules, as reported in table \ref{tab:2}. The Ramond sector (denoted $V^{f\natural}_{tw}$) is given by $(v,0) \oplus (0,v) \oplus (c,s) \oplus (s,c)$, where the $(-1)^F$ operator acts with eigenvalue +1 on $(v,0)$ and $(0,v)$ and $-1$ on $(c,s)$ and $(s,c)$. 
As reviewed in section \ref{s:conj}, if we select a $c=6$ $\mathcal N=4$ SCA, where the $\widehat{su}(2)_1$ is contained in the $\widehat{so}(16)_1$, the graded trace \eqref{Vfnat:graded_part_func} should reproduce the elliptic genus of K3.

Let us choose our four plane to lie in an $\widehat{su}(2)_1 \subset \widehat{so}(4)_1 \subset \widehat{so}(16)_1$ preserved by $g$.
Therefore, we can grade the $\widehat{so}(16)_1$ by the Cartan of the $\widehat{su}(2)_1$ such that $\ch^{(16)}_\rho(\tau,z):= \Tr_{\rho^{(16)}}y^{J_0^3}q^{L_0-1/3}$ for $\rho \in \{0,v,s,c\}$. The explicit form of the corresponding characters is given in \eqref{ch04plane}--\eqref{chsc4plane}.
The formula for the elliptic genus \eqref{Vfnat:graded_part_func} can be expressed in terms of these characters as
 \begin{equation}
 \begin{split}
\label{newelliptic}\phi(V^{f\natural},\tau,z)=&\ch^{(16)}_v(\tau,z)\ch^{(8)}_0(\tau)+\ch^{(16)}_0(\tau,z)\ch^{(8)}_v(\tau)-\ch^{(16)}_c(\tau,z)\ch^{(8)}_s(\tau)-\ch^{(16)}_s(\tau,z)\ch^{(8)}_c(\tau)\\
 =& {2\over y} + 20 + 2y +\left (\frac{20}{y^2}-\frac{128}{y}+216-128 y +20 y^2\right )q + \ldots
\end{split}
\end{equation}
Now we consider the corresponding defect twined genera $\phi^\CN(V^{f\natural},\tau,z)$ for $\CN=\CN^{(1)}_g,\CN^{(2)}_g$. 
The formula will be given by \eqref{eq:ZdefectR} after including the $y^{J_0^3}$ grading. In this case, the twined $\widehat{so}(16)_1$ character will no longer vanish; it is given by
\be
\ch^{(16),S_\CN}_v(\tau,z):=\Tr_v(S_\CN y^{J^3_0}q^{L_0-1/3})= {1\over 2}{\left(\theta_3^2(\tau,z)\theta_3^2(\tau)\theta_4^4(\tau) - \theta_4^2(\tau,z)\theta_4^2(\tau)\theta_3^4(\tau)\right )\over \eta^8(\tau)},
\ee
and thus we find
\bea\nonumber
\phi^{\CN^{(1)}_g}(V^{f\natural},\tau,z)&=&\sqrt 2\ch^{(16),S_{ \CN^{(1)}_g}}_v(\tau,z)\ch^{(8),O_{\CN^{(1)}_g}}_0(\tau)\\\label{twinZ2dual+}
&=&\sqrt 2 {\left ( \theta_3^2(\tau,z)\theta_4^2 - \theta_4^2(\tau,z)\theta_3^2\right )(\theta_3^3 + \theta_4^3)(\theta_3\theta_4)^{5/2}\over 4\eta^{12}(\tau)}
\\\nonumber
&=& \sqrt 2\left (\frac{2}{y}-4 + 2y+\left (-\frac{4}{y^2}+\frac{36}{y}-64+36 y -4 y^2\right ) q + \ldots\right )
\eea
and
\bea\nonumber
\phi^{\CN^{(2)}_g}(V^{f\natural},\tau,z)&=&\sqrt 2\ch^{(16),S_{\CN^{(2)}_g}}_v(\tau,z)\ch^{(8),O_{\CN^{(2)}_g}}_0(\tau)\\\label{twinZ2dual-}
&=&\sqrt 2 {\left ( \theta_3^2(\tau,z)\theta_4^2 - \theta_4^2(\tau,z)\theta_3^2\right )(\theta_3 + \theta_4)(\theta_3\theta_4)^{7/2}\over 4\eta^{12}(\tau)}
\\ \nonumber
&=& \sqrt 2\left (\frac{2}{y}-4 + 2y+\left (-\frac{4}{y^2}+\frac{4}{y}+4 y -4 y^2\right ) q + \ldots\right ).
\eea

In the next section we will construct a $\ZZ_2$ duality defect in a K3 NLSM whose defect twined elliptic genus matches $\phi^{\CN^{(2)}_g}(V^{f\natural},\tau,z)$.

\section{A K3  model with $\ZZ_2$ duality defect preserving $\CN=(4,4)$} \label{s:Z2dualityK3}

In this section we construct a K3 NLSM with a $\mathbb Z_2$ Tambara-Yamagami fusion category symmetry which preserves the $\CN=(4,4)$ superconformal algebra. Our construction generalizes the case of the free boson on a circle $S^1$ that was considered in, e.g.,  \cite{Thorngren:2021yso}, and that we now briefly review.

 Suppose that the target space circle $S^1$ has radius $\sqrt{2}R_{sd}$, where $R_{sd}$ is the self-dual radius under T-duality (at which the $u(1)$ current algebra is enhanced to $\widehat{su}(2)_1$).
In \cite{Thorngren:2021yso}, it was shown that, at this particular radius, the $S^1$ model admits some topological defects generating a $\ZZ_2$ Tambara-Yamagami category. In particular, the $\ZZ_2$ invertible defect in this category corresponds to a half-period shift of the free boson along the circle. The orbifold of the $S^1$ model by such a $\ZZ_2$ symmetry gives again a free boson on a circle, but with half of the original radius $R'=R_{sd}/\sqrt{2}$. Therefore, the original model and its $\ZZ_2$ orbifold are equivalent by T-duality. This implies the existence of a duality defect, and also suggests how such a defect acts on the operators of this theory.

Inspired by this example, we will now construct such a K3 model with $TY(\mathbb{Z}_2)$ topological defects as the orbifold $T^4/\ZZ_2$ of a particular sigma model on $T^4$. Let $\calR$ be the $\ZZ_2$ reflection symmetry of $T^4$ acting by $X^k\to -X^k$ on the $T^4$ coordinates and by $\psi^k\to -\psi^k$, $\bar\psi^k\to -\bar\psi^k$ on their fermionic superpartners. Let $g$ be a half-period shift along one of the cycles of the torus.   The orbifold of any $T^4$ model by a half-period shift $g$ is again a sigma model on a torus $T^4/g$, obtained by identifying the points under the shift $g$. Furthermore, $g$ commutes with $\calR$, and gives rise to a $\ZZ_2$ symmetry with Frame shape $1^82^8$ on the K3 model $T^4/\calR$.  

 Now, the idea is to choose the geometry of the torus in such a way that, for a suitable half-period shift $g$, the original $T^4$ and the quotient $T^4/g$ are related by T-duality in the $4$-directions. This means that, as a CFT, the sigma model on such a  $T^4$ is self-orbifold with respect to the group $\langle g\rangle\cong \ZZ_2$. It is clear that the corresponding duality defect $\CN$ acts in the same way as  T-duality when moved across any $g$-invariant local operator. Because $4$-fold T-duality acts trivially on the $\CN=(4,4)$ superconformal algebra,\footnote{$4$-fold T-duality acts trivially on $\partial X^k$, $\psi^k$ and by $\bar\partial X^k\to -\bar\partial X^k$, $\bar\psi^k\to-\bar\psi^k$. The $\CN=4$ holomorphic and anti-holomorphic algebras can be constructed, respectively, in terms of $\psi^j\partial X^k$, $\psi^j\psi^k$ and $\bar\psi^j\bar\partial X^k$, $\bar\psi^j\bar\psi^k$, and are therefore invariant.} it follows that $\CN$ preserves such an algebra. Finally, both $g$ and $\CN$ can be chosen so that they commute with the $\ZZ_2$ reflection symmetry $\calR$. Therefore, they give rise to topological defects in the K3 model $T^4/\calR$, generating a Tambara-Yamagami $\ZZ_2$ category that preserves the $\CN=(4,4)$ superconformal algebra.

\subsection{The $\mathbb Z_2$ duality defect}\label{s:K3Z2duality}

Let us now describe the sigma model on $T^4$ in more detail. In order to define such a model, it is sufficient to specify its set of vertex operators $\CV_{(\lambda_L,\lambda_R)}(z,\bar z)$, where $\lambda_L\equiv (\lambda_L^1,\ldots,\lambda_L^4)$ and $\lambda_R\equiv (\lambda_R^1,\ldots,\lambda_R^4)$ are the vectors of eigenvalues under the $U(1)$ currents $i\partial X^k$ and $i\bar\partial X^k$, respectively. The conformal weights $(h_L,h_R)$ of $V_{(\lambda_L,\lambda_R)}(z,\bar z)$ are given by
\be h_L=\frac{\lambda_L\cdot\lambda_L}{2}\ ,\qquad h_R=\frac{\lambda_R\cdot\lambda_R}{2}\ .
\ee The set of allowed vectors $(\lambda_L,\lambda_R)$, with quadratic form
\be \lambda_L\cdot\lambda_L-\lambda_R\cdot\lambda_R
\ee forms an even self-dual lattice $\Gamma^{4,4}$ of signature $(4,4)$, the Narain lattice of winding-momenta.

Let us now consider the $T^4$ target space to be given by the product of four mutually orthogonal circles, with vanishing B-field; three such circles are at the self-dual radius $R_i=R_{sd}$, $i=2,3,4$, while the last $S^1$ has radius $R_1=\sqrt{2}R_{sd}$.

For this geometry and B-field, the Narain lattice of winding-momenta is given by $(\lambda_L,\lambda_R)$, with
\be \lambda_L=(\frac{n_1}{2}+w_1,\frac{n_2+w_2}{\sqrt{2}},\frac{n_3+w_3}{\sqrt{2}},\frac{n_4+w_4}{\sqrt{2}}),\quad \lambda_R=(\frac{n_1}{2}-w_1,\frac{n_2-w_2}{\sqrt{2}},\frac{n_3-w_3}{\sqrt{2}},\frac{n_4-w_4}{\sqrt{2}}),
\ee where $w\equiv (w_1,\ldots,w_4)\in \ZZ^4$ and $n\equiv (n_1,\ldots,n_4)\in \ZZ^4$ are, respectively, the winding and momentum quantum numbers. 

We consider the symmetry $g$ corresponding to a half-period shift along the first circle. This means that it acts trivially on all $\partial X^k$, $\bar\partial X^k$ and their fermionic superpartners $\psi^k$, $\bar\psi^k$, while it multiplies the vertex operators $\CV_{(\lambda_L,\lambda_R)}\equiv \CV_{n,w}$ by a sign
\be g(\CV_{n,w})=(-1)^{n_1}\CV_{n,w}\ .
\ee This $g$ is an element in the $U(1)^4\times U(1)^4$ group of symmetries acting by $\CV_{n,w}\mapsto e^{2\pi i(\alpha\cdot  n-\beta \cdot w)}\CV_{n,w}$, with $\alpha,\beta\in (\RR/\ZZ)^4$. In particular, $g$ corresponds to $\alpha_g=(1/2,0,0,0)$ and $\beta_g=(0,0,0,0)$. 

The orbifold $T^4/\langle g\rangle$ is obtained by first including the $g$-twisted sector that contains 
vertex operators with fractional winding $w\in \alpha_g+\ZZ^4$, i.e. with $w_1\in \frac{1}{2}+\ZZ$ and $w_2,w_3,w_4\in \ZZ$; and then by projecting onto the subspace of (untwisted and $g$-twisted) vertex operators with even $n_1\in 2\ZZ$. Therefore, by defining $\tilde n_1:=n_1/2$, $\tilde w_1:=2w_1$, the Narain lattice in the orbifold model is given by
\be (\lambda_L,\lambda_R)=(\tilde{n}_1+\frac{\tilde w_1}{2},\frac{n_2+w_2}{\sqrt{2}},\frac{n_3+w_3}{\sqrt{2}},\frac{n_4+w_4}{\sqrt{2}},\ \tilde{n}_1-\frac{\tilde w_1}{2},\frac{n_2-w_2}{\sqrt{2}},\frac{n_3-w_3}{\sqrt{2}},\frac{n_4-w_4}{\sqrt{2}}),
\ee
with $\tilde n_1,\tilde w_1,n_i,w_i\in \ZZ$.
It is clear that the orbifold model $T^4/\langle g\rangle$ is equivalent to the original one via T-duality. Therefore, one can define a duality defect $\CN$ whose corresponding operator $\hat\CN$ acts in the same way as $4$-fold T-duality on the $g$-invariant operators:
\be\label{T4Z2duality}
\hat\CN:\begin{cases} \partial X^k,\psi^k\to \sqrt{2}\partial X^k,\sqrt{2}\psi^k\ ,\\
\bar \partial X^k,\bar\psi^k\to -\sqrt{2}\bar\partial X^k,-\sqrt{2}\bar\psi^k\ ,\\
\CV_{n,w}\to 0 & \text{if }n_1\text{ is odd},\\
\CV_{n,w}\to (-1)^{n\cdot w} \sqrt{2} \CV_{\tilde w,\tilde n} & \text{if }n_1\text{ is even},
\end{cases}
\ee where $\tilde w=(2w_1,w_2,w_3,w_4)$ and $\tilde n=(n_1/2,n_2,n_3,n_4)$.
As discussed above, such an action preserves the $\CN=(4,4)$ superconformal algebra. Furthermore, both $\hat\CN$ and $\hat g$ commute with the $\ZZ_2$ symmetry $\calR$, acting by
\be\label{T4refl} \calR:\begin{cases}
 \partial X^k,\psi^k\to -\partial X^k,-\psi^k\ ,\\
\bar \partial X^k,\bar\psi^k\to -\bar\partial X^k,-\bar\psi^k\ ,\\
\CV_{n,w}\to \CV_{-n,-w}\ .
\end{cases}
\ee
Therefore, $g$ and $\CN$ should induce topological defects with the same fusion rules in the K3 model obtained as the orbifold $T^4/\langle \calR\rangle$. 
Let us determine the action of the corresponding operators $g$ and $\hat\CN$ on this K3 model. On the untwisted sector, $\hat\CN $ and $g$ are obviously determined by their action on the torus model.

Let us consider the (unprojected) $\calR$-twisted sector $\Hh_{\calR}$. When acting on any $\calR$-twisted state $|\theta\rangle\in \Hh_{\calR}$, the operators $\partial X^k(z)$ and $\bar\partial X^k(\bar z)$ have half-integral modes; in particular, there are no zero modes for these currents. On the other hand, the fermionic operators $\psi^k(z)$ and $\tilde\psi^k(\bar z)$ are integral moded on the NS-NS $\calR$-twisted sector and half-integral moded on the R-R $\calR$-twisted sector. We define the $\calR$-twisted ground states $|\theta_{gr}\rangle\in \Hh_{\calR}$ to be the states annihilated by all (bosonic or fermionic) positive modes.

The vertex operators $\CV_\lambda(z,\bar z)$, $\lambda\in \Gamma^{4,4}_{w-m}$ are given as usual as
\be \CV_\lambda(z,\bar z)=c_\lambda :\! e^{i\lambda\cdot X}\!:(z,\bar z)\ .
\ee The operators $c_\lambda$, $\lambda\in \Gamma^{4,4}$ obey the algebra
\be c_\lambda c_\mu =\epsilon(\lambda,\mu)c_{\lambda+\mu}\ ,
\ee where $\epsilon(\lambda,\mu)\in \{\pm 1\}$ are such that
\be \epsilon(\lambda,\mu)\epsilon(\lambda+\mu,\nu)=\epsilon(\lambda,\mu+\nu)\epsilon(\mu,\nu),\qquad \epsilon(\lambda,\mu)=(-1)^{\lambda\cdot\mu}\epsilon(\mu,\lambda)\ .
\ee The operators $c_\lambda$ are necessary in order for the OPE of the vertex operators $\CV_\lambda$ to be local.

When acting on a twisted ground state, one has
\be \lim_{z\to 0} \CV_\lambda(z,\bar z)|\theta_{gr}\rangle =c_\lambda |\theta_{gr}\rangle\ .
\ee Therefore, the space of $\calR$-twisted ground states is a representation of the algebra of the operators $c_\lambda$.  
We can take a basis of $\Gamma^{4,4}$ given by vertex operators carrying a single unit of either winding or momentum along one of the circles. We denote by $c_{n_1},\ldots,c_{n_4},c_{w_1},\ldots,c_{w_4}$ the corresponding operators, obeying
\be c_{n_i}^2=1=c_{w_i}^2\ ,\quad c_{w_i}c_{n_j}=
(-1)^{\delta_{ij}}c_{n_j}c_{w_i}\ ,\quad c_{n_i}c_{n_j}=c_{n_j}c_{n_i}\ ,\quad c_{w_i}c_{w_j}=c_{w_j}c_{w_i},\ee with $c_{n_i}^2=1=c_{w_i}^2$. For any other $\lambda\in \Gamma^{4,4}$, corresponding to $n_1,\ldots,n_4,w_1,\ldots,w_4\in \ZZ$, we can simply set $c_\lambda:=c_{n_1}^{n_1}\cdots c_{n_4}^{n_4}c_{w_1}^{w_1}\cdots c_{w_4}^{w_4}$. An irreducible representation of this algebra is $16$-dimensional, and we can take a basis $|\theta_1,\ldots,\theta_4\rangle$, $\theta_i\in \{\pm 1\}$, of eigenvectors of $c_{w_1},\ldots, c_{w_4}$:
\be c_{w_i}|\theta_1,\ldots,\theta_4\rangle=\theta_i |\theta_1,\ldots,\theta_4\rangle\ ,
\ee while the operators $c_{n_i}$ act by 
\be c_{n_1}|\theta_1,\theta_2,\theta_3,\theta_4\rangle=|-\theta_1,\theta_2,\theta_3,\theta_4\rangle\ ,\qquad \ldots\qquad c_{n_4}|\theta_1,\theta_2,\theta_3,\theta_4\rangle=|\theta_1,\theta_2,\theta_3,-\theta_4\rangle\ .
\ee The space of R-R $\calR$-twisted ground states is therefore $16$-dimensional and carries such a representation of $c_\lambda$. On the other hand, the NS-NS $\calR$-twisted ground states form a $16^2$-dimensional space, since one needs to tensor with a $16$-dimensional irreducible representation of the Clifford algebra of the zero modes of the $8$ fermions $\psi^k(z)$, $\tilde\psi^k(\bar z)$. The whole $\calR$-twisted sector is obtained by acting on the ground states with the negative modes of the bosonic and fermionic oscillators.

The symmetry $g$ acts on the $\calR$-twisted sector by a unitary operator $\rho(g)$ such that $\rho(g)^2=1$. It must commute with all oscillators, and act on the operators $c_{n_i}$, $c_{w_i}$ as
\be \begin{aligned}
& \rho(g)c_{n_1}\rho(g)=-c_{n_1}\qquad \rho(g)c_{n_i}\rho(g)=c_{n_i}\ ,\quad i=2,3,4 \\
&\rho(g)c_{w_j}\rho(g)=c_{w_j}\ ,\qquad j=1,\ldots,4\ . \\ \end{aligned}
\ee These conditions are satisfied by
\be\label{rhog} \rho(g)\text{(oscillators)}|\pm,\theta_2,\theta_3,\theta_4\rangle=\pm \text{(oscillators)}|\pm,\theta_2,\theta_3,\theta_4\rangle\ ,
\ee and we take this as the definition of $\rho(g)$. 
The other possibility is to define $\rho(g)$ to act  on $\Hh_\calR$ with the opposite signs with respect to \eqref{rhog}. The two possibilities are related by composition with the quantum symmetry $\CQ$. In fact, the action of $g$ on the torus model $T^4$ induces two different symmetries on $T^4/\langle \calR\rangle$, and it is just a matter of convention which one we choose to call $g$ and which one $g\CQ$. 

The duality defect $\hat\CN$ must annihilate all states that are not $g$-invariant, so that
\be \hat\CN \text{(oscillators)}|-,\theta_2,\theta_3,\theta_4\rangle=0\ .
\ee
Furthermore, it must satisfy
\be \hat\CN c_{n_i}=c_{w_i}\hat\CN\ ,\qquad i=2,\ldots,4\ ,
\ee and it must commute with the left-moving oscillators and anticommute with the right-moving ones. Finally, it must satisfy $\hat\CN^2=2$ on the $g$-invariant states. Notice that $c_{w_1}$ acts trivially on the $g$-invariant ground states, so it necessarily commutes with $\hat\CN$. 

It follows that $\CN$ must map the eigenvectors of $c_{w_2},c_{w_3},c_{w_4}$ to eigenvectors of $c_{n_2},c_{n_3},c_{n_4}$  with the same eigenvalues:
\be \hat\CN |+,+,+,+\rangle =(const) \sum_{\theta_2,\theta_3,\theta_4\in \{\pm 1\}} |+,\theta_2,\theta_3,\theta_4\rangle
\ee
\be \hat\CN |+,-,+,+\rangle =(const) \sum_{\theta_3,\theta_4\in \{\pm 1\}} |+,+,\theta_3,\theta_4\rangle-|+,-,\theta_3,\theta_4\rangle
\ee and so on. The normalization can be fixed, up to an overall sign, by the requirement that $\hat\CN^2=2$ on these $g$-invariant states. More precisely, we find:
\be \hat\CN |+,+,+,+\rangle ={1\over 2} \sum_{\theta_2,\theta_3,\theta_4\in \{\pm 1\}} |+,\theta_2,\theta_3,\theta_4\rangle
\ee
\be \hat\CN |+,-,+,+\rangle ={1\over 2} \sum_{\theta_2,\theta_3,\theta_4\in \{\pm 1\}} (\theta_2)|+,\theta_2,\theta_3,\theta_4\rangle
\ee
and its 2 permutations,
\be \hat\CN |+,+,-,-\rangle ={1\over 2} \sum_{\theta_2,\theta_3,\theta_4\in \{\pm 1\}} (\theta_3\theta_4)|+,\theta_2,\theta_3,\theta_4\rangle
\ee
and its 2 permutations,
\be \hat\CN |+,-,-,-\rangle ={1\over 2} \sum_{\theta_2,\theta_3,\theta_4\in \{\pm 1\}} (\theta_2\theta_3\theta_4)|+,\theta_2,\theta_3,\theta_4\rangle\ .
\ee

\subsection{The defect-twined elliptic genus}

We want to compute
\be
\phi^{\mathcal N}(T^4/\langle \mathcal R \rangle,\tau, z)=\Tr_{RR} \mathcal N (-1)^F y^{J_L}q^{L_0-c/24}\bar q^{\bar L_0-c/24},
\ee
which corresponds to the elliptic genus of the model with the insertion of the $\mathbb Z_2$ duality defect $\hat{\mathcal N}$. Let $\mathcal H$ be the untwisted sector of the orbifold model $T^4/\langle \mathcal R \rangle$, and $\mathcal H_{\mathcal R}$ be the twisted sector. Then we can express the twined elliptic genus as a sum of four terms, 
\begin{align}
\phi^{\mathcal N}(T^4/\langle \mathcal R \rangle,\tau, z)=&{1\over 2}\Tr_{\mathcal H^{RR}}((1+\mathcal R) \hat{\mathcal N} (-1)^F y^{J_L}q^{L_0-c/24}\bar q^{\bar L_0-c/24})\\
&+{1\over 2}\Tr_{\mathcal H_{\mathcal R}^{RR}}((1+\mathcal R) \hat{\mathcal N} (-1)^F y^{J_L}q^{L_0-c/24}\bar q^{\bar L_0-c/24}),
\end{align}
corresponding to the projection of the twisted and untwisted RR Hilbert spaces onto $\mathcal R$-invariant states.

Firstly, in the twisted sector, $\hat{\mathcal N}$ acts on the 16 $\mathcal R$-twisted ground states with eight zero eigenvalues, and four eigenvalues each of $\pm 1$, as described in \S \ref{s:K3Z2duality}. Therefore the trace in the projected Hilbert space $\mathcal H_{\mathcal R}^{RR}$ will be zero, and the contribution of the twisted sector to the expression above will vanish.

In the untwisted sector, it is clear from \eqref{T4Z2duality} and \eqref{T4refl} that the action of $\mathcal R \mathcal N$ leaves the right-moving fermions invariant, and thus the additional insertion of $(-1)^F$ will cause the second term to vanish.

Thus we are just left with the insertion of $\hat{\mathcal N}$ in the (unprojected) untwisted sector as the only non-vanishing contribution to the twined elliptic genus:
\be
\phi^{\mathcal N}(T^4/\langle \mathcal R \rangle,\tau, z)={1\over 2}\Tr_{\mathcal H^{RR}}(\hat{\mathcal N} (-1)^F y^{J_L}q^{L_0-c/24}\bar q^{\bar L_0-c/24}).
\ee

We can express this as a product of 3 terms 
\be
\frac{\sqrt{2}}{2}\phi_{\mathcal N}^{gd}(\tau,z)\phi_{\mathcal N}^{osc}(\tau,z)\phi_{\mathcal N}^{w-m}(\tau),
\ee
where 
\be
\phi^{gd}_{\mathcal N}(\tau,z)=y^{-1}(1-\zeta_L y)(1-\zeta_L^{-1} y) (1-\zeta_R)(1-\zeta_R^{-1})
\ee
is the contribution from RR ground states with $\zeta_L=1$, $\zeta_R=-1$,
\begin{equation}
\begin{split}
\phi^{osc}_{\mathcal N}(\tau,z) \, & = \, \prod_{n=1}^\infty {(1-\zeta_L y q^n)(1-\zeta_L^{-1} y q^n) (1-\zeta_L y^{-1} q^n)(1-\zeta_L^{-1} y^{-1} q^n) \over (1-\zeta_L  q^n)^2(1-\zeta_L^{-1}  q^n)^2}
\end{split}
\end{equation}
is the contribution from left-moving bosonic and fermionic oscillators, and
\begin{equation}
\begin{split}
\phi_{\mathcal N}^{w-m}(\tau) \, & = \sum_{n \in \mathbb Z^4} (-1)^{n_2^2 + n_3^2 + n_4^2} q^{2n_1^2 + n_2^2 + n_3^2 + n_4^2} \, =\\
& = \, 1- 6q + 14 q^2- 20 q^3 + \ldots=\theta_4(2\tau)^3\theta_3(4\tau) \\
\end{split}
\end{equation}
is the contribution from the winding-momentum modes with the action of $\hat{\mathcal N}$ inserted.

This form arises from imposing $n_1= 2w_1$ and $n_i= w_i$ for $i=2,3,4$, which corresponds to tracing over vertex operators which are eigenstates of $\hat{\mathcal N}$ with eigenvalue $(-1)^{n\cdot w}= (-1)^{n_2^2 + n_3^2 + n_4^2}$ (as $n_1 \in 2\mathbb Z$).

Putting all of this together, we find 
\bea\label{eq:K3Z2genus}
\phi^{\mathcal N}(T^4/\langle \mathcal R \rangle,\tau, z)&=&-2\sqrt 2 {\theta_1(\tau,z)^2\theta_4(2\tau)^3\theta_3(4\tau)\over \eta^6(\tau)}\\
&=&\sqrt 2\left (\frac{2}{y}-4 + 2y+\left (-\frac{4}{y^2}+\frac{4}{y}+4 y -4 y^2\right ) q + \ldots\right )
\eea
that matches precisely with \eqref{twinZ2dual-}, as expected from our general conjecture.

\section{Duality defects for $\ZZ_3$}\label{sec:Z3duality}

In this section, we describe the duality defects related to a $\ZZ_3$ symmetry $g\in \Aut_\tau(V^{f\natural})$ with Frame shape $1^63^6$. 

From the general discussion in section \ref{sec:TY_categories}, it follows that there are potentially four different Tambara-Yamagami categories $TY(\ZZ_3,\chi,\epsilon)$ associated with a cyclic group $\ZZ_3$, corresponding to $\epsilon\in\{\pm 1\}$ and $\chi=\chi_+$ or $\chi_-$, with $\chi_\pm$ as in eq.\eqref{bicharZ3}. In this section, we will prove that, for each given $g$ with Frame shape $1^63^6$, there are five different $\CN=1$ preserving duality defects $\CN^{(b_+)}$, $\CN^{(b_-)}$, $\CN^{(a'_+)}$, $\CN^{(a'_-)}$, $\CN^{(a'_0)}$, up to conjugation by invertible elements in the normalizer $N_{Co_0}(\langle g\rangle )$ of $\langle g\rangle$ in $Co_0$. One has the relations $\CN^{(b_-)}=\xi \CN^{(b_+)}$ and $\CN^{(a'_-)}=\xi\CN^{(a'_+)}$ where $\xi$ is the central involution in $Co_0$, that acts trivially on the NS states and changes the sign of all Ramond states  (the defects $\CN^{(a'_0)}$ and $\xi\CN^{(a'_0)}$ are conjugate within $N_{Co_0}(\langle g\rangle)$). It turns out that, for each $k\in \{a'_\pm,a'_0,b_\pm\}$, the TY category and twining partition functions of $\CN^{(k)}$ and $\xi\CN^{(k)}$ are the same, so we can consider them together. We will compute the twining partition function and elliptic genera for all such defects, and show that $\CN^{(b_\pm)}$,  $\CN^{(a'_\pm)}$, $\CN^{(a'_0)}$ belong to categories TY$(\ZZ_3,\chi_+,-1)$, TY$(\ZZ_3,\chi_-,+1)$, TY$(\ZZ_3,\chi_-,-1)$, respectively. The category TY$(\ZZ_3,\chi_+,1)$ is not realized by any duality defect that preserves the $\CN=1$ supercurrent, at least for this particular Frame shape.

We have not yet found a K3 sigma model corresponding to the $V^{f\natural}$ duality defects in this section; Conjecture \ref{conj:K3relation} can be taken as a prediction that such models should exist.

\subsection{The Leech lattice endomorphism associated with a $\ZZ_3$-duality defect}

The general discussion about $\CN=1$ preserving defects in $V^{f\natural}$ predicts that any such $\CL$ is associated with an endomorphism $\rho(\CL):
\Lambda\to\Lambda$ of the Leech lattice $\Lambda$. In this section, we discuss the possible endomorphisms associated with a duality defect $\CN_g$ for a $\ZZ_3$ symmetry.

Let $g\in \Aut_\tau(V^{f\natural})\cong Co_0$ be a symmetry with Frame shape $1^63^6$. An element in this conjugacy class of $Co_0$ is known to fix a sublattice $\Lambda^g\subset\Lambda$ of the Leech lattice that is isomorphic to the $12$-dimensional Todd-Coxeter lattice $\Lambda^g\cong K_{12}$ (see \cite{ConwaySloane:1983}, and \cite{ConwaySloane}, chapter 4, section 9). Its orthogonal complement $\Lambda_g:=(\Lambda^g)^\perp\cap \Lambda$ is also isomorphic to the same lattice $\Lambda_g\cong K_{12}$.

   For each element $g\in Co_0$ in this class, there is a unique (up to inversion) element $g'\in Co_0$ (called the \emph{mate} of $g$ in \cite{Curtis:1980}) such that: $g'$ has also Frame shape $1^63^6$; $g$ and $g'$ commute; and the sublattice $\Lambda^{g'}\subset \Lambda$ fixed by $g'$ is exactly the orthogonal complement $\Lambda_g$ of the lattice fixed by $g$, and vice versa, i.e. $\Lambda^g=\Lambda_{g'}$ and $\Lambda^{g'}=\Lambda_g$.
   In the following, we will need to consider some subgroups of $\Aut(V^{f\natural})\cong Co_0$, whose elements $k$ either commute with $g$ and/or $g'$, or are in the normalizer groups of $\langle g\rangle$ and/or $\langle g'\rangle$ (i.e. $kgk^{-1}=g^{-1}$ or $kg'k^{-1}={g'}^{-1}$). More precisely, we consider the centralizer in $Co_0$ of $\langle g,g'\rangle$
 \be C_{Co_0}(\langle g,g'\rangle):=\{k\in Co_0\mid kg=gk\text{ and }kg'=g'k\}=\langle \xi, g,g'\rangle.PSU_4(\FF_3)\ ,
\ee  the centralizers of the single elements $g$ and $g'$
\be C_{Co_0}(g)= \langle \xi, g,g'\rangle.PSU_4(\FF_3).\langle a'\rangle\ ,\qquad C_{Co_0}(g')=\langle \xi,g,g'\rangle).PSU_4(\FF_3).\langle a\rangle\ ,
\ee  and the normalizers of $\langle g\rangle$ and of $\langle g'\rangle$, that coincide
\be N_{Co_0}(\langle g\rangle)=N_{Co_0}(\langle g'\rangle)=\langle \xi,g,g'\rangle.PSU_4(\FF_3).\langle a,a'\rangle\ ,
\ee
 Here, the group $C_{Co_0}(\langle g,g'\rangle)$  contains $\langle \xi, g,g'\rangle\cong \ZZ_2\times\ZZ_3\times\ZZ_3$ as a central subgroup, where $\langle \xi\rangle\cong \ZZ_2$ is the centre of $Co_0$, and the quotient of $C_{Co_0}(\langle g,g'\rangle)/\langle \xi, g,g'\rangle$ is isomorphic to a finite simple group $PSU_4(\FF_3)$ of order $2^7\cdot 3^6\cdot 5\cdot 7$.\footnote{This group is denoted as $U_4(3)$ in \cite{Atlas}. As the name suggests, the group $ PSU_4(\FF_3)$ can be described as the quotient $SU_4(\FF_3)/\langle i\mathbf{1}_4\rangle$ of the group $SU_4(\FF_3)$ of special (i.e. $\det=1$) unitary $4\times 4$ matrices over the finite field $\FF_3+i\FF_3\cong \FF_9$, divided by its center $\langle i\mathbf{1}_4\rangle\cong \ZZ_4$ \cite{Atlas} Here, $\FF_3\cong \ZZ/3\ZZ$ is the field with $3$ elements and $i^2=-1\in \FF_3$. We will not use such a description in this article.} The group $N_{Co_0}(\langle g\rangle)=N_{Co_0}(\langle g'\rangle)$ contains $C_{Co_0}(\langle g,g'\rangle)$ as a normal subgroup and the quotient $N_{Co_0}(\langle g\rangle)/C_{Co_0}(\langle g,g'\rangle)$ is isomorphic to $\ZZ_2\times \ZZ_2$. We denote by $a$, $a'$, and $b$ some elements in $N_{Co_0}(\langle g\rangle)$  satisfying, respectively, the conditions
 \be\label{adef} a^2=1,\qquad aga=g^{-1},\qquad ag'=g'a,
 \ee
 \be\label{apdef} a'^2=1,\qquad a'g=ga',\qquad a'g'a'=g'^{-1},
 \ee
 \be\label{Cdef} b^2=1,\qquad bgb=g^{-1},\qquad bg'b=g'^{-1}.
 \ee Any three such elements $a,a',b\in N_{Co_0}(\langle g\rangle)$ with these properties can be taken as representatives of the three non-trivial cosets in $N_{Co_0}(\langle g\rangle)/C_{Co_0}(\langle g,g'\rangle)\cong \ZZ_2\times\ZZ_2$.
 With the help of Magma \cite{Magma}, we explicitly constructed the subgroup $N_{Co_0}(\langle g\rangle)$, and verified that, up to conjugation in $N_{Co_0}(\langle g\rangle)$, there are three elements $a_+$, $a_0$, $a_-=\xi a_+\in N_{Co_0}(\langle g\rangle)$ satisfying \eqref{adef} with Frame shapes, respectively, $1^82^8$, $2^{12}$, $1^{-8}2^{16}$, three elements $a'_+$, $a'_0$, $a'_-=\xi a'_+$ satisfying \eqref{apdef} with the same Frame shapes, and two elements $b_+$ and $b_-=\xi b_+$ satisfying  \eqref{Cdef}, both with Frame shape $2^{12}$.

The lattices $\Lambda^g$ and $\Lambda_g$ can be described as two isomorphic $6$-dimensional complex lattices, for example as 
{\footnotesize{
\be \Lambda^g=\Lambda_{g'}:=\{\frac{1}{\sqrt{3}}(z_1,\ldots,z_6)\in \CC^6\mid z_i\in \ZZ[\omega],\  z_1\equiv\ldots\equiv z_6 \bmod \theta  \ZZ[\omega],\ \sum_i z_i=0\bmod 3 \ZZ[\omega]\}
\ee
\be \Lambda_g=\Lambda^{g'}:=\{\frac{1}{\sqrt{3}}(w_1,\ldots,w_6)\in \CC^6\mid w_i\in \ZZ[\omega],\  w_1\equiv\ldots\equiv w_6 \bmod \theta  \ZZ[\omega],\ \sum_i w_i=0\bmod 3 \ZZ[\omega]\}
\ee
}}
where $\omega:=e^{\frac{2\pi i}{3}}$, $ \ZZ[\omega]= \ZZ+\omega\ZZ$, and $\theta:=\omega-\bar\omega=\sqrt{3} i$; see \cite{ConwaySloane:1983} for alternative descriptions and embeddings in the Leech lattice $\Lambda$. The corresponding real lattices are obtained via the standard maps $\CC^6\to \RR^{12}$, such as $(z_1,\ldots,z_6)\mapsto (\Re(z_1),\Im(z_1),\ldots,\Re(z_6),\Im(z_6))\in \RR^{12}$. In terms of these complex coordinates the action of $g$ and $g'$ is very simple
\be g(z_k)=z_k\ ,\qquad g(w_k)=\omega w_k\ ,
\ee
\be g'(z_k)=\omega z_k\ ,\qquad g'(w_k)=w_k\ .
\ee
More generally, the centralizer $C_{Co_0}(\langle g,g'\rangle)\cong \langle \xi,g,g'\rangle.PSU_4(\FF_3)$ acts by complex unitary transformations on the complex lattices $\Lambda^g$ and $\Lambda_g$ as
\be\label{UAhUBh} h(z_k)=\sum_l U^A(h)_{kl}z_l\ ,\qquad h(w_k)=\sum_l U^B(h)_{kl}w_l\ ,\qquad h\in C_{Co_0}(\langle g,g'\rangle)\ ,
\ee for suitable $6\times 6$ special unitary matrices $U^A(h),U^B(h)\in SU(6)$. In particular, $(U^A(g),U^B(g))=(\mathbf{1}_6,\omega \mathbf{1}_6)$, $(U^A(g'),U^B(g'))=(\omega \mathbf{1}_6,\mathbf{1}_6)$, and $(U^A(\xi),U^B(\xi))=(-\mathbf{1}_6,- \mathbf{1}_6)$  correspond to matrices in the centre of $SU(6)$.
On the other hand, an element $a'$ satisfying \eqref{apdef} acts by complex conjugation on the variables $z_1,\ldots,z_6$ and by a unitary transformations on $w_1,\ldots,w_6$. The fact that $a'_+$, $a'_0$ and $a'_-$ have Frame shapes, respectively, $1^82^8$, $2^{12}$ and $1^{-8}2^{16}$ implies that the unitary action on $w_1,\ldots,w_6$ have eigenvalue multiplicities $(-1)^1,1^5$ for $a'_+$, $(-1)^3,1^3$ for $a'_0$, and $(-1)^5,1^1$ for $a'_-$; in all cases, the determinant is $-1$. It is convenient to parametrize this action as
\be\label{apelem} a'_j(z_k)=\sum_l U^A(a'_j)_{kl}\bar z_l\ ,\qquad a'_j(w_k)=i\sum_l U^B(a'_j)_{kl}w_l\ ,\quad j\in\{+,0,-\},
\ee for suitable $U^A(a'_j),U^B(a'_j)\in SU(6)$, where $U^B(a'_1)$, $U^B(a'_2)$, $U^B(a'_3)$ have eigenvalues $(i)^1,(-i)^5$ for $a_+'$, $(i)^3,(-i)^3$ for $a_0'$, and $(i)^5,(-i)^1$ for $a_-'$. Analogously,
\be\label{aelem} a_j(z_k)=i\sum_l U^A(a_j)_{kl} z_l\ ,\qquad a_j(w_k)=\sum_l U^B(_j)_{kl}\bar w_l\ ,\qquad j\in\{+,0,-\},
\ee with $U^A(a_j),U^B(a_j)\in SU(6)$ and $U^A(a_j)$ having the same eigenvalues as $U^B(a'_j)$. Notice, in particular, that
\be\label{strangesquare} U^A(a_j)^2=-\mathbf{1}_6\ ,\qquad U^B(a'_j)^2=-\mathbf{1}_6\ .
\ee

Clearly, the quotient $N_{Co_0}(g)/\langle g\rangle$ acts faithfully on the real lattice $\Lambda^g\subset \RR^{12}$ by lattice automorphisms. In fact, since $|\Aut(K_{12})|=2^{10}\cdot 3^7\cdot 5\cdot 7=|N_{Co_0}(g)/\langle g\rangle|$ we have an isomorphism
\be \Aut(\Lambda^g)=\Aut(\Lambda_{g'})=N_{Co_0}(g)/\langle g\rangle\cong \langle \xi, g'\rangle.PSU_4(\FF_3).\langle a',a\rangle\ ,
\ee and similarly
\be \Aut(\Lambda^{g'})=\Aut(\Lambda_g)=N_{Co_0}(g')/\langle g'\rangle\cong \langle \xi, g\rangle.PSU_4(\FF_3).\langle a',a\rangle\ .
\ee 

The index $2$ subgroup $\Aut^+(\Lambda^g):=  \Aut(\Lambda^g)\cap U(6)$ that preserves the complex structure is 
\be \Aut^+(\Lambda^g)=C_{Co_0}(g)/\langle g\rangle\cong \langle \xi, g'\rangle.PSU_4(\FF_3).\langle a\rangle\subset U(6)\subset SO(12)\ ,
\ee and similarly
\be \Aut^+(\Lambda_g)=C_{Co_0}(g')/\langle g'\rangle\cong \langle \xi, g\rangle.PSU_4(\FF_3).\langle a'\rangle\subset U(6)\subset SO(12)\ .
\ee

Let us now run an argument similar to the one for the $\ZZ_2$ duality defect in section \ref{s:Z2endom}. The chain of lattice inclusions
\be \Lambda^g\oplus \Lambda_g\subset \Lambda\subset (\Lambda^g)^*\oplus (\Lambda_g)^*\ ,
\ee implies that every $\lambda\in\Lambda$ can be written as $\lambda=(\lambda^g,\lambda_g)\in (\Lambda^g)^*\oplus (\Lambda_g)^*$. Self-duality of $\Lambda$ implies that for every $\mu^g\in (\Lambda^g)^*$, there exists $\mu_g\in (\Lambda_g)^*$ such that $(\mu^g,\mu_g)\in \Lambda$. Furthermore, vectors of the form $(\lambda^g,0)$ are in $\Lambda$ if and only if $\lambda^g\in \Lambda^g\subset (\Lambda^g)^*$. Now, the fusion relation $\rho(\CN)^2=\sum_{k=1}^3 g^k$ and $\rho(\CN)^t=\rho(\CN)$ imply that
\be \rho(\CN)^2(\lambda^g,\lambda_g)=(3\lambda^g,0)\ ,\qquad \forall (\lambda^g,\lambda_g)\in\Lambda\ ,
\ee and therefore
\be \rho(\CN)(\lambda^g,\lambda_g)=(\sqrt{3}O_\CN\lambda^g,0)\in \Lambda\ ,\qquad \forall (\lambda^g,\lambda_g)\in\Lambda\ ,
\ee for some $12\times 12$ real matrix $O_\CN$ with \be\label{ONinvol} O_\CN^t=O_\CN,\qquad  O_{\CN}^2=1\ .\ee In particular, $O_\CN\in O(12,\RR)$. Because $(\sqrt{3}O_\CN\lambda^g,0)$ is in $\Lambda$, it follows that $\sqrt{3}O_\CN\lambda^g\in \Lambda^g$ for all $\lambda^g\in (\Lambda^g)^*$. Therefore, we have an injective map
\be\label{threemodular} \sqrt{3}O_\CN:(\Lambda^g)^*\to \Lambda^g\ .
\ee The observation that $|(\Lambda^g)^*/\Lambda^g|=3^6=|\det(\sqrt{3}O_\CN)|$ implies that this is surjective. We conclude that the lattice $\Lambda^g$ is $3$-modular, i.e. it is isomorphic $\Lambda^g\cong \sqrt{3}(\Lambda^g)^*$ to a rescaling of its dual lattice by $\sqrt{3}$.  It is well-known that the Todd-Coxeter lattice has this property: in the description of $\Lambda^g\cong K_{12}$ as a complex lattice, a similarity of the form \eqref{threemodular} is simply given by rescaling by $\theta=\sqrt{3}i$ \cite{ConwaySloane:1983}
\begin{align} \Theta:(\Lambda^g)^*&\to \Lambda^g\ ,\label{Thetadef}\\
v&\mapsto \sqrt{3}iv\ .\notag
\end{align} The map $\Theta$ is a $3$-similarity, i.e. an isomorphism of $\ZZ$-modules that rescales the quadratic form by $3$. Every other similarity between $(\Lambda^g)^*$  and $\Lambda^g$ is obtained by composing $\Theta$ from the left or from the right by lattice automorphisms
\be k_1\Theta k_2:(\Lambda^g)^*\to \Lambda^g\ ,
\ee for any $k_1,k_2\in  \Aut(\Lambda^g)=\Aut((\Lambda^g)^*)\subset O(12,\RR)$. Notice that $\Theta$ commutes with all $k\in \Aut^+(\Lambda^g)$, while for the elements $h\in \Aut(\Lambda^g)$ that involve complex conjugation, such as $a'$, it satisfies
\be\label{Thetaconjcomm} \Theta h=-h\Theta \ ,\qquad h\in \Aut(\Lambda^g)\setminus \Aut^+(\Lambda^g)\ .
\ee

The similarity $\sqrt{3}O_\CN$ in \eqref{threemodular} associated with a duality defect $\CN$ must also satisfy \eqref{ONinvol}, and therefore one cannot simply define $\sqrt{3} O_{\CN}^{(12)}=\Theta$ because $\Theta^2=-3$ has the wrong sign. Instead, by \eqref{Thetaconjcomm},  one can take $ \sqrt{3}O_{\CN}^{(12)}=\Theta h\ ,$ where $h$ is any element in $\Aut(\Lambda^g)\setminus \Aut_+(\Lambda^g)$ such that $h^2=1$.\footnote{In principle, one could also look for an element $h\in \Aut_+(\Lambda^g)$ such that $h^2=-1$. Indeed, given that all elements of $\Aut_+(\Lambda^g)$ commute with $\Theta$, one would then obtain $(\Theta h)^2=\Theta^2h^2=3$. However, there exists no element in $\Aut_+(\Lambda^g)$ that squares to $-1$, as can be checked from the character table of this group \cite{Atlas} or by a direct calculation (e.g. in Magma).} Recalling that $\Aut(\Lambda^g)=N_{Co_0}(\langle g\rangle)/\langle g\rangle$, we see that the elements $h\in \Aut(\Lambda^g)\setminus \Aut_+(\Lambda^g)$ with $h^2=1$ correspond exactly to elements in $N_{Co_0}(\langle g\rangle)$ satisfying the conditions \eqref{apdef} or \eqref{Cdef}. Thus, up to conjugation in $N_{Co_0}(\langle g\rangle)$, there are only five possibilities for $\sqrt{3}O_\CN$, namely
\be \Theta a'_\pm\ ,\qquad \Theta a'_0  \ ,\qquad \Theta b_\pm\ .
\ee  We will show that, for each of these five possibilities, there is a corresponding duality defect $\CN_g$ for $g$ preserving the $\CN=1$ supercurrent.

\subsection{The affine algebra fixed by $\ZZ_3$}

It is useful to consider a description of the NS sector of $V^{f\natural}$ as the lattice SVOA given by $12$ chiral free bosons on the odd unimodular lattice $D_{12}^+$ in \eqref{D12plus}. Up to conjugation in $Spin(24)$, one can take $g$ of Frame shape $1^63^6$ to be a symmetry fixing all free bosons and acting on the vertex operators $\CV_\lambda$, $\lambda\in D_{12}^+$, by a phase
\be g(\CV_\lambda(z))=e^{2\pi i \lambda\cdot \alpha}\CV_\lambda(z)\ ,\qquad \lambda\in D_{12}^+
\ee where 
\be\label{Z3alpha} \alpha=(0,0,0,0,0,0;\frac{1}{3},\frac{1}{3},\frac{1}{3},-\frac{1}{3},-\frac{1}{3},-\frac{1}{3})\ .
\ee

Thus, the $g$-invariant vertex operators $\CV_\lambda$ are the ones for $\lambda$ in the sublattice
\be\label{Z3fixedD12} (D_{12}^+)^g:=\{\lambda\in D_{12}^+\mid \lambda\cdot\alpha\in \ZZ\}\ .\ee In particular, the affine subalgebra of $\widehat{so}(24)_1$ that is preserved by $g$ is \be\label{Z3fixedaffine} \widehat{so}(12)_1^A\oplus \widehat{su}(6)^B_1\oplus \hat u(1)^B_3\ ,\ee corresponding to a lattice VOA based on the lattice
\be D_6\oplus A_5\oplus \sqrt{6}\ZZ\ \subset\ (D_{12}^+)^g\ ,
\ee  
where
\be D_6=\{(x_1,\ldots,x_{6},0,\ldots,0)\mid x_i\in \ZZ,\  \sum_{i=1}^{6}x_i\in 2\ZZ\}\ ,
\ee
\be\label{A5lattice} A_5=\{(0,\ldots,0,x_7,\ldots,x_{12})\mid x_i\in \ZZ,\  x_7+x_8+x_9-x_{10}-x_{11}-x_{12}=0\}\ ,
\ee
\be \sqrt{6}\ZZ=\{n(0,\ldots,0,1,1,1,-1,-1,-1),\ n\in\ZZ\}\ .
\ee

Analogously, the $g'$-fixed affine subalgebra is 
\be \hat u(1)^A_3\oplus \widehat{su}(6)^A_1\oplus  \widehat{so}(12)_1^B\ , \ee where $\hat u(1)^A_3\oplus \widehat{su}(6)^A_1\subset \widehat{so}(12)_1^A$ and $\hat u(1)^B_3\oplus \widehat{su}(6)^B_1\subset \widehat{so}(12)_1^B$. Finally, the affine algebra fixed by $\langle g,g'\rangle\cong \ZZ_3\times \ZZ_3$ is 
\be\label{ggpaffine} \hat u(1)^A_3\oplus \widehat{su}(6)^A_1\oplus \widehat{su}(6)^B_1\oplus \hat u(1)^B_3\ .
\ee

Recall that the affine algebra $\widehat{su}(6)_1$ has six irreducible unitary representations corresponding to the six cosets in $A_5^*/A_5\cong\ZZ/6\ZZ$. These representations can be labeled by ${\bf 1}$, ${\scalebox{0.5}{ \ydiagram{1}}}$, ${\scalebox{0.5}{ \ydiagram{1,1}}}$, ${\scalebox{0.5}{ \ydiagram{1,1,1}}}$,  $\overline{{\scalebox{0.5}{ \ydiagram{1,1}}}}$  $\overline{{\scalebox{0.5}{ \ydiagram{1}}}}$. For each such representation, the corresponding coset in $A_5^*/A_5$, the lowest conformal weight $h$, and the number of states with conformal weight $h$ (ground states) are as follows:
\begin{align}
{\bf 6},\bar{\bf 6}\text{ or } & {\scalebox{0.5}{ \ydiagram{1}}},\overline{{\scalebox{0.5}{ \ydiagram{1}}}} && (0,\ldots,0,\ \pm \frac{5}{6},\mp\frac{1}{6},\mp\frac{1}{6},\ \pm\frac{1}{6},\pm\frac{1}{6},\pm\frac{1}{6})+A_5&& h=\frac{5}{12}\ ,\quad \text{6 ground states}\ ,\\
{\bf 15},\bar{\bf 15} \text{ or }&{\scalebox{0.5}{ \ydiagram{1,1}}},\overline{{\scalebox{0.5}{ \ydiagram{1,1}}}}&& (0,\ldots,0,\pm \frac{4}{6} , \pm \frac{4}{6},\mp\frac{2}{6},\pm\frac{2}{6},\pm\frac{2}{6},\pm\frac{2}{6})+A_5&& h=\frac{2}{3}\ ,\quad \text{15 ground states}\ ,\\
{\bf 20} \text{ or } &{\scalebox{0.5}{ \ydiagram{1,1,1}}}&& (0,\ldots,0,\frac{3}{6}, \frac{3}{6},\frac{3}{6},\frac{3}{6},\frac{3}{6},\frac{3}{6})+A_5&& h=\frac{3}{4}\ ,\quad \text{20 ground states}\ .
\end{align} The algebra $\hat u(1)_3$ has also six representations labeled by $n_{u(1)}\in \ZZ/6\ZZ$, corresponding to the cosets $\frac{n_{u(1)}}{\sqrt{6}}+\sqrt{6}\ZZ$. For $n_{u(1)}=0,\pm 1,\pm 2,3$ the conformal weights are, respectively, $\frac{n_{u(1)}^2}{12}=0,\frac{1}{12},\frac{1}{3},\frac{3}{4}$; all such representations have a unique ground state, except for $n_{u(1)}=3$ that has two of them. 

It is convenient to write the decomposition of $\widehat{so}(24)_1$ representations with respect to the various $g$- and $g'$-fixed subalgebras in two steps, using the inclusions
\be \widehat{so}(24)_1\supset \widehat{so}(12)^A_1\oplus \widehat{so}(12)^B_1\ee and \be  \widehat{so}(12)^A_1\supset \widehat{su}(6)^A_1\oplus \hat u(1)^A_3\qquad \widehat{so}(12)^B_1\supset \widehat{su}(6)^B_1\oplus \hat u(1)^B_3\ .
\ee The decomposition of the representations $0,v,s,c$ of $\widehat{so}(24)_1$ with respect to $\widehat{so}(12)_1\oplus \widehat{so}(12)_1$ is
\begin{align}
&0 \to (0,0)+(v,v)\ ,&&  v\to (v,0)+(0,v)\\&s \to (s,s)+(c,c)\ ,&&  c\to (s,c)+(c,s)\ .
\end{align}
In turn, the four representations of $\widehat{so}(12)_1$ decompose as follows with respect to the representations of $\widehat{su}(6)_1\oplus \hat u(1)_3$ (the $\hat u(1)_3$ representation is indicated by a subscript $n\in\{0,\pm1,\pm2,
\pm3\}$): 
\begin{align}
&0 \to {\bf 1}_0+{\bf 15}_2+{\bf \bar{15}}_{-2}={\bf 1}_0+{{\scalebox{0.5}{ \ydiagram{1,1}}}}_2+\overline{{\scalebox{0.5}{ \ydiagram{1,1}}}}_{-2}\ ,&& h=0\ \text{(1 state),}\quad h=1\ \text{(15+15=30 states),}\label{so12tosu6u1adj}\\
& v\to {\bf 6}_1+\bar {\bf 6}_{-1}+{\bf 20}_{\pm 3}={{\scalebox{0.5}{ \ydiagram{1}}}}_1+\overline{{\scalebox{0.5}{ \ydiagram{1}}}}_{-1}+{{\scalebox{0.5}{ \ydiagram{1,1,1}}}}_{\pm 3} && \begin{matrix} & h=\frac{1}{2}\  \text{(6+6=12 states)},\\ & h=\frac{3}{2}\ \text{($2\cdot 20=40$ states)}  \end{matrix} \\&s \to{\bf 6}_{-2}+{\bf 20}_0+\bar{\bf 6}_2 ={{\scalebox{0.5}{ \ydiagram{1}}}}_{-2}+\overline{{\scalebox{0.5}{ \ydiagram{1}}}}_{2}+{{\scalebox{0.5}{ \ydiagram{1,1,1}}}}_0 \ , && h=\frac{3}{4}\  \text{(6+6+20=32 states)}\\&c \to {\bf 1}_{\pm 3}+{\bf 15}_{-1}+\bar{\bf 15}_1={\bf 1}_{\pm 3}+{{\scalebox{0.5}{ \ydiagram{1,1}}}}_{-1}+\overline{{\scalebox{0.5}{ \ydiagram{1,1}}}}_{1}\ , && h=\frac{3}{4}\  \text{(2+15+15=32 states)}\ .
\end{align}
Here, for each irreducible representation of $\widehat{su}(6)_1\oplus \hat u(1)_3$, we report the lowest conformal weight $h$ and the number of states at that conformal weight in the given representation. Notice that out of the $66$ currents generating $\widehat{so}(12)_1^B$, $30$ are the ground states of ${{\scalebox{0.5}{ \ydiagram{1,1}}}}_2$ and $\overline{{\scalebox{0.5}{ \ydiagram{1,1}}}}_{-2}$, and $36$ are descendants of ${\bf 1}_0$.  The decomposition of the $g$-eigenspaces in the $NS\pm$ and $R\pm$  sectors of $V^{f\natural}$ into representations of $\widehat{so}(12)^A_1\oplus \widehat{su}(6)_1\oplus \hat u(1)_3$ are given in table \eqref{tab:4}.
{\scriptsize{
		\begin{table}[h!]
			\centering
			\begin{tabular}{|c|c|c|c|c|}
				\hline
				untwisted & $\mathbf{so(24)}$ & $g=1$ & $g=\omega$ & $g=\omega^2$  \\
				\hline
				NS+ & 0 & {\scriptsize{$ (0 , {\bf 1}_0) \oplus(v,{\scalebox{0.4}{ \ydiagram{1,1,1}}}_{3}) $}} & {\scriptsize{$ ( 0, \overline{{\scalebox{0.4}{ \ydiagram{1,1}}}}_{-2}  ) \oplus (v, {\scalebox{0.5}{ \ydiagram{1}}}_{1} )  $}} & {\scriptsize{$(0, {\scalebox{0.3}{ \ydiagram{1,1}}}_{2})\oplus  (v, \overline{{\scalebox{0.4}{ \ydiagram{1}}}}_{-1}) $}} \\
				\hline
				NS- & s & {\scriptsize{$\left( s, {\scalebox{0.4}{ \ydiagram{1,1,1}}}_{0}\right) \oplus \left( c,{\bf 1}_{\pm 3}   \right)$}} 
                & {\scriptsize{$\left( s, {\scalebox{0.5}{ \ydiagram{1}}}_{-2} \right) \oplus \left( c, \overline{{\scalebox{0.4}{ \ydiagram{1,1}}}}_{+1}  \right)$}} 
                & {\scriptsize{$\left( s, \overline{{\scalebox{0.5}{ \ydiagram{1}}}}_{2} \right) \oplus \left( c, {\scalebox{0.4}{ \ydiagram{1,1}}}_{-1}   \right)$}} \\
				\hline
				R+ & v & {\scriptsize{$(v,{\bf 1}_0) $}} & {\scriptsize{$(0, {\scalebox{0.5}{ \ydiagram{1}}}_{1} ) $}} & {\scriptsize{$ (0, \overline{ \scalebox{0.5}{\ydiagram{1}}}_{-1})$}}  \\
				\hline
				R- & c & {\scriptsize{$\left( c, {\scalebox{0.4}{ \ydiagram{1,1,1}}}_{0}\right) \oplus \left( s,{\bf 1}_{\pm 3}   \right)$}} 
                & {\scriptsize{$\left( c, {\scalebox{0.5}{ \ydiagram{1}}}_{-2} \right) \oplus \left( s, \overline{{\scalebox{0.4}{ \ydiagram{1,1}}}}_{+1}  \right)$}} 
                & {\scriptsize{$\left( c, \overline{{\scalebox{0.5}{ \ydiagram{1}}}}_{2} \right) \oplus \left( s, {\scalebox{0.4}{ \ydiagram{1,1}}}_{-1}   \right)$}} \\
				\hline
			\end{tabular}
			\caption{\small{Decomposition of the $g$-eigenspaces of $V^{f \natural}$ into representations of the $g$-preserved subalgebra $\widehat{so}(12)_1 \oplus \widehat{su}(6)_1 \oplus \hat u(1)_3$.}}
			\label{tab:4}
		\end{table}
}}

The $g$-twisted and $g^2$-twisted sectors also admit a lattice VOA description corresponding respectively to the translates $\alpha+(D_{12}^+)^g$ and $-\alpha+ (D_{12}^+)^g$ of the lattice in \eqref{Z3fixedD12}. The vector $\alpha$ in \eqref{Z3alpha} represents a state in the representation $(0,{\bf 1}_{+2})$ of $\widehat{so}(12)_1\oplus\widehat{su}(6)_1\oplus \hat u(1)_3$. 
Thus, all the $\widehat{so}(12)\oplus\widehat{su}(6)_1\oplus \hat u(1)_3$ representations in the $g$-twisted and $g^2$-twisted sectors can be obtained by taking the fusion of  each representation in the untwisted sector with $(0,{\bf 1}_{+2})$ and $(0,{\bf 1}_{-2})$, respectively, i.e by shifting the $\hat u(1)_3$ reps by $+2$ or $-2$ ($\bmod\ 6$). The action of $g$ on the twisted sector is in general ambiguous, as it can be multiplied by a cubic root of unity. As usual, one can fix the ambiguity by requiring $g^k$ to coincide with $e^{2\pi i L_0}$ on the $g^k$-twisted sector, for $k=1,2$. The decompositions of each $g$-eigenspace in the $g$-twisted and $g^2$-twisted sectors into $\widehat{so}(12)_1\oplus\widehat{su}(6)_1\oplus \hat u(1)_3$ representations is reported in tables \ref{tab:5} and \ref{tab:6}.

{\scriptsize{
		\begin{table}[h!]
			\centering
			\begin{tabular}{|c|c|c|c|}
				\hline
				$g$-twisted  & $g=1$ & $g=\omega$ & $g=\omega^2$  \\
				\hline
				NS+ & {\scriptsize{$(0, {\scalebox{0.3}{ \ydiagram{1,1}}}_{-2})\oplus  (v, \overline{{\scalebox{0.4}{ \ydiagram{1}}}}_{1}) $}} & {\scriptsize{$ (0 , {\bf 1}_2) \oplus(v,{\scalebox{0.4}{ \ydiagram{1,1,1}}}_{-1}) $}} & {\scriptsize{$ ( 0, \overline{{\scalebox{0.4}{ \ydiagram{1,1}}}}_{0}  ) \oplus (v, {\scalebox{0.5}{ \ydiagram{1}}}_{3} )  $}}  \\
				\hline
				NS-  & {\scriptsize{$\left( s, \overline{{\scalebox{0.5}{ \ydiagram{1}}}}_{-2} \right) \oplus \left( c, {\scalebox{0.4}{ \ydiagram{1,1}}}_{1}   \right)$}} & {\scriptsize{$\left( s, {\scalebox{0.4}{ \ydiagram{1,1,1}}}_{2}\right) \oplus \left( c,{\bf 1}_{-1}   \right)$}} 
                & {\scriptsize{$\left( s, {\scalebox{0.5}{ \ydiagram{1}}}_{0} \right) \oplus \left( c, \overline{{\scalebox{0.4}{ \ydiagram{1,1}}}}_{3}  \right)$}} 
                 \\
				\hline
				R+ & {\scriptsize{$ (0, \overline{ \scalebox{0.5}{\ydiagram{1}}}_{1})$}}  & {\scriptsize{$(v,{\bf 1}_2) $}} & {\scriptsize{$(0, {\scalebox{0.5}{ \ydiagram{1}}}_{3} ) $}}  \\
				\hline
				R- & {\scriptsize{$\left( c, \overline{{\scalebox{0.5}{ \ydiagram{1}}}}_{-2} \right) \oplus \left( s, {\scalebox{0.4}{ \ydiagram{1,1}}}_{1}   \right)$}} & {\scriptsize{$\left( c, {\scalebox{0.4}{ \ydiagram{1,1,1}}}_{2}\right) \oplus \left( s,{\bf 1}_{-1}   \right)$}} 
                & {\scriptsize{$\left( c, {\scalebox{0.5}{ \ydiagram{1}}}_{0} \right) \oplus \left( s, \overline{{\scalebox{0.4}{ \ydiagram{1,1}}}}_{3}  \right)$}} 
                 \\
				\hline
			\end{tabular}
			\caption{\small{Decomposition of the $g$-eigenspaces of the $g$-twisted sector into representations of the $g-$preserved subalgebra $\widehat{so}(12)_1 \oplus \widehat{su}(6)_1 \oplus \hat u(1)_3$.}}
			\label{tab:5}
		\end{table}
}}

{\scriptsize{
		\begin{table}[h!]
			\centering
			\begin{tabular}{|c|c|c|c|}
				\hline
				$g^2$-twisted & $g=1$ & $g=\omega$ & $g=\omega^2$  \\
				\hline
				NS+ & {\scriptsize{$ ( 0, \overline{{\scalebox{0.4}{ \ydiagram{1,1}}}}_{2}  ) \oplus (v, {\scalebox{0.5}{ \ydiagram{1}}}_{-1} )  $}} &  {\scriptsize{$ (0 , {\bf 1}_-2) \oplus(v,{\scalebox{0.4}{ \ydiagram{1,1,1}}}_{1}) $}}  & {\scriptsize{$(0, {\scalebox{0.3}{ \ydiagram{1,1}}}_{0})\oplus  (v, \overline{{\scalebox{0.4}{ \ydiagram{1}}}}_{3}) $}} \\
				\hline
				NS- & {\scriptsize{$\left( s, {\scalebox{0.5}{ \ydiagram{1}}}_{2} \right) \oplus \left( c, \overline{{\scalebox{0.4}{ \ydiagram{1,1}}}}_{-1}  \right)$}} &  {\scriptsize{$\left( s, {\scalebox{0.4}{ \ydiagram{1,1,1}}}_{-2}\right) \oplus \left( c,{\bf 1}_{1}   \right)$}} 
                & {\scriptsize{$\left( s, \overline{{\scalebox{0.5}{ \ydiagram{1}}}}_{0} \right) \oplus \left( c, {\scalebox{0.4}{ \ydiagram{1,1}}}_{3}   \right)$}} \\
				\hline
				R+ & {\scriptsize{$(0, {\scalebox{0.5}{ \ydiagram{1}}}_{-1} ) $}} &  {\scriptsize{$(v,{\bf 1}_{-2}) $}}  & {\scriptsize{$ (0, \overline{ \scalebox{0.5}{\ydiagram{1}}}_{3})$}}  \\
				\hline
				R- & {\scriptsize{$\left( c, {\scalebox{0.5}{ \ydiagram{1}}}_{2} \right) \oplus \left( s, \overline{{\scalebox{0.4}{ \ydiagram{1,1}}}}_{-1}  \right)$}}
                &  {\scriptsize{$\left( c, {\scalebox{0.4}{ \ydiagram{1,1,1}}}_{-2}\right) \oplus \left( s,{\bf 1}_{1}   \right)$}} 
                & {\scriptsize{$\left( c, \overline{{\scalebox{0.5}{ \ydiagram{1}}}}_{0} \right) \oplus \left( s, {\scalebox{0.4}{ \ydiagram{1,1}}}_{3}   \right)$}} \\
				\hline
			\end{tabular}
			\caption{\small{Decomposition of the $g$-eigenspaces of the $g^2$-twisted sector into representations of the $g-$preserved subalgebra $\widehat{so}(12)_1 \oplus \widehat{su}(6)_1 \oplus \hat u(1)_3$.}}
			\label{tab:6}
		\end{table}
}}

The zero modes of the $g$-invariant and $\langle g,g'\rangle$-invariant currents generate the groups
\be
Spin(12)^A\times SU(6)^B\times U(1)^B\ ,
\ee and
\be\label{USUSUU}
U(1)^A\times SU(6)^A \times SU(6)^B\times U(1)^B\ ,
\ee respectively. These groups do not act faithfully on $V^{f\natural}$.  In particular, the central subgroup
\be\label{kerrep} \langle (\bar\omega,\omega \mathbf{1}_6,\mathbf{1}_6,1)\rangle\times \langle(1,\mathbf{1}_6,\omega \mathbf{1}_6,\bar\omega)\rangle\times\langle(-1,-\mathbf{1}_6,-\mathbf{1}_6,-1)\rangle \cong \ZZ_3\times \ZZ_3\times \ZZ_2
\ee acts trivially on all the fields in $V^{f\natural}$. Here, in a lattice description of $\hat u(1)_3$ and its representations, the element $e^{i\alpha}\in U(1)$, $\alpha\in \RR/2\pi\ZZ$, is the symmetry multiplying the vertex operator $\CV_{\frac{m}{\sqrt{6}}}(z)$, $m\in \ZZ$, by $e^{i\alpha m}$.

The fact that the action is not faithful leads to some ambiguity in identifying $g$ and $g'$ with elements in \eqref{USUSUU}. A convenient choice is\footnote{If one requires that $g$ and $g'$ have the expected action on the $g$- and $g'$-twisted sectors, then the correct identifications are $g=(1,\mathbf{1}_6,\bar\omega \mathbf{1}_6,\bar\omega)$, $g'=(\bar\omega,\bar\omega\mathbf{1}_6, \mathbf{1}_6,1)$. The elements that we write here are instead $gQ$ and $g'Q'$, where $Q$ and $Q'$ are the quantum symmetries acting by third roots of unity on the $g$- and $g'$-twisted sectors. The difference is not relevant for what follows.}
\be\label{gZ3} g=(1,\mathbf{1}_6,\omega \mathbf{1}_6,1)\in U(1)^A\times SU(6)^A \times SU(6)^B\times U(1)^B\ ,
\ee
\be\label{gpZ3} g'=(1,\omega\mathbf{1}_6, \mathbf{1}_6,1)\in U(1)^A\times SU(6)^A \times SU(6)^B\times U(1)^B\ .
\ee The element $\omega \mathbf{1}_6\in SU(6)$ acts by $\omega^d$ on the $\widehat{su}(6)_1$-representation labeled by $d$ boxes.

Let us now discuss in which representations of the $\langle g,g'\rangle$-invariant affine algebra \eqref{ggpaffine} is the supercurrent $\tau(z)$ sitting. The representation $s$ of $\widehat{so}(24)_1$ decomposes as $(s,s)\oplus (c,c)$ with respect to $\widehat{so}(12)_1^A\oplus \widehat{so}(12)_1^B$, so that the $\CN=1$ supercurrent must be a linear combination of a state in $(s,s)$ and  a state in $(c,c)$. In turn, the decomposition of $(s,s)$ and $(c,c)$ in terms of $u(1)^A_3\oplus \widehat{su}(6)^A_1\oplus \widehat{su}(6)^B_1\oplus \hat u(1)^B_3$ is
\be (s,s)\to \left( {\scalebox{0.5}{ \ydiagram{1}}}_{-2} \oplus  \overline{{\scalebox{0.5}{ \ydiagram{1}}}}_{2} \oplus {\scalebox{0.4}{ \ydiagram{1,1,1}}}_{0}, {\scalebox{0.5}{ \ydiagram{1}}}_{-2} \oplus  \overline{{\scalebox{0.5}{ \ydiagram{1}}}}_{2} \oplus {\scalebox{0.4}{ \ydiagram{1,1,1}}}_{0}\right)
\ee
\be (c,c)\to \left( {\bf 1}_{\pm3}  \oplus {\scalebox{0.4}{ \ydiagram{1,1}}}_{-1} \oplus \overline{{\scalebox{0.4}{ \ydiagram{1,1}}}}_{+1}  ,{\bf 1}_{\pm3}  \oplus {\scalebox{0.4}{ \ydiagram{1,1}}}_{-1} \oplus \overline{{\scalebox{0.4}{ \ydiagram{1,1}}}}_{+1}  \right)\ .
\ee  

The fact that $\tau$ is invariant under both $g$ and $g'$ in \eqref{gZ3} and \eqref{gpZ3} implies that the number of boxes with respect to both $\widehat{su}(6)_1^A$ and $\widehat{su}(6)_1^B$ must be a multiple of $3$. The only representations with this property in the $(s,s)$ and $(c,c)$ components are  \be \Bigl( {\scalebox{0.4}{ \ydiagram{1,1,1}}}_{0}, {\scalebox{0.4}{ \ydiagram{1,1,1}}}_{0}\Bigr)\subset (s,s) \quad  \text{(400 ground states)}\qquad \quad   ({\bf 1}_{\pm3},{\bf 1}_{\pm3})\subset (c,c)\quad \text{(4 ground states)} .\ee
In fact, if we decompose the supercurrent $\tau(z)$ as the sum
\be \tau(z)=\tau_s(z)+\tau_c(z)\ ,\qquad\tau_s\in \Bigl( {\scalebox{0.4}{ \ydiagram{1,1,1}}}_{0}, {\scalebox{0.4}{ \ydiagram{1,1,1}}}_{0}\Bigr),\quad \tau_c\in ({\bf 1}_{\pm3},{\bf 1}_{\pm3})
\ee
of a component $\tau_s\in (s,s)$ and one component $\tau_c\in (c,c)$, then one can prove that both $\tau_s$ and $\tau_c$ must be non-zero. We will not use this fact in the following.

The finite group $C_{Co_0}(g,g')\subset Co_0$ of elements of $Co_0$ that commute with both $g$ and $g'$ can be identified with a subgroup of $U(1)^A\times SU(6)^A \times SU(6)^B\times U(1)^B$ in \eqref{USUSUU}. In particular, one has
\begin{align} C_{Co_0}(g,g')&\to U(1)^A\times SU(6)^A \times SU(6)^B\times U(1)^B\\   h&\mapsto (1,\ U^A(h),\ U^B(h),\ 1) \ ,\label{centrtoLie}
\end{align} where $U^a(h)$ and $U^B(h)$ are the $SU(6)$ matrices in \eqref{UAhUBh}. Notice that $C_{Co_0}(g,g')$ is actually contained in the $SU(6)^A \times SU(6)^B$ subgroup of \eqref{USUSUU}. Because the four states with weight $3/2$ in $({\bf 1}_{\pm3},{\bf 1}_{\pm3})$ are in the trivial representation of $SU(6)^A \times SU(6)^B$, it follows that they are all automatically fixed by  $C_{Co_0}(g,g')$. On the other hand, using the character table of $C_{Co_0}(g,g')$, one can check that the trivial representation of $C_{Co_0}(g,g')$ appears with multiplicity $1$ in $\Bigl( {\scalebox{0.4}{ \ydiagram{1,1,1}}}_{0}, {\scalebox{0.4}{ \ydiagram{1,1,1}}}_{0}\Bigr)$. Such a representation must be spanned by the component $\tau_s$ of the supercurrent.

Each involution $a_j\in Co_0$ such that $a_jga_j=g^{-1}$ and $a_jg'a_j=g'$, as in \eqref{adef}, must act on the currents in $\widehat{su}(6)_1^B\oplus \hat u(1)_3^B$ by charge conjugation outer automorphisms, while it acts by an inner automorphism on $\widehat{su}(6)_1^A\oplus \hat u(1)_3^A$. We can represent such an element as
\be\label{aconj} a_j\mapsto (i,\ U^A(a_j),\ C_{SU(6)^B}(a_j),\ C_{U(1)^B})\ ,
\ee where $U^A(a_j)$ is the $SU(6)$ matrix in \eqref{aelem}, and $C_{SU(6)^B}(a_j)$, $C_{U(1)^B}$ are charge conjugation automorphisms. (This is an abuse of notation, as \eqref{aconj} is not in the Lie group \eqref{USUSUU}, but is an outer automorphism of the corresponding Lie algebra). Notice that different $a_j$'s satisfying \eqref{adef} are related by multiplication by some $h\in C_{Co_0}(g,g')$, and therefore by \eqref{centrtoLie} the outer $C_{U(1)^B}$ is the same for all such $a_j$. Recalling that $a_j$ is an involution in $Co_0$, and that $U^A(a_j)^2=-\mathbf{1}_6$ (see eq.\eqref{strangesquare}), we deduce that the charge conjugation automorphisms must satisfy 
\be C_{SU(6)^B}(a_j)^2=-\mathbf{1}_6\ ,\qquad C_{U(1)^B}^2=-1\ ,\ee so that
\be a^2_j\mapsto (-1,\ -\mathbf{1}_6,\ -\mathbf{1}_6,\ -1)\ee acts trivially on all the fields in $V^{f\natural}$, see \eqref{kerrep}.
Similarly, the involutions $a'_j\in Co_0$ such that $a'_jga'_j=g$ and $a'_jg'a'_j={g'}^{-1}$ as in \eqref{apdef} are associated with
\be\label{apconj} a'_j\mapsto (C_{U(1)^A},\ C_{SU(6)^A}(a'_j),\ U^B(a'_j),\ i)\ ,
\ee where $U^B(a'_j)$ is as in \eqref{apelem}, and
\be U^B(a'_j)^2=-\mathbf{1}_6,\qquad  C_{SU(6)^A}^2(a'_j)=-\mathbf{1}_6\ ,\qquad C_{U(1)^A}^2=-1\ .
\ee 
Finally, the involutions $b_\pm$ such that $b_\pm gb_\pm =g^{-1}$ and $b_\pm g'b_\pm ={g'}^{-1}$ as in \eqref{Cdef} are associated with
\be\label{bconj} b_\pm \mapsto (\pm C_{U(1)^A}i,\ C_{SU(6)^A}(b),\ C_{SU(6)^B}(b),\ \pm iC_{U(1)^B})\ ,
\ee  where $C_{SU(6)^A}(b)$ and $C_{SU(6)^B}(b)$ are charge conjugation automorphisms with
\be C_{SU(6)^A}^2(b)=-\mathbf{1}_6\ ,\qquad C_{SU(6)^B}^2(b)=-\mathbf{1}_6\ ,
\ee
while the central involution $\xi$ of $Co_0$ corresponds to
\be\label{xi} \xi\mapsto (-1,\ \mathbf{1}_6,\ \mathbf{1}_6,\ -1)\ .
\ee 
There is only one linear combination of the four ground states in $({\bf 1}_{\pm3},{\bf 1}_{\pm3})$ that is invariant under all $a_j$, $a'_k$, and $b_\pm $  -- this follows immediately from the fact that $a_j$, $a'_k$ and $b_\pm $ act as three commuting involutions with trace $0$ on this $4$-dimensional space. This invariant state must therefore be proportional to the component $\tau_c$ of the supercurrent.

\subsection{An explicit description of the $\ZZ_3$ duality defects}\label{s:Z3defects}

Let us now describe the main properties of a duality defect $\CN_g$ for the symmetry $g$ and its action on the fields of $V^{f\natural}$. The linear operator $\frac{1}{\sqrt{3}}\hat\CN_g$ must annihilate the fields that are not $g$-invariant, and must act on the $g$-invariant affine algebra $\widehat{so}(12)_1^A\oplus \widehat{su}(6)^B_1\oplus \hat u(1)^B_3$ in \eqref{Z3fixedaffine} by an automorphism of order $2$. In particular, for a Tambara-Yamagami category $TY(\ZZ_3,\chi,\epsilon)$, such an automorphism must exchange the $\widehat{so}(12)_1^A\oplus \widehat{su}(6)^B_1\oplus \hat u(1)^B_3$ representations that are $g$-eigenspaces with $g=\omega^k$, $k=1,2$, with the $g$-invariant representations in the $g^l$-twisted sector, $l=1,2$, where $k$ and $l$ are related by $\chi(l,1)=\omega^k$. By comparing tables \ref{tab:4}, \ref{tab:5}, and \ref{tab:6}, this means that:
\begin{itemize}
    \item For $\chi=\chi_+$ (see eq.\eqref{bicharZ3}), one exchanges the $g=\omega^k$ eigenspace with the $g^k$-twisted sector. This means that $\frac{1}{\sqrt{3}}\hat\CN_g$ acts by a outer charge conjugation automorphism of $\widehat{su}(6)^B_1$ and by inner automorphisms on $\widehat{so}(12)_1^A$ and  $\hat u(1)^B_3$.
    \item For $\chi=\chi_-$ (see eq.\eqref{bicharZ3}), one exchanges the $g=\omega^k$ eigenspace with the $g^{-k}$-twisted sector. This means that $\frac{1}{\sqrt{3}}\hat\CN_g$ acts by a outer charge conjugation automorphism on $\hat u(1)^B_3$, and by inner automorphisms on $\widehat{so}(12)_1^A$ and   $\widehat{su}(6)^B_1$.
\end{itemize}
Finally, $\hat\CN_g$ must act trivially on the supercurrent $\tau$.

In order to define such defects, let us first consider a linear operator $\Theta$ on the space of states of $V^{f\natural}$ that annihilates all the states that are not $g$-invariant, and such that $\frac{1}{\sqrt{3}}\Theta$ acts on the algebra $ \hat u(1)^A_3\oplus \widehat{su}(6)^A_1\oplus\widehat{su}(6)^B_1\oplus \hat u(1)^B_3\subset \widehat{so}(12)_1^A\oplus \widehat{su}(6)^B_1\oplus \hat u(1)^B_3$ by the automorphism
\be \frac{1}{\sqrt{3}}\Theta\mapsto (i,\ \mathbf{1}_6,\ \mathbf{1}_6,\ C_{U(1)^B})
\ee where $C_{U(1)^B}$ is the charge conjugation element in \eqref{aconj}. The restriction of this operator $\Theta$ to the $g$-invariant Ramond ground states is exactly as in eq.\eqref{Thetadef}.
For each two possible choice of $\chi$, and up to conjugation in $N_{Co_0}(g)$, the duality defects satisfying all required properties are given by
\be
\hat\CN_g^{(a'_j)}=\Theta a'_j\ , \qquad j\in \{+,0,-\}\ ,\qquad \text{for }\chi=\chi_-\ ,
\ee
\be \hat\CN_g^{(b_\pm)}=\Theta b_\pm\ ,\qquad \text{for }\chi=\chi_+\ .\ee  This implies 
\be\label{N31aut} \frac{1}{\sqrt{3}}\hat\CN_g^{(b_\pm )}\mapsto (\mp C_{U(1)^A},\ C_{SU(6)^A}(b),\ C_{SU(6)^B}(b),\ \mp i)\ ,\ee
\be\label{N32aut} \frac{1}{\sqrt{3}}\hat{\CN}_g^{(a'_j)}\mapsto (iC_{U(1)^A},\ C_{SU(6)^A}(a'_j),\ U^B(a'_j),\ C_{U(1)^B}i)
\ee and it is clear that such automorphisms exchange the representations of $\widehat{so}(12)_1^A\oplus \widehat{su}(6)^B_1\oplus \hat u(1)^B_3$ in the expected way.

Let us prove that $\frac{1}{\sqrt{3}}\hat\CN_g^{(b_\pm)}$ and $\frac{1}{\sqrt{3}}\hat{\CN}_g^{(a'_j)}$ act trivially on the supercurrent. Since $b_\pm$, $\xi$ and $a'_j$ are all in $\Aut_\tau(V^{f\natural})$, it is sufficient to show that the automorphism $\frac{1}{\sqrt{3}}\Theta$ preserves $\tau$. To show this, let us notice that $\frac{1}{\sqrt{3}}\Theta$ acts trivially on the representation
$\Bigl( {\scalebox{0.4}{ \ydiagram{1,1,1}}}_{0}, {\scalebox{0.4}{ \ydiagram{1,1,1}}}_{0}\Bigr)$, while it acts in the same way as $a_j\in \Aut_\tau(V^{f\natural})$ on the representation $({\bf 1}_{\pm3},{\bf 1}_{\pm3})$. Therefore, both components $\tau_s$ and $\tau_c$ of the supercurrent are fixed by $\frac{1}{\sqrt{3}}\Theta$.

In order to compute the twining partition functions and the twining genus, it is more useful to think of  $\frac{1}{\sqrt{3}}\hat\CN_g^{(k)}$, $k\in \{b_\pm,a'_0,a'_\pm\}$, as an automorphism of the $g$-invariant subalgebra $\widehat{so}(12)^A_1\oplus \widehat{su}(6)^B_1\oplus \hat u(1)^B_3$ as
\be \frac{1}{\sqrt{3}}\hat\CN_g^{(k)}\mapsto (O^{(k)},\ S^{(k)},\ T^{(k)})\ ,\qquad k\in \{b_\pm,a'_0,a'_\pm\}\ ,
\label{N_action:O-S-T}\ee where, by comparing with \eqref{N31aut} and \eqref{N32aut}, we have
\begin{align} \label{S-T:b_vectors} S^{(b_\pm )}&=C_{SU(6)^B}(b)\ ,& T^{(b_\pm )}&=\pm i\ ,\\
 \label{S-T:a_vectors} S^{(a'_j)}&=U^B(a'_j) \ ,& T^{(a'_j)}&=C_{U(1)^B}i\ .
\end{align} The automorphisms $O^{(k)}$ of $\widehat{so}(12)^A_1$ can be determined by requiring their action on the subalgebra $\hat u(1)^A_3\oplus \widehat{su}(6)^A_1$ to match with \eqref{N31aut} and \eqref{N32aut}. Because the group of automorphisms of $\widehat{so}(12)^A_1$ is $O(12)$, it is actually easier to determine $O^{(k)}$ by checking the action of $\hat\CN^{(k)}$ on the $g$-invariant Ramond ground states $\Lambda^g\otimes \RR$ of $V^{f\natural}$, comprising the vector representation of $O(12)$.
Notice that all $\frac{1}{\sqrt{3}}\hat\CN_g^{(k)}$ have trace $0$ on the $12$-dimensional space $\Lambda^g\otimes \RR$, because they all involve complex conjugation of the complex coordinates $z_1,\ldots,z_6$. Since these automorphisms have order $2$, it follows that the eigenvalues $+1$ and $-1$ of $\frac{1}{\sqrt{3}}\hat\CN_g^{(k)}$ (and therefore of $O^{(k)}\in O(12)$)  have the same multiplicity $6$ on $\Lambda^g\otimes \RR$. Thus, the affine subalgebra of $\widehat{so}(12)^A_1$ that is fixed by $\frac{1}{\sqrt{3}}\hat\CN_g^{(k)}$ is, in all cases, isomorphic to 
\be \widehat{so}(6)^{A+}_1\oplus \widehat{so}(6)^{A-}_1\subset \widehat{so}(12)^A_1\ .
\ee Notice that all $O^{(k)}$ are in $SO(12)\subset O(12)$, so they are inner automorphisms of $\widehat{so}(12)^A_1$.

As described in the introduction, in order to define the twining genus
\be \phi^{\CN^{(k)}} \left( V^{f \natural}, \tau, z \right) = Tr_{V^{f \natural}_{tw}} \left( \hat\CN^{(k)} \,(-1)^F q^{L_0 - 1/2} y^{J^3_0}\right)
    \label{N3_twininggenus}
\ee  one needs to choose a $4$-dimensional subspace $\Pi^\natural$ in the $6$-dimensional space of Ramond ground states that are fixed by $\frac{1}{\sqrt{3}}\hat\CN^{(k)}$. This choice determines a subalgebra $\widehat{su}(2)_1\oplus \widehat{su}(2)_1$ inside  $\widehat{so}(6)_1^{A+}\subset \widehat{so}(12)^{A}$. Then,  $J^3_0$ in \eqref{N3_twininggenus} is defined as the zero mode of a current in one of these $\widehat{su}(2)_1$. It is clear that the final form of the twining genera $\phi^{\CN^{(k)}}$ do not depend on the choices of $\Pi^\natural$ and of the current $J^3_0$.

The only representations of $\widehat{so}(12)^A_1\oplus \widehat{su}(6)^B_1\oplus \hat u(1)^B_3$ that contribute to the twining partition functions or twining genera are the ones that are $g$-invariant, see table \ref{tab:4}. Then, we can write
\begin{align} \frac{1}{\sqrt{3}}Z_{NS}^{\CN^{(k)},\pm}(\tau) &=\ch^{(12),O^{(k)}}_0(\tau)\ch^{(6),S^{(k)}}_{\bf{1}}(\tau)\ch^{(1),T^{(k)}}_{0}(\tau)+\ch^{(12),O^{(k)}}_v(\tau)\ch^{(6),S^{(k)}}_{{\scalebox{0.2}{ \ydiagram{1,1,1}}}}(\tau)\ch^{(1),T^{(k)}}_{3}(\tau)\notag\\
&\pm \ch^{(12),O^{(k)}}_s(\tau)\ch^{(6),S^{(k)}}_{{\scalebox{0.2}{ \ydiagram{1,1,1}}}}(\tau)\ch^{(1),T^{(k)}}_{0}(\tau)\pm\ch^{(12),O^{(k)}}_c(\tau)\ch^{(6),S^{(k)}}_{\bf{1}}(\tau)\ch^{(1),T^{(k)}}_{3}(\tau)
\end{align}
\begin{align} \frac{1}{\sqrt{3}}Z_{R}^{\CN^{(k)},\pm}(\tau) &=\ch^{(12),O^{(k)}}_v(\tau)\ch^{(6),S^{(k)}}_{\bf{1}}(\tau)\ch^{(1),T^{(k)}}_{0}(\tau)\notag\\
&\pm \ch^{(12),O^{(k)}}_c(\tau)\ch^{(6),S^{(k)}}_{{\scalebox{0.2}{ \ydiagram{1,1,1}}}}(\tau)\ch^{(1),T^{(k)}}_{0}(\tau)\pm\ch^{(12),O^{(k)}}_s(\tau)\ch^{(6),S^{(k)}}_{\bf{1}}(\tau)\ch^{(1),T^{(k)}}_{3}(\tau)
\end{align}
\begin{align} \frac{1}{\sqrt{3}}\phi^{\CN^{(k)}}(\tau,z)=&\ch^{(12),O^{(k)}}_v(\tau,z)\ch^{(6),S^{(k)}}_{\bf{1}}(\tau)\ch^{(1),T^{(k)}}_{0}(\tau)\notag\\
&- \ch^{(12),O^{(k)}}_c(\tau,z)\ch^{(6),S^{(k)}}_{{\scalebox{0.2}{ \ydiagram{1,1,1}}}}(\tau)\ch^{(1),T^{(k)}}_{0}(\tau)\\
&-\ch^{(12),O^{(k)}}_s(\tau,z)\ch^{(6),S^{(k)}}_{\bf{1}}(\tau)\ch^{(1),T^{(k)}}_{3}(\tau)\notag
\end{align} Here, $\ch^{(12),O^{(k)}}_M(\tau)$, $M\in\{0,v,s,c\}$, $\ch^{(6),S^{(k)}}_R$, $R\in\{\bf{1},{{\scalebox{0.2}{ \ydiagram{1,1,1}}}}\}$ and  $\ch^{(1),T^{(k)}}_{n}$, $n\in\{0,3\}$ denote the characters of a representation of the algebras $\widehat{so}(12)_1^A$, $\widehat{su}(6)_1^B$ and $\hat{u}(1)_3^B$, respectively, `twined' by their corresponding automorphism $O^{(k)}$, $S^{(k)}$ or $T^{(k)}$, i.e.
\be 
 \ch^{(12),O^{(k)}}_M(\tau)=\Tr_{M}(q^{L_0-\frac{6}{24}}O^{(k)})\ ,\qquad\ch^{(12),O^{(k)}}_M(\tau,z)=\Tr_{M}(q^{L_0-\frac{6}{24}}y^{J_0^3}O^{(k)})\ ,\ee
\be \ch^{(6),S^{(k)}}_R(\tau)=\Tr_{R}(q^{L_0-\frac{5}{24}}S^{(k)})\ ,\qquad \ch^{(1),T^{(k)}}_n(\tau)=\Tr_{n}(q^{L_0-\frac{1}{24}}T^{(k)})\ .
\ee 
Using the formulas for the twined characters in \eqref{eq:TrO0}, \eqref{eq:TrOcs} and \eqref{eq:TrOv}, with $2n=12$ and $\CO=\CO^{(k)}$ acting on the fermions with six eigenvalues equal to $+1$ and six equal to $-1$, we obtain the following simplifications:
\be\label{chvanish}  \ch^{(12),O^{(k)}}_s(\tau)= \ch^{(12),O^{(k)}}_c(\tau)= \ch^{(12),O^{(k)}}_s(\tau,z)= \ch^{(12),O^{(k)}}_c(\tau,z)=0\ .
\ee
\be \ch^{(12),O^{(k)}}_v(\tau)=0
\ee
and the following expression for the vacuum module:
\be \ch^{(12),O^{(k)}}_0(\tau)={\theta_3(\tau)^3\theta_4(\tau)^3\over \eta(\tau)^6}
\ee
Among the $6$ $O^{(k)}$-invariant free fermions, $4$ are charged under $J_0^3$ (the subspace $\Pi^\natural$) and two are neutral; the six $O^{(k)}=-1$ eigenstates are all neutral under $J_0^3$. Using this information, we can compute the twined character in the $v$ representation, graded with respect to this Cartan current
\be \ch^{(12),O^{(k)}}_v(\tau,z)={\theta_3(\tau,z)^2\theta_3(\tau)\theta_4(\tau)^3-\theta_4(\tau,z)^2\theta_4(\tau)\theta_3(\tau)^3\over 2\eta(\tau)^6}.
\ee
For the calculation of the other characters $\ch^{(6),S^{(k)}}_R$ and $\ch^{(1),T^{(k)}}_n$, it is useful to distinguish the cases $\CN^{(b_\pm)} $ and ${\CN}_g^{(a'_j)}$.

\paragraph{Defect $\CN^{(b_\pm)} $.}
    In this case, $S^{(b_\pm )}=C_{SU(6)^B}(b)$ is an outer charge conjugation automorphism of $\widehat{su}(6)_1$, and $T^{(b_\pm )}=\pm i$ an inner automorphism of $\hat u(1)_3$. We can use the description of $\hat u(1)_3$ in terms of a free boson, with the $n$ representation  corresponding to vertex operators $\CV_{\frac{m}{\sqrt{6}}}$ with $m\in n+6\ZZ$. Then, $T^{(b_\pm )}$ acts trivially on the bosonic oscillators, and  by
    \be T^{(b_\pm )}(\CV_{\frac{m}{\sqrt{6}}})=(\pm i)^m\CV_{\frac{m}{\sqrt{6}}}\ ,
    \ee on vertex operators. It follows that
    \be \ch_0^{(1),T^{(b_\pm )}}(\tau)={1\over \eta(\tau)}\sum_{k \in \ZZ}(\pm i)^{6k}q^{{1\over 12}(6k)^2}={\theta_4(6\tau)\over \eta(\tau)}
    \ee
    \be \ch_3^{(1),T^{(b_\pm )}}(\tau)={1\over \eta(\tau)}\sum_{k \in \ZZ}(\pm i)^{6k+3}q^{{1\over 12}(6k+3)^2}=0
    \ee 
    Let us now discuss the automorphism $S^{(b_\pm)}$ of $\widehat{su}(6)^B_1$. Kac's theorem (\cite{Kac}, Theorem 8.6) ensures that, up to conjugation, there are only two outer automorphisms of order $2$ of the (finite) Lie algebra $su(6)$, one fixing a subalgebra $so(6)\cong su(4)$ of dimension $15$ and one fixing a subalgebra $sp(6)$ of dimension $21$. In order to determine to which of these two cases $S^{(b_\pm )}=C_{SU(6)^B}(b)$ belongs, we notice that this element is in the same conjugacy class as the automorphism $C_{SU(6)^A}(b)$ of $\widehat{su}(6)^A_1$ in eq.\eqref{N31aut}. This follows because $g$ and $g'$ are conjugate within $Co_0$, since they have the same Frame shape, and that any element $k\in Co_0\subset Spin(24)$ such that $kgk^{-1}=g'$ must exchange the subalgebras $\widehat{su}(6)^A_1$ and $\widehat{su}(6)^B_1$; furthermore, $kbk^{-1}$ must be again an element of $N_{Co_0}(\langle g\rangle)=N_{Co_0}(\langle g'\rangle)$ satisfying \eqref{Cdef}. The automorphism $C_{SU(6)^A}(b)$ is induced by the automorphism $O^{(b_\pm )}$ of $\widehat{so}(12)_1^A$, that fixes the $15+15=30$ currents generating  $\widehat{so}(6)_1^{A+}\oplus \widehat{so}(6)_1^{A-}$ out of $66$ in $\widehat{so}(12)_1^A$. If we decompose the adjoint (vacuum) representation of $\widehat{so}(12)^A_1$ with respect to the subalgebra $\hat u(1)^A_3\oplus \widehat{su}(6)^A_1$ as in eq.\eqref{so12tosu6u1adj}, it is clear that $15$ of these $30$ $O^{(b_\pm )}$-preserved currents come from the representations ${{\scalebox{0.5}{ \ydiagram{1,1}}}}_2+\overline{{\scalebox{0.5}{ \ydiagram{1,1}}}}_{-2}$. Indeed, because $O^{(b_\pm )}$ exchanges ${{\scalebox{0.5}{ \ydiagram{1,1}}}}_2$ and $\overline{{\scalebox{0.5}{ \ydiagram{1,1}}}}_{-2}$, the $O^{(b_\pm )}$-invariant currents are given by symmetric linear combinations of each current in ${{\scalebox{0.5}{ \ydiagram{1,1}}}}_2$ and its image in $\overline{{\scalebox{0.5}{ \ydiagram{1,1}}}}_{-2}$. On the other hand, the current of $\hat u(1)^A_3$ is not fixed by $O^{(b_\pm )}$, because the action is by charge conjugation automorphism $C_{U(1)^A}$. Thus, the remaining $15$ $O^{(b_\pm )}$-fixed currents must be the ones in $\widehat{su}(6)^A_1$ that are fixed by $C_{SU(6)^A}(b)$. Combining this counting with Kac's theorem, we conclude that the subalgebra of $\widehat{su}(6)^A_1$ that is fixed by $C_{SU(6)^A}(b)$ is $\widehat{su}(4)_2\cong \widehat{so}(6)_2$; the level can be deduced by matching the central charge of the Sugawara stress tensor of $\widehat{su}(4)_2$ and $\widehat{su}(6)_1$. Analogously, the subalgebra of $\widehat{su}(6)_1^B$ preserved by the automorphism $S^{(b_\pm )}=C_{SU(6)^B}(b)$, must be isomorphic to $\widehat{su}(4)_2$.  

Using the description of $su(6)_1^B$ in terms of five chiral free bosons on the lattice $A_5$, the order-two automorphism we are interested in corresponds to a $\mathbb{Z}_2$ symmetry of the theory acting as a sign flip on the free bosons, $X^i \mapsto - X^i$, and flipping the sign of the momenta of the vertex operators $\mathcal{V}_{\lambda} \mapsto \mathcal{V}_{- \lambda}$ for $\forall \lambda \in A_5^{\ast}$. This symmetry implements charge conjugation, exchanging the fundamental and anti-fundamental representations, ${\scalebox{0.5}{ \ydiagram{1}}} \leftrightarrow \overline{{\scalebox{0.5}{ \ydiagram{1}}}}$, and ${\scalebox{0.5}{ \ydiagram{1,1}}} \leftrightarrow \overline{{\scalebox{0.5}{ \ydiagram{1,1}}}}$. Moreover, it leaves invariant exactly 15 currents, given by the symmetric combinations of vertex operators associated with the 30 roots $\lambda \in A_5$:
\begin{equation}
    \mathcal{V}_{\lambda} + \mathcal{V}_{- \lambda} \, . 
\end{equation}
The only contributions to the $\mathbb{Z}_2$-twining partition function come from oscillators, yielding to the following character:

 \begin{equation}
 \ch^{(6),S^{(b_\pm)}}_{\bf{1}}(\tau)={\eta(\tau)^5\over \eta(2\tau)^5}.
 \end{equation}

\paragraph{Defect ${\CN}_g^{(a'_j)}$.}
    The automorphism ${T}^{(a'_j)}$ acts by a minus sign on the bosonic oscillator of $\hat u(1)_3$ and maps states with $u(1)$ charge $m$ to states with charge $-m$.  Therefore, only the states with $u(1)$ charge $m=0$ contribute to the character $\ch_n^{(1),{T}^{(a'_j)}}(\tau)$. In particular,
    \be \ch_3^{(1),{T}^{(a'_j)}}(\tau)=0\ ,\qquad j\in\{0,+,-\}\ ,
    \ee  and 
    \be \ch_0^{(1),{T}^{(a'_j)}}(\tau)={\eta(\tau)\over \eta(2\tau)}\ ,\qquad j\in\{0,+,-\}\ .
    \ee
    The inner automorphism ${S}^{(a'_j)}$ on $\widehat{su}(6)_1^B$ is given by the $SU(6)$ matrix $U^B(a'_j)$, that has eigenvalues $+i,-i$ with multiplicities $(1,5)$, $(3,3)$ and $(5,1)$, respectively, for $j=+,0,-$. Let us use a description of $\widehat{su}(6)_1$ and its representations in terms of $A_5$ lattice VOA and cosets in $A_5^*/A_5$. If the $A_5$ lattice is described as in eq.\eqref{A5lattice}, then the dual $A_5^*$ is generated by the six (linearly dependent!) vectors
    \begin{align}\label{A5stargens}
        &(\frac{5}{6},-\frac{1}{6},-\frac{1}{6},\frac{1}{6},\frac{1}{6},\frac{1}{6}),\quad
        (-\frac{1}{6},\frac{5}{6},-\frac{1}{6},\frac{1}{6},\frac{1}{6},\frac{1}{6}),\quad
        (-\frac{1}{6},-\frac{1}{6},\frac{5}{6},\frac{1}{6},\frac{1}{6},\frac{1}{6}),\\
        &(-\frac{1}{6},-\frac{1}{6},-\frac{1}{6},-\frac{5}{6},\frac{1}{6},\frac{1}{6}),\quad
        (-\frac{1}{6},-\frac{1}{6},-\frac{1}{6},\frac{1}{6},-\frac{5}{6},\frac{1}{6}),\quad
        (-\frac{1}{6},-\frac{1}{6},-\frac{1}{6},\frac{1}{6},\frac{1}{6},-\frac{5}{6}),\notag
    \end{align} corresponding to the six ground states in the $\widehat{su}(6)_1$ representation ${\scalebox{0.5}{ \ydiagram{1}}}$. Then, up to conjugation in $SU(6)$, we can take the automorphism ${S}^{(a'_j)}$ to act on $\CV_\lambda$, $\lambda\in A_5^*$, by 
    \be {S}^{(a'_j)}(\CV_\lambda)=e^{2\pi i\beta_j\cdot\lambda}\CV_\lambda\ ,\qquad j\in\{0,+,-\}\ ,
    \ee with 
    \be \beta_\pm=\pm (\frac{5}{4},-\frac{1}{4},-\frac{1}{4},\frac{1}{4},\frac{1}{4},\frac{1}{4})\in \frac{1}{2}A_5^*\ ,
    \ee or
    \be \beta_0={3\over 4}(1,1,1,1,1,1)
    \ee One can easily verify that, with this choice of $\beta_j$, the vertex operators $\CV_\lambda$, where $\lambda$ are the six vectors in \eqref{A5stargens}, are eigenvectors of ${S}^{(a'_j)}$ with the same eigenvalues as $U^B(a'_j)$.
    A direct calculation using the lattice description yields
    \bea
 \ch^{(6),S^{(a'_\pm)}}_{\bf{1}}(\tau) & = & {\theta_4(6\tau)g(\tau)\over \eta(\tau)^5}\\
\nonumber &= & {1\over \eta(\tau)^5}(1+10 q-30 q^2+20 q^3-50 q^4+120 q^5-110 q^6+\ldots) 
    \eea
    where
    $$g(\tau)={1\over 4}(9\theta_4(6\tau)^4-5\theta_4(2\tau)^4)\ ,$$
and
    \bea
\ch^{(6),S^{(a'_0)}}_{\bf{1}}(\tau)&=&{h(\tau)\theta_4(2\tau)^2\over \eta(\tau)^5}\\
\nonumber &=&{1\over \eta(\tau)^5}(1-6 q+18 q^2-44 q^3+78 q^4-72 q^5+66 q^6+\ldots) 
    \eea
    where
    \be
h(\tau):=\theta_4(6\tau)^3+ {1\over 4}(\theta_4(2\tau/3)-\theta_4(6\tau))^3.
    \ee
 
Combining all these results, we get the following twining functions of $\CN^{(b_{\pm})}$
\begin{align}\nonumber
Z_{NS}^{\CN^{(b_\pm )},\pm}(\tau)=&\sqrt{3}\ch^{(12),O^{(b_{\pm})}}_0(\tau)\ch^{(6),S^{(b_{\pm})}}_{\bf{1}}(\tau)\ch^{(1),T^{(b_{\pm})}}_{0}(\tau)
=\sqrt 3{\theta_3(\tau)^3\theta_4(\tau)^3\theta_4(6\tau)\over\eta(\tau)^2\eta(2\tau)^5}\\
=&\sqrt 3\left (\frac{1}{q^{\frac{1}{2}}}-10 q^\frac{1}{2}+46 q^{\frac{3}{2}}-142 q^{\frac{5}{2}}+377 q^{\frac{7}{2}}-954 q^{\frac{9}{2}}+2238
   q^{\frac{11}{2}}
   +O(q^{\frac{13}{2}})\right)\\
   Z_{R}^{\CN^{(b_\pm )},\pm}(\tau)=&\sqrt{3}\ch^{(12),O^{(b_{\pm})}}_v(\tau)\ch^{(6),S^{(b_{\pm})}}_{\bf{1}}(\tau)\ch^{(1),T^{(b_{\pm})}}_{0}(\tau)=0
   \end{align}
 and the following twining genus 
   \begin{equation}
   \begin{split}
\phi^{\CN^{(b_\pm )}}(\tau,z)=&\sqrt{3}\ch^{(12),O^{(b_{\pm})}}_v(\tau,z)\ch^{(6),S^{(b_{\pm})}}_{\bf{1}}(\tau)\ch^{(1),T^{(b_{\pm})}}_{0}(\tau) \\
=&\sqrt 3{(\theta_3(\tau,z)^2\theta_3(\tau)\theta_4(\tau)^3-\theta_4(\tau,z)^2\theta_4(\tau)\theta_3(\tau)^3)\theta_4(6\tau)\over 2\eta(\tau)^2\eta(2\tau)^5} \, . 
\label{Nbpmgenus}
\end{split}
\end{equation}
For the twining functions and genus of $\CN^{(a'_\pm)}$ we get\footnote{Here and in the next equations the following identities have been used: $\theta_3(\tau)\theta_4(\tau)=\theta_4(2\tau)^2$ and $\eta(2\tau)\theta_4(2\tau)=\eta(\tau)^2$}, 
\begin{align}Z_{NS}^{\CN^{(a'_\pm)},\pm}(\tau)=&\sqrt{3}\ch^{(12),O^{(a'_\pm)}}_0(\tau)\ch^{(6),S^{(a'_\pm)}}_{\bf{1}}(\tau)\ch^{(1),T^{(a'_\pm)}}_{0}(\tau)=\sqrt 3{\theta_4(2\tau)^7\theta_4(6\tau)g(\tau)\over \eta(\tau)^{12}} \\ Z_{R}^{\CN^{(a'_\pm)},\pm}(\tau)=&\sqrt{3}\ch^{(12),O^{(a'_\pm)}}_v(\tau)\ch^{(6),S^{(a'_\pm)}}_{\bf{1}}(\tau)\ch^{(1),T^{(a'_\pm)}}_{0}(\tau)=0 \\
 \phi^{\CN^{(a'_\pm)}}(\tau,z)&=\sqrt{3}\ch^{(12),O^{(a'_\pm)}}_v(\tau,z)\ch^{(6),S^{(a'_\pm)}}_{\bf{1}}(\tau)\ch^{(1),T^{(a'_\pm)}}_{0}(\tau)\nonumber\\\label{Na'pmgenus}
&=\sqrt 3{(\theta_3(\tau,z)^2\theta_3(\tau)\theta_4(\tau)^3-\theta_4(\tau,z)^2\theta_4(\tau)\theta_3(\tau)^3)\theta_4(6\tau)g(\tau)\over 2\eta(\tau)^{10}\eta(2\tau)}.
\end{align}
 Finally, for the twining functions and genus of $\CN^{(a'_0)}$ we have 
\begin{align}Z_{NS}^{\CN^{(a'_0)},\pm}(\tau)=&\sqrt{3}\ch^{(12),O^{(a'_0)}}_0(\tau)\ch^{(6),S^{(a'_0)}}_{\bf{1}}(\tau)\ch^{(1),T^{(a'_0)}}_{0}(\tau)=\sqrt 3{\theta_4(2\tau)^9h(\tau)\over \eta(\tau)^{12}}\end{align}

\begin{align} Z_{R}^{\CN^{(a'_0)},\pm}(\tau)=&\sqrt{3}\ch^{(12),O^{(a'_0)}}_v(\tau)\ch^{(6),S^{(a'_0)}}_{\bf{1}}(\tau)\ch^{(1),T^{(a'_0)}}_{0}(\tau)=0
\end{align}
\begin{align} \phi^{\CN^{(a'_0)}}(\tau,z)&=\sqrt{3}\ch^{(12),O^{(a'_0)}}_v(\tau,z)\ch^{(6),S^{(a'_0)}}_{\bf{1}}(\tau)\ch^{(1),T^{(a'_0)}}_{0}(\tau)\nonumber\\
&=\sqrt 3{(\theta_3(\tau,z)^2 \theta_4 (\tau)^2-\theta_4(\tau,z)^2 \theta_3 (\tau)^2) \theta_4(2\tau)^5 h(\tau)\over 2\eta(\tau)^{12}}.\label{Na0genus}
\end{align}
These twining genera, reported in equations \eqref{Nbpmgenus}, \eqref{Na'pmgenus}, and \eqref{Na0genus}, are predictions for the twining genera of K3 NLSMs with $\ZZ_3$ duality defects, according to Conjecture \ref{conj:K3relation}.

Finally, we can determine the corresponding TY categories of each duality defect from the spin selection rules, by
considering the S-transform of the NS twining functions. In the case of $\CN^{(b_\pm )}$, we find
\begin{align}\nonumber
Z_{\CN^{(b_\pm )},NS}^{\pm}(\tau)=&\sqrt 3{\theta_3(-1/\tau)^3\theta_4(-1/\tau)^3\theta_4(-6/\tau)\over\eta(-1/\tau)^2\eta(-2/\tau)^5}={4\theta_3(\tau)^3\theta_2(\tau)^3\theta_2(\tau/6)\over\eta(\tau)^2\eta(\tau/2)^5}\\
=&64 q^{5/24}+64 q^{3/8}+768 q^{17/24}+704 q^{7/8}+5056 q^{29/24}+4288 q^{11/8}+ \ldots, 
\end{align} so that the conformal weights of the $\CN^{(b_\pm )}$-twisted sector take values in
$ {1\over 2} \ZZ+ \{{5\over 24}, -{1\over 8}\}$, as expected for a category TY$(\ZZ_3,\chi_+,-1)$ (see Table \ref{t:spinselectZ3}).

Similarly, in the case of $\CN^{(a'_\pm)}$ we find,
\begin{align}\nonumber
Z_{\CN^{(a'_\pm)},NS}^{\pm}(\tau)=&\sqrt 3{\theta_4(-2/\tau)^7\theta_4(-6/\tau)g(-1/\tau)\over \eta(-1/\tau)^{12}}={\theta_2(\tau/2)^7\theta_2(\tau/6)(\theta_2(\tau/6)^4-5\theta_2(\tau/2)^4)\over 256\eta(\tau)^{12}}\\
=&16 q^{1/24}+80 q^{3/8}+352 q^{13/24}+736 q^{7/8}+2768 q^{25/24}+4352 q^{11/8} + \ldots, 
\end{align}
so that the conformal weights of the $\CN^{(a'_\pm)}$-twisted sector take values in
$ {1\over 2} \ZZ+ \{{1\over 24}, -{1\over 8}\}$, as expected for TY$(\ZZ_3,\chi_-,+1)$.

Finally, for $\CN^{(a'_0)}$, we find
\begin{align}\nonumber
Z_{\CN^{(a'_0)},NS}^{\pm}(\tau)=&\sqrt 3{\theta_4(-2/\tau)^9h(-1/\tau)\over\eta(-1/\tau)^{12}}={\theta_2(\tau/2)^9 \left[ \theta_2(\tau/6)^3+ {1\over 4}(3\theta_2(3\tau/2)-\theta_2(\tau/6))^3 \right] \over 192\eta(\tau)^{12}}\\
=&16 q^{1/8}+96 q^{7/24}+256 q^{5/8}+1056 q^{19/24}+1872 q^{9/8}+6624 q^{31/24} + \ldots, 
\end{align}
such that the conformal weights of the $\CN^{(a'_0)}$-twisted sector take values in
$ {1\over 2} \ZZ+ \{-{5\over 24}, {1\over 8}\}$, as expected for TY$(\ZZ_3,\chi_-,-1)$.

\section{Duality defects for non-cyclic groups} \label{s:duality_def_noncyclic}

In this section, we consider some cases of Tambara-Yamagami category $TY(A)$ of topological defects in $\cTop$ for non-cyclic abelian groups $A$. In particular, after some general discussion for groups of the form $A=\ZZ_n\times\ZZ_n$, we will explicitly describe an example of duality defect for $A=\ZZ_2\times \ZZ_2$ in $V^{f\natural}$. Finally, we will describe a K3 model with the same TY category of defects and show the matching of the corresponding twining genera.

\subsection{Duality defects from orbifolds}

In \cite{Bhardwaj:2017xup} it was shown that a large class of TY categories $TY(A)$ can be obtained by an orbifold procedure similar to the one that was discussed in section 4 of \cite{Angius:2025zlm}. This construction can only work when the duality defect has integral dimension, i.e. when the order $|A|$ of the abelian group is a perfect square.

Let us briefly review the general construction in section 4 of \cite{Angius:2025zlm}. Suppose a holomorphic (super)VOA $V$ admits a finite group of symmetries $G$, and let $H\subseteq G$ be a subgroup. For simplicity, we assume here that $G$ has trivial 't Hooft anomaly. Then, the orbifold $V':=V/H$ is a well defined holomorphic VOA,\footnote{Strictly speaking, one needs to assume that the $H$-fixed subalgebra is regular. This is conjecturally true for any holomorphic VOA $V$ and finite group $H$, but it has been rigorously proved only when $H$ is solvable \cite{Carnahan:2016guf}.} and one can consider the category of topological defects in $V'$ preserving the $G$-fixed VOA $V^G\subseteq V^H$, which is a common subVOA both in $V$ and $V'$. As described in \cite{Angius:2025zlm}, such defects can be labeled by a pair $(HgH,\rho)$, where $HgH$ is a double coset in $H\backslash G/H$, and $\rho\in Irr(H)$ an irreducible representation of $H$. The twined and twisted partition functions for the defect $\CL_{HgH,\rho}$ in $V'=V/H$ are given by
\be\label{twinHgH} Z^{HgH,\rho}(V/H,\tau)=\frac{1}{|H|}\sum_{g'\in HgH}\sum_{\substack{h\in H\\ hg'=g'h}}Tr_{\rho}(h)^*Z^{g'}_{h^{-1}}(V,\tau)\ ,
\ee and
\be Z_{HgH,\rho}(V/H,\tau)=\frac{1}{|H|}\sum_{g'\in HgH}\sum_{\substack{h\in H\\ hg'=g'h}}Tr_{\rho}(h)^*Z_{g'}^h(V,\tau),
\ee where $Z_g^h(V,\tau)$ is the $g$-twisted $h$-twining partition function in $V$. See \cite{Angius:2025zlm} for an explanation of these formulae. In particular, for the trivial coset $HeH$, one gets the Wilson lines associated with the irreducible representations $\rho$ of the finite group $H$ that has been `gauged'. Under some mild conditions, this construction can be generalized to the case of SVOAs, and in particular when $V'=V^{f\natural}$ and $V^G$ contains the supercurrent $\tau$. If the defects $\CL_{HgH,\rho}$ obtained in this way also preserve a four dimensional subspace of the ground states in $V^{f\natural}_{tw}$, then one can define the corresponding twining genus in terms of the twisted-twining genera $\phi_g^h$ in the parent theory $V$.

\bigskip

Let us now apply this general construction to define some duality defects for an abelian group $A$ whose order is a perfect square. Let us  specialize this discussion to the case where the abelian group $A$ is isomorphic to $\ZZ_n\times \ZZ_n$, for some $n$.

Suppose that the VOA $V$ contains a group of symmetries $G\cong(\ZZ_n\times\ZZ_n)\rtimes \ZZ_2$ of order $2n^2$ with generators and relations 
\be G=\langle a,b,g\mid a^n=1=b^n\ ,\quad ab=ba\ ,\quad g^2=1\ ,\quad gag=b\ ,\quad gbg=a\rangle\ .
\ee This means that $a,b$ generate the normal abelian subgroup $\ZZ_n\times \ZZ_n$, while conjugation by the involution $g$ exchanges the two generators $a$ and $b$.  Let $H\subset G$ be the cyclic subgroup $H=\langle b\rangle\cong \ZZ_n$ generated by $b$. 

The group $G$ has $n(n+1)/2+n$ conjugacy classes, with representative
\be [a^jb^k]=[a^kb^j]\ ,\qquad 0\le j\le k\le n-1\ ,
\ee
\be [ga^j]=[ga^kb^{j-k}]\ ,\qquad 0\le j\le n-1\ .
\ee The classes in the first line contain either one (if $j=k$) or two (if $j<k$) elements, while each class $[ga^j]$ in the second line  contains the $n$ elements $ga^kb^{j-k}$, $0\le k\le n-1$. There are $2n$ irreducible $1$-dimensional representations $\rho_q,\tilde\rho_q\in Irr(G)$, $q=0,\ldots,n-1$, with
\be \rho_q(a)=\rho_q(b)=e^{\frac{2\pi i q}{n}}\ ,\qquad \rho_q(g)=1\ ,
\ee
\be \tilde\rho_q(a)=\tilde\rho_q(b)=e^{\frac{2\pi i q}{n}}\ ,\qquad \tilde\rho_q(g)=-1\ ,
\ee and $n(n-1)/2$ irreducible two-dimensional representations $\rho_{q_1,q_2}$, $0\le q_1<q_2\le n-1$, given by
\be \rho_{q_1,q_2}(a)=\begin{pmatrix}
	e^{\frac{2\pi i q_1}{n}} & 0\\ 0 & e^{\frac{2\pi i q_2}{n}}
\end{pmatrix}\ ,\quad \rho_{q_1,q_2}(b)=\begin{pmatrix}
e^{\frac{2\pi i q_2}{n}} & 0\\ 0 & e^{\frac{2\pi i q_1}{n}}
\end{pmatrix}\ ,\quad \rho_{q_1,q_2}(g)=\begin{pmatrix}
0 & 1\\ 1 & 0
\end{pmatrix}.
\ee One can check that the sum of the dimensions squared equals $2n^2$, so that the list is complete. 

The centralizers in $H$ and in $G$ of an element $b^j$, $j\neq 0$, are $C_H(b^j)=\langle b\rangle\cong\ZZ_n$ and $C_G(b^j)=\langle a,b\rangle\cong\ZZ_n\times \ZZ_n$, respectively. We denote by $\sigma_{q_a,q_b}$ the representation of $\langle a,b\rangle$ given by $\sigma_{q_a,q_b}(a)=e^{\frac{2\pi i q_a}{n}}$ and $\sigma_{q_a,q_b}(b)=e^{\frac{2\pi i q_b}{n}}$. 

Recall that the irreducible representations of the $G$-invariant subVOA $V^G$ of a holomorphic $V$ are labeled by a pair $([g],\rho)$, where $[g]$ is a conjugacy class in $G$ and $\rho\in Irr(C_G(g))$ is an irreducible representation of the centralizer $C_G(g)$ of $g$ in $G$. In particular, the $g$-twisted sector $V_g$ contains all representations of the form $([g],\rho)$, with $[g]$ the conjugacy class of $g$. The theory $V'=V/H$ contains the following $V^G$-primary operators
\be \begin{aligned} & \Phi_{1,\rho_{0}}\ ,\qquad \Phi_{1,\tilde \rho_{0}}\ ,\qquad \Phi^0_{1,\rho_{0,q_2}},\, && 1\le q_2\le n-1,\\ & \Phi_{b^j,\sigma_{q_a,0}},\,  && 1\le j\le n-1,\quad 0\le q_a\le n-1\ , \\ \end{aligned}
\ee each one with multiplicity one. Here, $\Phi_{1,\rho}$, $\rho\in Irr(G)$, denote the operators in the untwisted sector $V^H$, with $\Phi^0_{1,\rho_{0,q_2}}$ being the only $H$-invariant $V^G$-primary operator in representation $\rho_{0,q_2}$. Furthermore, $\Phi_{b^j,\sigma_{q_a,0}}$ is the $V^G$-primary operator in the $b^j$-twisted sector in representation $\sigma_{q_a,0}\in Irr(C_G(b^j))$.

The double quotient $H\backslash G/H$  contains $n$ cosets with $n$ elements each, given by
\be Ha^kH=\{a^kb^j,\ j=0,\ldots,n-1\}\ ,\qquad k=0,\ldots,n-1\ ,
\ee and one coset with $n^2$ elements, namely
\be HgH=\{ b^jgb^k\ ,\ j,k=0,\ldots,n-1\}\ .
\ee Because $H$ is abelian, the irreducible $H$ representation are $1$-dimensional and the corresponding Wilson lines are invertible, forming the quantum symmetry group $\langle Q\rangle\cong \ZZ_n$. More generally the cosets $Ha^kH$, $k=0,\ldots, n-1$, give rise to $n^2$ invertible defects $\CL_{a^jQ^k}$, $j,k=0,\ldots,n-1$, in the orbifold theory $V/H$, forming a group $\ZZ_n\times \ZZ_n$ generated by $a$ and by the quantum symmetry $Q$.

On the other hand, the coset $HgH$ corresponds to defects $\CL_{HgH,\rho}$, with $\rho\in Irr(H)$, with twisted and twining partition functions
\be Z^{HgH,\rho} =\frac{1}{n}\sum_{j,k=0}^{n-1}Z^{gb^{j+k}}=\sum_{j=0}^{n-1}Z^{gb^j}
\ .
\ee
\be Z_{HgH,\rho} =\frac{1}{n}\sum_{j,k=0}^{n-1}Z_{gb^{j+k}}=\sum_{j=0}^{n-1}Z_{gb^j}
\ee where we use that $b^jgb^{k}$ is conjugate to $gb^{j+k}$.

Notice that these partition functions are the same for all $\rho$, so all $\CL_{HgH,\rho}$ can be identified with a single defect of dimension $n$ that we denote by $\CN=\CL_{HgH,\rho}$. Because
\be \sum_{j,k=0}^{n-1} \langle \CL_{a^jQ^k}\rangle^2 +\langle \CN\rangle^2=n^2\cdot 1^2+n^2=2n^2=|G|\ ,
\ee we conclude that $\CL_{a^jQ^k}$ and $\CN$ are the only simple defects preserving the subalgebra $V^G$ in the orbifold theory $V/H$. 
From the twining partition functions, one can read off the eigenvalues 
\be \hat\CL(\Phi_{[h],R})=c_{[h],R} \Phi_{[h],R}\ ,
\ee
of a simple defect $\CL$ on the $V^G$-primary operator  $\Phi_{[h],R}$, $R\in Irr(C_G(h))$. They are given by
\be \begin{array}{c|cccc}
 & c_{1,\rho_0} & c_{1,\tilde\rho_0} & c_{1,\rho_{0,q_2}} & c_{b^j,\sigma_{q_a,0}}\\
 \hline
 \CL_{a^rQ^s} & 1 & 1 & e^{\frac{2\pi i rq_2}{n}} & e^{\frac{2\pi i (rq_a+sj)}{n}}\\
 \CN & n & -n & 0 & 0 
\end{array}
\ee

This implies that $\CN$ is unoriented and that
\be \CN\CL_{a^jQ^k}=\CL_{a^jQ^k}\CN=\CN\ ,\qquad \CN^2=\sum_{r,s=0}^{n-1}\CL_{a^rQ^s}\ .
\ee It follows that $\CN$ is a duality defect for $\ZZ_n\times \ZZ_n$, i.e. that the category of $V^G$-preserving defects in $V/H$ is a Tambara-Yamagami category $TY(\ZZ_n\times \ZZ_n)$. This suggests that the theory $V':=V/H$ should be self-orbifold with respect to the group $\langle a,Q\rangle\cong\ZZ_n\times \ZZ_n$, i.e. that
\be V'\cong V'/\langle Q,a\rangle\ .
\ee Indeed, we can easily prove that
\be V'/\langle Q,a\rangle=(V'/\langle Q\rangle)/\langle a\rangle=V/\langle a\rangle\cong V/\langle b\rangle=V'.
\ee In the first equality, we observed that the orbifold $V'/\langle Q,a\rangle$ can be completed in two steps, by dividing first by $\langle Q\rangle \cong\ZZ_n$ and then by $\langle a\rangle\cong\ZZ_n$. Then, because $Q$ is the quantum symmetry of $V'=V/\langle b\rangle$, taking the orbifold $V'/\langle Q\rangle$ simply gives back the theory $V$. Finally, we notice that $a=gbg$, i.e. $a$ and $b$ are conjugate within the symmetry group $G$, so that the orbifold of $V$ by $\langle a\rangle$ or $\langle b\rangle$ must give rise to isomorphic theories. 

By following the chain of isomorphisms, one can check that the eigenspace of $V'$ with eigenvalues $Q=e^{\frac{2\pi i k}{n}}$, $a=e^{\frac{2\pi i s}{n}}$ mapped to the $(a^kQ^s)$-twisted sector in $V'/\langle a,Q\rangle$. this means that the symmetric bicharacter $\chi: \langle a,Q\rangle\times \langle a,Q\rangle\to U(1)$ is given by
\be \chi(a,a)=1=\chi(Q,Q)\ ,\qquad \chi(a,Q)=e^{\frac{2\pi i}{n}}\ .
\ee The fact that the group $G$ is anomaly free implies that $\epsilon=+1$, so that the Tambara-Yamagami category generated by these defects is $TY(\ZZ_n\times \ZZ_n,\chi,+1)$. Vice versa, because the orbifold procedure is invertible, one obtains that any category if a VOA $V'$ has a category $TY(\langle a,Q\rangle,\chi,+1)$ for some group of symmetries $\langle a,Q\rangle\cong \ZZ_n\times\ZZ_n$ and $\chi$ defined as above, the the orbifold by $V=V'/\langle Q\rangle$ has a non anomalous group of symmetries $G\cong (\ZZ_n\times\ZZ_n)\rtimes \ZZ_2$.

\bigskip

One can obtain more general TY categories from the same group $G$ with non-trivial 't Hooft anomaly (of course, the anomaly must be trivial when restricted to the subgroup $H$). In particular, all TY categories of the groups $A=\ZZ_2\times \ZZ_2$ (four inequivalent ones) and $A=\ZZ_4$  (four inequivalent ones) can be obtained from an orbifold of a subgroup $H\cong \ZZ_2$ within a non-abelian group $G\cong (\ZZ_2\times \ZZ_2)\rtimes \ZZ_2\cong D_8$, upon choosing a suitable anomaly $\alpha\in H^3(G,U(1))$ that becomes trivial when restricted to $H$ \cite{Bhardwaj:2017xup}. In the next section, we will provide an example of this general construction for a $\ZZ_2\times \ZZ_2$ group of $\CN=1$-preserving symmetries in $V^{f\natural}$.

\subsection{Tambara-Yamagami for $\ZZ_2\times\ZZ_2$ in $V^{f\natural}$}\label{s:Z2Z2Vfnat}

In this section, we consider a particular example of the construction described in the previous section, where $V\cong V'\cong V^{f\natural}$ and $A=\langle a,Q\rangle \cong \ZZ_2\times \ZZ_2\subset Co_0$ is a group where all non-trivial elements $a$, $Q$, and $aQ$ have Frame shape $1^82^8$.

As a starting point, let us consider a non-abelian subgroup  of order $8$ of $Co_0$ of the form $G\cong (\ZZ_2\times \ZZ_2)\rtimes \ZZ_2\cong D_8$ with generators and relations 
\be G=\langle a,b,g\mid a^2=1=b^2\ ,\quad ab=ba\ ,\quad g^2=1\ ,\quad gag=b\ ,\quad gbg=a\rangle\ .
\ee
The group $G$ has 5 conjugacy classes 
\begin{align}
&[a]=[b]\,,[1]\,, [ab]\,,\\
&[g]=[gab]\,,[ga]=[gb]\,.
\end{align}
We assume that $a$, $b$, and $ab$ all have Frame shape $1^8 2^8$. This means that the  subgroup $\langle a,b\rangle\cong \ZZ_2\times \ZZ_2$ is anomaly free. Now, as discussed in section \ref{section:TDLs_Vfnat}, the orbifold $V':=V^{f\natural}/\langle b\rangle$ by an element $b$ of Frame shape $1^82^8$ is again isomorphic to $V^{f\natural}$, and the quantum symmetry $Q$ has again Frame shape $1^82^8$. Furthermore, the fact that the group $\langle a,b\rangle\cong \ZZ_2\times \ZZ_2$ in $V$ is not anomalous implies that in the orbifold theory $V'$ the group generated by $a$ and $Q$ is isomorphic to $\langle a,Q\rangle\cong \ZZ_2\times\ZZ_2$, i.e. it is a trivial central extension of $\langle a\rangle\cong \ZZ_2$ by the quantum symmetry $\langle Q\rangle$. Vice versa,  the group $\langle a,Q\rangle$ in the theory $V'$ must be non-anomalous, since the symmetry induced by this group on theory $V=V'/\langle Q\rangle$ is $\langle a,b\rangle\cong\ZZ_2\times\ZZ_2$ that is a trivial central extension of $\langle a\rangle$ by the $Q$-quantum symmetry $\langle b\rangle$. In fact, one can explicitly show (see below) that $a$, $Q$ and $aQ$ all have Frame shape $1^82^8$ in $V'$.

One can prove that any subgroup of order $2^r$ in $Co_0$, for any power $r$, is contained in a subgroup of the form $\ZZ_2^{12}:M_{24}\subset Co_0$. One can always find a suitable orthonormal basis of $\Lambda\otimes\RR\cong \RR^{24}$ that simultaneously diagonalizes $a$ and $b$; let us assume that $g$ acts on the basis vectors by a permutation in the Mathieu group $M_{24}$.  The set of eigenvalues of $a$ with Frame shape $1^82^8$, up to re-orderings of the basis vectors, is given by
\begin{equation}\label{aZ2Z2}
    a = (1,1,1,1,\ 1,1,1,1,\ 1,1,1,1,\ 1,1,1,1,\ -1,-1,-1,-1,\ -1,-1,-1,-1)\,,
\end{equation}
and the requirement that also $b$ and $ab$ have Frame shape $1^8 2^8$ implies that 
\begin{equation}
    b = (-1,-1,-1,-1,\ 1,1,1,1,\ 1,1,1,1,\ 1,1,1,1,\ -1,-1,-1,-1,\ 1,1,1,1)\,,
\end{equation}
up to permutations of the first $16$ eigenvalues among themselves and  the last $8$ among themselves. Notice that on the $16$ $b$-fixed vectors, that represent the untwisted Ramond ground states in the orbifold $V/\langle b\rangle$, the symmetry $a$ has $4$ eigenvalues $-1$ and $12$ eigenvalues $+1$. This means that both $a$ and $aQ$ must have Frame shape $1^82^8$ in the orbifold theory, since this is the only possibility with so many fixed vectors.

The element $g$ acts as a permutation on this lists of eigenvalues so as to swap $a \leftrightarrow b$. If we assume that $g$ is an involution in $M_{24}$, then the only possibilities for the Frame shape of $g$ is $1^82^8$ ($8$ transpositions and $8$ fixed points) or $2^{12}$ ($12$ transpositions).

We can simply list all possible permutations $g$, up to conjugation by permutations in $S_{24}$ that commute with $a$ and $b$. Every permutation $g$ of order $2$ must contain $4$ transpositions exchanging the four $(a,b)$-eigenvectors with eigenvalues $(1,-1)$ with the four eigenvectors with eigenvalues $(-1,1)$. The remaining transpositions must exchange eigenvectors with the same $(a,b)$-eigenvalues.
For examples, we can then take $g\equiv p_1$, with
\be p_1=(1,21)(2,22)(3,23)(4,24)(9,13)(10,14)(11,15)(12,16)\ ,
\ee
so that 
\begin{equation}
    b = p_1 a p_1\,.
\end{equation}
In this case, $p_1$ has Frame shape $1^8 2^8$, and
we have $(p_1a)^4 = (p_1b)^4 = 1$, with eigenvalues of $p_1 a$ and $p_1 b$ 
\begin{equation}
  (-1,-1,-1,-1,-1,-1,-1,-1,i,i,i,i,-i,-i,-i,-i,1,1,1,1,1,1,1,1)\,, 
\end{equation}
implying a Frame shape $2^4 4^4$ ($4D$).
We have $(p_1 a b)^2=1$, with eigenvalues of $p_1 a b$
\begin{equation}
    (-1,-1,-1,-1,-1,-1,-1,-1,1,1,1,1,1,1,1,1,1,1,1,1,1,1,1,1)\,,
\end{equation}
implying a Frame shape $1^8 2^8$ ($2A$). The element $p_1$ is a representative of the set of $M_{24}$ permutations with cycle shape $1^82^8$ such that, besides the $4$ transpositions that exchange  the $(a,b)=(-1,1)$ eigenvectors with the $(a,b)=(1,-1)$ eigenvectors, contains $4$ transpositions that exchange $(a,b)=(1,1)$ eigenvectors among themselves. 

Another possible choice is that two transpositions in $g$ exchange eigenvectors of eigenvalues $(-1,-1)$ among themselves. For example, one can take $g=p_0$ with
\be p_0=(1,21)(2,22)(3,23)(4,24)(13,16)(14,15)(17,20)(18,19)
\ee
with $p_0 a$ and $p_0 b$ having the eigenvalues
\begin{equation}
\left(-1,-1,-1,-1,i,i,i,i,-i,-i,-i,-i,1,1,1,1,1,1,1,1,1,1,1,1\right)\,.
\end{equation}
However, $p_0b$ would have Frame shape $1^8 4^4$, which is not one of the allowed Frame shapes in $Co_0$. This is the signal that $p_0$ is not a permutation in $M_{24}$, and therefore we should not consider it.

The last possible choice for $g$ of Frame shape $1^82^8$ is that $1$ transposition exchanges eigenvectors of $(a,b)$-eigenvalues $(-1,-1)$ among themselves, whereas other $3$ exchange eigenvectors of eigenvalues $(1,1)$, namely
\be p_2=(1,21)(2,22)(3,23)(4,24)(5,8)(6,7)(13,14)(19,20)
\ee
with $p_2 a$ and $p_2 b$ having the eigenvalues
\begin{equation}
    \left(-1,-1,-1,-1,-1,-1,i,i,i,i,-i,-i,-i,-i,1,1,1,1,1,1,1,1,1,1\right)\,,
\end{equation}
consistent with Frame shape $1^4 2^2 4^4$ ($4C$). $p_2 a b$ still has Frame shape $1^8 2^8$ ($2A$).\\

Finally, when $g$ has Frame shape $2^{12}$, there is essentially only one possibility, namely $g=p_3$ with
\be p_3=(1,21)(2,22)(3,23)(4,24)(5,8)(6,7)(9,13)(10,14)(11,15)(12,16)(17,20)(18,19)
\ee
with Frame shape $2^{12}$, so that 
\begin{equation}
    b = p_3 a p_3\,.
\end{equation} 
$p_3 a$ and $p_3 b$ have Frame shape $2^4 4^4$ ($4D$); $p_3 a b$ has a Frame shape $2^{12}$.

Notice that the three cases that we considered, namely $g=p_1$, $g=p_2$, and $g=p_3$, correspond to subgroups $G\subset Co_0$ with different anomalies. Indeed, for $g=p_2$ the group $G$ is non-anomalous (see table \ref{tbl:anomalous} for the list of anomalous Frame shapes). For $g=p_1$, the anomaly is trivial when restricted to $\langle g\rangle\cong \ZZ_2$, but the groups $\langle ga\rangle\cong \ZZ_4\cong \langle gb\rangle$ have an anomaly of order $2$. Finally, when $g=p_3$, both $\langle g\rangle\cong \ZZ_2$ and $\langle ga\rangle\cong \ZZ_4$ have an order $2$ anomaly. As a consequence, the orbifold construction leads to inequivalent Tambara-Yamagami categories $TY(A,\chi,\epsilon)$ with different twining partition functions and genera for the duality defect $\CN$. More precisely:

 \begin{itemize}
 \item $g=p_1$. 
    \be\label{bichZ2Z2-} \chi(Q,Q)=1,\qquad \chi(a,a)=\chi(a,Q)=-1,\qquad \qquad \epsilon=+1
    \ee
    \begin{align}\label{eq:Z2Z2-+}
        \mathcal{Z}^{\CN_{p_1}}=\mathcal{Z}^{1^82^8}+\mathcal{Z}^{2^44^4}\qquad \phi^{\CN_{p_1}}=\phi^{1^82^8}+\phi^{2^44^4}
    \end{align}
     This is the same category as $TY(\ZZ_2\times\ZZ_2,\chi_s,+1)\cong Rep(H_8)$, the category of representations of the Hopf algebra $H_8$ \cite{Tambara:1998}.
    \item $g=p_2$. 
    \be\label{bichZ2Z2+} \chi(Q,Q)=1=\chi(a,a),\qquad \chi(Q,a)=-1,\qquad \qquad \epsilon=+1\ ,
    \ee
    \begin{align}\label{eq:Z2Z2++}
        \mathcal{Z}^{\CN_{p_2}}=\mathcal{Z}^{1^82^8}+\mathcal{Z}^{1^42^24^4}\qquad \phi^{\CN_{p_2}}=\phi^{1^82^8}+\phi^{1^42^24^4}\ .
    \end{align}
    This TY category is equivalent to the category $TY(\ZZ_2\times\ZZ_2,\chi_a,+1)\cong Rep(D_8)$ of representations of the binary dihedral group $D_8$.
    \item $g=p_3$. 
    \be \chi(Q,Q)=1,\qquad \chi(a,a)=\chi(a,Q)=-1\qquad \qquad \epsilon=-1
    \ee
    \begin{align}\label{eq:Z2Z2--}
        \mathcal{Z}^{\CN_{p_3}}=\mathcal{Z}^{2^{12}}+\mathcal{Z}^{2^44^4}\qquad \phi^{\CN_{p_3}}=\phi^{2^{12}}+\phi^{2^24^4}
    \end{align}
      This category $TY(\ZZ_2\times\ZZ_2,\chi_s,-1)$ is not arising as a category of representations of any Hopf algebra.
 \end{itemize} In each case, the correct choices of $\chi$ and $\epsilon$  are determined by matching that the modular properties of the $\CN$-twisted partition function with the expected spins of the fields in the $\CN$-twisted sector for a given Tambara-Yamagami category, as described in table \ref{t:spinselectZ2Z2} in section \ref{s:spinselect}. The only ambiguity concerns the choice of $\epsilon$ when $g=p_1$. This ambiguity can be fixed by observing that $\epsilon=-1$ would imply that a circle with defect $\CN$ encircling the vacuum on the sphere would bring a factor $-2$ rather than $+2$. Since the whole group $G$ we started from is not anomalous in the case $g=p_1$, this minus sign cannot occur, so the correct choice in this case must be $\epsilon=+1$.

 A similar analysis can be performed starting from a subgroup $\langle a,b',g\rangle \subset Co_0$, $\langle a,b',g\rangle\cong D_8$, with the usual relations $a^2={b'}^2=g^2=1$, $ab'=b'a$ and $gag=b'$, where $a$ is as in eq.\eqref{aZ2Z2}, while $b'$ is given by \begin{equation}
    b' = (-1,-1,-1,-1,\ -1,-1,-1,-1,\ 1,1,1,1,\ 1,1,1,1,\ 1,1,1,1,\ 1,1,1,1)\,.
\end{equation} This means that both $a$ and $b'$ have Frame shape $1^82^8$, but now $ab'$ has Frame shape $1^{-8}2^{16}$; the group $\langle a,b'\rangle$ is still non-anomalous. If we take $g$ to be a permutation with Frame shape $1^82^8$, then the only possibility to get $gag=b'$ is that the $8$ transpositions of $g$ exchange the eight  $-1$-eigenvalues of $a$ with the eight $-1$-eigenvalues of $b'$. An example of such a permutation is $g=p_4$ with
\be p_4=(1,17)(2,18)(3,19)(4,20)(5,21)(6,22)(7,23)(8,24)\ ,
\ee and $p_4b'$, $p_4a$ have eigenvalues
\be (1,1,1,1,1,1,1,1,i,i,i,i,i,i,i,i,-i,-i,-i,-i,-i,-i,-i,-i)
\ee corresponding to a Frame shape $1^82^{-8}4^8$. Thus the twining partition function and genus are
\be \label{VfnatRepD8} \CZ^{\CN_{p_4}}=\CZ^{1^82^8}+\CZ^{1^82^{-8}4^8}\qquad \phi^{\CN_{p_4}}=\phi^{1^82^8}+\phi^{1^82^{-8}4^8}\ .
\ee This implies that the conformal weights of $\CN$-twisted states take values in $\frac{1}{2}\ZZ$ and $\frac{1}{4}+\frac{1}{2}\ZZ$, and the relevant category is $TY(\ZZ_2\times \ZZ_2,\chi_a,+1)\cong Rep(D_8)$ with $\chi_a(a,a)=1$.

On the other hand, if we consider $g=p_5$ with Frame shape $2^{12}$, then the only choice is essentially
\be p_5=(1,17)(2,18)(3,19)(4,20)(5,21)(6,22)(7,23)(8,24)(9,10)(11,12)(13,14)(15,16) .
\ee In this case, $p_5b$ and $p_5a$ have eigenvalues
\be (1,1,1,1,-1,-1,-1,-1,i,i,i,i,i,i,i,i,-i,-i,-i,-i,-i,-i,-i,-i)
\ee corresponding to the Frame shape $2^{-4}4^8$. Thus the twining partition function and genus are
\be \label{VfnatRepQ8}\CZ^{\CN_{p_5}}=\CZ^{2^{12}}+\CZ^{2^{-4}4^8}\qquad \phi^{\CN_{p_5}}=\phi^{2^{12}}+\phi^{2^{-4}4^8}\ .
\ee This implies that the conformal weights of $\CN$-twisted states take values in $\frac{1}{2}\ZZ$ and $\frac{1}{4}+\frac{1}{2}\ZZ$, and the relevant category is $TY(\ZZ_2\times \ZZ_2,\chi_a,-1)$ with $\chi_a(a,a)=1$. This category is equivalent to the category $Rep(Q_8)$ of representations of the quaternionic group $Q_8$. This can be shown using the explicit formulae in section 5.6 of \cite{Bhardwaj:2017xup}, upon identify the generators of $D_8$ here with $X_{\rho,\sigma}$ and $Y_{\rho,\sigma}$ in \cite{Bhardwaj:2017xup} by $X_{0,0}=1$, $X_{1,0}=a$, $X_{0,1}=b'$ $X_{1,1}=ab'$, $Y_{0,0}=g$, $Y_{0,1}=bg$, $Y_{1,0}=ag$, $Y_{1,1}=abg$. Then, by (5.44) in \cite{Bhardwaj:2017xup}, one has that the $3$-cocycle $\alpha$ determining the anomaly of $D_8$ satisfies $\alpha(g,g,g)\equiv \alpha(Y_{0,0},Y_{0,0},Y_{0,0}) =\epsilon$. Therefore, $\epsilon=+1$ or $\epsilon=-1$ depending on whether the anomaly of $\langle g\rangle\cong \ZZ_2$ vanishes or not.

\subsection{Duality defect for non-cyclic groups in the GTVW K3 model} \label{s:Z2Z2_GTVW}

In this section, we provide two examples of Tambara-Yamagami categories $TY(\ZZ_2\times\ZZ_2,\chi,+1)$ in a non-linear sigma model on K3, such that all defects in the category preserve the $\CN=(4,4)$ algebra and spectral flow. The two categories will be equivalent to $Rep(H_8)$ and $Rep(D_8)$, i.e. the bicharacters $\chi$ will be as in \eqref{bichZ2Z2-} and \eqref{bichZ2Z2+}, respectively. 

We consider the K3 model $\calC_{GTVW}$, first discussed in \cite{Gaberdiel:2013psa}, which can be seen as the $\CN=(4,4)$ sigma model having a specific $T^4/\ZZ_2$ target, $T^4$ being the maximal $\text{Spin}(8)$ torus $T^4=\RR^4/D_4$. The orbifold $T^4/\ZZ_2$ is distinguished by having  the symmetry group $ G_{GTVW}=\ZZ_2^8\rtimes M_{20}$ with maximal order among K3 models. The SCFT $\calC_{GTVW}$ is most easily understood as an extension of  $\mathcal{A}=\hat{\mathfrak{su}}(2)_1^6$ RCFT. $\hat{\mathfrak{su}}(2)_1$ has two unitary highest weight representations of conformal weight $0$ and $1/4$, corresponding to singlet and doublet representations of the global $SU(2)$ symmetry. The fusion ring of $\hat{\mathfrak{su}}(2)_1$ has the structure of the cyclic group $\ZZ_2$ of order $2$ so that a $\hat{\mathfrak{su}}(2)_1^6$-module $M_{a_1,\ldots,a_6}$  is labeled by the elements $a_i\in \{0,1\}$ in $\ZZ_2$. The Hilbert space of GTVW model is therefore given by
\begin{equation}
\label{eq:HilbertGTVW}
    \mathcal{H}^{\text{GTVW}}= \bigoplus_{[a_1,\ldots,a_6;b_1,\ldots,b_6]\in {A} }M_{a_1,\ldots, a_6}\otimes \bar M_{b_1,\ldots, b_6}\, ,
\end{equation}
where the bar denotes right-movers, and ${A} \subset \ZZ_2^6 \times \ZZ_2^6$ is the subgroup defined by $a_i = b_i +x$ for $i=1, \ldots, 6 $, where $x\in \ZZ_2$ is independent of $[a_1,\ldots,a_6;b_1,\ldots,b_6]$ and can be either $0$ or $1$.The space of states \eqref{eq:HilbertGTVW} is $(\ZZ_2 \oplus \ZZ_2)$-graded,  with the grading $a \equiv \sum_{i=1,\ldots,6}a_i$ denoting the NS-NS and R-R sectors according to whether $a=0,1$ respectively, and the $x$ grading distinguishing the bosonic and fermionic states according to the identification $e^{i \pi x} = (-1)^{F_L + F_R}$, where $F_{L,R}$ are left- and right-moving fermion numbers. 
The GTVW model contains several copies of the $\CN=(4,4)$ related to each other by symmetries. We make the choice to identify the first factor in the (anti-)chiral algebra $\hat{\mathfrak{su}}(2)_1^6$ with the $\hat{\mathfrak{su}}(2)_1$ subalgebra of the (anti-)holomorphic $\CN=4$. The four supercurrents are then suitably chosen ground states in $M_{1, 1, 1, 1, 1,1}\otimes \bar M_{0,0,0,0,0,0}$.

The full group of (invertible) symmetries of this K3 model is a semidirect product $(SU(2)^6\times SU(2)^6)\rtimes S_6$, where $SU(2)^6\times SU(2)^6$ is the Lie group generated by the zero modes of the holomorphic and anti-holomorphic currents, and $S_6$ acts diagonally by permutations of the six $\widehat{su}(2)_1$ both in chiral and antichiral algebra. Strictly speaking, this group does not act faithfully on the CFT, since a subgroup $Z_0\cong \ZZ_2^5$ of the centre $\ZZ_2^6\times \ZZ_2^6$ of $SU(2)^6\times SU(2)^6$ acts trivially on all fields of the theory. Let $t_i,\tilde t_i$, $i=1,\ldots, 6$ be the generators of the central subgroup $\ZZ_2^6\times \ZZ_2^6\subset SU(2)^6\times SU(2)^6$, acting on the fields in the representation $[a_1,\ldots,a_6;b_1,\ldots,b_6]$ by multiplication by $(-1)^{a_i}$ (for $t_i$) or $(-1)^{b_i}$ (for $\tilde t_i$).  Then, all elements of the form $\prod_{i=1}^6 (t_i\tilde t_i)^{r_i}$ with $\sum_i r_i\in 2\ZZ$, act trivially, and generate $Z_0\cong \ZZ_2^5$. 
It can be proved that the category of topological defects that preserve the whole $\widehat{su}(2)^6\oplus \widehat{su}(2)^6$ bosonic chiral and antichiral algebra of the K3 model is group-like, i.e. it is generated by \emph{invertible} simple lines (Verlinde lines), and can be identified with the quotient $(\ZZ_2^6\times \ZZ_2^6)/Z_0$ of the central subgroup of $SU(2)^6\times SU(2)^6$ \cite{Angius:2024evd}.

The subgroup $G_{GTVW}$ of $((SU(2)^6\times SU(2)^6)/Z_0)\rtimes S_6$ that preserves the $\CN=(4,4)$ algebra and the spectral flow is finite and isomorphic to $\ZZ_2^8\rtimes M_{20}$, an extension of the Mathieu group $M_{20}$ by $\ZZ_2^8$. The intersection of $G_{GTVW}$ with $(\ZZ_2^6\times \ZZ_2^6)/Z_0\cong \ZZ_2^7$ is isomorphic to $\ZZ_2^4$ and is generated by $t_2t_i$, $i=3,\ldots, 6$ (modulo $Z_0$).

Let us first consider a TY category $TY(A)$ where  $A=\langle t_2t_5,t_2t_6\rangle\cong \ZZ_2\times \ZZ_2$ is a subgroup of $G_{GTVW}\cap (\ZZ_2^6\times \ZZ_2^6)/Z_0$. All non-trivial elements of $A$ have Frame shape $1^82^8$ when acting on the R-R ground states, so they are non-anomalous and their twining genera match the ones considered in section \ref{s:Z2Z2Vfnat} for $V^{f\natural}$. Since $A$ acts trivially on the chiral and antichiral algebras  $\mathcal{A}\times\bar{\mathcal{A}}=\widehat{su}(2)^6\oplus \widehat{su}(2)^6$, the duality defect $\frac{1}{2}\hat\CN$ must act on $\mathcal{A}\times\bar{\mathcal{A}}$ by an automorphism of order $2$. The group of automorphisms of $\mathcal{A}\times\bar{\mathcal{A}}$ is the product
\be\label{su26aut} \Aut(\widehat{su}(2)^6)\times \Aut(\widehat{su}(2)^6)=((SU(2)/\ZZ_2)^6\rtimes S_6)\times ((SU(2)/\ZZ_2)^6\rtimes S_6)\ ,
\ee
with the first factor acting on the holomorphic currents and the second on the antiholomorphic ones. Let us now argue that the condition that $\CN$ is a simple non-invertible defect in this K3 model necessarily implies that the algebra automorphism $\frac{1}{2}\hat\CN$ cannot be extended to a symmetry $g$ of the whole CFT. Indeed, if such a symmetry $g$ existed, then the fusion $\CN\CL_{g^{-1}}$ would be a topological defect of dimension $2$ acting trivially on all holomorphic and antiholomorphic currents. But we know that all such topological defects are superpositions of invertible Verlinde lines, so $\CN\CL_{g^{-1}}=\CL_{h_1}+\CL_{h_2}$ for some symmetries $h_1,h_2$. Therefore, $\CN=\CL_{h_1g}+\CL_{h_2g}$ would not be simple, contradicting our hypothesis.

The only automorphisms in \eqref{su26aut} that do not lift to symmetries of the CFT are the ones that permute in a different way (non-diagonally) the holomorphic and the antiholomorphic currents. Indeed, such elements would map representations $[a_1,\ldots,a_6;b_1,\ldots,b_6]$ of $\CA\times\bar\CA$ that are in the CFT spectrum of the theory to representations that are not. In order for $\CN$ to be well-defined, we only need the automorphism $\frac{1}{2} \hat \CN\equiv (\rho_L,\rho_R)\in ((SU(2)\ZZ_2)^6\rtimes S_6)^2$ to preserve the subset of $A$-invariant representations of $\CA\times\bar\CA$, but not necessarily the whole set of representations in which $\Hh$ decomposes.

We also want to impose that $\CN$ preserves the $\CN=(4,4)$ superconformal algebra and spectral flow. To this aim we have to require that the action $\rho_L$ of $\frac{1}{2}\hat\CN$ on the holomorphic fields is the same as an element $g_L\in G_{GTVW}$, and $\rho_R$ on the antiholomorphic fields is the same as another element $g_R\in G_{GTVW}$. The condition that $\frac{1}{2}\hat\CN$ does not lift to a symmetry of the CFT implies that $g_L$ and $g_R$ are not the same element of $G_{GTVW}$, and in particular they permute the currents in a different way. 

The $A$-invariant representations $[a_1,\ldots,a_6;b_1,\ldots,b_6]$ are the ones satisfying
\be a_2=a_5=a_6\ ,\qquad\text{and}\qquad b_2=b_5=b_6\ .
\ee We can arrange the set of $A$-invariant representations as follows
\begin{align*}
    \Omega_1&=\{[000000;000000],\ [111111;000000],\ [000000;111111],\ [111111;111111] \}\ ,\\
    \Omega_2&=\{[101000;101000],\ [010111;101000],\ [101000;010111],\ [010111;010111] \}\ ,\\
    \Omega_3&=\{[100100;100100],\ [011011;100100],\ [100100;011011],\ [011011;011011] \}\ ,\\
    \Omega_4&=\{[001100;001100],\ [110011;001100],\ [001100;110011],\ [110011;110011] \}\ ,
\end{align*} for the NS-NS sector, and
\begin{align*}
    \tilde\Omega_1&=\{[100000;100000],\ [011111;100000],\ [100000;011111],\ [011111;011111] \}\ ,\\
    \tilde\Omega_2&=\{[001000;001000],\ [110111;001000],\ [001000;110111],\ [110111;110111] \}\ ,\\
    \tilde\Omega_3&=\{[000100;000100],\ [111011;000100],\ [000100;111011],\ [111011;111011] \}\ ,\\
    \tilde\Omega_4&=\{[101100;101100],\ [010011;101100],\ [101100;010011],\ [010011;010011] \}\ ,
\end{align*} for the R-R sector. Here, each set $\Omega_i$ or $\tilde \Omega_i$ contains the representations that are related with each other by the holomorphic and antiholomorphic supercurrents in the $\CN=(4,4)$ algebra.

An example of a duality defect for this group was provided in appendix C of \cite{Angius:2024evd}, where it was denoted as $\CN_{256}$. The corresponding automorphism $\frac{1}{2}\hat\CN\equiv (\rho_L,\rho_R)\in (SU(2)^6\rtimes S_6)^2$ was given by
\be\label{rhoLrhoR} \rho_{L}= (11\Omega \Omega^\dagger 1 1)(26)(34)\in SU(2)^6\rtimes S_6, ~~ \rho_{R}=(111111)(34)(56)\in SU(2)^6\rtimes S_6,\ee where
\be \Omega=\begin{pmatrix}
    \frac{1-i}{2} & \frac{1+i}{2}\\ \frac{1-i}{2} & \frac{1+i}{2}
\end{pmatrix}\in SU(2)\ ,
\ee and $1$ denotes the identity element in $SU(2)$. By comparing with the generators of $G_{GTVW}$, as described in \cite{Angius:2024evd}, one can check that $\rho_L$ and $\rho_R$ correspond to the action on, respectively, the holomorphic and antiholomorphic fields of the CFT of two elements $g_L,g_R\in G_{GTVW}$. It is easy to verify that the combined action of $(\rho_L,\rho_R)$ has order $2$ and preserves the subset of representations $[a_1,\ldots,a_6;b_1,\ldots,b_6]$ that are $A$-invariant.

Let us compute the twining elliptic genus $\phi^{\CN_{256}}(\tau,z)$. The only non-vanishing contributions come from Ramond-Ramond states with right-moving conformal weight $\bar h=1/4$, thanks to the usual cancelations due to supersymmetry. All such states and their superconformal descendants are contained in the representations in the set $\tilde \Omega_1$. The fact that the contribution from the other sets vanishes can also be easily checked by a direct computation. The twining elliptic genus is therefore given by
\begin{align} \label{GTVWRepH8} \phi^{\CN_{256}}(\tau,z)&=\Tr_{\rm RR}(\hat\CN (-1)^F q^{L_0-c/24}\bar q^{\bar L_0-\bar c/24}y^{J_L})\\\nonumber
=&2\sum_{[a_1,\ldots, b_6]\in \tilde \Omega_1}(-1)^{a_1+b_1}\chi^1_{a_1}(\tau,z)\chi^1_{a_2}(2\tau)\chi^1_{a_3}(2\tau)\chi^1_{a_5}(\tau)\overline{\chi^1_{b_1}(\tau)\chi^1_{b_2}(\tau)\chi^1_{b_3}(2\tau)\chi^1_{b_5}(2\tau)}\\\nonumber
=&{2\over \eta^2(\tau)\eta^2(2\tau)}\left(\theta_2(2\tau,2z)\theta_3(2\tau)\theta_3^2(4\tau)-\theta_3(2\tau,2z)\theta_2(2\tau)\theta_2^2(4\tau)\right)\\\nonumber
=&\phi^{1^82^8}(\tau,z)+\phi^{2^44^4}(\tau,z).
\end{align}
 In this equation, $\chi^1_0$ and $\chi^1_1$ are the characters of the two  $\widehat{su}(2)_1$ representations $[0]$ and $[1]$ (see \eqref{su2kcharz} and \eqref{su2kchar}), and we used that $(-1)^F$ acts by multiplication by $(-1)^{a_1+b_1}$ on the representation $[a_1,\ldots,a_6,b_1,\ldots,b_6]$. This result matches perfectly with the twining genus $\phi^{\CN_{p_1}}$ in eq.\eqref{eq:Z2Z2-+}, in agreement with our conjecture. This also shows that the TY category generated by $\CN$ and the invertible defect in $A$ is the one equivalent to $Rep(H_8)$.

As discussed in \cite{Angius:2024evd}, there are other $\ZZ_2\times \ZZ_2$ subgroups of $G_{GTVW}$ under which the GTVW model is self-orbifold, and whose duality defect preserves the $\CN=(4,4)$ algebra and spectral flow. An example is given by the defects denoted as $\CN_{34,56}$ in appendix C of \cite{Angius:2024evd}; the corresponding $\ZZ_2\times \ZZ_2$ group $A':=\langle t_3t_4, t_5t_6\rangle$ contains two symmetries with Frame shape $1^82^8$ and one element ($t_3t_4t_5t_6$) with Frame shape $1^{-8}2^{16}$. These are the same Frame shapes as in the subgroups of $\Aut_\tau(V^{f\natural})\cong Co_0$ denoted by $\langle a,b'\rangle$ in section \ref{s:Z2Z2Vfnat}. The duality defect $\CN_{34,56}$ acts on the holomorphic and antiholomorphic $\widehat{su}(2)^6\oplus \widehat{su}(2)^6$ currents by $\frac{1}{2}\hat\CN=(\rho'_L,\rho'_R)\in ((SU(2)^6/\ZZ_2)\rtimes S_6)^2$
\be
  \rho'_L= (11111 1)(34)(56), ~~ \rho'_R=1. \ee
  This automorphism has order $2$ and preserves the representations $[a_1,\ldots,a_6;b_1,\ldots,b_6]$ that are invariant under $A'$, namely the ones satisfying
  \be a_3=a_4,\qquad a_5=a_6,\qquad b_3=b_4,\qquad b_5=b_6\ .
  \ee We can arrange the R-R $A'$-invariant representations as
  \begin{align*}
    \tilde\Omega_1&=\{[100000;100000],\ [011111;100000],\ [100000;011111],\ [011111;011111] \}\ ,\\
    \tilde\Omega_2&=\{[010000;010000],\ [101111;010000],\ [010000;101111],\ [101111;101111] \}\ ,\\
    \tilde\Omega_3&=\{[101100;101100],\ [010011;101100],\ [101100;010011],\ [010011;010011],\\
     \tilde\Omega_4&=\{[100011;100011],\ [011100;100011],\ [100011;011100],\ [011100;011100]
    \}\ .
\end{align*}
The twining elliptic genus is 
\begin{align} \phi^{\CN_{34,56}}(\tau,z)=&2\sum_{[a_1,\ldots, b_6]\in \tilde \Omega_1\cup\tilde \Omega_2}(-1)^{a_1+b_1}\chi^1_{a_1}(\tau,z)\chi^1_{a_2}(\tau)\chi^1_{a_3}(2\tau)\chi^1_{a_5}(2\tau)\nonumber\\
&\qquad\qquad\qquad\qquad\qquad\qquad\cdot\overline{\chi^1_{b_1}(\tau)\chi^1_{b_2}(\tau)\chi^1_{b_3}(\tau)^2\chi^1_{b_4}(\tau)^2}\nonumber\\
&={2\over \eta^2(\tau)\eta^2(2\tau)}\left(\theta_2(2\tau,2z)\theta_3(2\tau)\theta_3^2(4\tau)-\theta_3(2\tau,2z)\theta_2(2\tau)\theta_2^2(4\tau)\right)\nonumber\\
&\quad +{2\over \eta^2(\tau)\eta^2(2\tau)}\left(\theta_3(2\tau,2z)\theta_2(2\tau)\theta_3^2(4\tau)-\theta_2(2\tau,2z)\theta_3(2\tau)\theta_2^2(4\tau)\right)\nonumber\\
&=\phi^{1^82^8}(\tau,z)+\phi^{1^82^{-8}4^8}(\tau,z)\ ,\label{twinN3456}
\end{align} where we used the fact that the sets $\tilde\Omega_3$ and $\tilde \Omega_4$ do not contain any Ramond-Ramond states with $\bar h=1/4$, nor their supersymmetric descendants, and therefore they contribute $0$ to the twining genus. This result matches perfectly with the function $\phi^{\CN_{p_4}}$ in \eqref{VfnatRepD8}, thus providing further evidence for Conjecture \ref{conj:K3relation}, and showing that the relevant TY category in this case is the one isomorphic to $Rep(D_8)$.

\section{Fibonacci defects and icosians} \label{s:Fibonacci}

In this section, we prove that the category $\cTop$ of $V^{f\natural}$ admits some $\CN=1$-preserving topological defect $W$ generating a Fibonacci fusion category, i.e. such that
\be W^*=W\ ,\qquad W^2=\CI+W\ ,
\ee and more generally $W^n=F_{n}W+F_{n-1}\CI$, where $F_n$ is the $n$-th Fibonacci number. In a unitary theory, a Fibonacci defect has quantum dimension $\langle W\rangle=\frac{1+\sqrt{5}}{2}$, and the possible eigenvalues of $\hat W$ are $\frac{1\pm\sqrt{5}}{2}$.  

In fact, we will prove a stronger result. One of the maximal subgroups of $Co_0$ is isomorphic to $2.(A_5\times J_2):2$, where $A_5$ is the group of even permutations of $5$ objects, $J_2$ is the Hall-Janko group, a finite sporadic simple group of order $|J_2|=2^7\cdot 3^3\cdot 5^2\cdot 7=604800$, $2.(A_5\times J_2)$ denotes a $\ZZ_2$-central extension of the direct product $A_5\times J_2$, and $2.(A_5\times J_2):2$ is a semidirect product of $2.(A_5\times J_2)$ and a group $\ZZ_2$ acting on both $A_5$ and $J_2$ by an outer automorphism \cite{Atlas}. We will show that, for each choice of a maximal subgroup of $\Aut_\tau(V^{f\natural})\cong Co_0$ isomorphic to $2.(A_5\times J_2):2$, there is  a fusion subcategory of $\cTop$ generated by the invertible defects $\CL_g$, $g\in 2.(A_5\times J_2):2$, and a pair of Fibonacci defects $W^A$ and $W^B$, that commute with each other and with the normal subgroup $2.(A_5\times J_2)$, while they are exchanged by conjugation by the other elements of $2.(A_5\times J_2):2$, i.e.
\be \label{e:Fib_commutativity_conj} W^AW^B=W^BW^A\ ,\qquad \CL_gW^A\CL_g^*=\begin{cases}
	W^A & \text{if }g\in 2.(A_5\times J_2)\\W^B & \text{if }g\in 2.(A_5\times J_2):2,\ g\notin 2.(A_5\times J_2)\ .
\end{cases}
\ee
Examples of Fibonacci defects in K3 sigma models will be described in section \ref{sec:338}.

\subsection{The icosian construction of the Leech lattice}

If there exists a TDL $W\in \cTop$ generating a Fibonacci category, then the corresponding lattice endomorphism $\rho(W):\Lambda\to\Lambda$ must be symmetric $\rho(W)=\rho(W)^t$, and satisfy $\rho(W)^2=1+\rho(W)$. In order to find such an endomorphism, it is useful to consider the construction of the Leech lattice $\Lambda$ using  the ring of \emph{icosians} (see \cite{ConwaySloane}, Chap. 8 sec. 2 or \cite{Wilson:1986}). 

Let us start by considering a real algebra $\HH\oplus \ssi \HH$, where $\HH=\RR\oplus\RR\ii\oplus\RR\jj\oplus\RR\kk$ is the algebra of quaternions with the usual relations $\ii^2=\jj^2=\kk^2=-1=\ii\jj\kk$, and $\ssi$ commutes with all quaternions and satisfies
\be \ssi^2=\ssi+1\ ,\qquad \ssi\ii=\ii\ssi\ ,\qquad \ssi\jj=\jj\ssi\ ,\qquad \ssi\kk=\kk\ssi\ .
\ee (In \cite{ConwaySloane}, they restrict to quaternions with rational coefficients, and identify $\ssi$ with the irrational real number $\frac{1-\sqrt{5}}{2}$, the shortest solution of $x^2=x^2+1$. Here, we will want to work with quaternions with real coefficients, so in order to avoid confusion we prefer to leave $\ssi$ as an abstract algebra generator.)
We define complex conjugation as the $\RR$-linear involution $x\mapsto \bar x$ on  $\HH+\ssi\HH$ given by \be \bar\ii=-\ii\ ,\qquad \bar\jj=-\jj\ ,\qquad\bar\kk=-\kk\ ,\qquad\bar\ssi=\ssi\ .\ee
Within this algebra, we define the icosian group $I_{120}$ as the set of elements $\alpha+\ii \beta+\jj \gamma+\kk \delta\in \HH+\ssi\HH$, where  $(\alpha,\beta,\gamma,\delta)\in (\RR\oplus \ssi\RR)^4$ are given by
\be \frac{1}{2}(\pm 2,0,0,0) \qquad \frac{1}{2}(\pm 1,\pm 1,\pm 1,\pm1)\qquad \frac{1}{2}(0,\pm1,\pm \ssi,\pm(1-\ssi))\ ,
\ee for all possible choices of signs and for all even permutations of the entries. 
This set of $120$ elements forms a multiplicative group $I_{120}$ that is isomorphic to the binary icosahedral group,  a double covering of the group $A_5$ of even permutations of $5$ objects, $I_{120}/\{\pm 1\}\cong A_5$. The icosian ring $\CJ$ is the  group ring $\mathcal{J}:=\ZZ I_{120}$ given by finite integral linear combinations $n_1 x_1+\ldots+n_rx_r $, with $n_i\in \ZZ$ and $x_i\in I_{120}$.

We can define two different norms on $\HH+\ssi\HH$ and therefore on $\CJ$. One is the quaternion norm
\be QN(q+\ssi q'):=(\bar q+\frac{1-\sqrt{5}}{2}\bar q') (q+\frac{1-\sqrt{5}}{2}q')\in \RR_{\ge 0}\ ,\qquad q,q'\in \HH
\ee obtained by replacing $\ssi$ with the real number $\frac{1-\sqrt{5}}{2}$ to map $\HH\oplus\ssi\HH\to \HH$,  and then taking the usual identification $\HH\cong \RR^4$ with $\{1,\ii,\jj,\kk\}$ being an orthonormal basis of $\HH$.  Alternatively, we can define the Euclidean norm  by taking $\{1,\ii,\jj,\kk,\ssi,\ssi\ii,\ssi\jj,\ssi\kk\}$ to be an orthonormal basis of $\HH+\ssi\HH$
\be EN(q+\ssi q'):=\bar q q+\bar q'q'\in \RR_{\ge 0}\ ,\qquad q,q'\in \HH\ .
\ee 
The space $\CJ\otimes_\ZZ \RR$ with the Euclidean norm $EN$ is an $8$-dimensional real vector space  isometric to the Euclidean space $\RR^8$.

The free $\ZZ$-module $\CJ\subset  \CJ\otimes\RR \cong \RR^8$, with the Euclidean norm $EN$, is an $8$-dimensional lattice isomorphic to $E_8(\frac{1}{2})$, a rescaled version of $E_8$ with the shortest vectors having square length $1$. In particular, the $240$ roots of $E_8$ are given by the $120$ elements of the icosian group $I_{120}$, that have all Euclidean norm $1$, and by the $120$ elements $\ssi q$, $q\in I_{120}$.

The Leech lattice can  be constructed as  a $\CJ$-module  of rank $3$, and in particular as a submodule of $\CJ\oplus\CJ\oplus\CJ\subset \RR^{24}$. Consider the elements
\be h:=\ssi +\frac{1}{2}(-1+\ii+\jj+\kk)\in \CJ\ ,\qquad
\bar h:=\ssi +\frac{1}{2}(-1-\ii-\jj-\kk)\in \CJ\ ,
\ee and let
\be \CJ h:=\{qh\mid q\in \CJ\}\subset \CJ\ ,\qquad
\CJ \bar h:=\{q\bar h\mid q\in \CJ\}\subset\CJ\ ,
\ee be two left ideals for the icosian ring $\CJ$. Notice that $h\bar h=\bar h h=2$ and that the intersection $\CJ h\cap \CJ \bar h$ is the bilateral ideal
\be \CJ h\cap \CJ \bar h=2\CJ=\{2q\mid q\in \CJ\}\ .
\ee
We define the Leech lattice $\Lambda$ as
\be \Lambda:= \{(x_1,x_2,x_3)\in \CJ\oplus\CJ\oplus\CJ\mid x_i- x_j\in \CJ h,\ x_1+x_2+x_3\in \CJ\bar h\}\ .
\ee It is clear that it is a left $\CJ$-module, with
\be y(x_1,x_2,x_3):=(y x_1,y x_2,y x_3)\in \Lambda\ ,\qquad \forall y\in \CJ,\quad (x_1,x_2,x_3)\in \Lambda\ .
\ee Because $\CJ$ itself is a free $\ZZ$-module of rank $8$, $\Lambda$ is a free $\ZZ$-module of rank $24$. One can define the quaternionic and euclidean norm on $\Lambda$, by
\be QN(x_1,x_2,x_3):=\sum_i QN(x_i)\ ,\qquad EN(x_1,x_2,x_3):=\sum_i EN(x_i)\ .
\ee It can be proved that, with respect to the Euclidean norm, $\Lambda$ is an even self-dual lattice whose shortest vectors have norm $4$, so that $\Lambda$ is indeed isomorphic to the Leech lattice. 

The subgroup of $\Aut(\Lambda)\cong Co_0$ that preserves the quaternionic norm, besides the Euclidean one, is $2.(A_5\times J_2)$, the extension of $A_5\times J_2$ by the centre $\{\pm 1\}$ of $ Co_0$. Here, $J_2$ is the Janko-Hall group, a sporadic finite simple group of order $|J_2|=2^7\cdot 3^3\cdot 5^2\cdot 7=604800$ \cite{Atlas}. The group $2.J_2$  is generated by the following $3\times 3$ matrices with entries in $ \HH+\ssi \HH$ 
\be\label{J2gens}\begin{gathered}
	 \begin{pmatrix}
	0 & 1 & 0 \\
	1 & 0 & 0\\
	0 & 0 & 1
\end{pmatrix}\ ,\qquad \begin{pmatrix}
	0 & 1 & 0 \\
	0 & 0 & 1\\
	1 & 0 & 0
\end{pmatrix}\\ 
 \begin{pmatrix}
	1 & 0 & 0 \\
	0 & \ii & 0\\
	0 & 0 & \jj
\end{pmatrix}\ ,\qquad \begin{pmatrix}
\ii & 0 & 0 \\
0 & \jj & 0\\
0 & 0 & \kk
\end{pmatrix}\ ,\qquad \frac{1}{2}\begin{pmatrix}
0 & h & h \\
-\bar h & 1 & -1\\
-\bar h & -1 & 1
\end{pmatrix}\ ,
\end{gathered}\ee that act on $(x_1,x_2,x_3)\in \Lambda\subset (\HH+\ssi\HH)^3$ by multiplication from the right. Notice that the first two matrices generate the group $S_3$ of permutations of the three components of $(x_1,x_2,x_3)$. The group $2.A_5$ is just the icosian group $I_{120}$, and acts on $(x_1,x_2,x_3)$ by left multiplication by elements $y\in I_{120}$. Clearly, $2.A_5$ commutes with $2.J_2$, so that the subgroup of $Co_0$ preserving the quaternionic norm is
\be (2.A_5\times 2.J_2)/\ZZ_2\cong 2.(A_5\times J_2)\ ,
\ee where the quotient is due to the fact that the non-trivial central elements of $2.A_5$ and $2.J_2$ act in the same way \be (x_1,x_2,x_3)\mapsto (-x_1,-x_2,-x_3)\ ,
\ee on every lattice vector $(x_1,x_2,x_3)\in \Lambda$,
 and therefore should be identified.

In this description, we can define the lattice endomorphism $\rho(W)\equiv \rho(W^A)$
\begin{align} \rho(W^A):\qquad\Lambda\qquad&\to \qquad\Lambda\\
	(x_1,x_2,x_3)&\mapsto \ssi (x_1,x_2,x_3)=(\ssi x_1,\ssi x_2,\ssi x_3)\ .
\end{align}
Then, because $\ssi^2=1+\ssi$ it is clear that $\rho(W)^2=1+\rho(W)$. There is a second endomorphism obeying the same equation, namely
\begin{align} \rho(W^B):\qquad\Lambda\qquad&\to\qquad  \Lambda\\
		(x_1,x_2,x_3)&\mapsto (1-\ssi) (x_1,x_2,x_3)\ ,
\end{align} and because $\ssi(1-\ssi)=-1$, one has
\be \rho(W^A)\rho(W^B)=\rho(W^B)\rho(W^A)=-1\ .
\ee Therefore, $\rho(W^A)$ and $\rho(W^B)$ are invertible as $\ZZ$-linear maps of $\Lambda$ to itself; however, they do not preserve the Euclidean norm, and therefore they are not elements of $\Aut(\Lambda)=Co_0$.

Notice that $\ssi\in \CJ$ can be written as the sum of two elements of the icosian group $I_{120}$, for example
\be \ssi= \frac{\ssi+0\ii+1\jj+(1-\ssi)\kk}{2}+\frac{\ssi-0\ii-1\jj-(1-\ssi)\kk}{2}\ ,
\ee which correspond to two elements $g,g^{-1}\in I_{120}\subset Co_0$ that are inverse of each other and with Frame shape $1^62^{-6}5^{-6}10^6$. Therefore,
\be \rho(W^A)=\rho(g)+\rho(g^{-1})=\rho(g)+\rho(g)^t\ ,
\ee so that both $\rho(W^A)$ and $\rho(W^B)=1-\rho(W^A)$ are symmetric and have trace $2\Tr\rho(g)=12$.

Since they commute and are symmetric, the operators $\rho(W^A)$ and $\rho(W^B)$ can be simultaneously diagonalized on $\Lambda\otimes \RR$,  and there is an orthogonal decomposition 
\be \Lambda\otimes \RR\cong (\RR^{12})^A\oplus_\perp  (\RR^{12})^B\ ,
\ee
where $(\RR^{12})^A$ and $(\RR^{12})^B$ are eigenspaces of $(\rho(W^A),\rho(W^B))$ with eigenvalues  $(\frac{1+\sqrt{5}}{2},\frac{1-\sqrt{5}}{2})$ and $(\frac{1-\sqrt{5}}{2},\frac{1+\sqrt{5}}{2})$, respectively. In particular, the eigenspace with $\rho(W^A)$-eigenvalue $\frac{1\pm \sqrt{5}}{2}$ is spanned by the vectors
\be (x_1,x_2,x_3)+\frac{1\pm \sqrt{5}}{2}\ssi( x_1, x_2, x_3)\qquad (x_1,x_2,x_3)\in\Lambda
\ee as can be checked by direct calculation:
\begin{align}
	& \nonumber \ssi\left[ (x_1,x_2,x_3)+\frac{1\pm \sqrt{5}}{2}\ssi( x_1, x_2, x_3)\right]=\ssi(x_1,x_2,x_3)+ \frac{1\pm \sqrt{5}}{2}(1+\ssi)( x_1, x_2, x_3)\\
	& \nonumber =\frac{1\pm \sqrt{5}}{2}\left[-\frac{1\mp \sqrt{5}}{2}\ssi(x_1,x_2,x_3)+ (1+\ssi)( x_1, x_2, x_3)\right] \\
    & =\frac{1\pm \sqrt{5}}{2}\left[(x_1,x_2,x_3)+\frac{1\pm \sqrt{5}}{2} \ssi( x_1, x_2, x_3)\right]\ .
\end{align} 

\subsection{Symmetries commuting with the Fibonacci endomorphism}

Because $\ssi$ commutes with all other icosians in $\CJ$, it is clear that both $\rho(W^A)$ and $\rho(W^B)$ commute with the whole group $2.(A_5\times J_2)\subset Co_0$. In fact, we will show here that $2.(A_5\times J_2)$ is the largest subgroup of $Co_0$ commuting with $\rho(W^A)$.

The smallest non-trivial representations of $I_{120}=2.A_5$ are self-conjugate and $2$-dimensional; up to isomorphisms, there are two such representations, that we denote by ${\bf 2}_{2A_5}$ and ${\bf 2}'_{2A_5}$. The group $2.A_5$ admits a non-trivial outer automorphism of order $2$ that exchanges ${\bf 2}_{2A_5}$ and ${\bf 2}'_{2A_5}$.

Similarly, the two smallest non-trivial representations of $2.J_2$, denoted by  ${\bf 6}_{2J_2}$ and ${\bf 6}'_{2J_2}$, are $6$-dimensional  and each of them is self-conjugate. Furthermore, they are exchanged by the only non-trivial outer automorphism of  $2.J_2$, that has order $2$ \cite{Atlas}.

Since $2.(A_5\times J_2)$  commutes with $\rho(W^A)$, each eigenspace $(\RR^{12})^A$ and $(\RR^{12})^B$ is a representation of $2.(A_5\times J_2)$ where both subgroups $2.A_5$ and $2.J_2$ act non-trivially. For dimensional reasons, the only possibility is that both $(\RR^{12})^A$ and $(\RR^{12})^B$ are tensor product of a $2$-dimensional representation of $2.A_5$ and a $6$-dimensional irrep of $2.J_2$.  Up to renaming the representations, we can assume that \be\label{Arep}(\RR^{12})^A= {\bf 2}_{2A_5}\otimes {\bf 6}_{2J_2} \ ,\ee as a $2.(A_5\times J_2)$-representation.  Then, one can show that \be\label{Brep} (\RR^{12})^B={\bf 2}'_{2A_5}\otimes {\bf 6}'_{2J_2}\ .\ee One way to show this is to compare the character of the $24$-dimensional representation ${\bf 24}_{Co_0}$ of $Co_0$ with the characters of  $2.A_5$ and $2.J_2$: such characters match only if ${\bf 24}_{Co_0}$ decomposes as \be {\bf 24}_{Co_0}=({\bf 2}_{2A_5}\otimes {\bf 6}_{2J_2})\oplus ({\bf 2}'_{2A_5}\otimes {\bf 6}'_{2J_2})\ .\ee 

The group $2.(A_5\times J_2)$ has index $2$ in one of the  maximal subgroup $2.(A_5\times J_2)\rtimes \ZZ_2$ (or, more shortly, $2.(A_5\times J_2):2$) of $Co_0$, namely
\be 2.(A_5\times J_2)\subset 2.(A_5\times J_2):2\subset Co_0\ .
\ee In the semidirect product $2.(A_5\times J_2)\rtimes \ZZ_2$, the $\ZZ_2$ factor acts on the normal $2.(A_5\times J_2)$ by outer automorphisms on both $2.A_5$ and $2.J_2$. This means that each element $k\in 2.(A_5\times J_2)\rtimes \ZZ_2$, such that $k\notin 2.(A_5\times J_2)$, exchanges the representations $({\bf 2}_{2A_5}\otimes {\bf 6}_{2J_2})$ and  $({\bf 2}'_{2A_5}\otimes {\bf 6}'_{2J_2})$, i.e. exchanges the eigenspaces $(\RR^{12})^A$ and $(\RR^{12})^B$. As a consequence, we have
\be\label{FibonacciConj} k\rho(W^A)k^{-1}=\rho(W^B) \ ,
\ee for all $k\in 2.(A_5\times J_2)\rtimes \ZZ_2$, $k\notin 2.(A_5\times J_2)$. This means that if we can prove that $W^A$ is a well-defined topological defect preserving the $\CN=1$ current, then one can obtain $W^B$ simply by conjugation $W^B=\CL_kW^A\CL_k^*$  with the invertible defect $\CL_k$.

The group $2.(A_5\times J_2)\rtimes \ZZ_2$ preserves the orthogonal decomposition $\Lambda\otimes \RR=(\RR^{12})^A\oplus (\RR^{12})^B$. Because it is maximal in $Co_0$, it follows that any element in $Co_0$  that is not in $2.(A_5\times J_2)\rtimes \ZZ_2$ does not preserve such decomposition, and in particular does not commute with $\rho(W^A)$. Indeed, if such an element existed, the whole $Co_0$ would preserve the decomposition, which is not true. We conclude that the subgroup of $Co_0$ that commutes with $\rho(W^A)$ and $\rho(W^B)$ is exactly $2.(A_5\times J_2)$.

The fact that $2.(A_5\times J_2)$ acts separately on the eigenspaces $(\RR^{12})^A$ and $(\RR^{12})^B$, implies that there are homomorphisms 
\be \pi^A:2.(A_5\times J_2)\to SO(12)^A\ ,
\ee
\be \pi^B:2.(A_5\times J_2)\to SO(12)^B\ ,
\ee into group of orthogonal transformations in each eigenspace. We can be describe such homomorphisms quite explicitly. Recall that $2.A_5\cong I_{120}$ was defined as a subset of the algebra $\HH+\ssi\HH$. The subspace $(\RR^{12})^A\subset (\HH+\ssi \HH)^3$ is an eigenspace of $\ssi$ with eigenvalue $\frac{1+\sqrt{5}}{2}$, so we can make the replacement
\be\label{replaceA} \ssi \mapsto \frac{1+\sqrt{5}}{2} \qquad\text{on }(\RR^{12})^A\ ,
\ee that turns every element of $2.A_5\cong I_{120}\subset \HH+\ssi\HH$ into a quaternion $q\in \HH$ of unit norm $q\bar q=1$. The set of unit quaternions forms a group isomorphic to $SU(2)$
\be \HH^{(1)}:=\{q\in \HH\mid q\bar q=\bar q q=1\}\cong SU(2)\ ,
\ee with the isomorphism given by replacing $1,\ii,\jj,\kk\in \HH^{(1)}$ by, respectively, the $2\times 2$ identity and Pauli matrices $\mathbf{1},\sigma_1,\sigma_2,\sigma_3$. This means that $\pi^A$ maps $I_{120}$ to a subgroup $SU(2)^A_{diag}\subset SO(12)^A$, with each $q\in SU(2)^A_{diag} \cong \HH^{(1)}$ acting on a generic vector $(x_1,x_2,x_3)\in (\RR^{12})^A\subset (\HH+\ssi\HH)^3$ by left multiplication
\be (x_1,x_2,x_3)\mapsto (qx_1,qx_2,qx_3)\ .
\ee (We use the subscript $diag$ to stress that $q$ acts diagonally on the three components of $(x_1,x_2,x_3)$).  Similarly, the replacement \eqref{replaceA} turns the matrices \eqref{J2gens} generating $2.J_2$ (in particular, the last one), into $3\times 3$ matrices $A$ with entries in $\HH$ rather than in $\HH+\ssi\HH$, and satisfying a unitarity condition $\bar A^t A=A\bar A^t=1$. Thus, $\pi^A$ maps $2.J_2$ to a subgroup of 
$USp(3)\cong Sp(6,\RR)$, the group of $3\times 3$ quaternionic unitary matrices, which is isomorphic to the group $Sp(6,\RR)$ of $6\times 6$ symplectic matrices. Altogether, $\pi^A$ maps $2.(A_5\times J_2)$ into a Lie subgroup of $SO(12)^A$
\be\label{Aembed} \pi^A(2.(A_5\times J_2))\subset (SU(2)^A_{diag}\times Sp(6,\RR)^A)/\ZZ_2\subset SO(12)^A\ ,
\ee where the $\ZZ_2$ quotient is again due to the fact that the non-trivial central elements of $SU(2)^A_{diag}$ and $Sp(6,\RR)^A$ act on the same way on $(\RR^{12})^A$.  Similarly, $\pi^B$ provides an embedding 
\be\label{Bembed} \pi^B(2.(A_5\times J_2))\subset (SU(2)^B_{diag}\times Sp(6,\RR)^B)/\ZZ_2\subset SO(12)^B\ ,
\ee with the important difference that the mapping $\HH+\ssi\HH\to \HH$ is now given by
\be\label{replaceB} \ssi \mapsto \frac{1-\sqrt{5}}{2} \qquad\text{on }(\RR^{12})^B\ .
\ee The different signs in the replacements \eqref{replaceA} and \eqref{replaceB} are the ultimate reason why $(\RR^{12})^A$ and $(\RR^{12})^B$ afford different representations of $2.(A_5\times J_2)$, see \eqref{Arep} and \eqref{Brep}.

\subsection{The affine algebra preserved by $W^A$}

From now on, we focus on $W^A$ and denote it simply as 
\be W\equiv W^A\ .
\ee  Our goal is to extend the definition of $\rho(W)$ to the whole $V^{f\natural}$ and $V^{f\natural}_{tw}$ so as to obtain a well defined topological defect of quantum dimension $\langle W\rangle =\frac{1+\sqrt{5}}{2}$ that preserves the supercurrent. 

Let us first determine the subalgebra of $\widehat{so}(24)_1$ that is preserved by $W$, i.e. such that $\hat W$ acts with eigenvalue equal to the quantum dimension. The decomposition $V^{f\natural}_{tw}(1/2)=(\RR^{12})^A\oplus (\RR^{12})^B$ into eigenspaces of $\rho(W)$ determines a conformally embedded subalgebra
\be \widehat{so}(12)_1^A\oplus \widehat{so}(12)_1^B\subset \widehat{so}(24)_1\ ,
\ee	where $ \widehat{so}(12)_1^A$ (respectively, $ \widehat{so}(12)_1^B$) is generated by bilinears in the $12$ Ramond ground fields in $(\RR^{12})^A$ (resp., $(\RR^{12})^B$).
	Because $W$ preserves the  fields in $(\RR^{12})^A$, it must also preserve the corresponding subalgebra $\widehat{so}(12)_1^A\subset \widehat{so}(24)_1$. The algebra $\widehat{so}(24)_1$ decomposes into the adjoint and the $(v,v)$ representation of $\widehat{so}(12)_1^A\oplus \widehat{so}(12)_1^B$. 
 The $12\times 12=144$ currents in the $(v,v)$ are obtained from the OPE of a field in $(\RR^{12})^B$ with a field in $(\RR^{12})^A$. Because the latter is transparent to the defect $W$, it follows that these $144$ currents must have the same $\hat W$-eigenvalue as $(\RR^{12})^B$, i.e.  $\frac{1- \sqrt{5}}{2}$, and therefore are not preserved by $W$. 
 
Thus the only currents with unknown $\hat W$-eigenvalues are the ones in the $\widehat{so}(12)_1^B$ algebra. In order to find a well-defined TDL $W$, we use some guesswork, and then verify a posteriori that all consistency conditions are satisfied. We expect $W$ to commute with the subgroup $2.(A_5\times J_2)\subset Co_0$ preserving the $\CN=1$ supercurrent, and therefore also with the image $\pi^B(2.(A_5\times J_2))$ in $SO(12)^B$. A natural ansatz for $W$ is that it commutes with the whole Lie group $(SU(2)^B_{diag}\times Sp(6,\RR)^B)/\ZZ_2\subset SO(12)^B$ containing $\pi^B(2.(A_5\times J_2))$, see \eqref{Bembed}. This means that $W$ preserves the corresponding affine algebra of currents\footnote{Via the identification $sp(2)\cong su(2)$, this can be recognized as a particular case ($k=3,n=1$) of the well known series of maximal conformal embeddings $\widehat{sp}(2k)_n\oplus \widehat{sp}(2n)_k\subset \widehat{so}(4nk)_1$.}
\be \widehat{su}(2)_3^B\oplus \widehat{sp}(6)_1^B \subset \widehat{so}(12)_1^B\ .
\ee  Altogether, our ansatz is that the currents of $V^{f\natural}$ that are preserved by $W$ generate the affine algebra
\be\label{Aansatz} \widehat{so}(12)_1^A\oplus \widehat{su}(2)_3^B\oplus \widehat{sp}(6)_1^B 
\ee which is conformally embedded in $\widehat{so}(24)_1$, i.e. the Sugawara stress tensor of the preserved subalgebra \eqref{Aansatz} coincides with the stress tensor of $\widehat{so}(24)_1$ and therefore of $V^{f\natural}$. Similarly, we expect the defect $W^B$ to preserve the subalgebra
\be \widehat{su}(2)_3^A\oplus \widehat{sp}(6)_1^A\oplus \widehat{so}(12)_1^B\ ,
\ee whose current zero modes generate the Lie group $(SU(2)^A_{diag}\times Sp(6,\RR)^A)/\ZZ_2\times SO(12)^B$.

The levels in $\widehat{su}(2)_3^B\oplus \widehat{sp}(6)_1^B$ can be understood as follows. Consider the subgroup \be SU(2)^{l,1,B}\times SU(2)^{l,2,B}\times SU(2)^{l,3,B}\subset SO(12)^B\ ,\ee
 acting on $(x_1,x_2,x_3)\in (\RR^{12})^B\subset (\HH+\ssi\HH)^3$ by left multiplications
 \be (x_1,x_2,x_3)\mapsto (q_1x_1,q_2x_2,q_3x_3) \ ,
 \ee by unit quaternions $q_1,q_2,q_3\in \HH^{(1)}$. There is an analogous  subgroup \be SU(2)^{r,1,B}\times SU(2)^{r,2,B}\times SU(2)^{r,3,B}\subset SO(12)^B\ ,\ee acting on $(x_1,x_2,x_3)\in (\RR^{12})^B\subset (\HH+\ssi\HH)^3$ by right multiplications by unit quaternions. The groups 
 \be Spin(4)^{i,B}=SU(2)^{l,i,B}\times SU(2)^{r,i,B}\ ,\qquad i=1,2,3\ ,
 \ee are generated by the zero modes of pairwise commuting subalgebras 
 \be \widehat{so}(4)^{i,B}_1=\widehat{su}(2)^{l,i,B}_1\oplus \widehat{su}(2)^{r,i,B}_1\ ,\qquad i=1,2,3\ ,
 \ee of $\widehat{so}(12)_1^B$. It is clear that there is a diagonal embedding of groups
 \be SU(2)_{diag}^B\subset SU(2)^{l,1,B}\times SU(2)^{l,2,B}\times SU(2)^{l,3,B}\ ,
 \ee and this must correspond to a diagonal embedding of the corresponding affine algebras
 \be \widehat{su}(2)_3^B\subset \widehat{su}(2)_1^{l,1,B}\oplus \widehat{su}(2)_1^{l,2,B}\oplus \widehat{su}(2)_1^{l,3,B}\ ,
 \ee that implies that the affine algebra associated with $SU(2)_{diag}^B$ has level $3$. Similarly, the embedding of groups
 \be SU(2)^{r,1,B}\times SU(2)^{r,2,B}\times SU(2)^{r,3,B}\subset USp(3)\cong Sp(6,\RR)^B\ ,
 \ee where the LHS is identified with the group of \emph{diagonal} unitary $3\times 3$ quaternionic matrices, leads to a conformal embedding of the corresponding affine algebras
 \be \widehat{su}(2)_1^{r,1,B}\oplus \widehat{su}(2)_1^{r,2,B}\oplus \widehat{su}(2)_1^{r,3,B}\subset  \widehat{sp}(6)_1^B\ ,
 \ee that is compatible with the level of $\widehat{sp}(6)_1^B$ being $1$.

\subsection{The Fibonacci defect $W^A$: complete description and twining functions}\label{s:FibDef}

Given the ansatz \eqref{Aansatz} for the preserved affine algebra, we can now explicitly determine the Fibonacci defect $W\equiv W^A$, and verify that it satisfies all consistency conditions for a topological defect in $V^{f\natural}$. The Fibonacci defect $W^B$ can be obtained by conjugation $W^B=\CL_k W^A\CL_{k^{-1}}$ with a suitable $k\in Co_0$ (see \eqref{FibonacciConj}), and therefore it satisfies similar properties.

Recall that $\widehat{su}(2)_k$ has central charge $c=\frac{3k}{k+2}$ and its primary fields are labeled by $\ell\in \{0,\ldots,k\}$ and have conformal weight $\frac{\ell(\ell+2)}{4(k+2)}$ with respect to the Sugawara stress-tensor of $\widehat{su}(2)_k$. The S-matrix is
\be S^{\widehat{su}(2)_k}_{\ell,\ell'}=\sqrt{\frac{2}{k+2}}\sin\left[\frac{\pi}{k+2}(\ell+1)(\ell'+1) \right] \qquad \ell,\ell'\in\{0,\ldots,k\}\ .\ee
In particular, $\widehat{su}(2)_{3}$ has $c=\frac{9}{5}$ and four primary fields $[0],[1],[2],[3]$ with conformal weights $0$, $\frac{3}{20}$, $\frac{2}{5}$, $\frac{3}{4}$, respectively. 

 On the other hand, $sp(2r)$ has rank $r$, dimension $2r^2+r$, and dual Coxeter number $r+1$, so that $\widehat{sp}(2r)_k$ has central charge $\frac{k(2r^2+r)}{k+r+1}$. In particular, $\widehat{sp}(6)_1$ has dimension $21$, central charge $c=\frac{21}{5}$ and four unitary representations, corresponding to the affine Dynkin labels $[\lambda_0;\lambda_1,\lambda_2,\lambda_3]$ being $[1;0,0,0]$ (vacuum), $[0;1,0,0]$, $[0;0,1,0]$, $[0;0,0,1]$ with conformal weights $0$, $\frac{7}{20}$, $\frac{3}{5}$, and $\frac{3}{4}$, respectively. In general $\widehat{sp}(2k)_n$ and  $\widehat{sp}(2n)_k$ have the same number of irreducible representations and the same S-matrix -- this is an example of level-rank duality. In particular, using the correspondence
 \be\label{levelrank}\begin{gathered}
{} 	[0]\leftrightarrow [1;0,0,0]\ ,\qquad [1]\leftrightarrow [0;1,0,0]
 	\\ 
 	[2]\leftrightarrow [0;0,1,0]\ ,\qquad [3]\leftrightarrow [0;0,0,1]
 \end{gathered} 
 \ee between representations of $\widehat{su}(2)_3=\widehat{sp}(2)_3$ and $\widehat{sp}(6)_1$  the $4\times 4$ S-matrix for these two algebras is the same, namely
\be S_{\ell,\ell'}=\sqrt{\frac{2}{5}}\sin\left[\frac{\pi}{5}(\ell+1)(\ell'+1) \right]\ ,\qquad \ell,\ell'\in \{0,1,2,3\}\ ,
	\ee
	leading to the fusion rules
	\be \begin{array}{c|cccc}
		& 0 & 1 & 2 & 3\\ \hline
		0& 0 & 1 & 2 & 3\\
		1 & 1 & 0+2 & 1+3 & 2\\
		2 & 2 & 1+3 & 0+2 & 1 \\
		3 & 3 & 2 & 	1 & 0
	\end{array}
	\ee
	It is useful to consider an analogy with a diagonal $\widehat{su}(2)_3$ (or $\widehat{sp}(6)_1$) WZW model. In this CFT, the Verlinde lines $W_\ell$ are labeled by $\ell\in \{0,1,2,3\}$ and act on the primary state $|\ell'\rangle$ by
	\be \hat W_\ell |\ell'\rangle=\frac{S_{\ell,\ell'}}{S_{\ell,0}}|\ell'\rangle =\frac{\sin\left[\frac{\pi}{5}(\ell+1)(\ell'+1) \right]}{\sin\left[\frac{\pi}{5}(\ell+1) \right]}|\ell'\rangle\ .
	\ee
	Using the identities
	\be \frac{\sin\left(2\frac{\pi}{5} \right)}{\sin\left(\frac{\pi}{5} \right)}=\frac{1+\sqrt{5}}{2}\ ,\qquad \frac{\sin\left(\frac{\pi}{5} \right)}{\sin\left(2\frac{\pi}{5} \right)}=-\frac{1-\sqrt{5}}{2}
	\ee
	one gets the following actions of the Verlinde lines 
	\be \begin{array}{c|cccc}
		& |0\rangle & |1\rangle & |2\rangle & |3\rangle\\ \hline
		\hat W_0 & 1 & 1 & 1 & 1\\
		\hat W_1 & \zeta & -\sigma & \sigma & -\zeta \\
		\hat W_2 & \zeta & \sigma & \sigma & \zeta \\
		\hat W_3 & 1 & -1 & 	1 & -1
	\end{array}
	\ee where $\zeta =\frac{1+\sqrt{5}}{2}$ and $\sigma=\frac{1-\sqrt{5}}{2}$. Then, it is clear that the fusion category of such Verlinde lines is of the form $Fib\otimes \ZZ_2$, i.e. it is generated by a Fibonacci and a $\ZZ_2$ line that commute with each other. In particular, $W\equiv W_2$ is a Fibonacci line preserving the vacuum and the $[3]$ representation and acting non-trivially on $[1]$ and $[2]$. On the other hand, $\eta:=W_3$ is an invertible $\ZZ_2$ line, acting by $(-1)^\ell$ on $[\ell]$. Finally, $W_1$ is just the product $\eta W$.
	
	With this example in mind, we (tentatively) define $W$ on $V^{f\natural}$ to be a line with quantum dimension $\langle W\rangle=\frac{1+\sqrt{5}}{2}$, preserving the subalgebra \eqref{Aansatz} of $\widehat{so}(24)_1$, and such that $\hat W$ has eigenvalue $\langle W\rangle=\frac{1+\sqrt{5}}{2}$ on all fields in the $[0]$ and $[3]$ representations of $\widehat{su}(2)_3^B$, and eigenvalue $\frac{1-\sqrt{5}}{2}$ on the fields in the $[1]$ and $[2]$ representations. 
	
	In order to compute the $W$-twining partition functions, we need to decompose $V^{f\natural}$ and $V^{f\natural}_{tw}$ into representations of $\widehat{so}(12)_1^A\oplus \widehat{su}(2)_3^B\oplus \widehat{sp}(6)_1^B$. We first notice that, with respect to the subalgebra $\widehat{so}(12)_1^A\oplus\widehat{so}(12)_1^B$ of $\widehat{so}(24)_1$, $V^{f\natural}$ and $V^{f\natural}_{tw}$ decompose as
	\be V^{f\natural}=(0,0)\oplus (v,v)\oplus (s,s)\oplus (c,c)
	\ee
	\be V^{f\natural}_{tw}=(v,0)\oplus (0,v)\oplus (s,c)\oplus (c,s)\ ,
	\ee
	where $0,v,s,c$ denote the four irreducible representations of $\widehat{so}(12)_1$ with conformal weights $0$, $1/2$, $3/4$, $3/4$, respectively. Next, the decompositions of the $\widehat{so}(12)^B_1$ representations into irreducible representations of $\widehat{sp}(6)_1\oplus \widehat{su}(2)_3$ can be found, for example, in \cite{Verstegen:1990at}, and read:
    \begin{align}
		(0)&= [0_{sp},0_{su}]\ \oplus\  [2_{sp},2_{su}]=(0,0)\ \oplus\  (\frac{3}{5},\frac{2}{5}) \\
		(v)&=[1_{sp},1_{su}]\ \oplus\   [3_{sp},3_{su}]=(\frac{7}{20},\frac{3}{20})\ \oplus\  (\frac{3}{4},\frac{3}{4}) \\
		(s)&=[3_{sp},0_{su}]\ \oplus\  [1_{sp},2_{su}]=(\frac{3}{4},0)\ \oplus\  (\frac{7}{20},\frac{2}{5})\\
		(c)&=  [2_{sp},1_{su}]\ \oplus\  [0_{sp},3_{su}]=(\frac{3}{5},\frac{3}{20})\ \oplus\  (0,\frac{3}{4})
	\end{align}
    where the irreps of $\widehat{sp}(6)_1\oplus \widehat{su}(2)_3$ are denoted either by $[\ell_{sp},\ell_{su}]$, with $\ell_{sp}\in \{0,1,2,3\}$ corresponding to the $\widehat{sp}(6)_1$ Dynkin labels as in \eqref{levelrank},  or by the conformal weights $(h_{\widehat{sp}(6)_1},h_{\widehat{su}(2)_3})$ with respect to the Sugawara stress-tensors of $\widehat{sp}(6)_1$ and $\widehat{su}(2)_3$.

	In particular, the $24$ Ramond fields of weight $1/2$ in $V^{f\natural}_{tw}(1/2)$ belong to representations
	\begin{align} &(v,[0_{sp},0_{su}]) && (12\text{ fields})\\
		&(0,[1_{sp},1_{su}]) && (12\text{ fields})
	\end{align} so that $W$ preserves the first $12$ and acts non-trivially on the second $12$, as expected. On the other hand, the $276$ NS fields of weight $1$ (currents) in $V^{f\natural}$ belong to
	\begin{align} &(0,[0_{sp},0_{su}]) && (66+21+3=90\text{ fields})\\
	&(0,[2_{sp},2_{su}]) && (14\times 3=42\text{ fields})\\
		&(v,[1_{sp},1_{su}]) && (12\times 6\times 2=144 \text{ fields})
\end{align} and we see that only the $90$ currents generating the algebra \eqref{Aansatz} are preserved.

The $W$-twining partition functions are\footnote{Note that replacing all occurrences of $\zeta,\sigma$ in the above formulas with 1 yields the well-known characters of $V^{f\natural}$ in the four spin structures, as expected.} (the argument $\tau$ is everywhere omitted)
    \begin{align} Z_{NS}^{W,\pm}(\tau)=&\ch^{(12)}_0(\zeta\chi^{sp}_{0}\chi^{su}_{0}+ \sigma\chi^{sp}_{2}\chi^{su}_{2})+\ch^{(12)}_v(\sigma\chi^{sp}_{1}\chi^{su}_{1}+ \zeta\chi^{sp}_{3}\chi^{su}_{3})\\
		&\pm 
		\ch^{(12)}_s(\zeta\chi^{sp}_{3}\chi^{su}_{0}+ \sigma\chi^{sp}_{1}\chi^{su}_{2})\pm\ch^{(12)}_c(\sigma\chi^{sp}_{2}\chi^{su}_{1}+ \zeta\chi^{sp}_{0}\chi^{su}_{3})\notag\\
		Z_{R}^{W,\pm}(\tau)=&\ch^{(12)}_v(\zeta\chi^{sp}_{0}\chi^{su}_{0}+ \sigma\chi^{sp}_{2}\chi^{su}_{2})+\ch^{(12)}_0(\sigma\chi^{sp}_{1}\chi^{su}_{1}+ \zeta\chi^{sp}_{3}\chi^{su}_{3})\\
		&\pm 
		\ch^{(12)}_c(\zeta\chi^{sp}_{3}\chi^{su}_{0}+ \sigma\chi^{sp}_{1}\chi^{su}_{2})\pm\ch^{(12)}_s(\sigma\chi^{sp}_{2}\chi^{su}_{1}+ \zeta\chi^{sp}_{0}\chi^{su}_{3}\notag
        )
	\end{align}
	where $\zeta=\frac{1+\sqrt{5}}{2}$ and $\sigma=\frac{1-\sqrt{5}}{2}$, $\ch^{(12)}_\rho(\tau)$ with $\rho\in\{0,v,s,c\}$ are the characters of $\widehat{so}(12)_1$, and $\chi^{sp}_{\ell_{sp}}\chi^{su}_{\ell_{su}}$ are the characters of the representations $[\ell_{sp},\ell_{su}]$ of  $\widehat{sp}(6)_1\oplus \widehat{su}(2)_3$ (see appendix \ref{a:minimal} for explicit formulae). In particular,
    \begin{align}
        Z_{NS}^{W,\pm}=&\frac{1}{2}Z_{NS}^\pm+\frac{\sqrt{5}}{2}(\frac{1}{\sqrt{q}}-96 q^{1/2}\mp 896 q-5016 q^{3/2}\mp 21888 q^2-81980 q^{5/2}
        +\ldots)
    \end{align}
    \begin{align} Z_R^{W,+}=&\frac{1}{2}Z_R^++\sqrt{5}( -896 q-21888 q^2-274560 q^3-2402816 q^4-16600320 q^5
    +\ldots)
    \end{align}
	By a computer calculation, we verified that 
	\be Z_{R}^{W,-}=12\ ,\ee
	is a constant (up to order $O(q^{40})$), thus providing strong evidence of the fact that $W$ preserves the $\CN=1$ supercurrent. This will be proved rigorously in the next section.
	The S-transformation of these expressions provide the $W$-twisted partition functions:
    \begin{align}
     Z_{W,NS}^{\pm}(\tau)=&\ch^{(12)}_0(\chi^{sp}_{0}\chi^{su}_{2}+\chi^{sp}_{2}\chi^{su}_{0}+ \chi^{sp}_{2}\chi^{su}_{2})+\ch^{(12)}_v(\chi^{sp}_{1}\chi^{su}_{1}+ \chi^{sp}_{1}\chi^{su}_{3}+ \chi^{sp}_{3}\chi^{su}_{1})\\
		&\pm 
		\ch^{(12)}_s(\chi^{sp}_{1}\chi^{su}_{0}+ \chi^{sp}_{3}\chi^{su}_{2}+ \chi^{sp}_{1}\chi^{su}_{2})\pm\ch^{(12)}_c(\chi^{sp}_{0}\chi^{su}_{1}+\chi^{sp}_{2}\chi^{su}_{1}+ \chi^{sp}_{2}\chi^{su}_{3})\ ,\notag\\
		Z_{W,R}^{\pm}(\tau)=&\ch^{(12)}_v(\chi^{sp}_{0}\chi^{su}_{2}+\chi^{sp}_{2}\chi^{su}_{0}+ \chi^{sp}_{2}\chi^{su}_{2})+\ch^{(12)}_0(\chi^{sp}_{1}\chi^{su}_{1}+ \chi^{sp}_{1}\chi^{su}_{3}+ \chi^{sp}_{3}\chi^{su}_{1})\\
		&\pm 
		\ch^{(12)}_c(\chi^{sp}_{1}\chi^{su}_{0}+ \chi^{sp}_{3}\chi^{su}_{2}+ \chi^{sp}_{1}\chi^{su}_{2})\pm\ch^{(12)}_s(\chi^{sp}_{0}\chi^{su}_{1}+\chi^{sp}_{2}\chi^{su}_{1}+ \chi^{sp}_{2}\chi^{su}_{3})\ .\notag 
    \end{align}
	Notice that they have an integral decomposition into irreps of the preserved subalgebra $\widehat{so}(12)_1^A\oplus \widehat{sp}(6)_1\oplus \widehat{su}(2)_3$, thus confirming that $W^A$ is a well defined topological defect in $V^{f\natural}$. 
    Explicitly, we have
    \begin{equation*} Z_{W,NS}^{\pm}(\tau)=q^{-\frac{1}{2}}(3 q^{2/5}+14 q^{3/5}\pm 64 q^{9/10}+186 q\pm 192 q^{11/10}+606 q^{7/5}\pm 1472 q^{3/2}
   +\ldots)\ ,
    \end{equation*}
     \begin{equation*} Z_{W,R}^{+}(\tau)=2(6+64 q^{2/5}+192 q^{3/5}+1472 q+3648 q^{7/5}+6912 q^{8/5}+35520 q^2+63744 q^{12/5}
   +\ldots) ,
     \end{equation*} with $Z_{W,R}^{-}(\tau)=12$. Notice that the conformal weights of the bosonic NS states in the $W$-twisted sector take values in $\pm \frac{2}{5}+\ZZ$ and in $\ZZ$, as predicted by the spin selection rules in \cite{Chang_2019}.
    The partition functions of the $W^B$ defect are the same, as expected since the two defects are related by conjugation in $Co_0$.

We would now like to evaluate $\phi^W(V^{f\natural},\tau,z)=Z^{W,-}_R(\tau,z)$ where $z$ is a chemical potential for an $\widehat{su}(2)_1\subset \widehat{so}(4)_1 \subset \widehat{so}(12)_1^A$, belonging to  a $c=6$, $\mathcal N=4$ SCA in $V^{f\natural}$. 
Firstly, before the insertion of $W$, we find a new expression for the function $\phi(V^{f\natural},\tau,z):=Z^-_R(\tau,z)$ (equal to the K3 elliptic genus) which respects the structure of the affine algebra preserved by W in \eqref{Aansatz}. From a straightforward generalization of the discussion around eq.\eqref{newelliptic}, we get
\be\nonumber
\phi(V^{f\natural},\tau,z)&=&\ch^{(12)}_v(\tau,z)(\chi^{sp}_{0}\chi^{su}_{0}+ \chi^{sp}_{2}\chi^{su}_{2})+\ch^{(12)}_0(\tau,z)(\chi^{sp}_{1}\chi^{su}_{1}+ \chi^{sp}_{3}\chi^{su}_{3})\\
		&-&
		\ch^{(12)}_c(\tau,z)(\chi^{sp}_{3}\chi^{su}_{0}+ \chi^{sp}_{1}\chi^{su}_{2})-\ch^{(12)}_s(\tau,z)(\chi^{sp}_{2}\chi^{su}_{1}+ \chi^{sp}_{0}\chi^{su}_{3}).
\ee  where the `flavoured' characters $\ch^{(12)}_\rho(\tau,z):= \Tr_{\rho^{(12)}}y^{J_0^3}q^{L_0-1/3}$,  with $\rho\in\{0,v,c,s\}$, are given in \eqref{ch04plane}--\eqref{chsc4plane}. 
Moreover, once we insert $W$, we then find the twining genus,
\be\nonumber
\phi^W(V^{f\natural},\tau,z)&=&\ch^{(12)}_v(\tau,z)(\zeta\chi^{sp}_{0}\chi^{su}_{0}+ \sigma\chi^{sp}_{2}\chi^{su}_{2})+\ch^{(12)}_0(\tau,z)(\sigma\chi^{sp}_{1}\chi^{su}_{1}+ \zeta\chi^{sp}_{3}\chi^{su}_{3})\\\nonumber
		&-&
		\ch^{(12)}_c(\tau,z)(\zeta\chi^{sp}_{3}\chi^{su}_{0}+ \sigma\chi^{sp}_{1}\chi^{su}_{2})-\ch^{(12)}_s(\tau,z)(\sigma\chi^{sp}_{2}\chi^{su}_{1}+ \zeta\chi^{sp}_{0}\chi^{su}_{3}),
\ee
which can equivalently be written as, 
\be\label{eq:FibTwining}
\phi^W(V^{f\natural},\tau,z)= {1\over 2}\phi(V^{f\natural},\tau,z)+{\sqrt 5}{\theta_1^2(\tau,z)\over \eta^6(\tau)}f(\tau)
\ee
where
\be
f(\tau):= {\eta(\tau)^5\over \eta(5\tau)} + 25{\eta(5\tau)^5\over \eta(\tau)},
\ee
is the weight 2 theta series of the $A_4$ lattice.

Using the modular properties of $\eta$ (see appendix \ref{a:minimal}),
it is clear that the S-transform of $f(\tau)$ is 
\be
f(-1/\tau)= {\sqrt{-i\tau}^4\over \sqrt 5} \left({\eta(\tau/5)^5\over \eta(\tau)} + 5{\eta(\tau)^5\over \eta(\tau/5)}\right )
\ee
and thus the S-transform of equation \eqref{eq:FibTwining} has integral coefficients.

	\subsection{$W^A$ preserves the $\CN=1$ supercurrent}
	
	In this section, we will prove that the topological defect $W^A$ preserves the $\CN=1$ supercurrent $\tau(z)$. The strategy we use is the following. We know that $W^A$ commutes with the group $2.(A_5\times J_2)$, since the latter is contained in the Lie group of symmetries generated by the zero modes of the preserved affine algebra $ \widehat{so}(12)_1^A\oplus \widehat{sp}(6)_1^B\oplus \widehat{su}(2)_3^B$. We will show that, in the space $V^{f\natural}(3/2)$ of spin $3/2$ NS fields, the subspace that is fixed by $2.(A_5\times J_2)\subset Co_0$ is $2$-dimensional, and is generated by two fields in the representations $[0]$ and $[3]$ of $\widehat{su}(2)_3^B$. Because the supercurrent is invariant under the whole $Co_0$, it must belong to this fixed subspace; but $W^A$ preserves all fields in the $[0]$ and $[3]$ representations of $\widehat{su}(2)_3^B$, so it must in particular preserve $\tau$.
	
	The $2^{11}=2048$ fields in $V^{f\natural}(3/2)$ belong to the following representations of $ \widehat{so}(12)_1^A\oplus \widehat{sp}(6)_1^B\oplus \widehat{su}(2)_3^B$:
    \begin{align}
		&(s,[3_{sp},0_{su}]) && 32\times 14\text{ fields}\\
		&(s,[1_{sp},2_{su}]) && 32\times 6\times 3=32\times 18\text{ fields}\\
	&(c,[2_{sp},1_{su}]) && 32\times 14\times 2=32\times 28\text{ fields}\\
&(c,[0_{sp},3_{su}]) && 32\times 4\text{ fields}\ .
	\end{align}
    Notice that in all such representations the lowest conformal weight is exactly $3/2$. We make one further decomposition of the $\widehat{so}(12)^A$ representations with respect to its subalgebra $ \widehat{sp}(6)_1^A\oplus \widehat{su}(2)_3^A$. It is useful to denote the representations of $ \widehat{sp}(6)_1^A\oplus \widehat{su}(2)_3^A\oplus \widehat{sp}(6)_1^B\oplus \widehat{su}(2)_3^B$ by $[\ell^A_{sp},\ell^A_{su}][\ell^B_{sp},\ell^B_{su}]$, where $\ell^A_{sp},\ell^B_{sp}\in \{0,1,2,3\}$ correspond to the $\widehat{sp}(6)_1$ representations as in \eqref{levelrank}. Thus, we have the spin $3/2$ Ramond fields belong to

\be \begin{aligned} 
& \left[([3^A_{sp},0^A_{su}]\oplus [1^A_{sp},2^A_{su}])\  ([3^B_{sp},0^B_{su}]\oplus [1^B_{sp},2^B_{su}])\right] \\
& \oplus \left[([2^A_{sp},1^A_{su}] \oplus [0^A_{sp},3^A_{su}])\ ([2^B_{sp},1^B_{su}]\oplus [0^B_{sp},3^B_{su}])\right]\ . \\
\end{aligned}
\ee
Each element $y\in 2.A_5\subset 2.(A_5\times J_2)$ acts on a given field of $V^{f\natural}$ by a transformation $(\pi^A(y),\pi^B(y))\in SU(2)^A_{diag}\times SU(2)^B_{diag}$, while the elements $z\in 2.J_2\subset 2.(A_5\times J_2)$ act by $(\pi^A(z),\pi^B(z))\in Sp(6,\RR)^A\times Sp(6,\RR)^B$. Thus, the action on a given spin $3/2$ field depends only on which $ \widehat{sp}(6)_1^A\oplus \widehat{su}(2)_3^A\oplus \widehat{sp}(6)_1^B\oplus \widehat{su}(2)_3^B$ representation it belongs. In particular, for each $[\ell^A_{sp},\ell^A_{su}][\ell^B_{sp},\ell^B_{su}]$, we can determine in which representation of $2.(A_5\times J_2)$ the ground fields transform.

Let us denote by ${\bf 1}_+,{\bf 3}_+,{\bf 3'}_+,{\bf 4}_+,{\bf 5}_+,{\bf 2}_-,{\bf 2'}_-,{\bf 4}_-,{\bf 6}_-$ the irreducible representations of $2.A_5$, and by ${\bf 1}_+,{\bf 6}_-,{\bf 6'}_-,{\bf 14}_+,{\bf 14'}_+,{\bf 14}_-,\ldots$ the smallest irreducible representations of $2.J_2$ \cite{Atlas}. Here, the number indicates the dimension, and the subscript $\pm$ indicates whether the non-trivial central element in either $2.A_5$, or $2.J_2$ acts by $+1$ (i.e., trivially) or by $-1$. Representations with the same dimension and subscript are exchanged by the outer automorphism. 

Then, we have the following correspondences between the labels $\ell^A_{sp},\ell^A_{su},\ell^B_{sp},\ell^B_{su}$ and representations of $2.(A_5\times J_2)$:
\begin{align}
	& 0^A_{sp}\leftrightarrow ({\bf 1}_+)_{2.J_2} && 1^A_{sp}\leftrightarrow ({\bf 6}_-)_{2.J_2} && 2^A_{sp}\leftrightarrow ({\bf 14}_+)_{2.J_2} && 3^A_{sp}\leftrightarrow ({\bf 14}_-)_{2.J_2}\\
	& 0^B_{sp}\leftrightarrow ({\bf 1}_+)_{2.J_2} && 1^B_{sp}\leftrightarrow ({\bf 6'}_-)_{2.J_2} && 2^B_{sp}\leftrightarrow ({\bf 14'}_+)_{2.J_2} && 3^B_{sp}\leftrightarrow ({\bf 14}_-)_{2.J_2}\\
	& 0^A_{su}\leftrightarrow ({\bf 1}_+)_{2.A_5} && 1^A_{su}\leftrightarrow ({\bf 2}_-)_{2.A_5} && 2^A_{su}\leftrightarrow ({\bf 3}_+)_{2.A_5} && 3^A_{su}\leftrightarrow ({\bf 4}_-)_{2.A_5}\\
	& 0^B_{su}\leftrightarrow ({\bf 1}_+)_{2.A_5} && 1^B_{su}\leftrightarrow ({\bf 2'}_-)_{2.A_5} && 2^B_{su}\leftrightarrow ({\bf 3'}_+)_{2.A_5} && 3^B_{su}\leftrightarrow ({\bf 4}_-)_{2.A_5}
\end{align} The assignments of $1^A_{sp},1^A_{su},1^B_{sp},1^B_{su}$ are determined by recalling that the $24$ Ramond ground fields, that are in the $[1^A_{sp},1^A_{su}] [0^B_{sp},0^B_{su}]\oplus [0^A_{sp},0^A_{su}] [1^B_{sp},1^B_{su}]$ affine algebra representation, transform in the $({\bf 6}_-\otimes {\bf 2}_-)\oplus ({\bf 6'}_-\otimes {\bf 2'}_-)$ representation of $2.(A_5\times J_2)$. Then, all the other assignments are fixed by compatibility with the fusion rules and by dimensional arguments. Here and in the following, the decompositions of tensor products of $2.A_5$ and $2.J_2$ representations into irreducible ones were computed using the character table for these groups (we used GAP \cite{GAP4} for this purpose).

For example, the ground states of $[1^A_{sp},2^A_{su}] [3^B_{sp},0^B_{su}]$ transform as 
\be [1^A_{sp},2^A_{su}] [3^B_{sp},0^B_{su}] \, \, \leftrightarrow \, \,({\bf 6}_-\otimes {\bf 14}_-)_{2.J_2}\otimes  ({\bf 3}_+\otimes {\bf 1}_+)_{2.A_5}= ({\bf 14}_+\oplus {\bf 70}_+)_{2.J_2}\otimes  ({\bf 3}_+)_{2.A_5} .\ee As another example,
\be [2^A_{sp},1^A_{su}] [2^B_{sp},1^B_{su}] \leftrightarrow ({\bf 14}_+\otimes {\bf 14'}_+)_{2.J_2}\otimes  ({\bf 2}_-\otimes {\bf 2'}_-)_{2.A_5}=({\bf 36}_+\oplus {\bf 160}_+)_{2.J_2}\otimes ({\bf 4}_+)_{2.A_5}\ .
\ee It is easy to see that, among the affine algebra representations $[\ell^A_{sp},\ell^A_{su}][\ell^B_{sp},\ell^B_{su}]$ appearing in the decomposition of $V^{f\natural}(3/2)$, the only ones where the trivial representation of $2.(A_5\times J_2)$ occurs are
\be \begin{aligned}{} [3^A_{sp},0^A_{su}][3^B_{sp},0^B_{su}] \, \, \leftrightarrow \, \, & ({\bf 14}_-\otimes {\bf 14}_-)_{2.J_2}\otimes  ({\bf 1}_+\otimes {\bf 1}_+)_{2.A_5} \\
& = ({\bf 1}_+\oplus {\bf 21}_+\oplus {\bf 21'}_+\oplus {\bf 63}_+\oplus {\bf 90}_+)_{2.J_2}\otimes ({\bf 1}_+)_{2.A_5} \\
\end{aligned}
\ee and
\be \begin{aligned} {} [0^A_{sp},3^A_{su}] [0^B_{sp},3^B_{su}] \, \, \leftrightarrow \, \, &  ({\bf 1}_+\otimes {\bf 1}_+)_{2.J_2}\otimes  ({\bf 4}_-\otimes {\bf 4}_-)_{2.A_5} \\
& =({\bf 1}_+)_{2.J_2}\otimes ({\bf 1}_+\oplus {\bf 3}_+\oplus {\bf 3}_+\oplus {\bf 4}_+\oplus {\bf 5}_+)_{2.A_5}\ . \\ \end{aligned} \ee
The $\CN=1$ supercurrent must be a linear combination of the two fields in the trivial representation of $2.(A_5\times J_2)$. Such fields belong to the representations $[3^A_{sp},0^A_{su}][3^B_{sp},0^B_{su}]$ and  $[0^A_{sp},3^A_{su}] [0^B_{sp},3^B_{su}]$ of $ \widehat{sp}(6)_1^A\oplus \widehat{su}(2)_3^A\oplus \widehat{sp}(6)_1^B\oplus \widehat{su}(2)_3^B$, that are preserved by both $W^A$ and $W^B$. It follows that both such topological defects preserve the supercurrent.

\section{Spectral flow preserving defects in  Gepner models of K3}\label{s:Gepner}

In this section, we consider a family of exactly solvable points in the moduli space of K3 NLSMs, given by Gepner models. In \S \ref{s:N=2defects}, we first review the structure of supersymmetry--preserving TDLs in diagonal (A-type) $\mathcal N=2$ minimal models. Then in \S \ref{s:Gepnerdefects}, we discuss the generalization to Gepner models constructed from the orbifold of a tensor product of diagonal $\CN=2$ minimal models  \cite{Cordova:2023qei,Benjamin:2025knd}.
 In section (\ref{sec:338}), we apply these general methods to construct Fibonacci defect  lines in the $(3)^2(8)$ and $(2)(3)(18)$ Gepner models and non-invertible defects for $Rep(S_3)$ in the $(4)^3, (1)^2(2)^2$, and $(2)(4)(10)$ Gepner models.  We demonstrate that their defect twining genera  match that of the defect $W$ in $V^{f\natural}$ constructed in \S \ref{s:FibDef} in Eq. \eqref{eq:FibTwining}, and that of the defects for $Rep(S_3)$ computed in \cite{Angius:2025zlm}, respectively, providing further evidence for  Conjecture \ref{conj:K3relation}.

\subsection{Defects and $\CN=2$ minimal models}\label{s:N=2defects}

In this section, we briefly review the $\CN=2$ minimal models and their  TDLs which preserve the $\CN=2$ superconformal algebra and spectral flow.  The $\CN=2$ minimal series has central charge $c= {3k\over k+2}$ for $k=1,2,\ldots$ and bosonic subalgebra 
\be\label{minmodcoset}\frac{\widehat{su}(2)_{k}\hat{u}(1)_2}{\hat{u}(1)_{k+2}}.\ee
The unitary irreducible highest weight representations $M_{l,m,\epsilon}\equiv M^k_{l,m,\epsilon}$ of the bosonic algebra \eqref{minmodcoset} are labeled by three integers, $(l,m,\epsilon)$, with $l\in \{0,\ldots,k\},$ $m\in \ZZ/(2k+4),$ $\epsilon\in \ZZ/4$, subject to the condition $l+m+\epsilon= 0 \mod 2$. Furthermore, representations related by the transformation
$
(l,m,\epsilon) \sim (k-l, m+k + 2, \epsilon+2)
$ are isomorphic; i.e.
\be M^k_{l,m,\epsilon}\cong M^k_{k-l,m+k+2,\epsilon+2}\qquad l\in\{0,\ldots,k\},\ m\in \ZZ/(2k+4)\ZZ,\ \epsilon\in \ZZ/4\ZZ.
\ee

The primary operators $\phi_{l,m,\epsilon}$ of the representation $M^k_{l,m,\epsilon}$ have conformal dimension $h_{l,m,\epsilon}$ and $U(1)$ charge ($J_0$ eigenvalue) $Q_{l,m,\epsilon}$ given by
\be\label{weightN2minimal}
h_{l,m,\epsilon}={l(l+2)-m^2\over 4(k+2)}+{\epsilon^2\over 8}+\text{(integer)}, ~~~ Q_{l,m,\epsilon}={m\over k+2}-{\epsilon\over 2}.
\ee 
Let $\chi^k_{l,m,\epsilon}(\tau,z):=\Tr_{M^k_{l,m,\epsilon}} q^{L_0-c/24}y^{J_0}$ be the character of the representation $M^k_{l,m,\epsilon}$ with primary state $|l,m,\epsilon\rangle$ and field $\phi_{l,m,\epsilon}$; see appendix \ref{a:minimal} for explicit formulas. The modular transformations of the characters are
\be \label{eq:bosonicSmatrix} \chi^k_{l,m,\epsilon}\left(-\frac{1}{\tau},0\right)=\frac{1}{2}\sum_{l',m',\epsilon'} S^k_{lm\epsilon,l'm'\epsilon'}\chi^k_{l',m',\epsilon'}(\tau,0),
\ee where the sum is over all $l'\in\{0,\ldots,k\}$, $m'\in \ZZ/(2k+4)\ZZ$, $\epsilon'\in \ZZ/4\ZZ$ such that $l'+m'+\epsilon'\in 2\ZZ$, and where the S-matrix is given by
\be S^k_{lm\epsilon,l'm'\epsilon'}=\frac{1}{k+2}\sin \left ({\pi (l+1)(l'+1)\over k+2}\right ) e^{\pi i mm'\over k+2}e^{-\pi i \epsilon\epsilon'\over 2}\ .
\ee The overall factor of $1/2$ in \eqref{eq:bosonicSmatrix} keeps track of the fact that $\chi^k_{l',m',\epsilon'}(\tau,z)$ and $\chi^k_{k-l',m'+k+2,\epsilon'+2}(\tau,z)$ are characters of isomorphic representations. 

One can define a bosonic CFT with both the chiral and antichiral algebras isomorphic to the coset \eqref{minmodcoset} and with diagonal modular invariant partition function, which includes one primary operator $|l,m,\epsilon\rangle\otimes \overline{|l,m,\epsilon\rangle}$ for all inequivalent representations $M_{l,m,\epsilon}$. The simple topological defects that preserve the chiral and antichiral algebra are the Verlinde lines $\CL^k_{lm\epsilon}$, that are in one-to-one correspondence with the irreducible representations $M_{l,m,\epsilon}$, and that act on the primaries by
\be \hat\CL^k_{lm\epsilon}|l',m',\epsilon'\rangle\otimes \overline{|l',m',\epsilon'\rangle}=\frac{S^k_{l'm'\epsilon',lm\epsilon}}{S^k_{l'm'\epsilon',000}}|l',m',\epsilon'\rangle\otimes \overline{|l',m',\epsilon'\rangle}\ .
\ee

However, we are interested in theories with $\CN=2$ superconformal symmetry and spectral flow, and their corresponding symmetry-preserving TDLs are subject to further constraints. We begin by reviewing the $\CN=2$ minimal theories. All irreducible unitary representations  of the $\CN=2$ superconformal algebra at $c=\frac{3k}{k+2}$ can be obtained as direct sums $M^k_{l,m,\epsilon}\oplus M^k_{l,m,\epsilon+2}$ of representations of the bosonic algebra \eqref{minmodcoset}. Even $\epsilon$ corresponds to NS sector representations and odd $\epsilon$ corresponds to the Ramond sector.  
Define the sets $\mathcal S^{\NS}_l:=\{-l, -l+2,\ldots, l-2,l\}$ and $\mathcal S^{\R}_l:=\{-l-1, -l+1,\ldots, l-3,l-1\}$. Because of the identifications above, w.l.o.g. we can label $\CN=2$ NS/R sector by \be \CM^{k,\NS}_{l,m}=M^k_{l,m,0}\oplus M^k_{l,m,2}\ ,\qquad\qquad  \CM^{k,\R}_{l,m}=M^k_{l,m,-1}\oplus M^k_{l,m,1}\ ,
\ee where $l \in \{0,\ldots, k\}$ and $m\in \mathcal S^{\NS}_l$ or $m\in \mathcal S^{\R}_l$, respectively. The superconformal primaries $|l,m\rangle_{\NS}:= |l,m,0\rangle$ and $|l,m\rangle_{\R}:= |l,m,-1\rangle$ of $\CM^{k,\NS}_{l,m}$ and $\CM^{k,\R}_{l,m}$ are the ground states of $M_{l,m,0}$ and 
$M_{l,m,-1}$, and their conformal weight and $U(1)$ charge are as in \eqref{weightN2minimal} with vanishing $\text{integer}$. The component $M^k_{l,m,\epsilon}$ of $\CM_{l,m}^{k,X}$, $m\in \mathcal{S}^X_l$, has fermion parity ($(-1)^F$ eigenvalue)  $+1$ for $\epsilon\in\{0,1\}$ and $-1$ for  $\epsilon\in\{2,-1\}$.

The corresponding $\CN=2$ characters can be written as, 
\begin{alignat}{2}
\tilde \chi_{l,m}^{k,\NS,\pm}(\tau,z)&:=&\Tr_{\CM^{k,\NS}_{l,m}}(\pm 1)^F q^{L_0-c/24}y^{J_0} &=\chi^k_{l,m,0}(\tau,z) \pm \chi^k_{l,m,2}(\tau,z)\\
\tilde \chi_{l,m}^{k,\R,\pm}(\tau,z)&:=&\Tr_{\CM^{k,\R}_{l,m}}(\pm 1)^F q^{L_0-c/24}y^{J_0} &=\chi^k_{l,m,1}(\tau,z) \pm \chi^k_{l,m,-1}(\tau,z).
\end{alignat}
In the following, we will sometimes drop the superscript $k$ if it is clear from the context, and/or the superscripts $\NS$, $\R$, as they correspond to $l+m$ being even or odd. The characters obey the following modular transformation formulae: 
\begin{align*}
    \tilde \chi_{l,m}^{\NS,+}\left(-\frac{1}{\tau},0\right)&=\sum_{l'=0}^k\sum_{m'\in \mathcal{S}^{\NS}_{l'}}2S^k_{lm0,l'm'0}\tilde \chi_{l',m'}^{\NS,+}(\tau,0), &\chi_{l,m}^{\NS,-}\left(-\frac{1}{\tau},0\right)&=\sum_{l'=0}^k\sum_{m'\in \mathcal{S}^{\R}_{l'}}2S^k_{lm0,l'm'1}\tilde \chi_{l',m'}^{\R,+}(\tau,0)\ ,\\
    \tilde \chi_{l,m}^{\R,+}\left(-\frac{1}{\tau},0\right)&=\sum_{l'=0}^k\sum_{m'\in \mathcal{S}^{\NS}_{l'}}2S^k_{lm1,l'm'0}\tilde \chi_{l',m'}^{\NS,-}(\tau,0), &\tilde \chi_{l,m}^{\R,-}\left(-\frac{1}{\tau},0\right)&=\sum_{l'=0}^k\sum_{m'\in \mathcal{S}^{\R}_{l'}}2S^k_{lm1,l'm'1}\tilde \chi_{l',m'}^{\R,-}(\tau,0)\ .
\end{align*}
or equivalently
\be \begin{pmatrix}
    \tilde \chi_{l,m}^{\NS,+}\\
    \tilde \chi_{l,m}^{\NS,-}\\
    \tilde \chi_{l,m}^{\R,+}\\
    \tilde \chi_{l,m}^{\R,-}
\end{pmatrix}(-\frac{1}{\tau},0)=\sum_{l'=0}^k\sum_{m'\in \mathcal{S}^{\NS}_l\cup \mathcal{S}^{\R}_l}S^k_{lm;l'm'}\begin{pmatrix}
    \tilde \chi_{l',m'}^{\NS,+}\\
    \tilde \chi_{l',m'}^{\R,+}\\
    \tilde \chi_{l',m'}^{\NS,-}\\
    -i\tilde \chi_{l',m'}^{\R,-}
\end{pmatrix}(\tau,0) \ee where we set $\tilde \chi_{l,m}^{X,\pm}(\tau,z)=0$ if $m\notin \mathcal{S}^X_l$, $X\in \{\NS,\R\}$, and where the S matrix for the $\CN=2$ characters is given by 
\be\label{eq:minSmatrix}
S^k_{lm;l'm'}={2\over k+2}\sin \left ({\pi (l+1)(l'+1)\over k+2}\right ) e^{\pi i mm'\over k+2}
\ .
\ee

It is a well-known result that $\CN=2$ minimal theories with modular--invariant partition functions which preserve spectral flow  have an ADE classification \cite{Cappelli:1987xt,Gepner:1987qi,Gray:2008je}; from this point on, we restrict to the diagonal (A-type) solutions.
We will denote a non-chiral state  in a given $\CN=2$ minimal model   corresponding to $\Phi^{XX'}_{l,m;l'm'}=\phi^X_{l,m}\otimes \bar \phi^{X'}_{l',m'}$ by $|l,m;l',m'\rangle_{XX'}$.
The corresponding (diagonal) partition functions can be expressed as
\bea
Z^{k,\pm}_{XX}(\tau,z,\bar \tau,\bar z)&=&\Tr_{XX}(\pm 1)^Fq^{L_0-c/24}\bar q^{\bar L_0-c/24}y^{J_0}\bar y^{\bar J_0}
\\&=&\sum_{l=0}^k \sum_{m\in \mathcal S^{X}_l}\tilde \chi_{l,m}^{X,\pm}(\tau,z)\tilde{{\bar\chi}}_{l,m}^{X,\pm}(\bar\tau,\bar z),
\eea
for $X\in\{\NS,\R\}$. Partition functions in the $\NS\R$ or $\R\NS$ sectors can be obtained by spectral flow of the chiral or anti-chiral characters.

For a given $\mathcal N=2$ minimal model, the spectrum of topological defect lines which preserve the $\mathcal N=2$ superconformal algebra and the spectral flow can be determined by the familiar modular bootstrap approach \cite{Petkova:2000ip}. As we are restricting to the $A$-series, the defect lines are simply the Verlinde lines \cite{Verlinde:1988sn}.
The $k$th A-type minimal model has $(k+1)(k+2)/2$ $\CN=2$ primary operators and thus $(k+1)(k+2)/2$ Verlinde lines.\footnote{Note that unlike the discussion of TDLs in $\CN=2$ minimal models in \cite{Cordova:2023qei}, where they only require the defects to preserve the Virasoro symmetry, here we are requiring our Verlinde lines to preserve the full $\CN=2$ superconformal algebra.} Because the the $\CN=2$ primaries are in one-to-one correspondence with the Verlinde lines, we will label the Verlinde lines as $\CL^k_{lm}$, where $l\in\{0,\ldots, k\}, m\in \CS_l^\NS$. The quantum dimensions can be determined from the S-matrix \eqref{eq:minSmatrix} via the Verlinde formula and are given by 
\be
\langle \CL^k_{lm}\rangle={S^k_{00;lm}\over S^k_{00;00}},~~l\in \{0,\ldots k\}, m\in \CS^\NS_{l}.
\ee
From this expression, it is easy to determine that there exist $k+2$ invertible lines (with quantum dimension one); these arise from the $\ZZ_{k+2}$ invertible symmetry of the theory. In our convention, these invertible lines correspond to the defects $\CL^k_{00}$ (the identity line) and $\CL^k_{km},$ $m\in \CS_k^\NS$. This leaves $(k+2)(k-1)/2$ non-invertible topological defect lines which correspond to $\CL^k_{lm}$ for $l \in \{1,...,k-1\}$.

The Verlinde line $\CL^k_{lm}$ acts on the basis of $\NS\NS$ primary operators as
\be
\CL^k_{lm}|l',m';l',m'\rangle_\NSNS=\lambda^k_{lm;l'm'}|l',m';l',m'\rangle_\NSNS,
\ee
where the eigenvalues are given by 
\be\label{VerlindeEigs}
\lambda^k_{lm;l'm'}={S^k_{l'm';lm}\over S^k_{l'm';00}},~~l\in \{0,\ldots k\}, m\in \CS^\NS_{l}.
\ee
Notably, the eigenvalues for the invertible lines $\CL^k_{km}$ are given by
\be
\lambda^k_{km;l'm'}=(-1)^{l'}\omega_k^{mm'\over2},
\ee
where $\omega_k=e^{2\pi i\over {k+2}}$. The action on the $\R \R$ sector is determined by the one on the $\NS\NS$ sector up to a sign ambiguity
\be\label{RRaction}
\CL^k_{lm}|l',m';l',m'\rangle_{\R\R}=\pm\lambda^k_{lm;l'm'}|l',m';l',m'\rangle_{\R\R},
\ee which is related to the existence of a (invertible) symmetry acting by $-1$ on the RR sector and trivially on the $\NSNS$ sector. Here, $\lambda^k_{lm;l'm'}$ are given by the same formula \eqref{VerlindeEigs}, but with $m'\in \CS^\R_{l'}$.
It follows that one can compute the twined partition functions for a corresponding Verlinde line by the following formula,
\bea
Z^{\CL^k_{lm},k,\pm}_{\NSNS}(\tau,z,\bar \tau,\bar z)&=&\Tr_{\NSNS}\CL^k_{lm}(\pm 1)^Fq^{L_0-c/24}\bar q^{\bar L_0-c/24}y^{J_0}\bar y^{\bar J_0}
\\&=&\sum_{l'=0}^k \sum_{m'\in \mathcal S^{\NS}_{l'}}\lambda^k_{lm;l'm'}\tilde \chi_{l',m'}^{\NS,\pm}(\tau,z)\tilde{{\bar\chi}}_{l',m'}^{\NS,\pm}(\bar\tau,\bar z).
\eea
The formula for the RR sector is completely analogous, except for the overall sign ambiguity, as discussed around eq.\eqref{RRaction}.

We now consider the spectrum of Verlinde lines for the first few minimal models. The  $k=1$ minimal model has a $\ZZ_3$ (invertible) symmetry group and no non-invertible TDLs, $k=2$ has a $\ZZ_4$ symmetry group and two (oriented) non-invertible defects of quantum dimension $\sqrt{2}$ for a total of six lines, and the $k=3$ minimal model has a $\ZZ_5$ invertible symmetry group and a Fibonacci defect line $W:=\CL^3_{20}$ (according to the above notation) of quantum dimension $\zeta={1+\sqrt 5\over 2}$, which, when fused with the invertible lines, leads to a total of 10 simple TDLs. The action of $W$  on the basis of $\CN=2$ primaries in the NSNS sector can be determined from the formula \eqref{VerlindeEigs} and is given by
\be
W|l,m;l,m\rangle_\NSNS = \begin{cases}\zeta|l,m;l,m\rangle_\NSNS, ~~~l\in\{0,3\}\\
\sigma|l,m;l,m\rangle_\NSNS, ~~~l\in\{1,2\}
\end{cases}
\ee
where $\sigma={1-\sqrt 5\over 2}$ as in the previous section.

In section \ref{sec:338}, we  construct  Gepner models of K3 which have  Fibonacci lines built from this defect that preserve the $\CN=(4,4)$ superconformal symmetry. First, we briefly review Gepner models and some of their topological defects.

\subsection{Defects in Gepner models}\label{s:Gepnerdefects}

A Gepner model is an exactly solvable point in the moduli space of NLSMs on a Calabi-Yau $d$-fold described by an orbifold of a tensor product of $\CN=2$ minimal models such that the total central charge is $c=3d$ \cite{Gepner:1987qi}. If the model is a tensor product of A-type minimal models $(k_1)\otimes...\otimes (k_r)$, then one must orbifold by $\ZZ_N$, where $N=lcm(k_i+2)$, to impose charge integrality. The $\NSNS$ and $\HRR$ Hilbert spaces of the Gepner model are 
\be
\CH^{(k_1)\ldots(k_r)}_{XX}:=\bigoplus_{a\in \ZZ_N}\bigoplus_{l_i,m_i}{}'\bigotimes_{i=1}^r\left (\CM^{k_i,X}_{l_i,m_i}\otimes\bar\CM^{k_i,X}_{\sigma^a(l_i,m_i)}\right ),
\ee
where $X\in \{\NS,\R\}$, the sum is over $l_i\in\{0,\ldots,k_i\}$, $m_i\in \mathcal{S}^X_{l_i}$ satisfying the charge integrality condition
\be\label{GepnerProj} \sum_{i=1}^r\Bigl(\frac{m_i}{k_i+2}-\frac{\epsilon_i}{2}\Bigr)\in \ZZ
\qquad \Leftrightarrow \qquad \begin{cases}
   \sum_{i=1}^r\frac{m_i}{k_i+2} \in \ZZ & \text{NSNS sector}\\
   \sum_{i=1}^r\frac{m_i}{k_i+2} \in \frac{r}{2}+\ZZ & \text{RR sector}
\end{cases} 
\ee with $\epsilon_i\equiv 0\mod 2$ in the NS sector and $\epsilon_i\equiv 1\mod 2$ in the Ramond sector, and $\sigma$ is the (one unit) spectral flow transformation defined by
\be\label{specflowdef} \sigma(l_i,m_i)=\begin{cases}
    (l_i,m_i+2) & \text{for }m_i\in \mathcal{S}^X_{l_i},\ m_i\le l_i-2\ ,\\
    (k_i-l_i,m_i-k_i) & \text{for }m_i\in \mathcal{S}^X_{l_i},\ m_i> l_i-2\ .
\end{cases}
\ee
Each component
$ \bigotimes_{i=1}^r\left (\CM^{k_i,X}_{l_i,m_i}\otimes\bar\CM^{k_i,X}_{\sigma^a(l_i,m_i)}\right )$ decomposes into a sum of irreducible representations $\bigotimes_{i=1}^r (M^{k_i}_{l_i,m_i,\epsilon_i}\otimes\bar M^{k_i}_{\bar l_i,\bar m_i,\bar\epsilon_i})$ of the product of bosonic coset algebras. The fermion number $(-1)^F$ of the Gepner model is defined as $(-1)^F=(-1)^{J_0-\bar J_0}$, so that each such coset representation has definite fermion number
\be {(-1)^{F}}_{\rvert \bigotimes_{i=1}^r M^{k_i}_{l_i,m_i,\epsilon_i}\otimes\bar M^{k_i}_{\bar l_i,\bar m_i,\bar\epsilon_i}}=(-1)^{\sum_i (Q_{m_i,\epsilon_i}-Q_{\bar m_i,\bar\epsilon_i})}=(-1)^{\sum_{i=1}^r\Bigl(\frac{m_i-\bar m_i}{k_i+2}-\sum_i\frac{\epsilon_i-\bar\epsilon_i}{2}\Bigr)}\ .
\ee 
The orbifold partition function 
\be
Z^{(k_1)\ldots (k_r),\pm}_{XX}(\tau,z,\bar \tau,\bar z):= \Tr_{\CH_{XX}^{(k_1)\ldots(k_r)}} (\pm 1)^F q^{L_0-c/24}\bar q^{\bar L_0-c/24}y^{J_0}\bar y^{\bar J_0}\ee
 of a Gepner model, where $X\in\{\NS,\R\}$,  can be expressed in terms of minimal model characters as \cite{Kawai:1993jk}
\be
Z^{(k_1)\ldots (k_r),+}_{XX}(\tau,z,\bar \tau,\bar z)={1\over N}\sum_{a,b=0}^{N-1}\prod_{i=1}^r \sum_{l_i=0}^{k_i}\sum_{m_i\in S_{l_i}^X}e^{2\pi i b(\frac{m_i}{k_i+2}-\frac{\epsilon_i}{2})}\tilde\chi^{X,+}_{l_i,m_i}(\tau,z)\tilde\chi^{X,+}_{\sigma^a(l_i,m_i)}(\bar\tau,\bar z)
\ee
where $\epsilon_i=0$ for $X=\NS$ and $\epsilon_i=1$ for $X=\R$, and
\be
Z^{(k_1)\ldots (k_r),-}_{XX}(\tau,z,\bar \tau,\bar z)=Z^{(k_1)\ldots (k_r),+}_{XX}(\tau,z+\frac{1}{2},\bar \tau,\bar z+\frac{1}{2})\ .
\ee 
Furthermore, one can obtain the elliptic genus by considering the RR sector, setting $\bar z=0$, and keeping the $(-1)^F$ insertion; i.e. $Z^{(k_1)\ldots (k_r),-}_{RR}(\tau,z,\bar \tau,0)$ which can be evaluated using the identity 
\be
\tilde\chi^{R,-}_{l,m}(\tau,0)=\delta_{l+1,m}-\delta_{-l-1,m}.
\ee

We will now describe the full chiral algebra $\CA$ of the Gepner model. Let us use the notation
\be \CM_{{\bf l},{\bf m}}:=\CM^{k_1,X}_{l_1,m_1}\otimes \CM^{k_2,X}_{l_2,m_2}\otimes\cdots \otimes \CM^{k_r,X}_{l_r,m_r}
\ee
where $X\in\{\NS,\R\}$, and $({\bf l},{\bf m})$ are the elements of the sets
\be\label{setReps} K^X:=\Bigl\{({\bf l},{\bf m}):=((l_1,m_1),(l_2,m_2),\ldots)\mid l_i\in\{0,\ldots, k_i\},\ m_i\in \mathcal{S}_{l_i}^X,\ \sum_i\Bigl(\frac{m_i}{k_i+2}-\frac{\epsilon_i}{2}\Bigr)\in \ZZ \Bigr\}\ .
\ee
Spectral flow acts on each set $K^X$ in \eqref{setReps}, $X\in\{\NS,\R\}$, by permutations
\be \sigma^a({\bf l},{\bf m})=(\sigma^a(l_1,m_1),\sigma^a(l_2,m_2),\ldots)\ ,
\ee where $\sigma(l_i,m_i)$ is defined in \eqref{specflowdef}. Notice that $\sigma^N({\bf l},{\bf m})=({\bf l},{\bf m})$ for all $({\bf l},{\bf m})\in K^X$. We can decompose each set $K^X$ into orbits under spectral flow:
\be [{\bf l m}]:=\{ ({\bf l},{\bf m}), \sigma({\bf l},{\bf m}),\sigma^2({\bf l},{\bf m}),\ldots\}\subset K^X\ .
\ee For what follows, it is important to notice that there can be two kind of orbits in $K^X$. The long, or generic, orbits $[{\bf l m}]$ contain $N$ distinct elements $\sigma^a({\bf l},{\bf m})$, $a\in \ZZ_N$; for example, the vacuum orbit $[({\bf 0},{\bf 0})]$ is always long. Many Gepner models also contain short orbits $[{\bf l m}]$, such that $\sigma^{N/2}({\bf l},{\bf m})=({\bf l},{\bf m})$, and therefore consisting of $N/2$ distinct elements. It is easy to check that an orbit  $[{\bf l m}]$ is short if and only if $l_i=k_i/2$ for all $i$ such that $k_i+2$ does not divide $N/2$. 

Let us consider the case with $c=6$, i.e. K3 sigma models. The chiral algebra of the tensor product of $\CN=2$ minimal model is 
\be \CM_{{\bf 0},{\bf 0}}:=\CM^{k_1,\NS}_{0,0}\otimes\cdots \otimes \CM^{k_r,\NS}_{0,0}\ .
\ee 
The chiral superalgebra $\CA$ of the Gepner model is a simple current extension of $\CM_{{\bf 0},{\bf 0}}$ by the (one unit) spectral flow operators, and is given by the holomorphic fields in the Hilbert space $\CH_{\NS\NS}$, i.e.
\be \CA\cong \bigoplus_{a\in \ZZ_N} \bigotimes_{i=1}^r\CM^{k_i,\NS}_{\sigma^a(0,0)}=\bigoplus_{a\in \ZZ_N} \CM_{\sigma^a({\bf 0},{\bf 0})}\ ,
\ee where $N=lcm(k_i+2)$. The extended algebra $\CA$ contains the (small) $\CN=4$ superconformal algebra at $c=6$, while $\CM_{{\bf 0},{\bf 0}}$ only contains $\CN=2$.

The representations of the extended algebra $\CA$ that appear in the Hilbert space $\CH_{XX}$ are obtained by summing over all $\CM_{{\bf 0},{\bf 0}}$-representations in a given spectral flow orbit $[{\bf l m}]$, i.e.
\be R_{[{\bf l m}]}:= \bigoplus_{a\in \ZZ_N} \CM_{\sigma^a({\bf l},{\bf m})}\ .
\ee If $[{\bf l m}]$ is a long orbit, then $R_{[{\bf l m}]}$ is irreducible as an $\CA$-representation. On the other hand, if ${[{\bf l m}]}$ is short, then $R_{[{\bf l m}]}$ decomposes as a direct sum of two irreducible representations
\be R_{[{\bf l m}]}=R_{[{\bf l m}]_+}\oplus R_{[{\bf l m}]_-}\ ,\qquad [{\bf l m}]\text{ short.}
\ee This is a general phenomenon in simple current extensions of chiral algebras, where the representations that are stabilised by some non-trivial subgroup of the simple current symmetry (in this case, $\langle \sigma\rangle\cong\ZZ_N$) are known as `fixed points' of the extension \cite{Schellekens:1989uf,Schellekens:1990xy}.  Roughly speaking, for short orbits one can interpret $\sigma^{N/2}$ as an involution acting on $R_{[{\bf l m}]}$, and the irreducible components $R_{[{\bf l m}]_+}$ and $R_{[{\bf l m}]_-}$ are the corresponding eigenspaces \cite{Fuchs:1995tq}. $R_{[{\bf l m}]_+}$ and $R_{[{\bf l m}]_-}$ are isomorphic as representations of the smaller algebra $\CM_{{\bf 0},{\bf 0}}$ (in particular, they have the same character), but they are inequivalent as representations of $\CA$. In terms of representations of $\CA\times \bar \CA$, the Hilbert space $\CH_{XX}$ of the Gepner model decomposes as
\be \CH_{XX}=\bigoplus_{\text{long orbits}} R_{[{\bf l m}]}\otimes \bar R_{[{\bf l m}]}\ \oplus \ \bigoplus_{\text{short orbits}} (R_{[{\bf l m}]_+}\otimes \bar R_{[{\bf l m}]_+}\, \oplus\, R_{[{\bf l m}]_-}\otimes \bar R_{[{\bf l m}]_-})
\ee where the sum is over the spectral flow orbits in the set $K^X$, $X\in\{\NS,\R\}$.  Therefore, the Gepner model is the diagonal modular invariant for the chiral algebra $\CA$. More generally, there might exist other modular invariants for the same algebra $\CA$, but we will not consider them in this article.

The characters of the irreducible representations of $\CA$ can be easily obtained from the ones of the $\CN=2$ minimal models,
\be \chi_{R[{\bf l m}]}^{X,\pm}(\tau,z)=\sum_{a=0}^{N-1}\prod_{i=1}^r \tilde \chi^{k^i,X,+}_{\sigma^a(l_i,m_i)}(\tau,z+\frac{\delta_-}{2})\qquad [{\bf l m}]\text{ long},
\ee
\be \chi_{R[{\bf l m}]_+}^{X,\pm}(\tau,z)=\chi_{R[{\bf l m}]_-}^{X,\pm}(\tau,z)=\frac{1}{2}\sum_{a=0}^{N-1}\prod_{i=1}^r \tilde \chi^{k^i,X,+}_{\sigma^a(l_i,m_i)}(\tau,z+\frac{\delta_-}{2})\qquad [{\bf l m}]\text{ short},
\ee where $\delta_-=1$ if $(-1)^{F}$ is inserted and $\delta_-=0$ otherwise.
Therefore, the modular S-matrix $\mathbf{S}$ for the extended algebra $\CA$ is related to the S-matrices $S^{k_i}_{l_im_i}$.  In particular, if $[{\bf l m}]$ and $[{\bf l' m'}]$ are both long orbits, then
\be \mathbf{S}_{[{\bf l m}],[{\bf l' m'}]}
=\prod_{i=1}^r S^{k_i}_{l_im_i,l_i'm_i'}\ ,\qquad [{\bf l m}],[{\bf l' m'}]\text{ long,}
\ee where one can prove that the RHS does not depend on the choice of the representatives $({\bf l},{\bf m}),({\bf l'},{\bf m'})\in K^\NS \cup K^\R$ of the spectral flow orbits. When one of the orbits, for example $[{\bf l m}]$, is short, and $[{\bf l' m'}]$ is long, there is also a simple formula
\begin{align} &\mathbf{S}_{[{\bf l m}]_+,[{\bf l' m'}]}=\mathbf{S}_{[{\bf l m}]_-,[{\bf l' m'}]}
=\frac{1}{2}\prod_{i=1}^r S^{k_i}_{l_im_i,l_i'm_i'} && [{\bf l m}]\text{ short,}\quad [{\bf l' m'}]\text{ long,}
\end{align} and $\mathbf{S}_{[{\bf l' m'}],[{\bf l m}]_\pm}= \mathbf{S}_{[{\bf l m}]_\pm,[{\bf l' m'}]}$. On the other hand, when both $[{\bf l m}]$ and $[{\bf l' m'}]$ are short, the corresponding S-matrix elements cannot be directly expressed in terms of the ones of the minimal models. The fact that the modular transformations of the characters are not sufficient to uniquely determine the S-matrix is due to the fact that the representations $R_{[{\bf l m}]_+}$ and $R_{[{\bf l m}]_-}$ have the same character. The matrix $\mathbf{S}$ can be unambiguously defined in terms of the modular tensor category of $\CA$-representations, rather than just the characters.  The problem of determining the precise S-matrix for simple current extensions is known as the resolution of the fixed points. There are some general results when the stabilizer $\langle \sigma^{N/2}\rangle\cong \ZZ_2$ of short orbits has order $2$ (see for example \cite{Fuchs:1996dd}), which in the present case becomes
\be \mathbf{S}_{[{\bf l m}]_+,[{\bf l' m'}]_+}=\mathbf{S}_{[{\bf l m}]_-,[{\bf l' m'}]_-}\ ,\qquad \mathbf{S}_{[{\bf l m}]_+,[{\bf l' m'}]_-}=\mathbf{S}_{[{\bf l m}]_-,[{\bf l' m'}]_+}
\ee and
\be\label{SmatrixRelation} \mathbf{S}_{[{\bf l m}]_+,[{\bf l' m'}]_\pm}+\mathbf{S}_{[{\bf l m}]_-,[{\bf l' m'}]_\pm}=\frac{1}{2}\prod_{i=1}^r S^{k_i}_{l_im_i,l_i'm_i'}\ ,\qquad [{\bf l m}],[{\bf l' m'}]\text{ short.}
\ee In this article, we will not attempt to determine the single terms $\mathbf{S}_{[{\bf l m}]_+,[{\bf l' m'}]_\pm}$ and $\mathbf{S}_{[{\bf l m}]_-,[{\bf l' m'}]_\pm}$.

We are interested in determining the TDLs of the Gepner model that preserve the full extended algebra $\CA\times \bar\CA$. Since the CFT is a diagonal modular invariant, the simple TDLs are just the Verlinde lines, in one-to-one correspondence with the irreducible $\NS$ representations of $\CA$ (representations in the Ramond sector correspond to defects that act on $\CA$ by the fermion number). This means that for each long orbit $[{\bf l m}]\subset K^\NS$ there is one simple TDL
\be\label{eq:GepnerTDL} \CL_{[{\bf l m}]}:=\CL^{k_1}_{l_1,m_1}\otimes \cdots \otimes \CL^{k_r}_{l_r,m_r}\qquad [{\bf l m}]\subset K^{\NS}\text{ long,}
\ee that acts on \emph{all} (long or short) $\NSNS$ representations $R_{[{\bf l' m'}]}\otimes \bar R_{[{\bf l' m'}]}$ by
\be \CL_{[{\bf l m}]}|{\bf l'},{\bf m'};{\bf  l'},{\bf m'}\rangle_\NSNS=\prod_{i=1}^r \lambda^{k_i}_{l_i,m_i;l_i',m_i'}|{\bf l'},{\bf m'};{\bf  l'},{\bf m'}\rangle_\NSNS\ ,
\ee where $\lambda^{k_i}_{l_im_i;l'_im'_i}={S^{k_i}_{l'_im'_i;l_im_i}\over S^{k_i}_{l'_im'_i;00}}$ as in eq.\eqref{VerlindeEigs}. On the other hand, for each short orbit $[{\bf l m}]\subset K^\NS$ there are two different simple defects
\be \CL_{[{\bf l m}]_+},\qquad \CL_{[{\bf l m}]_-}\ ,\qquad [{\bf l m}]\subset K^{\NS}\text{ short.}
\ee These two defects act in the same way on \emph{long} orbit representations $R_{[{\bf l' m'}]}\otimes \bar R_{[{\bf l' m'}]}$ by
\be \CL_{[{\bf l m}]_\pm}|{\bf l'},{\bf m'};{\bf  l'},{\bf m'}\rangle_\NSNS=\frac{1}{2}\prod_{i=1}^r \lambda^{k_i}_{l_i,m_i;l_i',m_i'}|{\bf l'},{\bf m'};{\bf  l'},{\bf m'}\rangle_\NSNS\ ,\qquad [{\bf l' m'}]\text{ long.}
\ee However, $\CL_{[{\bf l m}]_+}$ and $\CL_{[{\bf l m}]_-}$ act in a different way on short orbit representations $R_{[{\bf l' m'}]_\pm}\otimes \bar R_{[{\bf l' m'}]_\pm}$. In order to find the precise eigenvalues one would need to resolve the fixed point and determine the exact S-matrix. However, for the identities above, one can show that the superposition of the two defects is given by
\be \CL_{[{\bf l m}]_+}+\CL_{[{\bf l m}]_-}=\CL^{k_1}_{l_1,m_1}\otimes \cdots \otimes \CL^{k_r}_{l_r,m_r}\qquad [{\bf l m}]\subset K^{\NS}\text{ short,}
\ee
which implies that on every representation $R_{[{\bf l' m'}]_\pm}\otimes \bar R_{[{\bf l' m'}]_\pm}$ the action is
\be (\CL_{[{\bf l m}]_+}+\CL_{[{\bf l m}]_-})|{\bf l'},{\bf m'};{\bf  l'},{\bf m'}\rangle_\NSNS=\prod_{i=1}^r \lambda^{k_i}_{l_i,m_i;l_i',m_i'}|{\bf l'},{\bf m'};{\bf  l'},{\bf m'}\rangle_\NSNS\ .
\ee The quantum dimensions are given by
\begin{align} \langle\CL_{[{\bf l m}]}\rangle&=\prod_{i=1}^r \lambda^{k_i}_{l_i,m_i;00}& \text{long orbit}\ ,\\ 
\langle \CL_{[{\bf l m}]_\pm}\rangle&=\frac{1}{2}\prod_{i=1}^r \lambda^{k_i}_{l_i,m_i;00}&\text{short orbit}\ .
\end{align}
The actions on the $\R\R$ representations are analogous, except for a sign ambiguity related to the possibility of multiplying the defect by the symmetry acting trivially on the $\NSNS$ sector and by $-1$ on the $\R\R$ sector (see eq.\eqref{RRaction}).

We are particularly interested in understanding the action of these defects on the four RR ground that are the generators of left- and right-moving half-unit spectral flow. Such fields are contained in the $\CA\times\bar\CA$ representation $R_{[(0,-1)^r]}\otimes \bar R_{[(0,-1)^r]}$, where 
\be R_{[(0,-1)^r]}:=\bigoplus_{a\in \ZZ_N} \bigotimes_{i=1}^r\CM^{k_i,\R}_{\sigma^a(0,-1)}\ .
\ee The orbit $[(0,-1)^r]$ is always long, and therefore we can easily compute the eigenvalues of every TDL  $\CL_{[{\bf l m}]_\pm}$ on $R_{[(0,-1)^r]}\otimes \bar R_{[(0,-1)^r]}$.  Using that $S^k_{lm,0,-1}=S^k_{lm,00}e^{-\pi i\frac{m}{k+2}}$, we obtain that the action of any $\CL_{[{\bf l m}]_\pm}$ (for both long or short orbit) on the primary states in $R_{[(0,-1)^r]}\otimes \bar R_{[(0,-1)^r]}$ is
\be \CL_{[{\bf l m}]_\pm}|(0,-1)^r;(0,-1)^r\rangle_{\R\R}=\pm e^{-\pi i\sum_{i=1}^r\frac{m_i}{k_i+2}}\langle \CL_{[{\bf l m}]_\pm}\rangle|(0,-1)^r;(0,-1)^r\rangle_{\R\R}\ .
\ee Notice that $e^{-\pi i\sum_{i=1}^r\frac{m_i}{k_i+2}}$ is a sign, because $\sum_{i=1}^r\frac{m_i}{k_i+2}$ is an integer. Therefore, using the possibility of multiplying any defect by the symmetry that acts by $-1$ on the RR sector, one can always choose any defect $\CL_{[{\bf l m}]_\pm}$ to preserve the spectral flow operators, i.e. to act by 
\be \CL_{[{\bf l m}]_\pm}|(0,-1)^r;(0,-1)^r\rangle_{\R\R}=\langle \CL_{[{\bf l m}]_\pm}\rangle|(0,-1)^r;(0,-1)^r\rangle_{\R\R}\ .
\ee Notice that if any two defects preserve the half-unit spectral flow operators, then  their fusion also has the same property. We conclude that the TDLs $\CL_{[{\bf l m}]_\pm}$ generate a category of defects preserving the $\CN=(4,4)$ superconformal algebra and spectral flow. We stress that these are not the most general defects with these properties; in particular, there might be TDLs that preserve $\CN=(4,4)$ but not $\CA\times \bar\CA$. Some examples,  described in \cite{Gaberdiel:2011fg}, are the invertible symmetries in the $(1)^6$ or in the $(2)^4$ Gepner models, acting by permutations of the various $\CN=2$ factors.

Finally, let us consider the twining partition functions and twining genera associated with these defects. Consider the defect $\CL\equiv \CL_{[{\bf l m}]}$ for $[{\bf l m}]$ a long orbit. 
The corresponding twined orbifold partition function  is given by
\begin{align}\label{eq:Gepnertwining}
&Z^{(k_1)\ldots (k_r),\CL,\pm}_{XX}(\tau,z,\bar \tau,\bar z):= \Tr_{\CH_{XX}^{(k_1)\ldots(k_r)}}\CL(\pm 1)^F q^{L_0-c/24}\bar q^{\bar L_0-c/24}y^{J_0}\bar y^{\bar J_0}
\\&=\frac1N\sum_{a,b=0}^{N-1}\prod_{i=1}^r \sum_{l_i=0}^{k_i}\sum_{m_i\in S_{l_i}^X}\lambda^{k_i}_{s_it_i;l_im_i} e^{2\pi i b(\frac{m_i}{k_i+2}-\frac{\epsilon_i}{2})}\tilde\chi^{X,+}_{l_i,m_i}(\tau,z+\tfrac{\delta_-}{2})\tilde\chi^{X,+}_{\sigma^a(l_i,m_i)}(\bar\tau,\bar z+\tfrac{\delta_-}{2})\notag
\end{align}
where $\delta_-=1$ if $(-1)^{F+\tilde F}$ is inserted and $\delta_-=0$ otherwise. As discussed above, the overall sign in the Ramond sector is chosen so that the half-unit spectral flow operators are invariant. For short orbits, even though we haven't determined the precise eigenvalues of the two defects $\CL_{[{\bf l m}]_+}$ and $\CL_{[{\bf l m}]_-}$ on the representations $R_{[{\bf l' m'}]_\pm}\otimes \bar R_{[{\bf l' m'}]_\pm}$, one can nevertheless prove that their twining partition functions are the same and given by 
\be Z^{(k_1)\ldots (k_r),\CL_{[{\bf l m}]_+}}(\tau,z)=Z^{(k_1)\ldots (k_r),\CL_{[{\bf l m}]_-}}(\tau,z)=\frac{1}{2}Z^{(k_1)\ldots (k_r),\CL}(\tau,z)\ ,
\ee
where $Z^{(k_1)\ldots (k_r),\CL}(\tau,z)$ is given by \eqref{eq:Gepnertwining} with
\be \CL:=\CL_{[{\bf l m}]_+}+\CL_{[{\bf l m}]_-}\ .\ee The reason is that each pair of representations $R_{[{\bf l' m'}]_+}\otimes \bar R_{[{\bf l' m'}]_+}$ and $R_{[{\bf l' m'}]_-}\otimes \bar R_{[{\bf l' m'}]_-}$ have the same character, and appear with multiplicity $1$ in the spectrum. Therefore, the twining partition function only depends on the sum of the eigenvalues on $R_{[{\bf l' m'}]_+}\otimes \bar R_{[{\bf l' m'}]_+}$ and on $R_{[{\bf l' m'}]_-}\otimes \bar R_{[{\bf l' m'}]_-}$, which by \eqref{SmatrixRelation} is $\frac{1}{2}\prod_{i=1}^r\lambda^{k_i}_{l_i,m_i;l_i',m_i'}$ for both $\CL_{[{\bf l m}]_+}$ and $\CL_{[{\bf l m}]_-}$.

\subsection{Examples of TDLs in Gepner models}\label{sec:338}
Our first goal is to construct K3 NLSMs which contain Fibonacci fusion categories, and moreover compare the defect-twined elliptic genus to the Fibonacci twining genus in $V^{f\natural}$ constructed in \S \ref{s:FibDef}. Our approach will be to use the fact that the $\CN=2$ minimal model with $k=3$ contains a Fibonacci fusion category generated by the Verlinde line $\CL^3_{20}$. Thus if we can build a Gepner model of K3 which contains the $k=3$ minimal model as one of its factors, we can construct a TDL in that theory of the form \eqref{eq:GepnerTDL} by tensoring  $\CL^3_{20}$ with the identity line for all other factors, and this TDL will generate a Fibonacci fusion category in the Gepner model.

There are 16 A-type Gepner constructions of K3 \cite{Gepner:1987vz}, and only two of them contain the $k=3$ minimal model as a component: the $(3)^2(8)$ Gepner model and the $(2)(3)(18)$ Gepner model.  In the $(3)^2(8)$ model, we will define the non-invertible TDL
\be
W^{(3)^2(8)}:=\CL^3_{20}\otimes \CL^3_{00}\otimes \CL^8_{00},
\ee
which is a Fibonacci defect line of quantum dimension $\zeta$. This defect corresponds to the spectral flow orbit $[(2,0),(0,0),(0,0)]\subset K^\NS$, which is a long orbit, since it contains $N=lcm(5,5,10)=10$ distinct elements. Short orbits in this model are given by $[(l_1,m_1),(l_2,m_2),(l_3,m_3)]$ with $l_3=k_3/2=4$. Note that $\CL^k_{00}$ always represents the identity line in the $k$th minimal model. Secondly, in the $(2)(3)(18)$ model, we define the non-invertible TDL
\be
W^{(2)(3)(18)}:= \CL^2_{00}\otimes \CL^3_{20}\otimes \CL^{18}_{00},
\ee
which is also a Fibonacci defect line of quantum dimension $\zeta$. Again, the orbit $[(0,0),(2,0),(0,0)]\subset K^\NS$ is long, since it contains $N=lcm(4,5,20)=20$ distinct elements. Note that in both cases, as $m_i=0$ for each Verlinde line appearing in the tensor product,  \eqref{GepnerProj} is satisfied. It follows that both $W^{(3)^2(8)}$ and $W^{(2)(3)(18)}$ preserve the $\CN=(4,4)$ superconformal algebra in the corresponding Gepner models.

Finally, we calculate the twining function in each case using equation \eqref{eq:Gepnertwining}. In both cases, we compute
\be\label{eq:GepnerFibgenus}
\phi^W(\tau,z)= {1\over 2}\phi(\tau,z)+{\sqrt 5}{\theta_1^2(\tau,z)\over \eta^6(\tau)}f(\tau),
\ee
where $W=W^{(3)^2(8)}$ or $W=W^{(2)(3)(18)}$, and $\phi(\tau,z)$ is equal to the elliptic genus of K3. This formula exactly reproduces the Fibonacci twining genus $\phi^W(V^{f\natural},\tau,z)$ in $V^{f\natural}$ in eq.\eqref{eq:FibTwining}, thus providing further support to our conjecture.

As another set of examples, note that in the $k=4$ minimal model, the defects $\CL^4_{00}, \CL^4_{20}$, and $\CL^4_{40}$ generate a $Rep(S_3)$ which has fusion rules
 $$ (\CL^4_{20})^2= \CL^4_{00}+\CL^4_{20} + \CL^4_{40}, ~~ (\CL^4_{40})^2= \CL^4_{00}.$$ Now consider the  Gepner model of type $X$, where 
$$X\in \{(4)^3,(1)^2(4)^2,  (2)(4)(10)\}.$$
Let  $\CL^X_\rho:=\CL^4_{40}\otimes \prod_{i\neq 1} \CL^{k_i}_{00}$ and $\mathcal X^X:= \CL^4_{20}\otimes \prod_{i \neq 1} \CL^{k_i}_{00}$, where our convention is to reorder the minimal model factors such that $k_1=4$. It follows that these lines, along with the identity line $\mathcal I:= \prod_{i} \CL^{k_i}_{00}$, form a $Rep(S_3)$ fusion category  if they are  all in distinct long orbits of spectral flow.
Then one can easily compute 
\be\label{RepS3twining}
\phi^{\CL^X_\rho}=\phi^{1^82^8}, ~~~ \phi^{{\cal X}^X}=\phi^{1^82^8} + \phi^{1^63^6},~~~~~ X\in\{(4)^3,(1)^2(4)^2,  (2)(4)(10)\}
\ee
exactly matching the results of the $Rep(S_3)$ category in $V^{f\natural}$ considered in section 5.2 in our previous article \cite{Angius:2025zlm}.

Analogous defects also occur in the other two Gepner models that admit a $k=4$ factor, namely $(1)^4(4)$ and $(1)(2)^2(4)$. 
However, for $X=(1)^4(4)$, we first notice that $\CL^X_\rho$ is in the same spectral flow orbit as the vacuum $\sigma^3((4,0),(0,0)^4)=(0,0)^5$, so that $\CL^X_\rho$ is just the identity defect. On the other hand, the orbit $[(2,0),(0,0)^4]$ associated with ${\cal X}^X$ is short for this model, so that ${\cal X}^X$ is the superposition of two simple invertible defects $\CL_{[(2,0),(0,0)^4]_+}$ and $\CL_{[(2,0),(0,0)^4]_-}$, with twining genus
\be
\phi^{\CL_{[(2,0),(0,0)^4]_+}}=\phi^{\CL_{[(2,0),(0,0)^4]_-}}=\phi^{1^63^6}.
\ee
This shows that the two invertible defects have order $3$. Therefore, in this case, the $Rep(S_3)$ category that is present in the tensor product of minimal models, does not survive the orbifold.

Finally, for  the $X=(1)(2)^2(4)$ Gepner model, we find that $\phi^{\CL^X_\rho}=\phi^{2^{16}/1^8}$ while $\phi^{{\cal X}^X}$ does not match with any of the twining genera that we computed in $V^{f\natural}$. It would be interesting to determine a $Rep(S_3)$ category of defects reproducing these results in $V^{f\natural}$.

\bigskip

{\bf Acknowledgments.}  
R.A. acknowledges support from the ERC Starting Grant QGuide-101042568- StG 2021, Deutsche Forschungsgemeinschaft under Germany’s Excellence Strategy EXC 2121 Quantum Universe 390833306  and Deutsche Forschungsgemeinschaft through the Collaborative Research Center 1624 "Higher Structures, Moduli Spaces and Integrability". S.M.H. is supported by the US DOE under grant DE-SC0024787. R.V. acknowledges support from CARIPARO Foundation Grant under grant n. 68079 and PRIN Project n. 2022ABPBEY.

\appendix

\section{Basis of the Leech lattice and some lattice endomorphisms}\label{a:Leechbasis}

The Leech lattice $\Lambda\subset \RR^{24}$ is generated by the rows of the following matrix \cite{ConwaySloane}
\be \arraycolsep=3pt \def\arraystretch{0.8}
    \frac{1}{\sqrt{8}} {\tiny{\left(
\begin{array}{cccccccccccccccccccccccc}
 8 & 0 & 0 & 0 & 0 & 0 & 0 & 0 & 0 & 0 & 0 & 0 & 0 & 0 & 0 & 0 & 0 & 0 & 0 & 0 & 0 & 0 & 0 & 0 \\
 4 & 4 & 0 & 0 & 0 & 0 & 0 & 0 & 0 & 0 & 0 & 0 & 0 & 0 & 0 & 0 & 0 & 0 & 0 & 0 & 0 & 0 & 0 & 0 \\
 4 & 0 & 4 & 0 & 0 & 0 & 0 & 0 & 0 & 0 & 0 & 0 & 0 & 0 & 0 & 0 & 0 & 0 & 0 & 0 & 0 & 0 & 0 & 0 \\
 4 & 0 & 0 & 4 & 0 & 0 & 0 & 0 & 0 & 0 & 0 & 0 & 0 & 0 & 0 & 0 & 0 & 0 & 0 & 0 & 0 & 0 & 0 & 0 \\
 4 & 0 & 0 & 0 & 4 & 0 & 0 & 0 & 0 & 0 & 0 & 0 & 0 & 0 & 0 & 0 & 0 & 0 & 0 & 0 & 0 & 0 & 0 & 0 \\
 4 & 0 & 0 & 0 & 0 & 4 & 0 & 0 & 0 & 0 & 0 & 0 & 0 & 0 & 0 & 0 & 0 & 0 & 0 & 0 & 0 & 0 & 0 & 0 \\
 4 & 0 & 0 & 0 & 0 & 0 & 4 & 0 & 0 & 0 & 0 & 0 & 0 & 0 & 0 & 0 & 0 & 0 & 0 & 0 & 0 & 0 & 0 & 0 \\
 2 & 2 & 2 & 2 & 2 & 2 & 2 & 2 & 0 & 0 & 0 & 0 & 0 & 0 & 0 & 0 & 0 & 0 & 0 & 0 & 0 & 0 & 0 & 0 \\
 4 & 0 & 0 & 0 & 0 & 0 & 0 & 0 & 4 & 0 & 0 & 0 & 0 & 0 & 0 & 0 & 0 & 0 & 0 & 0 & 0 & 0 & 0 & 0 \\
 4 & 0 & 0 & 0 & 0 & 0 & 0 & 0 & 0 & 4 & 0 & 0 & 0 & 0 & 0 & 0 & 0 & 0 & 0 & 0 & 0 & 0 & 0 & 0 \\
 4 & 0 & 0 & 0 & 0 & 0 & 0 & 0 & 0 & 0 & 4 & 0 & 0 & 0 & 0 & 0 & 0 & 0 & 0 & 0 & 0 & 0 & 0 & 0 \\
 2 & 2 & 2 & 2 & 0 & 0 & 0 & 0 & 2 & 2 & 2 & 2 & 0 & 0 & 0 & 0 & 0 & 0 & 0 & 0 & 0 & 0 & 0 & 0 \\
 4 & 0 & 0 & 0 & 0 & 0 & 0 & 0 & 0 & 0 & 0 & 0 & 4 & 0 & 0 & 0 & 0 & 0 & 0 & 0 & 0 & 0 & 0 & 0 \\
 2 & 2 & 0 & 0 & 2 & 2 & 0 & 0 & 2 & 2 & 0 & 0 & 2 & 2 & 0 & 0 & 0 & 0 & 0 & 0 & 0 & 0 & 0 & 0 \\
 2 & 0 & 2 & 0 & 2 & 0 & 2 & 0 & 2 & 0 & 2 & 0 & 2 & 0 & 2 & 0 & 0 & 0 & 0 & 0 & 0 & 0 & 0 & 0 \\
 2 & 0 & 0 & 2 & 2 & 0 & 0 & 2 & 2 & 0 & 0 & 2 & 2 & 0 & 0 & 2 & 0 & 0 & 0 & 0 & 0 & 0 & 0 & 0 \\
 4 & 0 & 0 & 0 & 0 & 0 & 0 & 0 & 0 & 0 & 0 & 0 & 0 & 0 & 0 & 0 & 4 & 0 & 0 & 0 & 0 & 0 & 0 & 0 \\
 2 & 0 & 2 & 0 & 2 & 0 & 0 & 2 & 2 & 2 & 0 & 0 & 0 & 0 & 0 & 0 & 2 & 2 & 0 & 0 & 0 & 0 & 0 & 0 \\
 2 & 0 & 0 & 2 & 2 & 2 & 0 & 0 & 2 & 0 & 2 & 0 & 0 & 0 & 0 & 0 & 2 & 0 & 2 & 0 & 0 & 0 & 0 & 0 \\
 2 & 2 & 0 & 0 & 2 & 0 & 2 & 0 & 2 & 0 & 0 & 2 & 0 & 0 & 0 & 0 & 2 & 0 & 0 & 2 & 0 & 0 & 0 & 0 \\
 0 & 2 & 2 & 2 & 2 & 0 & 0 & 0 & 2 & 0 & 0 & 0 & 2 & 0 & 0 & 0 & 2 & 0 & 0 & 0 & 2 & 0 & 0 & 0 \\
 0 & 0 & 0 & 0 & 0 & 0 & 0 & 0 & 2 & 2 & 0 & 0 & 2 & 2 & 0 & 0 & 2 & 2 & 0 & 0 & 2 & 2 & 0 & 0 \\
 0 & 0 & 0 & 0 & 0 & 0 & 0 & 0 & 2 & 0 & 2 & 0 & 2 & 0 & 2 & 0 & 2 & 0 & 2 & 0 & 2 & 0 & 2 & 0 \\
 -3 & 1 & 1 & 1 & 1 & 1 & 1 & 1 & 1 & 1 & 1 & 1 & 1 & 1 & 1 & 1 & 1 & 1 & 1 & 1 & 1 & 1 & 1 & 1 \\
\end{array}
\right).}}\label{Leechbasis}
\ee

\subsection{Explicit results for the $\mathbb{Z}_2$ duality defect}
Let $g_2$ be an element of $Co_0$ with Frame shape $1^82^8$. In the Leech lattice basis  \eqref{Leechbasis}, $g_2$  acts by $+1$ on the first $16$ rows and by $-1$ on the last $8$. Thus, the $g_2$-fixed sublattice $\Lambda_{16} \subset \Lambda$ is generated by the first $16$ rows of \eqref{Leechbasis}; this is the Barnes-Wall lattice considered in section \ref{s:Z2duality}.\\
In the same basis, the orthogonal matrix $\mathcal{O}_{\mathcal{N}}^{(16)}$ implementing the isomorphism \eqref{NgLambda16} in section \ref{s:Z2duality} is represented by the following block-diagonal matrix: 
\be
\arraycolsep=3pt \def\arraystretch{0.8}
 O^{(16)}_\CN = \frac{1}{\sqrt{2}} {\tiny{\left(
\begin{array}{cccccccccccccccc}
 1 & 1 & 0 & 0 & 0 & 0 & 0 & 0 & 0 & 0 & 0 & 0 & 0 & 0 & 0 & 0  \\
 1 & -1 & 0 & 0 & 0 & 0 & 0 & 0 & 0 & 0 & 0 & 0 & 0 & 0 & 0 & 0 \\
 0 & 0 & 1 & 1 & 0 & 0 & 0 & 0 & 0 & 0 & 0 & 0 & 0 & 0 & 0 & 0  \\
 0 & 0 & 1 & -1 & 0 & 0 & 0 & 0 & 0 & 0 & 0 & 0 & 0 & 0 & 0 & 0 \\
 0 & 0 & 0 & 0 & 1 & 1 & 0 & 0 & 0 & 0 & 0 & 0 & 0 & 0 & 0 & 0  \\
 0 & 0 & 0 & 0 & 1 & -1 & 0 & 0 & 0 & 0 & 0 & 0 & 0 & 0 & 0 & 0  \\
 0 & 0 & 0 & 0 & 0 & 0 & 1 & 1 & 0 & 0 & 0 & 0 & 0 & 0 & 0 & 0  \\
 0 & 0 & 0 & 0 & 0 & 0 & 1 & -1 & 0 & 0 & 0 & 0 & 0 & 0 & 0 & 0  \\
 0 & 0 & 0 & 0 & 0 & 0 & 0 & 0 & 1 & 1 & 0 & 0 & 0 & 0 & 0 & 0  \\
 0 & 0 & 0 & 0 & 0 & 0 & 0 & 0 & 1 & -1 & 0 & 0 & 0 & 0 & 0 & 0 \\
 0 & 0 & 0 & 0 & 0 & 0 & 0 & 0 & 0 & 0 & 1 & 1 & 0 & 0 & 0 & 0  \\
 0 & 0 & 0 & 0 & 0 & 0 & 0 & 0 & 0 & 0 & 1 & -1 & 0 & 0 & 0 & 0\\
 0 & 0 & 0 & 0 & 0 & 0 & 0 & 0 & 0 & 0 & 0 & 0 & 1 & 1 & 0 & 0  \\
 0 & 0 & 0 & 0 & 0 & 0 & 0 & 0 & 0 & 0 & 0 & 0 & 1 & -1 & 0 & 0 \\
 0 & 0 & 0 & 0 & 0 & 0 & 0 & 0 & 0 & 0 & 0 & 0 & 0 & 0 & 1 & 1  \\
 0 & 0 & 0 & 0 & 0 & 0 & 0 & 0 & 0 & 0 & 0 & 0 & 0 & 0 & 1 & -1
\end{array}
\right).}}\label{O16Lattice} 
\ee

\subsection{Explicit results for the $\mathbb{Z}_3$ duality defect}

An element $g_3\in Co_0$ with Frame shape $1^63^6$, written as a matrix in the basis in eq.\eqref{Leechbasis}, is given by (here and in the following matrices, we use the notation $\bar n$ for $-n\in \ZZ$):
\begin{equation*}\arraycolsep=3pt \def\arraystretch{0.8}
g_3={\tiny{\begin{pmatrix}
\bar 1 & 1 & 0 & 0 & 1 & 0 & \bar 1 & 1 & 1 & 0 & 0 & 0 & 1 & \bar 1 & 1 & \bar 1 & 0 & 0 & \bar 1 & 0 & \bar 1 & 1 & 0 & 0\\
\bar 1 & 1 & 0 & 0 & 1 & 0 & \bar 1 & 1 & 1 & 0 & 0 & 0 & 1 & \bar 1 & 0 & \bar 1 & 0 & 0 & 0 & 0 & \bar 1 & 1 & 0 & 0\\
5 & \bar 1 & \bar 2 & \bar 2 & \bar 1 & \bar 1 & \bar 1 & 1 & 0 & \bar 1 & \bar 1 & 1 & 0 & 0 & 1 & \bar 1 & 0 & 0 & 0 & \bar 1 & 1 & 0 & \bar 1 & 1\\
0 & 0 & 0 & 0 & 1 & 0 & \bar 1 & 1 & 1 & 0 & 0 & 0 & 1 & \bar 1 & 0 & \bar 1 & 0 & 0 & \bar 1 & 0 & 0 & 0 & 0 & 1\\
0 & 0 & \bar 1 & 0 & 1 & 0 & 0 & 0 & 0 & \bar 1 & 0 & 1 & 1 & 0 & 0 & \bar 1 & 0 & 1 & \bar 1 & 0 & 0 & 0 & 0 & 0\\
\bar 1 & 0 & 0 & 0 & 2 & 0 & 0 & 0 & 1 & \bar 1 & 0 & 1 & 1 & 0 & 0 & \bar 1 & 1 & 0 & \bar 1 & \bar 1 & \bar 1 & 0 & 0 & 1\\
1 & 0 & \bar 1 & \bar 1 & 0 & 0 & \bar 1 & 1 & 1 & 0 & 0 & 0 & 1 & \bar 1 & 1 & \bar 1 & 0 & 0 & 0 & 0 & 0 & 0 & \bar 1 & 1\\
6 & \bar 2 & \bar 3 & \bar 2 & 0 & \bar 1 & \bar 1 & 1 & 0 & \bar 2 & \bar 1 & 2 & 1 & 0 & 0 & \bar 2 & 0 & 1 & 0 & \bar 1 & 1 & \bar 1 & \bar 1 & 2\\
\bar 2 & 1 & 1 & 0 & 1 & 0 & 0 & 0 & 1 & 0 & 0 & 0 & 1 & \bar 1 & 0 & 0 & 0 & 0 & 0 & 0 & \bar 1 & 1 & 0 & 0\\
0 & 1 & 1 & 0 & \bar 1 & 0 & \bar 1 & 0 & \bar 1 & 1 & 1 & \bar 2 & \bar 1 & 0 & 1 & 1 & \bar 1 & 0 & 0 & 1 & 0 & 1 & 0 & \bar 1\\
0 & 0 & \bar 1 & \bar 1 & 1 & 0 & 0 & 1 & 1 & 0 & 0 & 1 & 1 & \bar 1 & 0 & \bar 1 & 1 & 0 & \bar 1 & \bar 1 & 0 & 0 & 0 & 1\\
1 & 0 & 0 & \bar 1 & 0 & 0 & \bar 1 & 1 & 0 & 0 & 0 & 0 & 0 & \bar 1 & 0 & 0 & 0 & 0 & 0 & 0 & 0 & 1 & 0 & 0\\
\bar 1 & 1 & 0 & 1 & 0 & 1 & 0 & \bar 1 & 0 & 0 & 0 & \bar 1 & 0 & 0 & 1 & 0 & 0 & 1 & \bar 1 & 0 & 0 & 0 & 0 & 0\\
1 & 0 & 0 & 0 & 0 & 0 & 0 & \bar 1 & \bar 1 & \bar 1 & 0 & 0 & 0 & 1 & 0 & 0 & 0 & 1 & 0 & 0 & 0 & 0 & 0 & 0\\
2 & \bar 1 & \bar 2 & \bar 1 & 0 & 0 & 0 & 0 & 1 & \bar 1 & 0 & 1 & 1 & 0 & 0 & \bar 1 & 1 & 1 & \bar 1 & \bar 1 & 1 & \bar 1 & \bar 1 & 2\\
0 & 0 & 0 & 1 & 0 & 1 & 0 & \bar 1 & \bar 1 & \bar 1 & 0 & 0 & 0 & 0 & 0 & 0 & \bar 1 & 2 & \bar 1 & 1 & 0 & 0 & 1 & \bar 1\\
0 & 0 & 0 & 0 & 1 & 0 & 0 & 0 & 1 & 0 & 0 & 0 & 0 & 0 & 0 & \bar 1 & 0 & 0 & \bar 1 & 0 & 0 & 0 & 0 & 1\\
4 & \bar 1 & \bar 1 & \bar 1 & \bar 1 & \bar 1 & 0 & 0 & \bar 1 & \bar 1 & 0 & 0 & \bar 1 & 1 & 0 & 0 & \bar 1 & 1 & 0 & 0 & 1 & 0 & 0 & 0\\
\bar 1 & 0 & 0 & 0 & 2 & 0 & 1 & \bar 1 & 1 & \bar 1 & 0 & 1 & 1 & 0 & \bar 1 & \bar 1 & 1 & 1 & \bar 1 & \bar 1 & 0 & \bar 1 & 0 & 2\\
\bar 1 & 0 & 0 & 0 & 1 & 0 & \bar 1 & 1 & 1 & 0 & 0 & 0 & 1 & \bar 1 & 0 & \bar 1 & 0 & 0 & 0 & 1 & \bar 1 & 1 & 0 & 0\\
4 & \bar 1 & \bar 2 & \bar 1 & 0 & 0 & 0 & \bar 1 & 0 & \bar 1 & \bar 1 & 1 & 0 & 0 & 0 & \bar 1 & 0 & 1 & 0 & \bar 1 & 2 & \bar 1 & \bar 1 & 2\\
2 & 0 & 0 & 0 & \bar 1 & 0 & 0 & \bar 1 & \bar 1 & 0 & 0 & \bar 1 & \bar 2 & 1 & 1 & 1 & 0 & 0 & 0 & 0 & 1 & 0 & 0 & 0\\
\bar 2 & 1 & 0 & 1 & 1 & 1 & 1 & \bar 2 & 1 & 0 & 0 & 0 & 0 & 0 & 0 & 0 & 1 & 1 & \bar 1 & \bar 1 & 0 & \bar 1 & 0 & 2\\
3 & \bar 1 & \bar 1 & 0 & \bar 1 & 0 & 0 & \bar 1 & \bar 1 & \bar 1 & 0 & 0 & \bar 1 & 1 & 0 & 0 & 0 & 1 & 0 & 0 & 1 & \bar 1 & 0 & 1
\end{pmatrix}}}\end{equation*}
Its `mate' $g_3'\in Co_0$ (see section \ref{sec:Z3duality}) is:
\begin{equation*}\arraycolsep=3pt \def\arraystretch{0.8}
g'_3={\tiny{\begin{pmatrix}
1 & \bar 1 & 0 & 1 & 0 & 0 & 0 & 0 & \bar 1 & 0 & 0 & 0 & 0 & 0 & \bar 1 & 0 & \bar 1 & 1 & 0 & 1 & 0 & \bar 1 & 1 & 0\\
2 & \bar 1 & \bar 1 & 0 & \bar 1 & 0 & 0 & 0 & \bar 1 & 0 & 0 & 0 & 0 & 0 & 0 & 0 & \bar 1 & 1 & 0 & 1 & 1 & \bar 1 & 0 & 0\\
\bar 5 & 1 & 2 & 2 & 2 & 1 & 0 & 0 & 1 & 1 & 1 & \bar 1 & 1 & \bar 1 & \bar 1 & 0 & 0 & 0 & \bar 1 & 1 & \bar 1 & 0 & 1 & 0\\
\bar 5 & 1 & 2 & 2 & 1 & 1 & 1 & \bar 1 & 0 & 1 & 1 & \bar 1 & 0 & 0 & \bar 1 & 1 & 0 & 0 & 0 & 1 & \bar 1 & 0 & 1 & \bar 1\\
1 & \bar 1 & 0 & 0 & 0 & 0 & 0 & 0 & 0 & 0 & 0 & 0 & 0 & 0 & 0 & 0 & 0 & 0 & 0 & 0 & 0 & \bar 1 & 0 & 1\\
1 & 0 & 0 & 0 & \bar 1 & 0 & \bar 1 & 1 & 0 & 1 & 0 & \bar 1 & 0 & \bar 1 & 0 & 0 & \bar 1 & 0 & 1 & 1 & 0 & 0 & 0 & 0\\
0 & 0 & 1 & 1 & \bar 1 & 0 & 0 & 0 & \bar 1 & 1 & 0 & \bar 1 & \bar 1 & 0 & 0 & 1 & \bar 1 & 0 & 0 & 1 & 0 & 0 & 1 & \bar 1\\
\bar 4 & 1 & 2 & 1 & 0 & 1 & 0 & 0 & 1 & 2 & 1 & \bar 2 & 0 & \bar 1 & 0 & 1 & 0 & \bar 1 & 0 & 1 & 0 & 0 & 0 & 0\\
3 & \bar 2 & \bar 2 & 0 & 0 & 0 & \bar 1 & 1 & 0 & 0 & 0 & 1 & 0 & 0 & 0 & \bar 1 & 0 & 0 & \bar 1 & 0 & 1 & \bar 1 & 0 & 1\\
1 & \bar 1 & \bar 1 & 0 & 1 & 0 & 1 & 0 & 0 & \bar 1 & \bar 1 & 2 & 1 & 0 & \bar 1 & \bar 1 & 0 & 1 & 0 & \bar 1 & 0 & \bar 1 & 0 & 1\\
2 & \bar 1 & 0 & 0 & \bar 1 & \bar 1 & \bar 1 & 1 & \bar 1 & 0 & 0 & 0 & 0 & 0 & 0 & 0 & \bar 1 & 0 & 1 & 1 & 0 & 0 & 0 & \bar 1\\
1 & \bar 1 & \bar 1 & 0 & 0 & 0 & 0 & 0 & 0 & 0 & 0 & 1 & 1 & 0 & 0 & \bar 1 & 0 & 0 & 0 & 0 & 1 & \bar 1 & \bar 1 & 1\\
\bar 1 & 0 & 0 & 0 & 1 & 0 & 0 & 1 & 0 & 0 & 0 & 1 & 1 & \bar 1 & \bar 1 & 0 & 0 & 0 & 0 & 0 & \bar 1 & 0 & 1 & 0\\
2 & \bar 1 & \bar 2 & \bar 1 & 0 & 0 & 0 & 1 & 1 & 0 & 0 & 1 & 1 & \bar 1 & 0 & \bar 1 & 0 & 0 & 0 & \bar 1 & 1 & \bar 1 & \bar 1 & 2\\
0 & 0 & 0 & 0 & 0 & 0 & \bar 1 & 1 & 0 & 1 & 0 & 0 & 0 & \bar 1 & 1 & 0 & 0 & \bar 1 & 0 & 0 & 0 & 0 & 0 & 0\\
2 & \bar 1 & \bar 1 & \bar 1 & 0 & 0 & 0 & 0 & 0 & 0 & 0 & 1 & 0 & 0 & 0 & 0 & 1 & \bar 1 & 0 & \bar 1 & 1 & \bar 1 & \bar 1 & 2\\
2 & \bar 1 & \bar 1 & 0 & \bar 1 & 0 & 0 & 0 & \bar 1 & 0 & 0 & 0 & 0 & 0 & 0 & 0 & 0 & 1 & 0 & 0 & 1 & \bar 1 & 0 & 0\\
0 & \bar 1 & \bar 1 & 0 & 1 & 1 & 0 & 0 & 1 & 0 & 0 & 1 & 1 & \bar 1 & 0 & \bar 1 & 1 & 0 & \bar 1 & \bar 1 & 1 & \bar 1 & \bar 1 & 2\\
2 & \bar 1 & \bar 1 & \bar 1 & \bar 1 & 0 & \bar 1 & 1 & 0 & 1 & 0 & 0 & 0 & \bar 1 & 1 & 0 & 0 & \bar 1 & 1 & 0 & 1 & 0 & \bar 1 & 0\\
6 & \bar 2 & \bar 2 & \bar 1 & \bar 3 & 0 & 0 & \bar 1 & \bar 2 & 0 & 0 & 0 & \bar 1 & 1 & 1 & 0 & \bar 1 & 1 & 0 & 0 & 3 & \bar 2 & \bar 1 & 1\\
\bar 3 & 0 & 0 & 0 & 1 & 1 & 0 & 1 & 1 & 1 & 1 & 0 & 1 & \bar 1 & 0 & 0 & 1 & \bar 1 & \bar 1 & 0 & 0 & 0 & 0 & 0\\
2 & \bar 1 & \bar 2 & \bar 1 & 0 & 0 & 0 & 1 & 0 & 0 & 0 & 1 & 1 & \bar 1 & 0 & \bar 1 & 0 & 0 & 0 & \bar 1 & 1 & 0 & \bar 1 & 1\\
6 & \bar 2 & \bar 3 & \bar 3 & \bar 1 & \bar 1 & \bar 1 & 2 & \bar 1 & 0 & \bar 1 & 2 & 0 & \bar 1 & 1 & \bar 1 & 0 & \bar 1 & 1 & \bar 1 & 2 & 0 & \bar 1 & 0\\
2 & 0 & \bar 1 & \bar 2 & \bar 1 & 0 & 0 & 0 & 0 & 1 & 0 & 0 & 0 & \bar 1 & 1 & 0 & 0 & \bar 1 & 1 & \bar 1 & 2 & 0 & \bar 2 & 1
\end{pmatrix}}}\end{equation*}
The generators of the normalizer $N_{Co_0}(\langle g_3\rangle)$ of the element $g_3$ in $Co_0$ are:
{\tiny{\begin{equation*}\arraycolsep=3pt \def\arraystretch{0.8}
\begin{pmatrix}
 \bar 5 & 2 & 3 & 1 & 1 & 1 & 0 & \bar 1 & 1 & 1 & 1 & \bar 2 & 0 & \bar 1 & \bar 1 & 1 & \bar 2 & 0 & 1 & 2 & \bar 1 & 1 &
1 & \bar 2 \\
 \bar 2 & 1 & 2 & 0 & 0 & 0 & 0 & 0 & 0 & 0 & 0 & \bar 1 & 0 & 0 & \bar 1 & 1 & \bar 1 & 0 & 1 & 1 & \bar 1 & 1 & 1 &
\bar 2 \\
 \bar 5 & 1 & 2 & 1 & 2 & 0 & \bar 1 & 1 & 2 & 1 & 1 & \bar 1 & 1 & \bar 1 & \bar 1 & 0 & 0 & \bar 1 & 0 & 1 & \bar 2 & 1 &
1 & \bar 1 \\
 \bar 3 & 1 & 2 & 1 & 0 & 1 & \bar 1 & 0 & 0 & 1 & 1 & \bar 2 & \bar 1 & 0 & 0 & 1 & \bar 1 & 0 & 0 & 2 & \bar 1 & 1 & 1 &
\bar 2 \\
 \bar 3 & 1 & 2 & 1 & 1 & 0 & 0 & 0 & 1 & 0 & 1 & \bar 1 & 0 & 0 & \bar 1 & 0 & \bar 1 & 0 & 0 & 1 & \bar 1 & 1 & 1 &
\bar 1 \\
 2 & 0 & 0 & 0 & \bar 1 & 0 & 0 & \bar 1 & \bar 1 & 0 & 0 & \bar 1 & \bar 1 & 0 & 0 & 0 & \bar 2 & 1 & 1 & 1 & 1 & 0 & 0 &
\bar 1 \\
 \bar 3 & 1 & 1 & 1 & 1 & 1 & 0 & \bar 1 & 1 & 0 & 1 & \bar 1 & 1 & \bar 1 & \bar 1 & 0 & \bar 1 & 1 & 0 & 1 & 0 & 0 & 0 &
0 \\
 \bar 2 & 0 & 1 & 1 & 1 & 0 & \bar 1 & 1 & 1 & 0 & 1 & \bar 1 & 0 & 0 & \bar 1 & 0 & 0 & 0 & \bar 1 & 1 & \bar 1 & 1 & 1 &
\bar 1 \\
 \bar 6 & 2 & 3 & 2 & 1 & 1 & 0 & 0 & 1 & 1 & 1 & \bar 2 & 0 & \bar 1 & \bar 1 & 1 & \bar 1 & 0 & 0 & 2 & \bar 2 & 1 & 2 &
\bar 2 \\
 \bar 2 & 1 & 1 & 0 & 1 & 1 & 0 & \bar 1 & 1 & 0 & 0 & 0 & 0 & \bar 1 & 0 & 0 & 0 & 0 & 0 & 0 & 0 & 0 & 0 & 0
\\
 \bar 1 & 0 & 1 & 0 & 0 & 0 & \bar 1 & 1 & 1 & 1 & 0 & \bar 1 & 0 & \bar 1 & 0 & 0 & \bar 1 & \bar 1 & 1 & 1 & 0 & 1 & 0 &
\bar 1 \\
 \bar 1 & 0 & 1 & 0 & 0 & 0 & \bar 2 & 2 & 1 & 1 & 0 & \bar 1 & 0 & \bar 1 & 0 & 0 & 0 & \bar 1 & 0 & 1 & \bar 1 & 1 & 1 &
\bar 1 \\
 \bar 3 & 1 & 1 & \bar 1 & 2 & 0 & 0 & 1 & 2 & 1 & 0 & 0 & 1 & \bar 2 & \bar 1 & 0 & 0 & \bar 1 & 1 & 0 & \bar 1 & 1 & 0 &
0 \\
 \bar 2 & 1 & 1 & 0 & 1 & 0 & 0 & 1 & 1 & 0 & 0 & 0 & 0 & \bar 1 & \bar 1 & 0 & 0 & 0 & 0 & 0 & \bar 1 & 1 & 1 &
\bar 1 \\
 \bar 5 & 1 & 2 & 1 & 2 & 0 & \bar 1 & 2 & 2 & 1 & 1 & \bar 1 & 1 & \bar 2 & \bar 1 & 0 & 0 & \bar 1 & 0 & 1 & \bar 2 & 2 &
1 & \bar 1 \\
 \bar 2 & 0 & 1 & 0 & 1 & 0 & \bar 1 & 2 & 2 & 1 & 1 & \bar 1 & 0 & \bar 1 & \bar 1 & 0 & 0 & \bar 1 & \bar 1 & 1 & \bar 1 & 1 &
1 & 0 \\
 0 & 1 & 1 & 0 & 0 & 0 & 0 & \bar 1 & 0 & 0 & 0 & \bar 1 & \bar 1 & 0 & 0 & 1 & \bar 1 & 0 & 1 & 0 & 0 & 0 & 0 & 0
\\
 \bar 3 & 1 & 1 & 1 & 2 & 0 & \bar 1 & 1 & 2 & 0 & 0 & 0 & 0 & \bar 1 & 0 & 0 & 1 & \bar 1 & \bar 1 & 0 & \bar 2 & 1 & 1 &
0 \\
 0 & 0 & 1 & 1 & 0 & 0 & \bar 1 & 1 & 0 & 0 & 0 & \bar 1 & \bar 1 & 0 & 0 & 0 & \bar 1 & 0 & 0 & 1 & \bar 1 & 1 & 1 &
\bar 1 \\
 1 & 0 & 0 & 0 & 0 & 0 & 0 & 0 & 0 & \bar 1 & 0 & 0 & 0 & 0 & \bar 1 & 0 & \bar 1 & 1 & 0 & 0 & 0 & 0 & 1 & 0
\\
 \bar 1 & 0 & 1 & 0 & 1 & \bar 1 & \bar 1 & 2 & 1 & 0 & 0 & 0 & 0 & 0 & \bar 1 & 0 & 1 & \bar 1 & 0 & 0 & \bar 2 & 1 & 1 &
0 \\
 \bar 1 & 1 & 0 & 0 & 1 & 0 & 0 & 0 & 1 & 0 & \bar 1 & 1 & 0 & \bar 1 & 0 & 0 & 1 & 0 & 0 & \bar 1 & \bar 1 & 0 & 0 &
1 \\
 2 & \bar 1 & 0 & \bar 1 & 0 & \bar 1 & \bar 1 & 2 & 0 & 0 & \bar 1 & 1 & 0 & \bar 1 & 0 & 0 & 0 & \bar 1 & 1 & 0 & \bar 1 & 1 &
0 & 0 \\
 4 & \bar 2 & \bar 2 & \bar 1 & 0 & \bar 1 & \bar 1 & 2 & 0 & \bar 1 & \bar 1 & 2 & 0 & 0 & 0 & \bar 1 & 1 & 0 & \bar 1 & \bar 1 & 0 &
0 & 0 & 1  
\end{pmatrix}
 \begin{pmatrix}
	  \bar 2 & 0 & \bar 1 & 1 & 2 & 2 & 2 & \bar 3 & 1 & \bar 1 & 0 & 1 & 1 & 0 & \bar 1 & \bar 1 & 1 & 2 & \bar 1 & \bar 2 & 1 & \bar 2 &
	 \bar 1 & 3 \\
	  \bar 2 & 0 & 0 & 1 & 1 & 2 & 1 & \bar 2 & 0 & 0 & 1 & 0 & 0 & 0 & \bar 1 & 0 & 0 & 1 & \bar 1 & 0 & 1 & \bar 1 & 0 &
	 1 \\
	  \bar 3 & 0 & 1 & 2 & 1 & 1 & 1 & \bar 2 & 1 & 0 & 1 & \bar 1 & 0 & 1 & \bar 1 & 0 & 0 & 1 & \bar 1 & 0 & 0 & \bar 1 & 0 &
	 1 \\
	  \bar 3 & 1 & 0 & 1 & 1 & 2 & 1 & \bar 2 & 1 & 0 & 0 & 0 & 0 & 0 & 0 & 0 & 1 & 1 & \bar 1 & \bar 1 & 0 & \bar 1 & 0 &
	 1 \\
	  3 & \bar 1 & \bar 2 & \bar 1 & 0 & 0 & 1 & \bar 2 & 0 & \bar 1 & 0 & 1 & 0 & 1 & 0 & \bar 1 & 1 & 1 & 0 & \bar 2 & 2 & \bar 2 &
	 \bar 2 & 3 \\
	  0 & 0 & \bar 1 & 0 & 1 & 1 & 0 & \bar 1 & 1 & 0 & 0 & 0 & 1 & \bar 1 & 0 & \bar 1 & 0 & 1 & 0 & \bar 1 & 1 & \bar 1 &
	 \bar 1 & 2 \\
	  0 & 0 & \bar 1 & 0 & 0 & 1 & 1 & \bar 1 & 0 & 0 & 0 & 0 & 0 & 0 & 0 & 0 & 0 & 1 & 0 & \bar 1 & 1 & \bar 1 & \bar 1 &
	 1 \\
	  \bar 1 & 0 & 0 & 0 & 0 & 1 & 0 & \bar 1 & 1 & 1 & 1 & \bar 1 & 0 & 0 & 0 & 0 & 0 & 0 & 0 & 0 & 1 & \bar 1 & \bar 1 &
	 1 \\
	  5 & \bar 2 & \bar 2 & \bar 1 & \bar 1 & 0 & 0 & \bar 1 & \bar 1 & \bar 1 & \bar 1 & 1 & 0 & 1 & 0 & \bar 1 & 0 & 1 & 0 & \bar 1 & 2 &
	 \bar 2 & \bar 1 & 2 \\
	  1 & 0 & \bar 1 & 0 & 1 & 1 & 1 & \bar 2 & 0 & \bar 1 & \bar 1 & 1 & 0 & 0 & 0 & \bar 1 & 0 & 1 & 0 & \bar 1 & 1 & \bar 1 &
	 \bar 1 & 2 \\
	  \bar 2 & 0 & \bar 1 & 1 & 2 & 1 & 1 & \bar 1 & 1 & \bar 1 & 0 & 1 & 1 & 0 & \bar 1 & \bar 1 & 1 & 1 & \bar 1 & \bar 1 & 0 & \bar 1 &
	 0 & 1 \\
	  1 & 0 & 0 & 1 & \bar 1 & 1 & 0 & \bar 2 & \bar 1 & 0 & 0 & \bar 1 & \bar 1 & 1 & 0 & 0 & \bar 1 & 1 & 0 & 1 & 1 & \bar 1 &
	 0 & 0 \\
	  3 & \bar 1 & \bar 2 & 0 & 0 & 1 & 2 & \bar 3 & \bar 1 & \bar 2 & \bar 1 & 2 & 0 & 1 & 0 & \bar 1 & 0 & 2 & 0 & \bar 2 & 2 & \bar 2 &
	 \bar 1 & 2 \\
	  8 & \bar 2 & \bar 3 & \bar 2 & \bar 2 & 0 & 0 & \bar 2 & \bar 2 & \bar 1 & \bar 1 & 1 & \bar 1 & 1 & 1 & \bar 1 & \bar 1 & 1 & 1 & \bar 1 & 4 &
	 \bar 2 & \bar 2 & 2 \\
	  3 & \bar 1 & \bar 1 & 0 & \bar 1 & 0 & 1 & \bar 2 & \bar 1 & \bar 1 & 0 & 0 & \bar 1 & 2 & 0 & 0 & 0 & 1 & 0 & \bar 1 & 2 & \bar 2 &
	 \bar 1 & 1 \\
	  5 & \bar 1 & \bar 2 & \bar 1 & \bar 2 & 0 & 0 & \bar 2 & \bar 1 & \bar 1 & \bar 1 & 1 & 0 & 1 & 1 & \bar 1 & 0 & 1 & 1 & \bar 1 & 2 &
	 \bar 2 & \bar 2 & 2 \\
	  \bar 3 & 0 & 0 & 1 & 2 & 1 & 1 & \bar 1 & 1 & 0 & 0 & 1 & 1 & 0 & \bar 1 & \bar 1 & 1 & 1 & \bar 1 & \bar 1 & 0 & \bar 1 &
	 0 & 1 \\
	  1 & \bar 1 & 0 & 0 & 0 & 0 & 0 & \bar 1 & 0 & 0 & 0 & 0 & 0 & 1 & 0 & \bar 1 & 0 & 0 & 0 & 0 & 1 & \bar 1 & \bar 1 &
	 1 \\
	  3 & \bar 1 & \bar 2 & \bar 1 & 0 & 0 & \bar 1 & 0 & 0 & 0 & \bar 1 & 1 & 0 & 0 & 1 & \bar 1 & 1 & 0 & 0 & \bar 1 & 1 & \bar 1 &
	 \bar 1 & 1 \\
	  4 & \bar 1 & \bar 1 & \bar 1 & \bar 2 & 0 & 0 & \bar 1 & \bar 2 & 0 & 0 & 0 & \bar 1 & 1 & 0 & 0 & \bar 1 & 1 & 1 & 0 & 2 & \bar 1 &
	 \bar 1 & 0 \\
	  2 & \bar 1 & 0 & 0 & \bar 1 & 0 & 0 & \bar 1 & \bar 1 & 0 & 0 & 0 & \bar 1 & 2 & 0 & 0 & 0 & 0 & 0 & 0 & 1 & \bar 1 & 0 &
	 0 \\
	  2 & 0 & 0 & 0 & \bar 1 & 0 & 0 & \bar 1 & \bar 2 & 0 & \bar 1 & 0 & \bar 1 & 1 & 1 & 0 & \bar 1 & 0 & 1 & 0 & 1 & 0 & 0 &
	 \bar 1 \\
	  0 & 0 & 0 & 0 & 1 & 0 & 0 & 0 & 0 & 0 & \bar 1 & 1 & 0 & 0 & 0 & 0 & 1 & \bar 1 & 0 & \bar 1 & 0 & 0 & 0 & 0
	 \\
	  0 & 1 & 1 & 0 & \bar 1 & 0 & \bar 1 & 0 & \bar 1 & 1 & 0 & \bar 1 & \bar 1 & 0 & 1 & 1 & \bar 1 & \bar 1 & 1 & 1 & 0 & 1 &
	 0 & \bar 2  
	 \end{pmatrix}
\end{equation*}
\begin{equation*}\arraycolsep=3pt \def\arraystretch{0.8}
\begin{pmatrix}
	2 & 0 & \bar 2 & \bar 1 & \bar 1 & \bar 1 & \bar 1 & 2 & 1 & 0 & 0 & 0 & 0 & 0 & 1 & 0 & 1 & \bar 1 & 0 & \bar 1 & 0 & 1 &
	\bar 1 & 0 \\
	\bar 1 & 1 & 0 & 0 & 1 & \bar 1 & 0 & 1 & 1 & 0 & 0 & 0 & 0 & 0 & 0 & 0 & 1 & \bar 1 & 0 & \bar 1 & \bar 1 & 1 & 0 &
	0 \\
	5 & \bar 1 & \bar 2 & \bar 1 & \bar 2 & \bar 1 & 0 & 0 & \bar 1 & \bar 1 & \bar 1 & 1 & \bar 1 & 1 & 1 & 0 & 0 & 0 & 1 & \bar 1 & 1 &
	0 & \bar 1 & 0 \\
	5 & \bar 1 & \bar 3 & \bar 1 & \bar 1 & \bar 1 & \bar 1 & 1 & 0 & \bar 1 & \bar 1 & 1 & 0 & 0 & 1 & \bar 1 & 0 & 0 & 0 & \bar 1 & 1 &
	0 & \bar 1 & 1 \\
	0 & 0 & \bar 1 & 0 & 0 & 0 & 0 & 1 & 0 & 0 & 0 & 0 & 0 & 0 & 0 & 0 & 0 & 0 & 0 & 0 & 0 & 1 & 0 & \bar 1
	\\
	0 & 0 & \bar 1 & 0 & 0 & 0 & 1 & 0 & 0 & \bar 1 & 0 & 1 & 0 & 1 & 0 & 0 & 1 & 0 & \bar 1 & \bar 1 & 0 & 0 & 0 & 0
	\\
	2 & \bar 1 & \bar 2 & \bar 1 & 0 & \bar 1 & 0 & 1 & 1 & \bar 1 & 0 & 1 & 1 & 0 & 0 & \bar 1 & 1 & 0 & 0 & \bar 1 & 0 & 0 &
	\bar 1 & 1 \\
	3 & \bar 1 & \bar 2 & 0 & 0 & \bar 1 & 1 & 0 & \bar 1 & \bar 2 & \bar 1 & 2 & 0 & 1 & 0 & \bar 1 & 0 & 1 & 0 & \bar 1 & 0 & 0 &
	0 & 0 \\
	0 & 1 & 0 & 0 & \bar 1 & 0 & 0 & 0 & 0 & 0 & 0 & \bar 1 & \bar 1 & 0 & 1 & 1 & 0 & 0 & 0 & 0 & 0 & 1 & 0 & \bar 1
	\\
	4 & \bar 1 & \bar 2 & \bar 1 & \bar 2 & \bar 1 & \bar 1 & 1 & \bar 1 & 0 & 0 & 0 & \bar 1 & 1 & 1 & 0 & 0 & 0 & 0 & 0 & 1 & 0 &
	0 & \bar 1 \\
	3 & 0 & \bar 2 & \bar 1 & \bar 1 & 0 & 0 & 0 & 0 & 0 & 0 & 0 & \bar 1 & 0 & 1 & 0 & 0 & 0 & 0 & \bar 1 & 1 & 0 & \bar 1 &
	1 \\
	5 & \bar 1 & \bar 2 & \bar 1 & \bar 1 & \bar 1 & 0 & 0 & \bar 1 & \bar 1 & \bar 1 & 1 & \bar 2 & 1 & 1 & 0 & 0 & 0 & 0 & \bar 1 & 1 &
	0 & 0 & 0 \\
	\bar 1 & 0 & 0 & 1 & 0 & 0 & 0 & 0 & 1 & 0 & 1 & \bar 1 & 0 & 1 & 0 & 0 & 0 & 0 & \bar 1 & 0 & 0 & 0 & 0 & 0
	\\
	0 & 0 & 0 & 1 & 0 & 0 & 1 & \bar 1 & \bar 1 & \bar 1 & 0 & 0 & \bar 1 & 2 & 0 & 0 & 0 & 1 & \bar 1 & 0 & 0 & 0 & 1 &
	\bar 1 \\
	3 & \bar 1 & \bar 1 & 0 & \bar 1 & 0 & 1 & \bar 1 & \bar 1 & \bar 1 & 0 & 0 & \bar 1 & 1 & 0 & 0 & \bar 1 & 1 & 0 & 0 & 1 & 0 &
	0 & 0 \\
	\bar 2 & 0 & 0 & 1 & 1 & 0 & 0 & 1 & 1 & 0 & 0 & 0 & 0 & 0 & 0 & 0 & 0 & 0 & \bar 1 & 0 & \bar 1 & 1 & 1 & \bar 1
	\\
	2 & 0 & \bar 1 & \bar 1 & \bar 1 & \bar 1 & \bar 1 & 1 & 0 & 0 & 0 & 0 & 0 & 0 & 1 & 0 & 0 & \bar 1 & 1 & 0 & 0 & 1 &
	\bar 1 & 0 \\
	5 & \bar 1 & \bar 2 & \bar 1 & \bar 2 & \bar 1 & 0 & 0 & \bar 2 & \bar 1 & \bar 1 & 1 & \bar 1 & 1 & 1 & 0 & \bar 1 & 1 & 1 & 0 & 1 &
	0 & 0 & \bar 1 \\
	2 & 0 & \bar 1 & 0 & \bar 1 & 0 & 1 & \bar 1 & \bar 1 & \bar 1 & 0 & 0 & \bar 1 & 1 & 0 & 0 & \bar 1 & 1 & 0 & 0 & 1 & 0 &
	0 & 0 \\
	\bar 1 & 0 & 0 & 0 & 1 & \bar 1 & 0 & 1 & 1 & 0 & 0 & 0 & 0 & 0 & 0 & 0 & 1 & \bar 1 & 0 & 0 & \bar 1 & 1 & 0 & 0
	\\
	1 & 0 & 0 & 1 & 0 & 0 & 1 & \bar 2 & \bar 1 & \bar 1 & 0 & 0 & \bar 1 & 1 & 0 & 0 & \bar 1 & 1 & 0 & 0 & 1 & 0 & 0 &
	0 \\
	2 & \bar 1 & \bar 1 & 0 & 0 & 0 & 0 & \bar 1 & 0 & \bar 1 & 0 & 1 & 0 & 1 & 0 & \bar 1 & 0 & 1 & \bar 1 & 0 & 1 & \bar 1 &
	0 & 1 \\
	\bar 3 & 1 & 2 & 2 & 0 & 1 & 1 & \bar 2 & 0 & 0 & 2 & \bar 2 & \bar 1 & 1 & \bar 1 & 1 & \bar 1 & 1 & \bar 1 & 1 & 0 & 0 &
	1 & 0 \\
	0 & \bar 1 & 0 & 1 & 1 & 0 & 1 & \bar 1 & 0 & \bar 1 & 0 & 1 & 0 & 1 & \bar 1 & \bar 1 & 0 & 1 & \bar 1 & 0 & 0 & \bar 1 &
	1 & 1  
\end{pmatrix}
\begin{pmatrix}
	\bar 1 & 1 & 0 & 1 & \bar 1 & 1 & 0 & \bar 1 & 0 & 1 & 1 & \bar 1 & 0 & \bar 1 & 1 & 0 & \bar 1 & 0 & 0 & 1 & 1 & 1 &
	\bar 1 & 0 \\
	0 & 0 & \bar 1 & 0 & 0 & 0 & 0 & 0 & 1 & 0 & 1 & 0 & 1 & \bar 1 & 0 & \bar 1 & 0 & 0 & 0 & 0 & 1 & 0 & \bar 1 & 1
	\\
	3 & \bar 1 & \bar 1 & 0 & \bar 1 & 0 & 1 & \bar 2 & \bar 1 & \bar 1 & 0 & 1 & 0 & 1 & 0 & \bar 1 & \bar 1 & 1 & 0 & 0 & 2 & \bar 1 &
	\bar 1 & 1 \\
	\bar 3 & 1 & 1 & 1 & 0 & 1 & 0 & 0 & 0 & 1 & 1 & \bar 1 & 1 & \bar 1 & 0 & 0 & \bar 1 & 0 & 0 & 1 & 0 & 1 & 0 &
	\bar 1 \\
	\bar 3 & 1 & 1 & 1 & 0 & 1 & 1 & \bar 1 & 0 & 1 & 1 & \bar 1 & 0 & \bar 1 & 0 & 1 & \bar 1 & 0 & 0 & 1 & 0 & 1 & 0 &
	\bar 1 \\
	\bar 1 & 1 & 1 & 1 & \bar 1 & 1 & 0 & \bar 1 & \bar 1 & 0 & 1 & \bar 1 & \bar 1 & 0 & 1 & 1 & \bar 1 & 0 & 0 & 1 & 0 & 1 &
	0 & \bar 1 \\
	0 & 0 & 0 & 0 & \bar 1 & 0 & \bar 1 & 1 & 0 & 1 & 0 & 0 & 0 & \bar 1 & 1 & 0 & 0 & \bar 1 & 0 & 1 & 0 & 1 & 0 &
	\bar 1 \\
	1 & \bar 1 & 0 & 0 & \bar 1 & 0 & 0 & 0 & \bar 1 & 0 & 1 & 0 & 0 & 0 & 0 & 0 & \bar 1 & 0 & 0 & 1 & 1 & 0 & 0 &
	\bar 1 \\
	1 & 0 & 0 & 0 & \bar 1 & 0 & \bar 1 & 0 & 0 & 1 & 0 & \bar 1 & 0 & \bar 1 & 1 & 0 & \bar 1 & \bar 1 & 1 & 1 & 1 & 1 &
	\bar 1 & 0 \\
	0 & 0 & 0 & 0 & \bar 1 & 0 & 0 & 0 & 0 & 1 & 1 & \bar 1 & 0 & 0 & 0 & 0 & \bar 1 & 0 & 0 & 1 & 1 & 0 & 0 & 0
	\\
	\bar 2 & 1 & 1 & 1 & 0 & 1 & 0 & \bar 1 & 0 & 1 & 1 & \bar 1 & 0 & \bar 1 & 0 & 0 & \bar 1 & 0 & 0 & 1 & 0 & 1 & 0 &
	0 \\
	0 & \bar 1 & 0 & 0 & 0 & 0 & 0 & 0 & 0 & 0 & 1 & 0 & 1 & 0 & \bar 1 & \bar 1 & \bar 1 & 0 & 0 & 1 & 1 & 0 & 0 & 0
	\\
	0 & 0 & 0 & 1 & \bar 1 & 1 & 0 & \bar 1 & 0 & 0 & 1 & \bar 1 & 0 & 0 & 0 & 0 & \bar 1 & 1 & \bar 1 & 1 & 1 & 0 & 0 &
	0 \\
	1 & \bar 1 & 0 & 0 & \bar 1 & 0 & 0 & 0 & 0 & 0 & 1 & \bar 1 & 0 & 0 & 0 & 0 & \bar 1 & 0 & 0 & 1 & 1 & 0 & 0 & 0
	\\
	2 & \bar 1 & 0 & 0 & \bar 1 & 0 & 0 & \bar 1 & \bar 1 & 0 & 0 & 0 & 0 & 0 & 0 & 0 & \bar 1 & 0 & 0 & 1 & 1 & 0 & 0 &
	0 \\
	\bar 4 & 0 & 2 & 2 & 0 & 1 & 0 & 0 & 0 & 1 & 2 & \bar 2 & 0 & 0 & \bar 1 & 1 & \bar 1 & 0 & \bar 1 & 2 & 0 & 1 & 1 &
	\bar 2 \\
	\bar 3 & 1 & 1 & 1 & 0 & 1 & 0 & \bar 1 & 1 & 1 & 1 & \bar 1 & 1 & \bar 1 & 0 & 0 & 0 & 0 & 0 & 1 & 0 & 0 & \bar 1 &
	1 \\
	1 & \bar 1 & 0 & 0 & \bar 1 & 0 & 0 & \bar 1 & 0 & 1 & 1 & \bar 1 & 0 & 0 & 0 & 0 & \bar 1 & 0 & 0 & 1 & 2 & \bar 1 &
	\bar 1 & 1 \\
	\bar 2 & 1 & 2 & 1 & \bar 1 & 1 & 0 & \bar 1 & \bar 1 & 1 & 1 & \bar 2 & 0 & \bar 1 & 0 & 1 & \bar 2 & 0 & 1 & 2 & 0 & 1 &
	0 & \bar 1 \\
	\bar 3 & 0 & 1 & 1 & 0 & 0 & \bar 1 & 1 & 1 & 1 & 1 & \bar 1 & 1 & \bar 1 & 0 & 0 & 0 & \bar 1 & 0 & 2 & \bar 1 & 1 & 0 &
	\bar 1 \\
	\bar 2 & 0 & 1 & 1 & 0 & 1 & 1 & \bar 2 & 0 & 0 & 1 & \bar 1 & 1 & 0 & \bar 1 & 0 & \bar 1 & 1 & 0 & 1 & 1 & \bar 1 & 0 &
	0 \\
	\bar 2 & 0 & 1 & 1 & 0 & 1 & 0 & \bar 1 & 1 & 1 & 1 & \bar 2 & 0 & 0 & 0 & 0 & 0 & 0 & \bar 1 & 1 & 1 & \bar 1 & 0 &
	1 \\
	1 & 0 & 0 & 1 & \bar 1 & 1 & 0 & \bar 2 & \bar 1 & 0 & 0 & \bar 1 & 0 & 0 & 0 & 0 & \bar 1 & 1 & 0 & 1 & 1 & \bar 1 & 0 &
	1 \\
	\bar 1 & \bar 1 & 1 & 1 & 0 & 0 & 0 & 0 & 0 & 0 & 1 & \bar 1 & 0 & 1 & \bar 1 & 0 & 0 & 0 & \bar 1 & 1 & 0 & \bar 1 & 1 &
	0  
\end{pmatrix}
\end{equation*}
\begin{equation*}\arraycolsep=3pt \def\arraystretch{0.8}
\begin{pmatrix}
	2 & \bar 1 & \bar 2 & \bar 2 & 1 & 0 & 0 & 1 & 2 & 0 & \bar 1 & 1 & 0 & \bar 1 & 1 & 0 & 2 & \bar 2 & \bar 1 & \bar 2 & 1 & 0 &
	\bar 1 & 2 \\
	\bar 2 & 0 & 0 & \bar 1 & 2 & 1 & 1 & 0 & 2 & 0 & 0 & 1 & 1 & \bar 1 & \bar 1 & 0 & 1 & \bar 1 & \bar 1 & \bar 1 & 0 & 0 &
	0 & 1 \\
	7 & \bar 2 & \bar 3 & \bar 2 & \bar 2 & \bar 1 & 0 & 0 & \bar 1 & \bar 1 & \bar 1 & 1 & \bar 1 & 1 & 1 & 0 & 0 & 0 & 0 & \bar 1 & 2 &
	\bar 1 & \bar 1 & 1 \\
	2 & \bar 1 & \bar 1 & \bar 2 & 0 & \bar 1 & 0 & 1 & 1 & 0 & 0 & 0 & 0 & 0 & 0 & 0 & 1 & \bar 1 & 0 & \bar 1 & 1 & 0 &
	\bar 1 & 1 \\
	\bar 1 & 0 & \bar 1 & 0 & 1 & 1 & 1 & \bar 1 & 1 & 0 & 0 & 0 & 0 & 0 & 0 & 0 & 1 & 0 & \bar 1 & \bar 1 & 1 & \bar 1 & 0 &
	1 \\
	2 & 0 & \bar 1 & \bar 1 & 0 & 0 & \bar 1 & 1 & 0 & 0 & \bar 1 & 0 & 0 & \bar 1 & 1 & 0 & 0 & \bar 1 & 0 & 0 & 0 & 1 & 0 &
	0 \\
	2 & \bar 1 & \bar 1 & \bar 1 & 0 & 0 & 0 & 0 & 1 & 0 & 0 & 0 & 0 & 0 & 0 & 0 & 1 & \bar 1 & \bar 1 & \bar 1 & 2 & \bar 1 &
	\bar 1 & 2 \\
	3 & \bar 1 & \bar 1 & \bar 1 & \bar 1 & 0 & 1 & \bar 1 & \bar 1 & \bar 1 & 0 & 0 & 0 & 1 & \bar 1 & 0 & \bar 1 & 1 & 0 & 0 & 2 &
	\bar 1 & 0 & 0 \\
	2 & \bar 1 & \bar 1 & \bar 1 & \bar 1 & 0 & \bar 1 & 1 & 1 & 1 & 0 & \bar 1 & 0 & \bar 1 & 1 & 0 & 0 & \bar 1 & 0 & 0 & 1 & 0 &
	\bar 1 & 1 \\
	1 & \bar 1 & \bar 1 & \bar 1 & 1 & 0 & 0 & 1 & 1 & \bar 1 & 0 & 1 & 0 & 0 & 0 & 0 & 1 & \bar 1 & \bar 1 & \bar 1 & 0 & 0 &
	0 & 1 \\
	1 & \bar 1 & \bar 2 & \bar 1 & 2 & 0 & 0 & 1 & 2 & \bar 1 & \bar 1 & 2 & 1 & \bar 1 & 0 & \bar 1 & 2 & \bar 1 & \bar 1 & \bar 2 & 0 &
	0 & \bar 1 & 2 \\
	4 & \bar 2 & \bar 2 & \bar 2 & 0 & \bar 1 & 0 & 1 & 1 & \bar 1 & 0 & 1 & 1 & 0 & \bar 1 & \bar 1 & 0 & 0 & 0 & \bar 1 & 1 & \bar 1 &
	\bar 1 & 2 \\
	0 & \bar 1 & 0 & 0 & 1 & 0 & 0 & 0 & 1 & 0 & 0 & 0 & 0 & 0 & 0 & 0 & 1 & \bar 1 & \bar 1 & 0 & 0 & 0 & 0 & 1
	\\
	\bar 3 & 0 & 1 & 1 & 1 & 1 & 0 & 0 & 1 & 0 & 1 & \bar 1 & 1 & 0 & \bar 1 & 0 & 0 & 0 & \bar 1 & 1 & \bar 1 & 0 & 1 &
	0 \\
	4 & \bar 2 & \bar 2 & 0 & \bar 1 & 0 & 0 & \bar 1 & 0 & \bar 1 & 0 & 0 & 0 & 1 & 0 & \bar 1 & 0 & 1 & \bar 1 & 0 & 2 & \bar 2 &
	\bar 1 & 2 \\
	2 & \bar 2 & \bar 1 & \bar 1 & 0 & \bar 1 & 0 & 1 & 1 & 0 & 0 & 0 & 1 & 0 & \bar 1 & \bar 1 & 0 & 0 & 0 & 0 & 1 & \bar 1 &
	0 & 1 \\
	\bar 1 & 0 & 0 & 0 & 1 & 0 & 1 & 0 & 1 & 0 & 0 & 0 & 0 & 0 & 0 & 0 & 1 & \bar 1 & \bar 1 & \bar 1 & 0 & 0 & 0 & 1
	\\
	2 & \bar 1 & \bar 1 & 0 & \bar 1 & 0 & 1 & \bar 1 & \bar 1 & \bar 1 & 0 & 0 & 0 & 1 & 0 & 0 & \bar 1 & 1 & 0 & 0 & 1 & \bar 1 &
	0 & 0 \\
	0 & 0 & 0 & 0 & 0 & 0 & 0 & 0 & 0 & 0 & 0 & \bar 1 & 0 & 0 & 0 & 0 & 0 & 0 & 0 & 0 & 0 & 0 & 0 & 0 \\
	2 & \bar 1 & \bar 1 & \bar 1 & 0 & 0 & 1 & \bar 1 & 1 & 0 & 0 & 0 & 1 & 0 & \bar 1 & \bar 1 & 0 & 0 & 0 & \bar 1 & 2 & \bar 2 &
	\bar 1 & 3 \\
	1 & \bar 1 & 0 & 0 & \bar 1 & 0 & 1 & \bar 1 & \bar 1 & 0 & 1 & \bar 1 & 0 & 1 & \bar 1 & 0 & \bar 1 & 1 & 0 & 1 & 1 & \bar 1 &
	0 & 0 \\
	\bar 2 & 0 & 1 & 1 & 0 & 0 & 0 & 0 & 0 & 0 & 1 & \bar 1 & 1 & 0 & 0 & 0 & 0 & 0 & 0 & 1 & \bar 1 & 0 & 0 & 0
	\\
	\bar 1 & 0 & 1 & 1 & 0 & 0 & 0 & 0 & 0 & 0 & 0 & \bar 1 & 0 & 0 & 0 & 0 & 0 & 0 & 0 & 1 & \bar 1 & 0 & 0 & 0
	\\
	2 & \bar 1 & 0 & 0 & \bar 1 & \bar 1 & 0 & 0 & \bar 1 & \bar 1 & 0 & 0 & 1 & 1 & \bar 1 & \bar 1 & \bar 1 & 1 & 1 & 1 & 0 & \bar 1 &
	0 & 0  
\end{pmatrix}
\begin{pmatrix}
	\bar 3 & 0 & 1 & 1 & 1 & 2 & 0 & \bar 1 & 0 & 0 & 2 & 0 & \bar 1 & 0 & \bar 1 & 1 & 0 & 1 & \bar 2 & 1 & 0 & 0 & 1 &
	\bar 1 \\
	\bar 1 & 0 & 0 & 0 & 1 & 1 & \bar 1 & 1 & 0 & 0 & 0 & 1 & 0 & \bar 1 & 0 & 0 & 0 & 0 & \bar 1 & 1 & \bar 1 & 1 & 1 &
	\bar 1 \\
	\bar 2 & 0 & 1 & 1 & 1 & 1 & 0 & \bar 1 & 0 & 0 & 1 & 0 & 0 & 0 & \bar 1 & 0 & \bar 1 & 1 & \bar 1 & 1 & 0 & 0 & 1 &
	\bar 1 \\
	2 & \bar 1 & \bar 1 & 0 & 0 & 1 & 0 & \bar 1 & \bar 1 & \bar 1 & 0 & 1 & \bar 1 & 1 & 0 & 0 & 0 & 1 & \bar 1 & 0 & 1 & \bar 1 &
	0 & 0 \\
	2 & \bar 2 & \bar 1 & 0 & 0 & 0 & 0 & 0 & 0 & \bar 1 & 1 & 1 & 0 & 1 & \bar 1 & \bar 1 & 0 & 1 & \bar 1 & 0 & 1 & \bar 1 &
	0 & 1 \\
	\bar 3 & 1 & 0 & 1 & 1 & 2 & 0 & \bar 1 & 1 & 0 & 1 & 0 & 0 & \bar 1 & 0 & 0 & 0 & 1 & \bar 1 & 0 & 0 & 0 & 0 & 0
	\\
	2 & \bar 1 & \bar 1 & 0 & \bar 1 & 1 & 0 & \bar 1 & \bar 1 & 0 & 1 & 0 & \bar 1 & 0 & 0 & 0 & \bar 1 & 1 & \bar 1 & 1 & 2 & \bar 1 &
	0 & 0 \\
	6 & \bar 2 & \bar 3 & \bar 1 & \bar 1 & 0 & \bar 1 & 0 & \bar 1 & \bar 1 & \bar 1 & 2 & 0 & 0 & 1 & \bar 2 & \bar 1 & 1 & 0 & 0 & 2 &
	\bar 1 & \bar 1 & 1 \\
	\bar 2 & 0 & 1 & 1 & 0 & 1 & \bar 1 & 0 & 0 & 1 & 1 & \bar 1 & \bar 1 & 0 & 0 & 1 & 0 & 0 & \bar 1 & 1 & 0 & 0 & 1 &
	\bar 1 \\
	3 & \bar 1 & \bar 1 & \bar 1 & \bar 1 & 0 & 0 & \bar 1 & \bar 1 & \bar 1 & 0 & 1 & \bar 1 & 1 & 0 & 0 & 0 & 1 & 0 & 0 & 1 & \bar 1 &
	0 & 0 \\
	\bar 3 & 0 & 1 & 0 & 2 & 1 & 0 & 0 & 2 & 1 & 1 & 0 & 0 & \bar 1 & \bar 1 & 0 & 1 & \bar 1 & \bar 1 & 0 & 0 & 0 & 0 &
	1 \\
	3 & \bar 1 & \bar 1 & \bar 1 & 0 & 0 & \bar 1 & 0 & 0 & 0 & \bar 1 & 1 & 0 & 0 & 0 & \bar 1 & 0 & 0 & 0 & 0 & 1 & \bar 1 &
	0 & 1 \\
	1 & \bar 1 & \bar 1 & \bar 1 & 1 & 1 & 1 & \bar 1 & 0 & \bar 1 & 0 & 2 & 0 & 0 & \bar 1 & 0 & 0 & 1 & \bar 1 & \bar 1 & 1 & \bar 1 &
	0 & 1 \\
	3 & \bar 1 & \bar 2 & \bar 1 & 0 & 0 & 0 & 0 & 0 & \bar 1 & \bar 1 & 2 & 0 & 0 & 0 & \bar 1 & 0 & 1 & 0 & \bar 1 & 1 & \bar 1 &
	0 & 1 \\
	3 & \bar 2 & \bar 1 & \bar 1 & 0 & 0 & 0 & 0 & 0 & 0 & 0 & 1 & 0 & 0 & \bar 1 & \bar 1 & \bar 1 & 0 & 0 & 0 & 2 & \bar 1 &
	0 & 1 \\
	5 & \bar 2 & \bar 2 & \bar 1 & \bar 1 & 0 & 0 & \bar 1 & \bar 1 & \bar 1 & \bar 1 & 2 & 0 & 1 & 0 & \bar 1 & 0 & 1 & 0 & \bar 1 & 2 &
	\bar 2 & \bar 1 & 2 \\
	\bar 1 & 1 & 1 & 1 & 0 & 1 & 0 & \bar 2 & \bar 1 & 0 & 1 & \bar 1 & \bar 1 & 0 & 0 & 1 & \bar 1 & 1 & 0 & 1 & 0 & 0 & 0 &
	0 \\
	7 & \bar 2 & \bar 2 & \bar 1 & \bar 2 & \bar 1 & \bar 1 & \bar 1 & \bar 2 & \bar 1 & \bar 1 & 1 & \bar 1 & 1 & 1 & \bar 1 & \bar 1 & 1 & 1 & 0 &
	2 & \bar 1 & \bar 1 & 1 \\
	2 & \bar 1 & \bar 1 & 0 & 0 & 0 & \bar 1 & 0 & 1 & 0 & 0 & 0 & 0 & 0 & 0 & \bar 1 & 0 & 0 & 0 & 0 & 1 & \bar 1 & \bar 1 &
	2 \\
	3 & \bar 1 & \bar 1 & 0 & \bar 1 & 0 & \bar 1 & 0 & \bar 1 & 0 & 0 & 0 & 0 & 0 & 0 & \bar 1 & \bar 1 & 1 & 0 & 1 & 1 & \bar 1 &
	0 & 1 \\
	3 & \bar 1 & \bar 1 & 0 & 0 & 0 & 0 & \bar 1 & \bar 1 & \bar 1 & \bar 1 & 1 & 0 & 1 & 0 & \bar 1 & \bar 1 & 1 & 0 & 0 & 1 & \bar 1 &
	0 & 1 \\
	2 & 0 & \bar 1 & 0 & 0 & 0 & 1 & \bar 2 & \bar 1 & \bar 1 & \bar 1 & 1 & \bar 1 & 1 & 0 & 0 & 0 & 1 & 0 & \bar 1 & 1 & \bar 1 &
	0 & 1 \\
	\bar 1 & 0 & 1 & 0 & 1 & 0 & 0 & 0 & 1 & 0 & 0 & 0 & 0 & 0 & \bar 1 & 0 & 0 & \bar 1 & 0 & 0 & 0 & 0 & 0 & 1
	\\
	4 & \bar 1 & \bar 2 & \bar 1 & 0 & \bar 1 & 0 & 0 & 0 & \bar 1 & \bar 2 & 2 & 1 & 0 & 0 & \bar 2 & 0 & 0 & 1 & \bar 1 & 1 & \bar 1 &
	\bar 1 & 2  
\end{pmatrix}
\end{equation*}}}

\section{Modular forms and character formulas}\label{a:minimal}
In this appendix, we collect some useful formulas for the modular forms and for the characters of affine algebras and cosets. These formulas are used in  various places in the main body of the article.

The Dedekind $\eta$-function $\eta(\tau)$ and the Jacobi theta functions $\theta_i(\tau,z)$ are defined by
\be\label{eta} \eta(\tau)=q^{\frac{1}{24}}\prod_{n=1}^\infty(1-q^n)\ ,
\ee
\begin{align} \theta_1(\tau,z)&=i\sum_{n\in \ZZ} (-1)^nq^{\frac{(n+\frac{1}{2})^2}{2}}y^{n+\frac{1}{2}}=q^{\frac{1}{8}}(iy^{\frac{1}{2}}-iy^{-\frac{1}{2}})\prod_{n=1}^\infty(1-q^{n}y)(1-q^{n}y^{-1})(1-q^n)\\
\theta_2(\tau,z)&=\sum_{n\in \ZZ} q^{\frac{(n+\frac{1}{2})^2}{2}}y^{n+\frac{1}{2}}=q^{\frac{1}{8}}(y^{\frac{1}{2}}+y^{-\frac{1}{2}})\prod_{n=1}^\infty(1+q^{n}y)(1+q^{n}y^{-1})(1-q^n)\\
\theta_3(\tau,z)&=\sum_{n\in \ZZ} q^{\frac{n^2}{2}}y^{n}=\prod_{n=1}^\infty(1+q^{n-\frac{1}{2}}y)(1+q^{n-\frac{1}{2}}y^{-1})(1-q^n)\\
\theta_4(\tau,z)&=\sum_{n\in \ZZ} (-1)^nq^{\frac{n^2}{2}}y^{n}=\prod_{n=1}^\infty(1-q^{n-\frac{1}{2}}y)(1-q^{n-\frac{1}{2}}y^{-1})(1-q^n)\ ,
\end{align} and we set $\theta_i(\tau):=\theta_i(\tau,0)$. 
The $\eta$ and $\theta_i$ are weight $1/2$ modular forms, satisfying
\begin{align}
\eta(\tau+1)&= e^{\frac{2\pi i}{24}}\eta(\tau) & \eta(-1/\tau)&= \sqrt{-i\tau}\eta(\tau)\\
\theta_2(\tau+1)&= e^{\frac{2\pi i}{8}}\theta_2(\tau) & \theta_2(-1/\tau)&= \sqrt{-i\tau}\theta_4(\tau)\\
\theta_3(\tau+1)&= \theta_4(\tau) & \theta_3(-1/\tau)&= \sqrt{-i\tau}\theta_3(\tau)\\
\theta_4(\tau+1)&= \theta_3(\tau) & \theta_4(-1/\tau)&= \sqrt{-i\tau}\theta_2(\tau).
\end{align}

The characters of the $\widehat{su}(2)_k$ algebra are given by (see \cite{DiFrancesco}, section 14.4.2)
\begin{align}\label{su2kcharz}
    \chi^{k}_l(\tau,z)&=\frac{\theta_{l+1,k+2}(\tau,z)-\theta_{-l-1,k+2}(\tau,z)}{\theta_{1,2}(\tau,z)-\theta_{-1,2}(\tau,z)}\end{align}
    where $l\in \{0,\ldots,k\}$ and the su(2) theta function $\theta_{m,k}$ is given by
\be\label{su2theta}
\theta_{m,k}(\tau,z):=\sum_{j\in \ZZ}q^{k(j+ {m\over 2k})^2}y^{k(j+{m\over 2k})}.
\ee
A useful formula for the limit $z\to 0$ is
    \begin{align}\label{su2kchar}
    \chi^{k}_l(\tau):=\chi^{k}_l(\tau,0)=\frac{q^{\frac{(l + 1)^2}{4 (k + 2)}}}{\eta(\tau)^3}\sum_{n\in \ZZ}(l + 1 + 2 n (k + 2)) q^{n (l + 1) + n^2 (k + 2)}\ .
\end{align} 
The characters of $\widehat{so}(n)_1$ for $n$ even are
\bea
\ch^{(n)}_0(\tau)&=&{1\over 2\eta(\tau)^{n/2}}(\theta_3^{n/2} \,(\tau) + \theta_4^{n/2}(\tau)) \label{ch_0_so(n)}\\
\ch^{(n)}_v(\tau)&=&{1\over 2\eta(\tau)^{n/2}} \, (\theta_3^{n/2}(\tau) - \theta_4^{n/2}(\tau)) \label{ch_v_so(n)}\\
\ch^{(n)}_s(\tau)&=& \ch^{(n)}_c(\tau) \, = \, {1\over 2\eta(\tau)^{n/2}}\theta_2^{n/2}(\tau) . \label{ch_s_so(n)}\ .
\eea
The $\CL$-twining genus $\phi^{\CL}(V^{f\natural},\tau,z)$ for a defect $\CL$ that preserves a four plane $\Pi^\natural\subset {}^\RR V^{f\natural}_{tw}(1/2)$ is often written in terms of $\widehat{so}(n)_1$ `flavoured' characters $\ch^{(n)}_\rho(\tau,z):= \Tr_{\rho^{(n)}}y^{J_0^3}q^{L_0-1/3}$  where $\rho\in\{0,v,c,s\}$, and $J_0^3$ is the zero mode of a current  in some algebra $\widehat{su}(2)_1\subset \widehat{so}(4)_1\subset \widehat{so}(n)_1$. These characters are given by
\bea\label{ch04plane}
\ch^{(n)}_0(\tau,z)&=& {1\over 2}{(\theta_3^2(\tau,z)\theta_3^{n/2-2}(\tau) + \theta_4^2(\tau,z)\theta_4^{n/2-2}(\tau))\over \eta^{n/2}(\tau)}\\\label{chv4plane}
\ch^{(n)}_v(\tau,z)&=& {1\over 2}{(\theta_3^2(\tau,z)\theta_3^{n/2-2}(\tau) - \theta_4^2(\tau,z)\theta_4^{n/2-2}(\tau))\over \eta^{n/2}(\tau)}\\\label{chsc4plane}
\ch^{(n)}_s(\tau,z)&=&\ch^{(n)}_c(\tau,z)={1\over 2}{\theta_2^{2}(\tau,z)\theta_2^{n/2-2}(\tau)\over \eta^{n/2}(\tau)}.
\eea
In section \ref{s:Fibonacci}, we use the character formulae for the representations of $\widehat{sp}(6)_1$. One can obtain them from the characters of $\widehat{so}(12)_1$ and of $\widehat{su}(2)_3$ by inverting the relations
\begin{align}
    \ch_0^{(12)}&=\chi^{sp}_0\chi^{su(2)_3}_0+\chi^{sp}_2\chi^{su(2)_3}_2\\
    \ch_v^{(12)}&=\chi^{sp}_1\chi^{su(2)_3}_1+\chi^{sp}_3\chi^{su(2)_3}_3\\
    \ch_s^{(12)}&=\chi^{sp}_3\chi^{su(2)_3}_0+\chi^{sp}_1\chi^{su(2)_3}_2\\
    \ch_c^{(12)}&=\chi^{sp}_2\chi^{su(2)_3}_1+\chi^{sp}_0\chi^{su(2)_3}_3
\end{align} 
to get
\begin{align}
\chi^{sp}_0(\tau) &= \frac{\ch_0^{(12)}\chi^{su(2)_3}_1 -\ch_c^{(12)}\chi^{su(2)_3}_2}{\chi^{su(2)_3}_0\chi^{su(2)_3}_1 -\chi^{su(2)_3}_3\chi^{su(2)_3}_2}\\
\chi^{sp}_1(\tau) &= \frac{\ch_v^{(12)}\chi^{su(2)_3}_0 -\ch_s^{(12)}\chi^{su(2)_3}_3}{\chi^{su(2)_3}_0\chi^{su(2)_3}_1 -\chi^{su(2)_3}_3\chi^{su(2)_3}_2}\\
\chi^{sp}_2(\tau) &= \frac{\ch_c^{(12)}\chi^{su(2)_3}_0-\ch_0^{(12)}\chi^{su(2)_3}_3 }{\chi^{su(2)_3}_0\chi^{su(2)_3}_1 -\chi^{su(2)_3}_3\chi^{su(2)_3}_2}\\
\chi^{sp}_3(\tau) &= \frac{\ch_s^{(12)}\chi^{su(2)_3}_1-\ch_v^{(12)}\chi^{su(2)_3}_2}{\chi^{su(2)_3}_0\chi^{su(2)_3}_1 -\chi^{su(2)_3}_3\chi^{su(2)_3}_2}\ .    \end{align}

In section \ref{s:Gepner}, we need the characters of $\CN=2$ minimal models. It is known that the unitary representations of the $\CN=2$ superconformal algebra at $c=\frac{3k}{k+2}$ decompose as representations  of  $\frac{\widehat{su}(2)_k\hat u(1)_2}{\hat u(1)_{k+2}}$.  The character of the $(l,m,\epsilon)$ representation of the bosonic coset is
\be
\chi_{l,m}^\epsilon(\tau,z):=\Tr_{M_{l,m}^\epsilon} q^{L_0-c/24}y^{J_0}=\sum_{j\in \ZZ}c^j_{m+4j-\epsilon}(\tau) q^{{k+2\over 2k}({m\over k+2}-{\epsilon\over 2}+2j)^2}y^{{m\over k+2}-{\epsilon\over 2}+2j},
\ee
where the string functions $c^l_m(\tau)$ can be determined from the relation
\be
\sum_{m\in \ZZ/2k\ZZ}c^l_m(\tau)\theta_{m,k}(\tau,z) = \chi^{su(2)_k}_l(\tau,z)
\ee
where $\chi^{su(2)_k}_l(\tau,z)$ is as in \eqref{su2kcharz} and $\theta_{m,k}$ as in \eqref{su2theta}.

 \printbibliography

\end{document}